\documentclass{article}
\usepackage[margin=0.9in]{geometry}
\usepackage[english]{babel}
\usepackage[utf8]{inputenc}

\usepackage{url}

\usepackage{subfiles}
\usepackage[dvips]{graphicx}
\usepackage{caption}
\usepackage{subcaption}
\usepackage{soul}
\usepackage{color}
\usepackage{float}
\usepackage{amsmath}
\usepackage{tikz}
\usetikzlibrary{shapes}
\usepackage{authblk}
\usepackage[nottoc]{tocbibind}

\usepackage{multirow}

\usepackage{pgfplotstable}
\pgfplotsset{compat=1.9}

\usepackage{xcolor}

\usepackage{amsmath}

\usepackage{xcolor}

\usepackage{multirow}

\usepackage{colortbl} 

\definecolor{rulecolor}{RGB}{103,68,133}
\definecolor{tableheadcolor}{RGB}{195,127,250}
\definecolor{NeutralColor}{RGB}{232,211,250}

\arrayrulecolor{rulecolor!70} 
\title{Quantifying the impact of persuasiveness, cautiousness and prior beliefs in (mis)information sharing on online social networks using Drift Diffusion Models}

\author[1]{Lucila G. Alvarez-Zuzek 
\thanks{lalvarezzuzek@fbk.eu}}
\author[2]{Lucio La Cava}
\author[3]{Jelena Gruji\'c}
\author[1]{Riccardo Gallotti
\thanks{rgallotti@fbk.eu}}
\affil[1]{ Complex Human Behavior Lab, Fondazione Bruno Kessler, Trento, Italy}
\affil[2]{ DIMES, Università della Calabria, Rende, Italy}
\affil[3]{ MLG, Université Libre de Bruxelles, Brussels, Belgium}

\begin{document}

    \maketitle

    \begin{abstract}

    Misleading newsletters can shape individuals' perceptions, and pose a threat to societies; as we witnessed by lowering the severity of follow-up stay-at-home orders and burdening a significant challenge to the fight against COVID-19. In this research, we study (mis)information spreading, reanalyzing behavioral data on online sharing, and analyzing decision-making mechanisms using the Drift Diffusion Model (DDM). We find that subjects display an increased instinctive inclination towards sharing misleading news, but rational thinking significantly curbs this reaction, especially for more cautious and older individuals. On top of network structures with similar characteristics as X, Mastodon, and Facebook, we use an agent-based model to expand this individual knowledge to a large scale where individuals are exposed to (mis)information through friends and share (or not) content with probabilities driven by DDM. We found that the natural shape of these social online networks provides a fertile ground for any news to rapidly become viral. Yet we have found that, for the case of X, limiting the number of followers of the most connected users proves to be an appropriate and feasible containment strategy.
    \end{abstract}

    \section*{Introduction}
        As individuals, we understand the world by constantly acquiring new information. The advent of the Internet has released numerous opportunities for information to spread faster and more expansive, reaching more people than ever before. However, the amount of information at our fingertips seems overwhelming, exposing us to a considerable amount of untrustworthy information \cite{lazer2018science}.  In recent times, this exposure has been potentially increased by the automatic use of highly persuasive large language models (LLMs) \cite{salvi2024conversational,hackenburg2024evidence}. This may lead to dangerous behaviors that for instance, may burden the acceptance of vaccines \cite{dube2013vaccine}, produce a negative effect on the public image of a political candidate through false narratives \cite{flynn2017nature,jerit2020political}, etc. The term {\it misinformation} refers to information that is false but not created with the intention of causing harm \cite{wardle2017information,west2021misinformation,wardle2018information}, for instance, individuals sharing online pieces of information without knowing that it is false and intending to be helpful. False information may have the capability of changing our perception of reality \cite {EU_Report} and its diffusion in a society may have serious consequences, such as compromising our ability to address hence fight the climate crisis \cite{treen2020online}, enhance public health systems \cite{suarez2021prevalence,swire2020public,ruggeri2023synthesis}, or maintain a stable democracy \cite{kuklinski2000misinformation,jerit2020political}. But why do people share false or unreliable information on social media? The scientific community has been studying this topic for decades. In \cite{pennycook2019lazy}, authors found that analytic thinking plays an important role when it comes to false political information. People tend to share it because they fail to think. Moreover, they suggested that susceptibility to fake news is driven more by lazy thinking rather than by partisan bias. On the contrary, in \cite{ceylan2023sharing} Ceylan, $et.al.$ found that the structure of online sharing in the platform is more important than critical reasoning or partisan bias. That is to say, users tend to form habits of sharing misinformation and react automatically to familiar platform cues (in their case of study Facebook). Habitual sharers do not take into consideration the informational consequences of what they share. Altogether, these pieces of research may suggest that the decision-making process of sharing false information is made rapidly and at a low cognitive level. 

\begin{figure}
            \centering
            \includegraphics[width=1\textwidth]{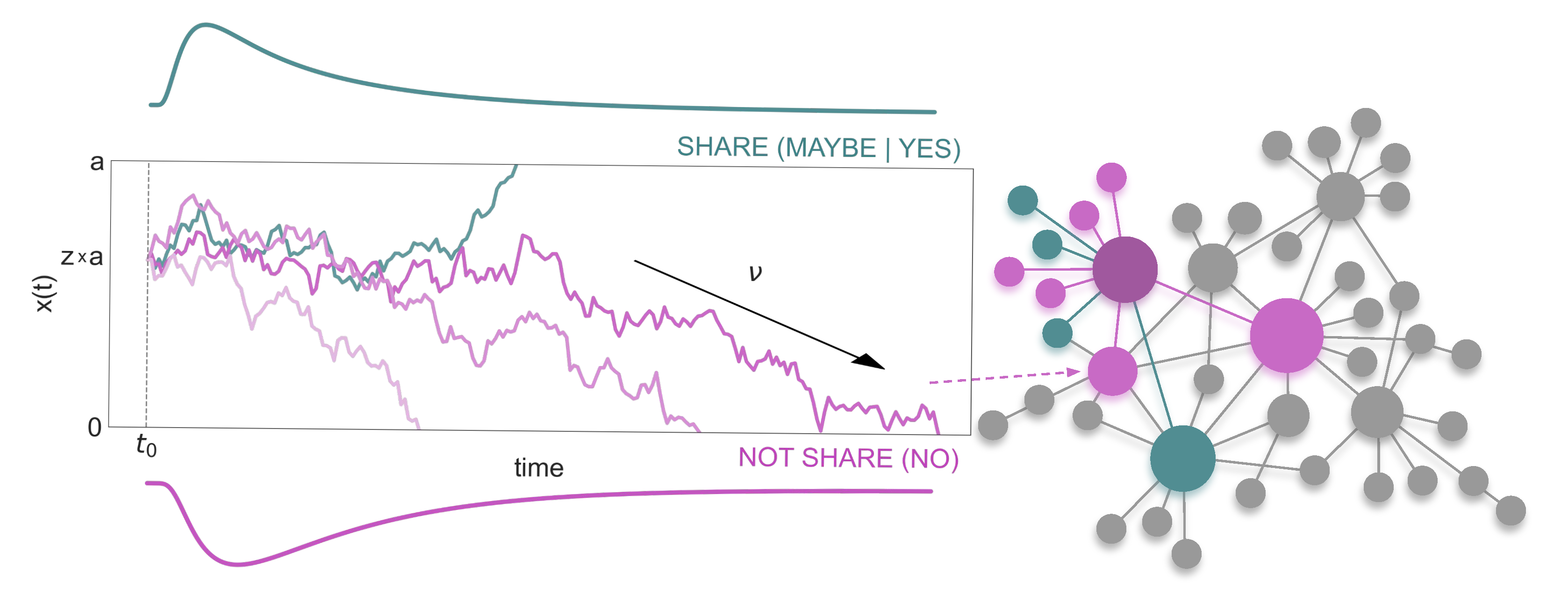}
            
            \caption{{\bf Information spreading in a society considering the neurocognitive mechanisms driving each individual response}. On the right side, there is the social network on top of which the spreading process occurs, $i.e.$, information diffuses in the social media platform where users interact. Every time an $S$ individual shares content (dark pink), its $NS$ neighbors will see it, and a single decision-making process is activated (left figure) for each individual to decide whether to share (pink) or not the content (green). The piece of news has not yet reached Grey nodes.}
            \label{model-network}
        \end{figure}

        Daily life activities involve the process of decision-making, whether to cross or not a street, which candidate to vote for, or when to take holidays. Some decisions are more important and require a deeper analysis. While others are made rapidly (almost instantaneously) and with less cognitive effort. Mathematically, cognitive processes that involve simple and rapid two-choice decisions can be seen as the noisy accumulation of evidence from a stimulus and can appropriately be described by {\it Drift Diffusion Models} (DDM) \cite{ratcliff2016diffusion}. These sequential sampling models have been extensively studied for decades, building a bridge between neuroscience and human behavior, taking advantage of collected data \cite{forstmann2016sequential}. Starting with the model proposed and developed by Ratcliff in \cite{ratcliff1978theory}, the DDM assumes a one-dimensional random walk representing the accumulation in our brain of noisy evidence in favor of one of the two alternative options. More in detail (see Fig.\ref{model-network}), the process is represented as the temporal evolution of a random variable $x(t)$ from a point $z$ named the bias. At each time step, as information is randomly collected in favor of one of the two choices, this accumulation is represented as increases or decreases in $x(t)$ up until a decision is made when one of the boundaries is reached $x=a$ or $x=0$ (being $a$ the boundary separation). The information accumulation rate is called the drift rate $\nu$ (see Methods for a more detailed mathematical description of the model). From a neuroscience perspective, these free parameters of the mathematical model have the following interpretation: (i) The bias ($z$): is related to the prior beliefs (the pre-existing ideas) the individual has about the options, (ii) The threshold or length of the barrier ($a$): psychologically quantifies cautiousness in response, (iii) The drift rate ($\nu$): the value of this quantity is related to the quality and velocity of the information extracted from the stimulus, also related to the task complexity. For tasks ``easy'' to understand, the drift rate has a high absolute value, and responses are fast and accurate on average. Conversely, when subjects find the tasks ``difficult'', the drift rate values are closer to zero, and responses are slower and less accurate on average. The accuracy of this model is measured with the response times (RTs) and response times distributions (PDF). By evaluating the predictions of the model with the shape of the response time data, we can obtain the free parameters and test the performance of the model. Previous studies have shown that decisions with smaller response times are more intuitive and less accurate, while longer decisions are more deliberate \cite{ratcliff1998modeling}; it is important to understand whether a person responds as quickly as possible or as accurately as possible. Hence, researchers generally focused on how response time probabilities change across experiment conditions (variations may appear among ages, genders, political orientations, etc.). Many variations to this model have been explored, such as adding non-decision time \cite{ratcliff1999connectionist}, across-trials variability of the rate of accumulation of information \cite{ratcliff1980note} and the starting point \cite{ratcliff1998modeling}, speed-accuracy optimality \cite{moran2015optimal,edwards1965optimal} and more \cite{ratcliff2008diffusion}. Additionally, more sophisticated models have been developed, mainly physiologically motivated by cases that involve more than two-option decisions \cite{bogacz2006physics}. For instance, the race model uses separate accumulators for each of the options, integrating evidence independently. 
        
        The universality of the Drift Diffusion Model allowed scientists to use it in a wide spectrum of different topics. In game theory, previous research used the model to describe and explore human cooperation-defection \cite{gallotti2019quantitative}. Finding that, in this context, individuals initially tend to cooperate, but rational deliberation quickly becomes dominant. Depending on the interaction with the environment, a positive or negative experience will be translated into a cooperative or defective answer. DDM models have also been used to describe human behavior in the context of altruism \cite{hutcherson2015neurocomputational}, food choices \cite{krajbich2015common}, moral judgments \cite{andrejevic2022response}, and more. In the context of online sharing misinformation, Lin. $et.al$ \cite{lin2023thinking} investigate accuracy and deliberation when sharing misinformation. They found that accuracy prompts increased sharing discernment by shifting peoples’ attention to accuracy while deciding what to share rather than increasing the amount of deliberation; this is interesting because they show that people can think better even without thinking more. However, little is known about this topic, and there is still much more to understand. For instance, can we take advantage of the drift-diffusion model to describe the decision-making process of sharing online misinformation? How well is the model's performance, and how does it vary across experimental conditions, such as age and other factors? In this research, our first aim is to explore and understand these addressed issues.
        
        At the moment, diffusion models link cognitive and decision-making processes at an individual level, providing a solid theoretical framework. But we know that social interactions play a fundamental role in human behavior, and interesting and unexpected outcomes can emerge. Widespread misinformation can lead to macroscopic clusters of people sharing misperceptions and biased collective opinions \cite{kuklinski2000misinformation}, for instance, leading pools of vaccine-hesitant individuals susceptible to a new epidemic outbreak \cite{alvarez2022spatial,tiu2022characterizing}. Further, it is well known that when certain features are present in the social patterns —namely homophily \cite{mcpherson2001birds}, polarization \cite{bessi2016homophily, la2024polarization}, echo chambers characteristics \cite{del2016spreading,boutyline2017social,quattrociocchi2016echo}— they have a strong impact for the spread of misinformation \cite{pierri2020topology}. Hence, our next step here is to expand these micro-level concepts to a macro-level. How can we link these neurocognitive mechanisms of decision-making taking place inside the brain of each individual (micro-level) to misleading content spreading and becoming viral on an online social platform (macro-level)?  
        
        The present research explores the decision-making process of sharing information online with two-fold goals. As a first step, we take the universality of the Drift Diffusion Model to describe and analyze data collected by  G. Pennycook \& D. G. Rand in \cite{pennycook2019lazy}. In this study, misleading and reliable content was presented to participants to then ask how prone they were to share it with response option {\it yes, maybe or not} (three-option process). To be able to apply DDM, we reduce the dimension of the problem by combining the positive options and analyzing the performance of the model. Data is disaggregated by content, content veracity —misleading and reliable— and age, and variation in the free parameters of the model are present across the different experiment conditions. As a second step, we explore how social contact patterns impact the outcomes of collective behavior. Hence, we consider different online social networks, spanning from more homogeneous to more heterogeneous: Twitter \cite{de2013anatomy}, Mastodon \cite{la2022information,bono2024exploration,cava2023drivers}, Facebook \cite{facebook_high_school} and \cite{facebook_copenhagen}. To describe the propagation of information, we propose a simplified version of the $SIR$ epidemic model \cite{tambuscio2015fact} in which individuals are divided into two compartments: Sharers ($S$) or Non-Sharers ($NS$). Every time an $S$ individual shares content will activate the decision-making process in their $NS$ neighbors. That is to say, a single DDM process runs, answering (share or not) to the stimulus. Finally, we adapt the previous mathematical framework described in \cite{newman2002spread} to build a bridge between the micro and macro levels to obtain the critical conditions the free parameters should satisfy in order for (mis)information to become viral; this leads us to propose an intervention strategy to prevent viralization.

    \section*{Results}
        In this Section, we explore if and how the Drift Diffusion Model can describe the decision-making dynamic of sharing misleading and reliable content. We use the open access dataset presented in \cite{pennycook2019lazy}, where authors carried out a large-scale experiment in which participants discerned the veracity of news headlines and expressed how prone they were to share them. Finally, we explore how contact patterns affect the possibility of viral content and the role played by the free parameters of the DDM.

        \subsection*{Drift Diffusion model is capable of capturing decision-making process of sharing information}\label{results-PDF}
        
\begin{figure}
             \centering
             \begin{subfigure}[b]{0.45\textwidth}
                 \centering
                 \includegraphics[width=\textwidth]{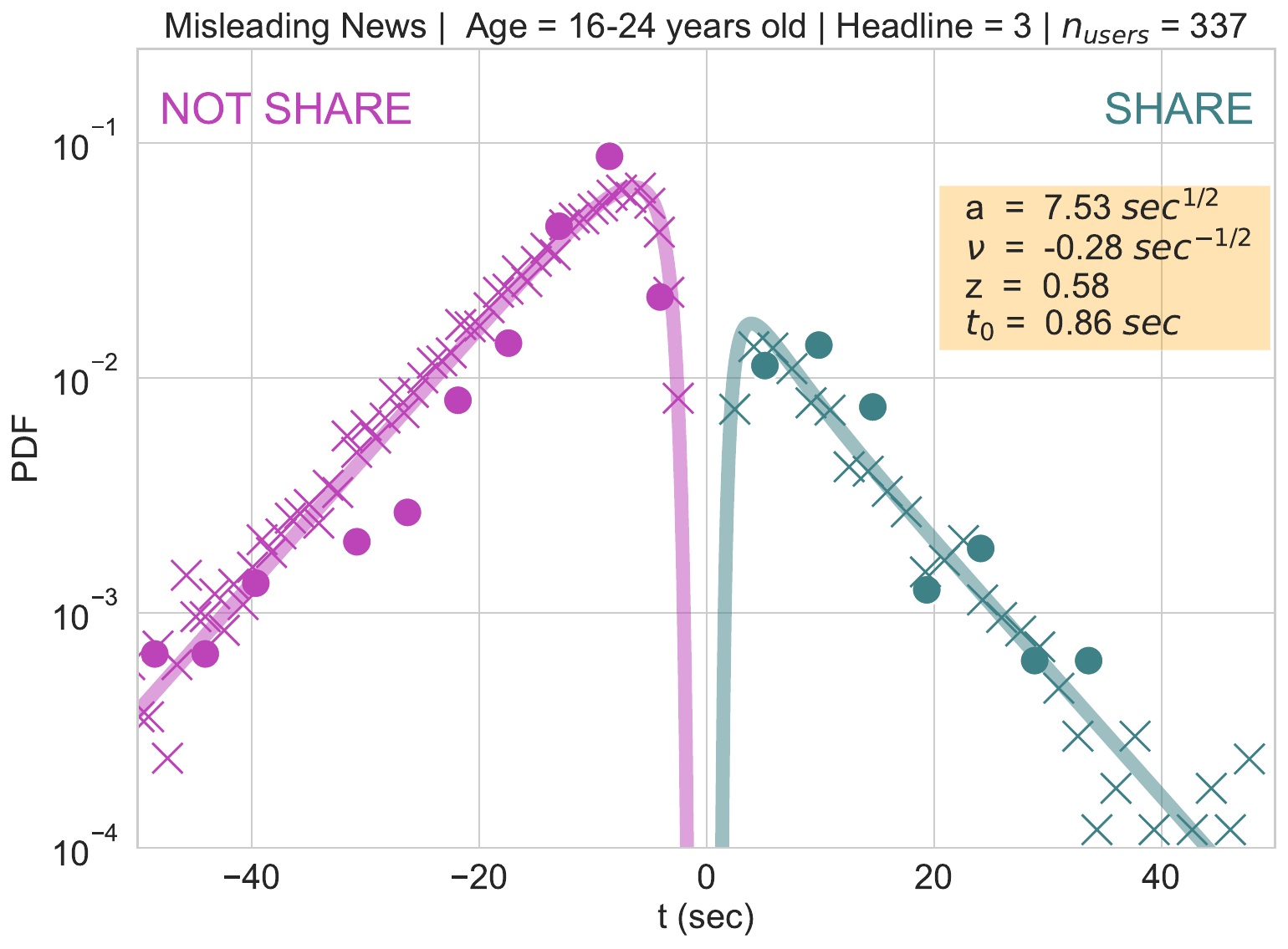}
                 \caption{}
                 \label{PDF-1a}
             \end{subfigure}
             \begin{subfigure}[b]{0.45\textwidth}
                 \centering
                 \includegraphics[width=\textwidth]{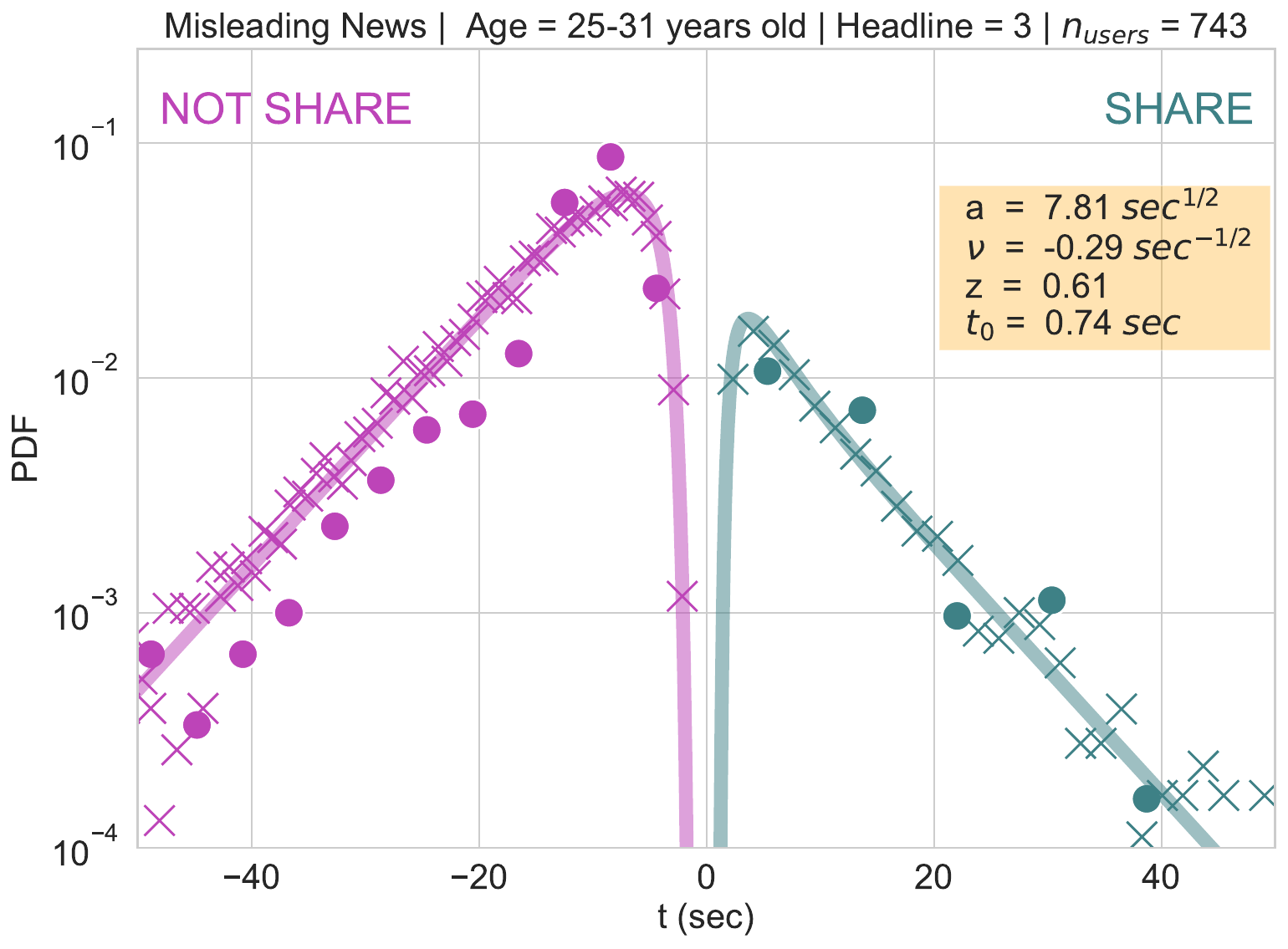}
                 \caption{}
                 \label{PDF-1b}
             \end{subfigure}
             \begin{subfigure}[b]{0.45\textwidth}
                 \centering
                 \includegraphics[width=\textwidth]{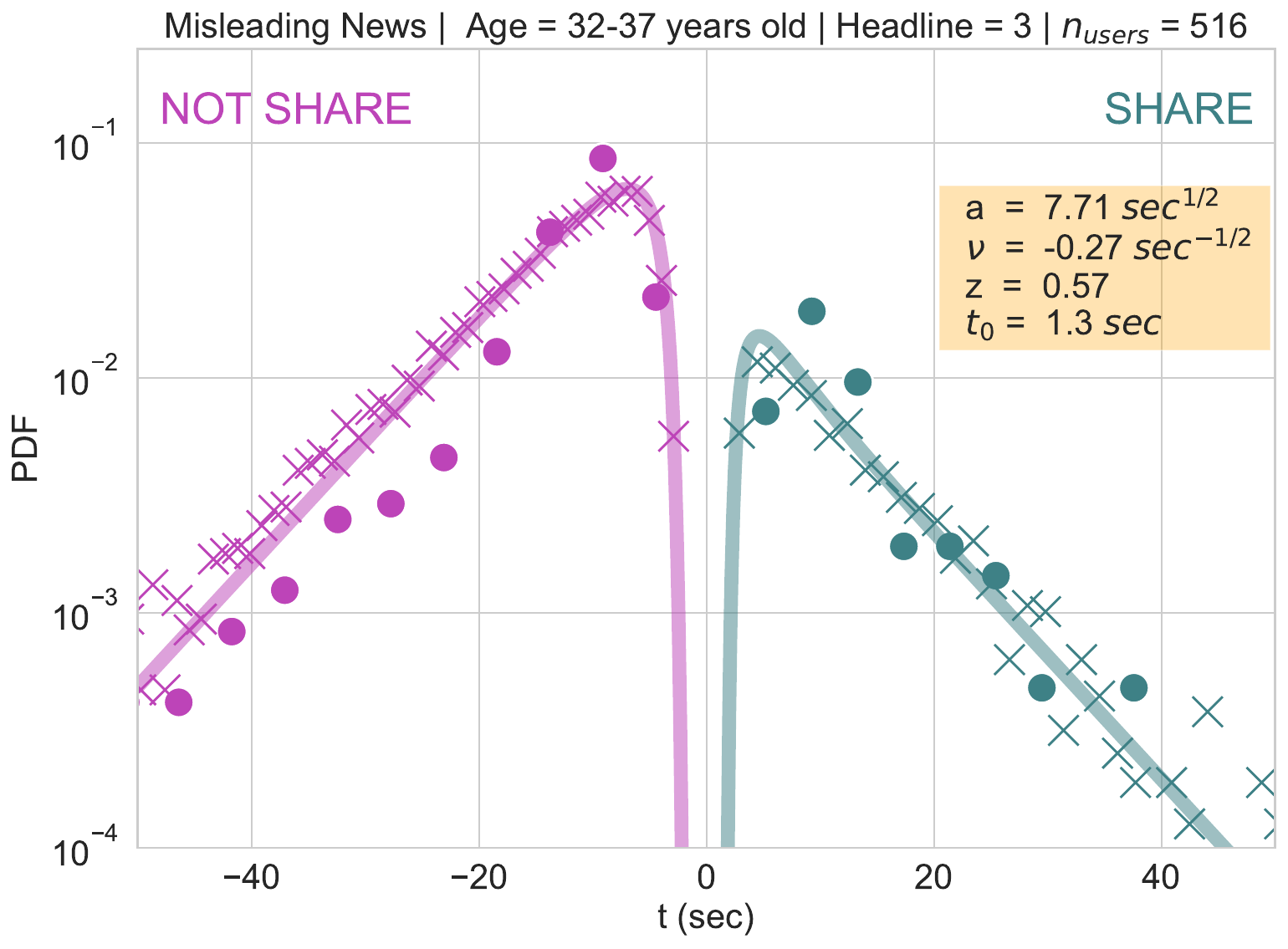}
                 \caption{}
                 \label{PDF-1c}
             \end{subfigure}\\
             \begin{subfigure}[b]{0.45\textwidth}
                 \centering
                 \includegraphics[width=\textwidth]{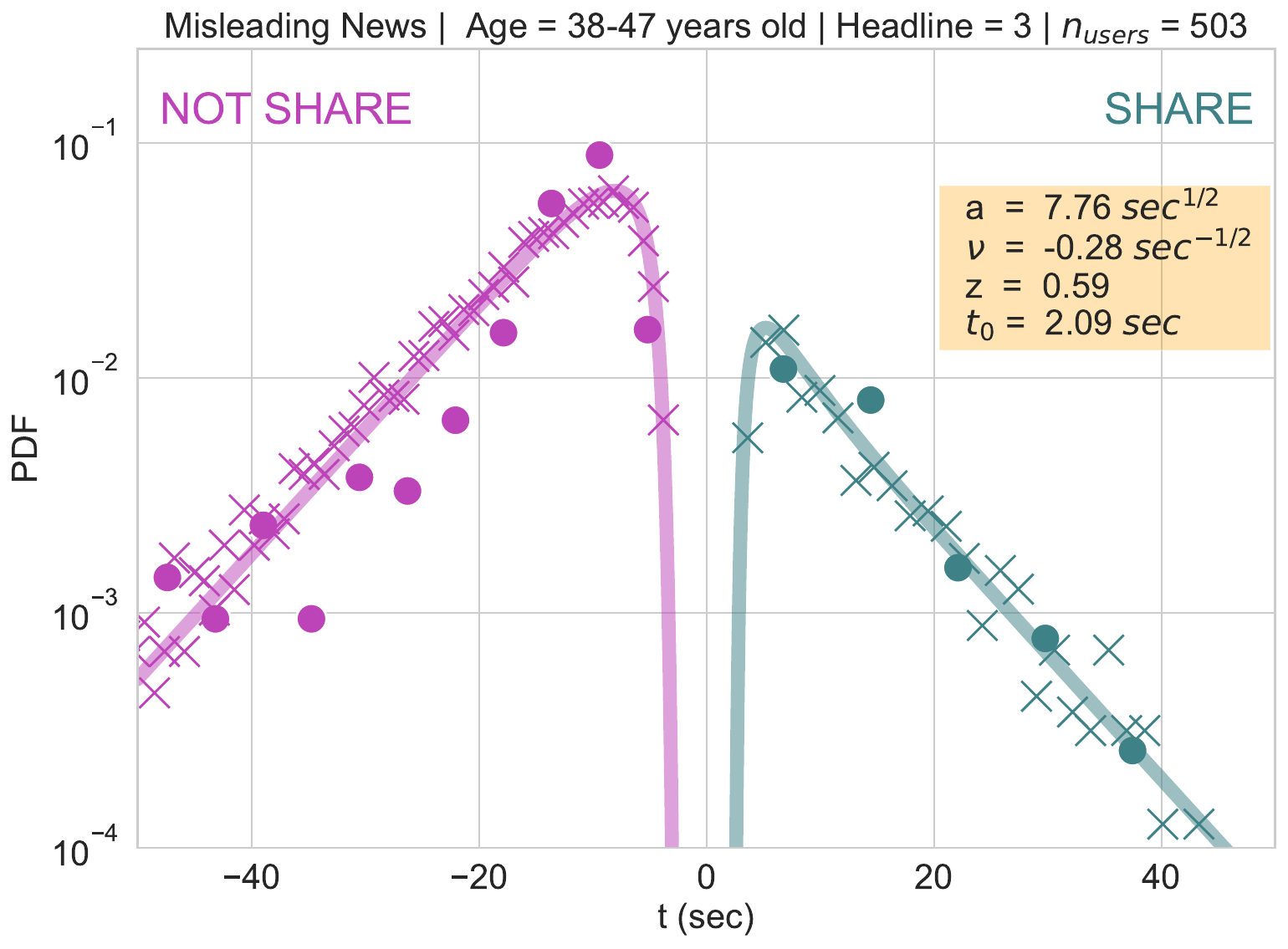}
                 \caption{}
                 \label{PDF-1d}
             \end{subfigure}
             \begin{subfigure}[b]{0.45\textwidth}
                 \centering
                 \includegraphics[width=\textwidth]{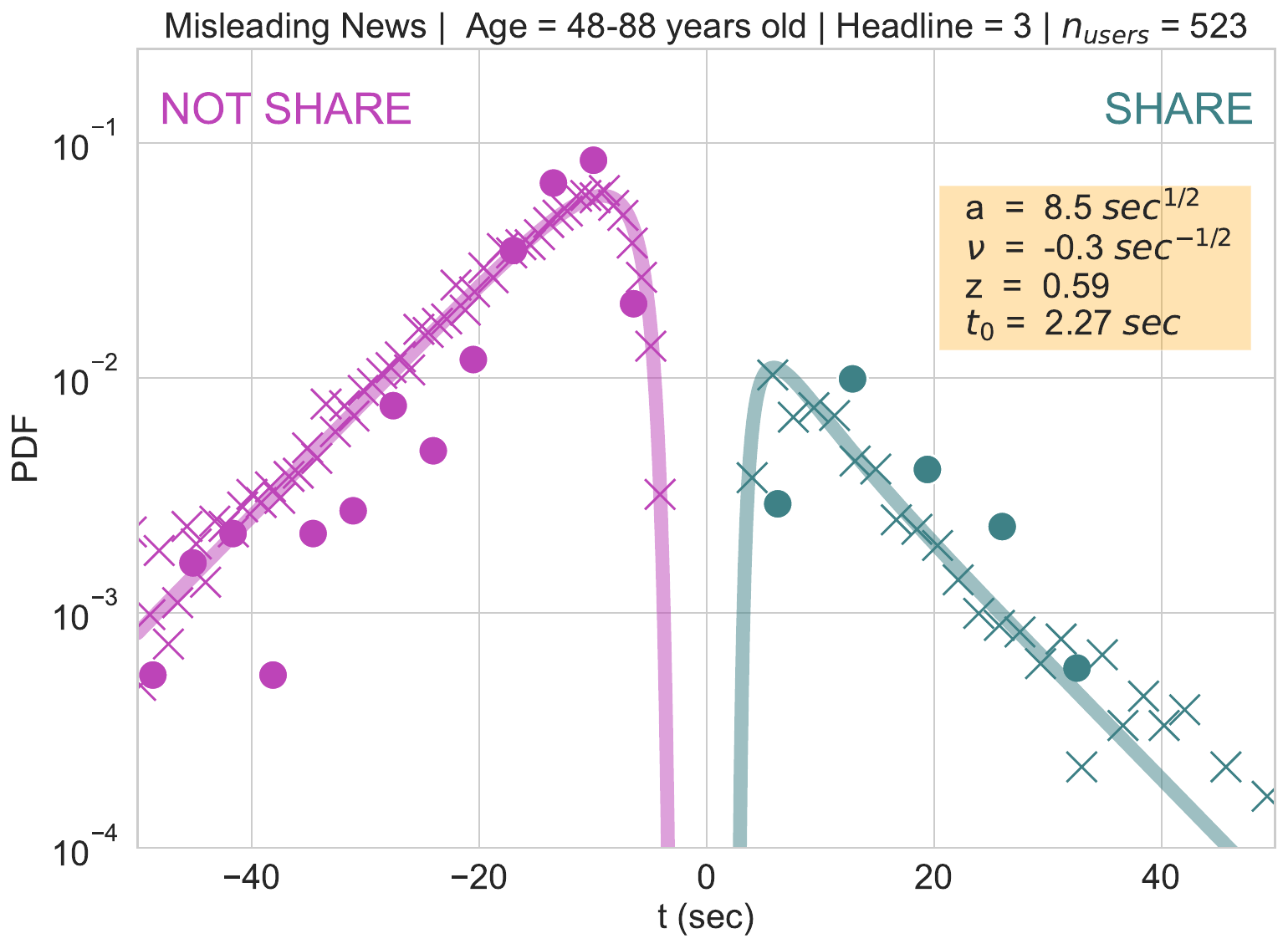}
                 \caption{}
                 \label{PDF-1e}
             \end{subfigure}
            \caption{{\bf Probability Distribution Function of the Response Times for the case of Misleading Information}. The estimation of the free parameters values (a, $\nu$, z and $t_0$ in the legend box) are obtained with the Python-based toolbox HDDM \cite{wiecki2013hddm} version=$0.6.0$. For simplicity, we display headline number $3$ (the remaining headlines are displayed in the Supplementary Information). Dots correspond to empirical data, lines are obtained using Eq.~\ref{RT-PDF} with the corresponding values of the free parameters, and crosses correspond to stochastic simulated data. Each panel corresponds to a different age bin: (a) $16-24$ years old, (b) $25-31$ years old, (c) $32-37$ years old, (d) $38-47$ years old, and (e) $48-88$ years old; and the corresponding number of participants for each case are displayed in the titles. The left side of each curve (violet) corresponds to data collected with not sharing answers, while the right side (blue) corresponds to sharing answers.}
            \label{PDF-misleading}
        \end{figure}

\begin{figure}
             \centering
             \begin{subfigure}[b]{0.45\textwidth}
                 \centering
                 \includegraphics[width=1\textwidth]{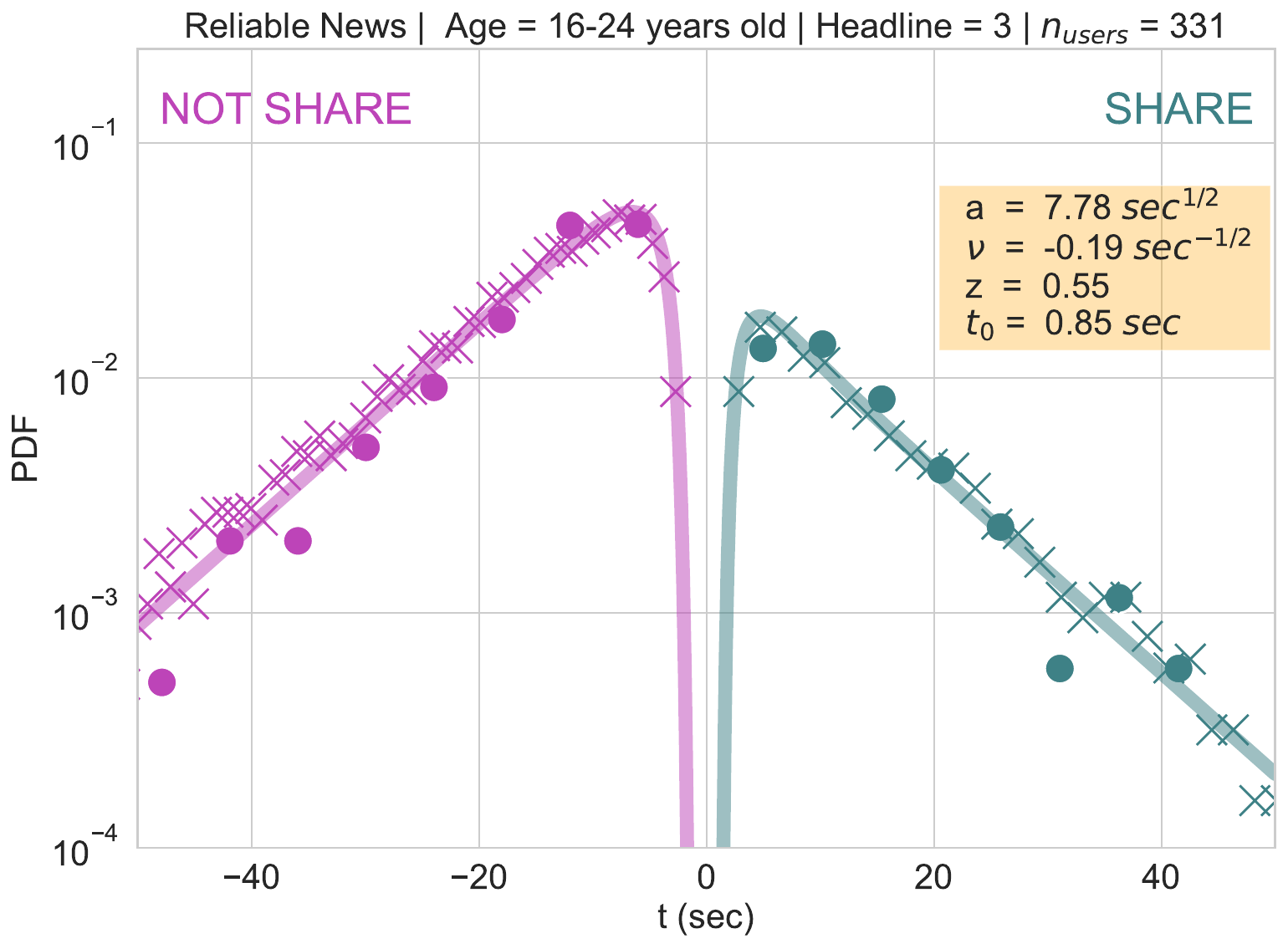}
                 \caption{}
                 \label{PDF-2a}
             \end{subfigure}
             \begin{subfigure}[b]{0.45\textwidth}
                 \centering
                 \includegraphics[width=1\textwidth]{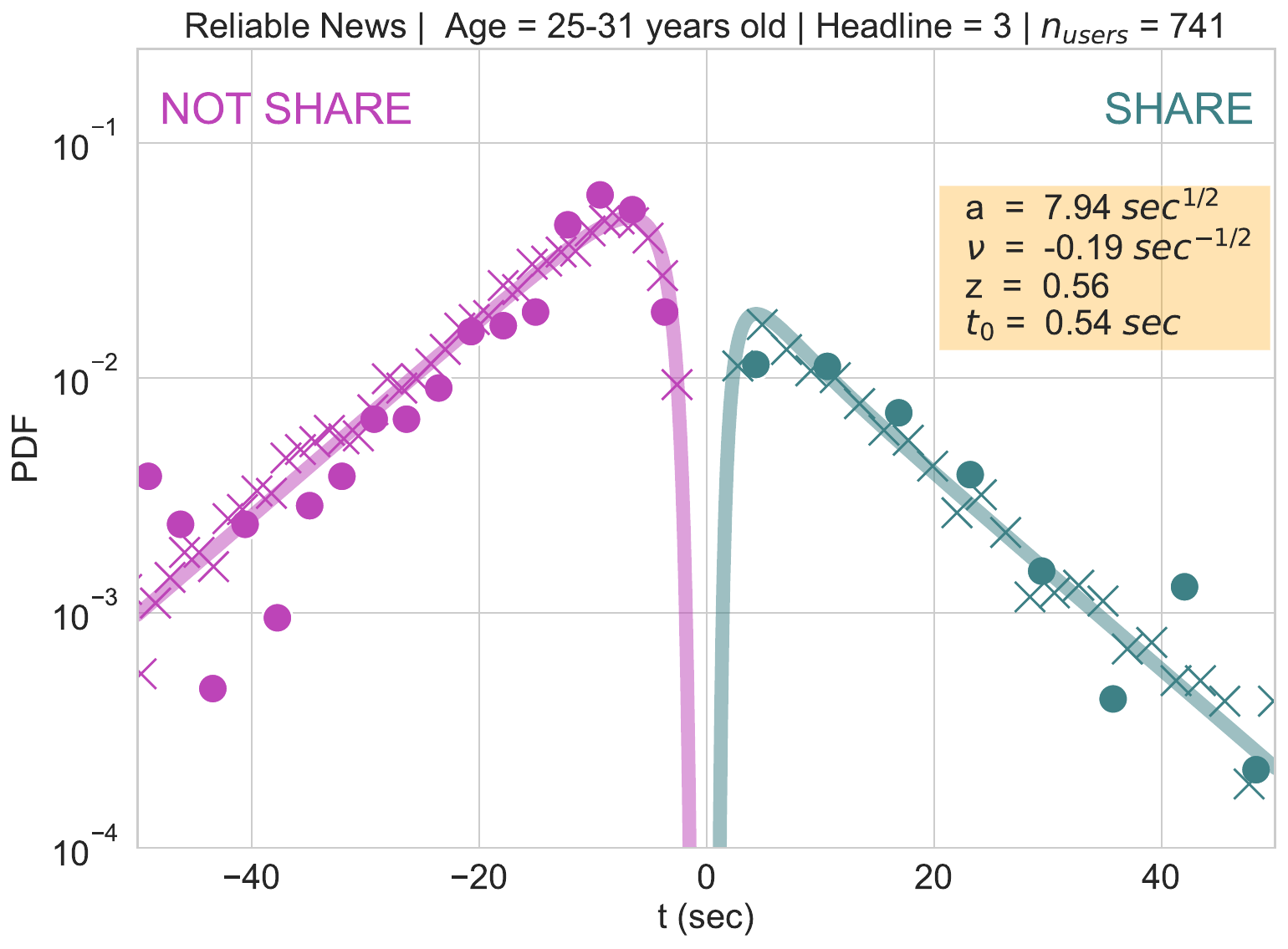}
                 \caption{}
                 \label{PDF-2b}
             \end{subfigure}
             \begin{subfigure}[b]{0.45\textwidth}
                 \centering
                 \includegraphics[width=1\textwidth]{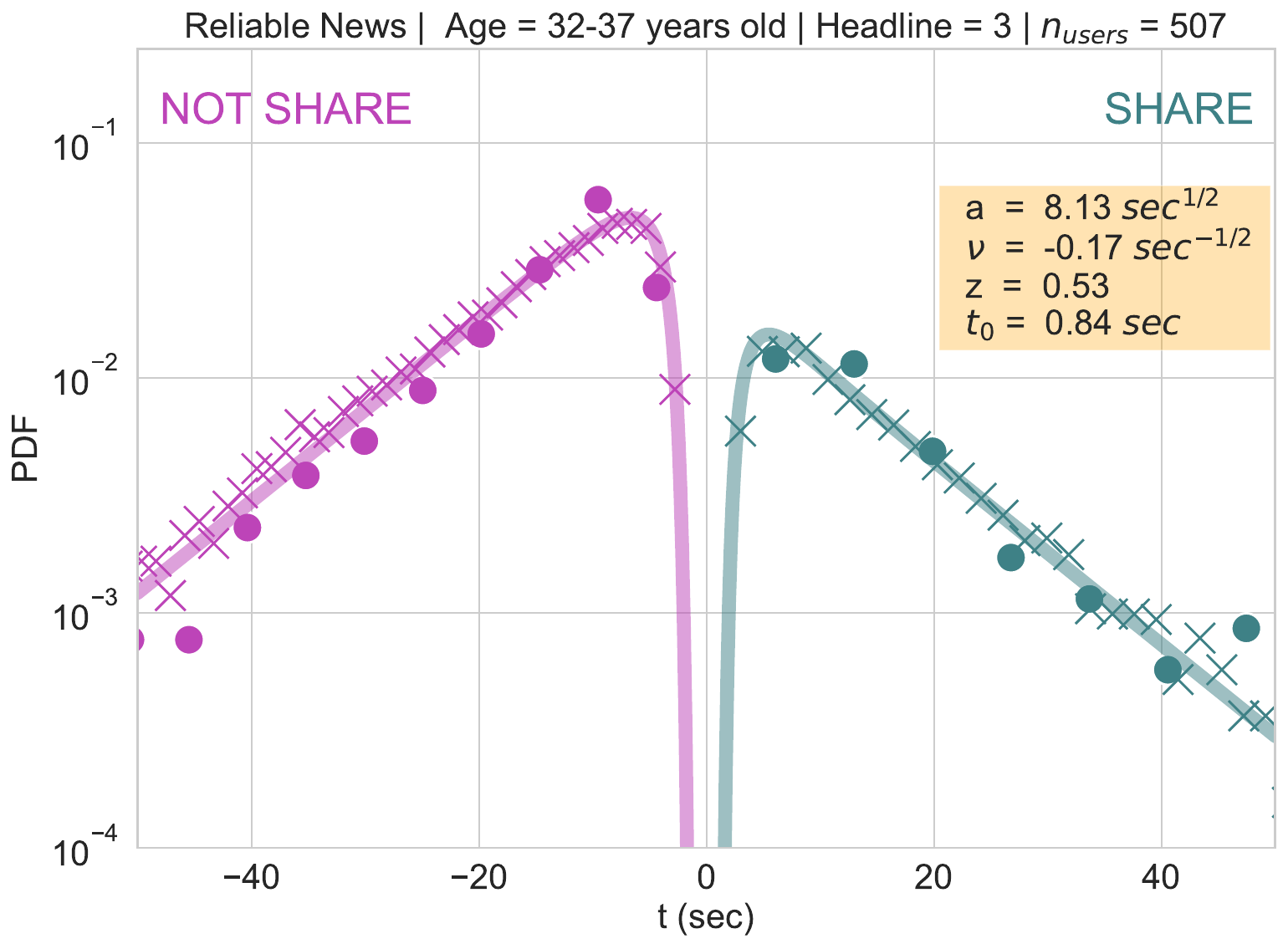}
                 \caption{}
                 \label{PDF-2c}
             \end{subfigure}\\
             \begin{subfigure}[b]{0.45\textwidth}
                 \centering
                 \includegraphics[width=1\textwidth]{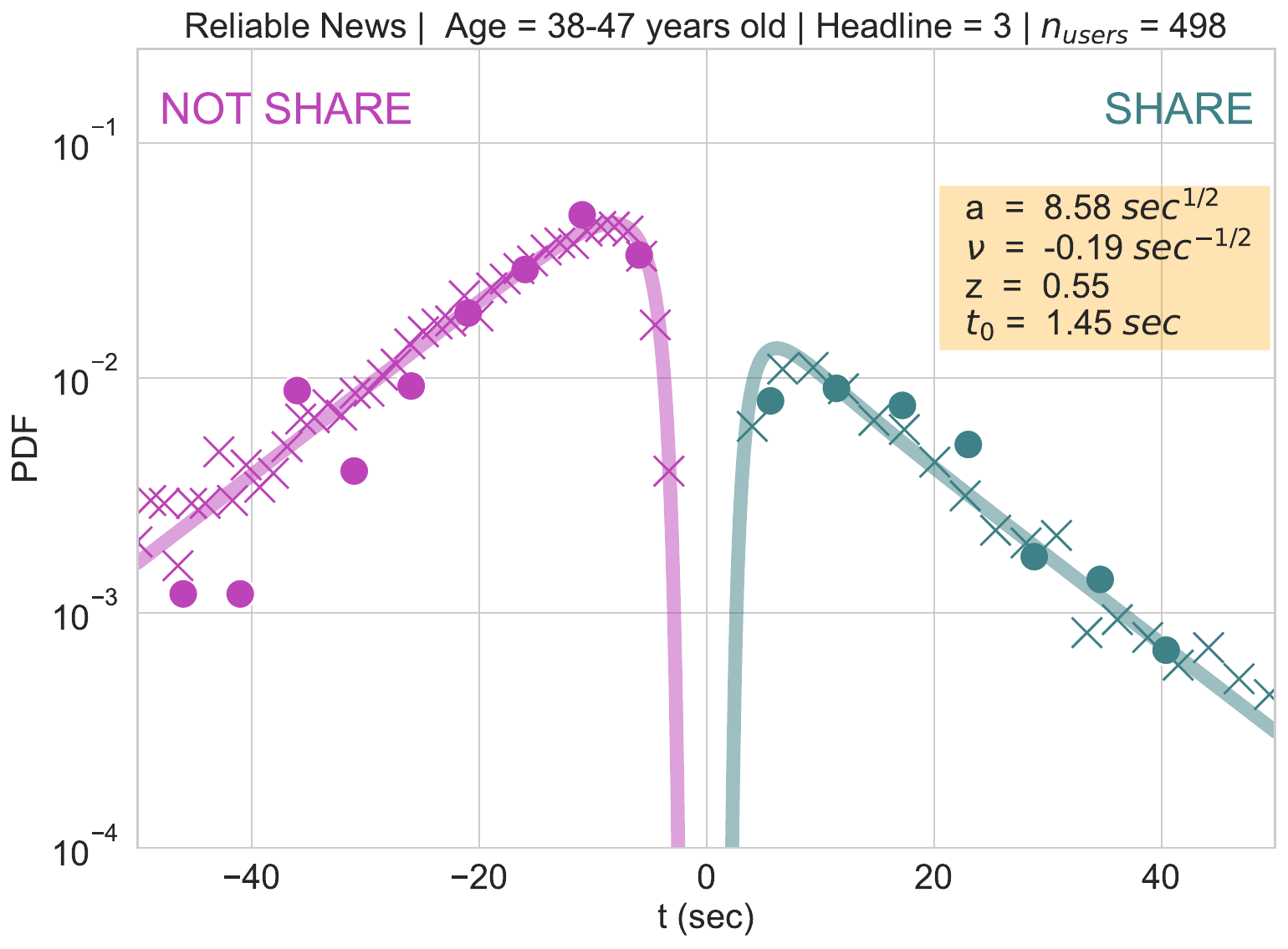}
                 \caption{}
                 \label{PDF-2d}
             \end{subfigure}
             \begin{subfigure}[b]{0.45\textwidth}
                 \centering
                 \includegraphics[width=1\textwidth]{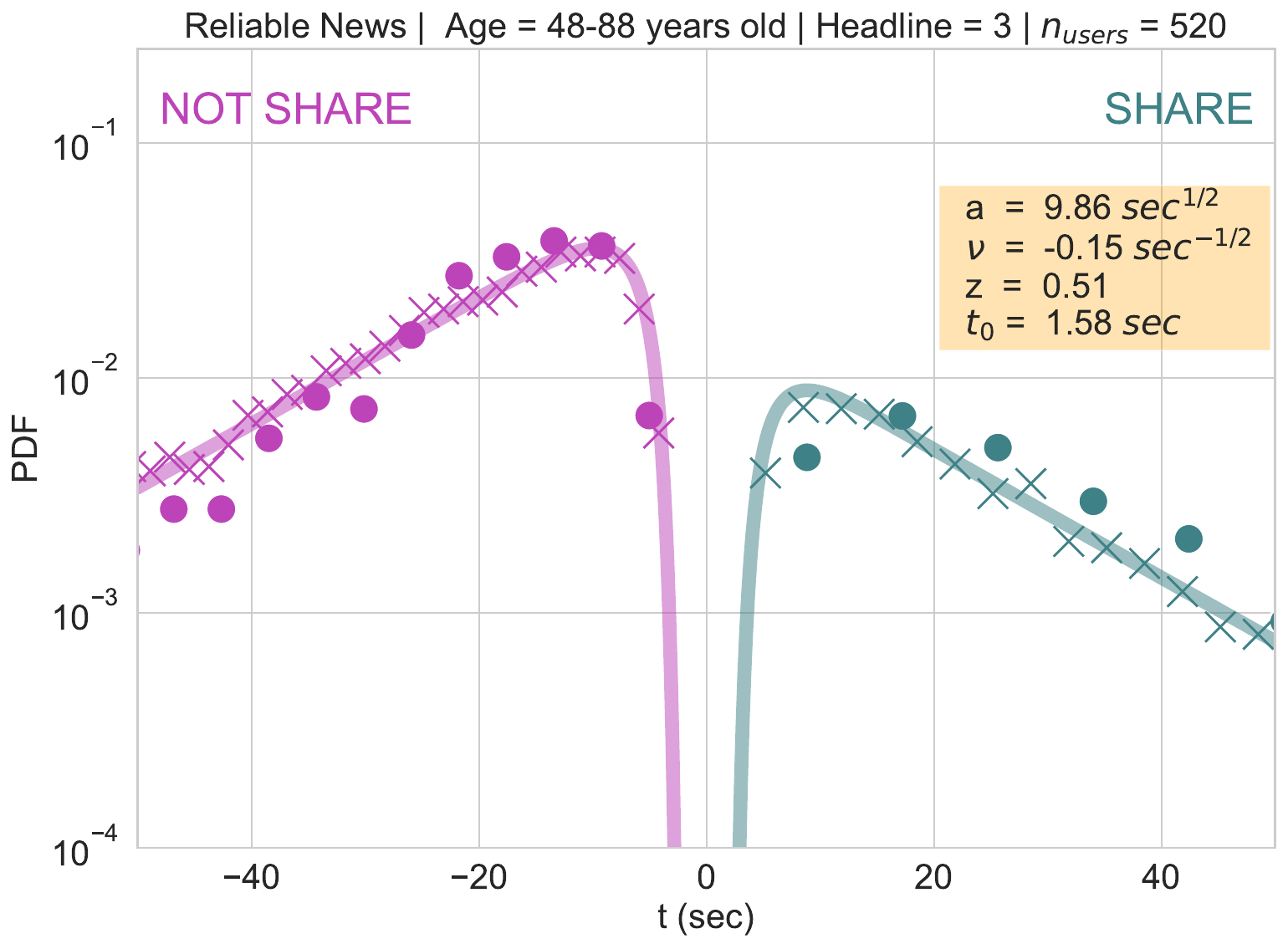}
                 \caption{}
                 \label{PDF-2e}
             \end{subfigure}
            \caption{{\bf Probability Distribution Function of the Response Times for the case of Reliable Information}. Details on the description are equal to Fig.~\ref{PDF-misleading} but considering the case of reliable headlines. By comparing both scenarios, we can observe that for reliable news responses tend to have longer response times and more they are more shared, as we can also observe in Fig.~\ref{statistics-share}.}
            \label{PDF-reliable}
        \end{figure}

        We applied the Hierarchical Drift Diffusion Model \cite{wiecki2013hddm} (described in the Methods Section) to obtain the set of parameters that better represent the data. Following, we display the probability density functions (PDF) of the response times (RTs) for the case of misleading in Fig.~\ref{PDF-misleading} and reliable content in Fig.~\ref{PDF-reliable}. A good agreement is obtained between empirical data (dots) and theoretical results (solid lines), which are derived by Eq.(\ref{RT-PDF}) and using the set of parameters in the legend (legend box, and see methods for further details on the dimension of the free parameters). We disaggregated our data by age range: $16-24$, $25-31$, $38-47$, and $48-88$ years old, and headlines: $12$ misleading and $12$ reliable; and we consider answers with $RTs < 100$ seconds (see the sensitivity analysis in Supplementary Fig.~\ref{RTs-analysis}. Besides, we classify the RTs according to individuals' responses, with a positive orientation for the case of sharing (orange left side) and a negative orientation for the case of not sharing (violet left side). For simplicity, we only display headline number $3$, but all headlines can be found in the Supplementary Information. In addition, Fig.\ref{statistics-share} displays the bar charts of the number of responses when sharing and not sharing for reliable and misleading content. Note that these results have already been analyzed in \cite{pennycook2019lazy}. From Figs.~\ref{PDF-misleading}-\ref{statistics-share}, we can see that the decision not to share is more trendy than the decision to share, and this is consistent among all ages and independent of the content. Besides, as already observed in \cite{pennycook2019lazy}, deciding not to share in general takes more time, $i.e.$, higher RTs have a higher probability for the not sharing scenario. And in a misleading scenario (Fig.~\ref{PDF-misleading}), this difference is more noticeable. An explanation for this could be that individuals prone to sharing may do it as a reflection or a habit, as suggested in \cite{ceylan2023sharing}; they do not take much time to think about the decision and do it automatically. While the decision not to share requires more analysis, longer times have more probability, meaning that individuals need to gather more evidence (probably) to discern the content's veracity and if they are prone to share it. Finally, in a reliable scenario (Fig.~\ref{PDF-reliable}), we can see that longer RTs now have a higher probability in both cases.
        \begin{figure}
            \centering
            \includegraphics[width=1\textwidth]{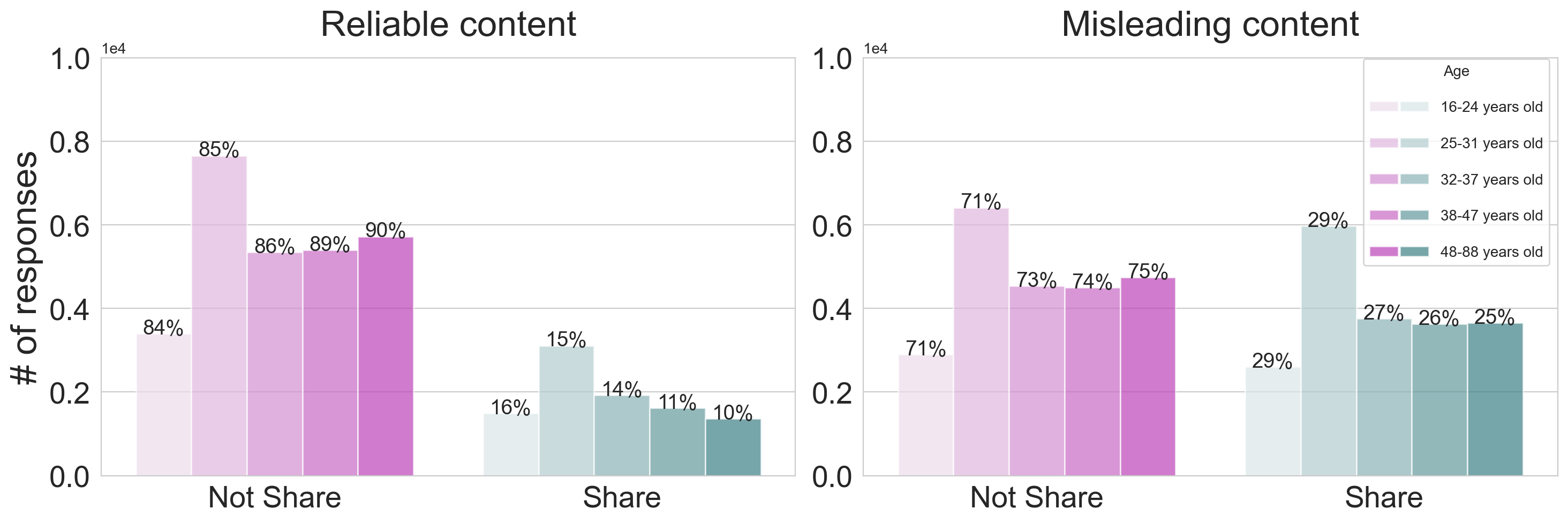}
            \caption{{\bf Number not share and share responses for the cases of reliable (left) and misleading (right) content}. Bars correspond to different age ranges, and responses are aggregated across all headlines. Older participants appear here slightly more reluctant to share both reliable and unreliable content.}
            \label{statistics-share}
        \end{figure}
        
        \subsection*{Neurocognitive mechanisms differ among ages and veracity of the content}\label{results-parameters}
        
\begin{figure}
            \centering
            \includegraphics[width=1\textwidth]{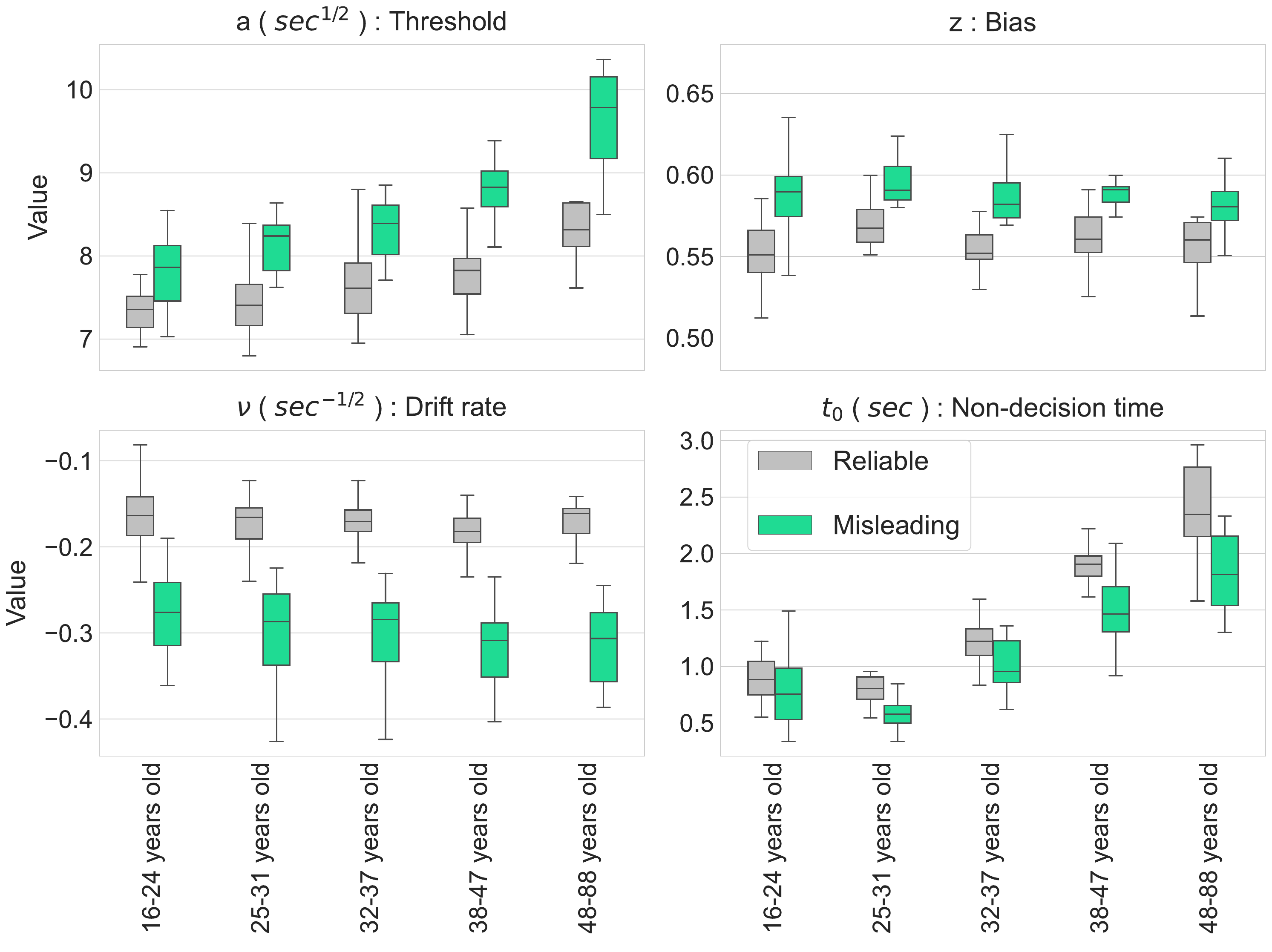}
            \caption{{\bf Free parameters of the drift-diffusion model}. Each plot corresponds to different free parameters: $a$ threshold (upper left), $\nu$ drift towards sharing the news (lower left), $z$ the a priori bias towards sharing (or not) information (upper right), and $t_0$ non-decision time before the brain starts the process of collecting information and deciding (lower right). See Methods for details on the dimensions of the free parameters. Each box plot shows the distribution of each age range averaging over $12$ headlines for the case of misleading (green) and reliable (grey) headlines.}
            \label{parameters-all}
        \end{figure}

        In this Section, we estimate the free parameters of the DDM. The fit of the parameters has been performed using a Hierarchical Bayesian Estimation to simultaneously fit the distribution of response times for both options --share and not share--. Fig.~\ref{parameters-all}, shows the results for the cases of (a) misleading news headlines and (b) reliable news headlines, where plots correspond to a different parameter, and each box plot shows the distribution of each range averaging over the $12$ headlines. Notice that even though the nature of both headlines differs, we can see similar trends in both cases --misleading and reliable--.
        
        {\bf Threshold ($a$)}: This parameter is also known as the ``criterion'' or the ``decision boundary'' and is related to the amount of evidence required to trigger a response. Speed-accuracy trade-off is modeled by this parameter. If accuracy is emphasized, boundary values tend to be higher and far from the starting point; in this way response times are slow, and accuracy is high. Conversely, when speed is emphasized, boundary values tend to be lower and closer to the starting point; in this way, individuals respond faster, leading to lowered accuracy due to the fact that processes that would have reached the correct boundary now are more likely to reach the wrong one randomly \cite{ratcliff1998modeling}. From Fig.~\ref{parameters-all}, we see a tendency in age to become more cautious when sharing any information independently of their nature. However, in the case of misleading content, the increase in cautiousness over age is considerably larger. From Fig.~\ref{statistics-share}, we can see that both --younger and older people-- are more prone not to share this type of content. This may imply that the parameter is greater for older generations because they take more time to discern the veracity of the content. 
        
        {\bf Bias towards sharing ($z$)}: The parameter indicates the starting point for the decision-making process, and it is related to the initial inclinations of the individuals; that is to say, the pre-existing ideas individuals have about the topic. In our case study, if initially subjects are more toward sharing, the starting point will be closer to the corresponding boundary $a=1$. Otherwise, it will be closer to $0$ (not sharing). From Fig.~\ref{parameters-all}, we can see that, in principle, individuals tend to be slightly biased toward sharing both types of content since $z \geq 0.55$ and have, therefore, an instinctive inclination to share content. The bias for reliable headlines, in particular, remains almost constant across ages and appears systematically larger for misleading headlines. This suggests that subjects may have an amplified intuitive inclination to share misleading news.
        
        {\bf Drift rate ($\nu$)}: This parameter is related to the information acquisition, and it measures the subjects' ability to gather evidence of the stimulus related to the task complexity. The module $|\nu|$ specifically indicates the rate at which individuals process the information given by the content to understand if they are willing to share it or not. From Fig.~\ref{parameters-all}, we can see that the values tend to be closer to zero, implying that the stimuli are ``difficult'' to understand and the information gathering is slow, particularly regarding reliable information. For misleading content, the task seems easier. On the other hand, the negative sign of the drift suggests individuals's rational thinking leads more toward not sharing information in general, independently of the veracity of the content and ages. Lastly, we can see that there is not much difference among ages; for younger people, it is slightly less challenging to understand the content, but the difference is not significant enough to draw strong conclusions. 
        
        {\bf Non-decision time ($t_0$)}: This parameter corresponds to the time before the decision process begins, is the component of RT that is not due to evidence accumulation, and is more related to the physiological components of the experiment. For instance, it could be related to the time individuals take to prepare themselves to start reading the headline (before they realize that the new round started) or the time used to press the bottom of the mouse. As expected, we can see almost no difference between the two types of content. Yet, there is a noticeable difference among ages: young people tend to be faster, while older people tend to be a bit slower and take more time to start with the task.
        
        In an unbiased scenario ($z=0.5$), when $\nu$ reaches high values, faster and more precise decisions are taken. In addition, high values of the $a$ reduce the role of noise, producing slower yet more accurate responses. Consequently, we can explore the speed/accuracy trade-off in our cases of study in a hypothetical absence of bias. In this context, we conclude that older generations tend to be better thinkers, especially in the case of misleading content. Their trade-off is superior to younger generations due to higher values of $a$ and $\nu$; they are faster towards not sharing and more accurate. Besides, we can also see in Supplementary Fig.~\ref{expected_share_fraction} that the expected fraction of individuals towards sharing is higher for younger people who appear to be more reckless or lazy thinkers. This raised point offers an alternative but aligned interpretation of the results obtained by \cite{pennycook2019lazy} on this same dataset.
        
        \subsection*{Contact patterns matters on going viral}\label{results-network}
        
\begin{figure}[H]
            \centering
             \begin{subfigure}[b]{1\textwidth}
                 \centering
                 \includegraphics[width=1\textwidth]{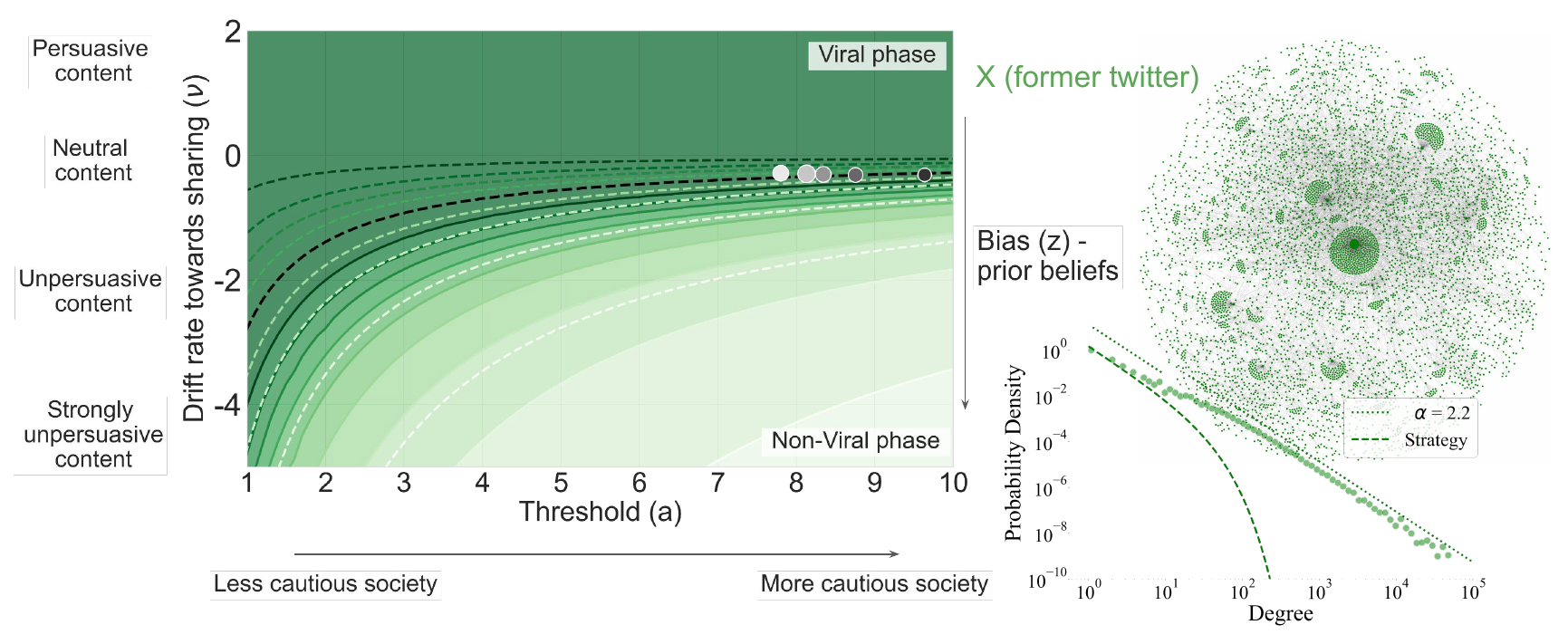}
                 \caption{}
                 \label{diagram1}
             \end{subfigure}
             \begin{subfigure}[b]{1\textwidth}
                 \centering
                 \includegraphics[width=1\textwidth]{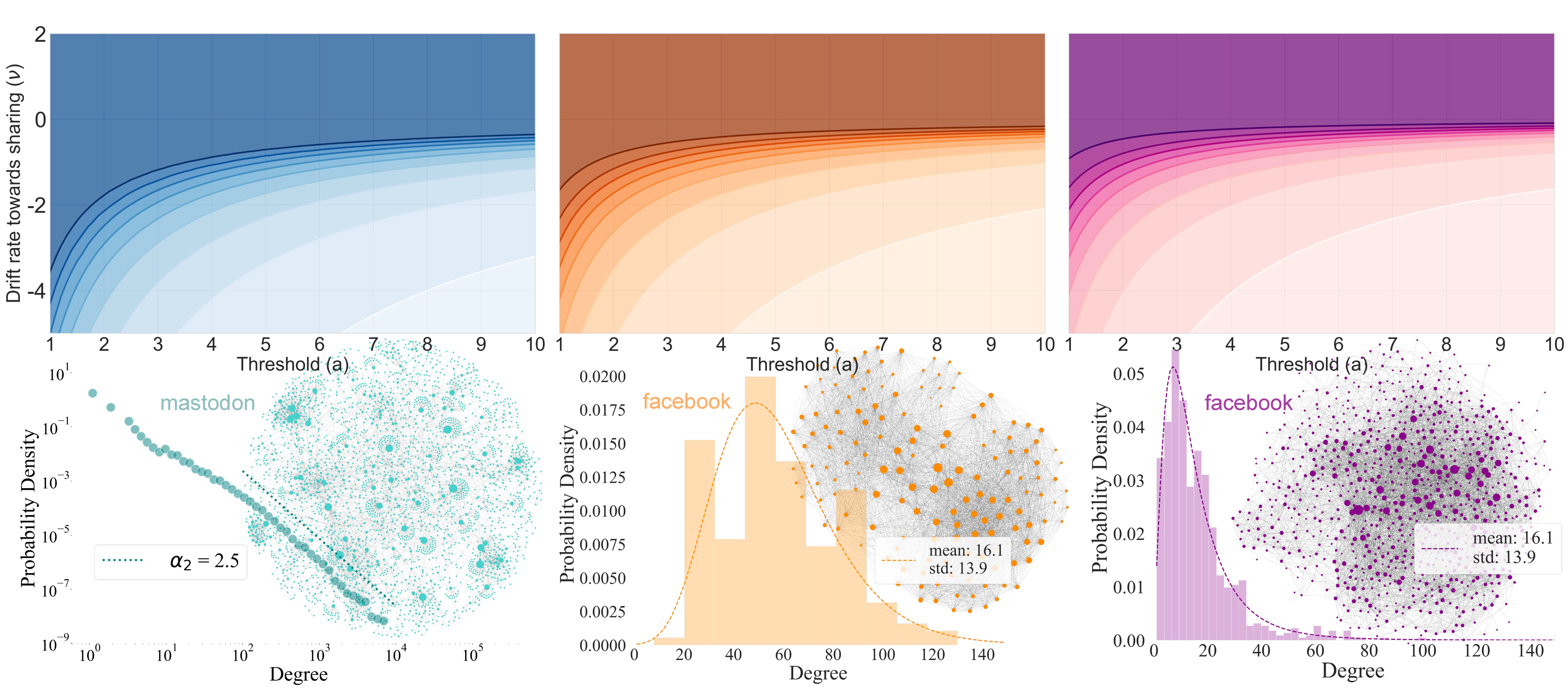}
                 \caption{}
                 \label{diagram2}
             \end{subfigure}
            \caption{{\bf Phase diagram for content to become viral according to the free parameters of DDM}. On the axis, we have the drift rate ($\nu$) as a function of the threshold ($a$), while each solid line corresponds to a different value of bias $z$ increasing from $0.1$ (top) to $z=0.9$ (bottom) with $\Delta z = 0.1$. The areas above the curves correspond to the combination of parameters that result in a phase of viral content, while below, the content does not go viral. Using synthetic networks with similar characteristics to empirical ones, we applied our model to obtain the displayed phase diagrams for X (upper panel), Mastodon (lower panel left), and two different datasets of Facebook (bottom middle/right). {\bf Upper panel}: X network exhibits a topology of a Scale-Free network $P(k) \sim k^{-\gamma}$ with $\gamma=2.2$, $k_{min}=1$ and $k_{min}=10^5$; dashed lines in this panel correspond to the proposed strategy of implementing a high-degree cutoff on users, thus $P(k) \sim k^{-\gamma}  \; \exp(-k/\kappa)$, with $\gamma = 2.2$ and cutoff $\kappa=20$. {\bf Lower left panel}: Mastodon network that also exhibits a topology of a Scale-Free network with $\gamma = 2.5$, $k_{min}=10^2$ and $k_{min}=10^4$. {\bf Lower middle and right panel}: the Facebook network that follows a Log-Normal degree distribution $P(k)=\frac{1}{x}\;e^{-\frac{(log(x)-\mu)^2}{2\sigma^2}}$ with $\langle k \rangle = 58$ and $\langle k \rangle = 16$ respectively. Each dot corresponds to the combination of values of the free parameters obtained from the empirical data for the different age ranges, considering the case of misleading content and averaging over headlines. The size of the dots corresponds to the value of $z$, being $\approx 0.55$ for all the cases.}
            \label{theory}
        \end{figure}

        How should a society be characterized to provide fertile ground for content to become viral? Even more, do the persuasive nature of the content play an important role in viralization? As a final step to answer these questions in this Section, we build a bridge between the individual-level concepts presented before and the behavior of the population (see Methods for details). Notice that, we assume all individuals in society are identical and follow a DDM. The condition for virality in this society is given by the expression given of Eq.~(\ref{criticality}) for critical threshold values of the free parameters of DDM: threshold $a_c$, drift $\nu_c$, and bias $z_c$. Besides, the non-decision times do not influence the mechanics behind the binary choice. In Fig.~\ref{theory}, we display the transition diagrams for the critical values of the free parameters of the DDM that divide two phases: the Viral content and the Non-Viral content phase. Meaning that all the possible combinations of parameters above the curves correspond to those scenarios in which the content becomes viral in that society. Conversely, in all cases below the curves, the content fades off. 
        
        In the x-axis of all panels in Fig.~\ref{theory}, we can see a clear division of societies in two: a {\it less cautious} one with low values of $a$ hence response times are shorter on average, and individuals tend to make faster decisions being more reckless; a {\it more cautious} one with high values of $a$, where response times are slow, subjects take time to gather evidence and thinking more before making a decision. On the other hand, the y-axis corresponds to the persuasiveness of the content. 
        
        For high values of the drift, the gathering of information is fast due to the stimuli being easy to understand, the content is clear and strongly persuasive (unpersuasive). Thus, subjects are more prone to not sharing (sharing), depending on the sign being negative (positive), independently of the veracity of the headline. Instead, for intermediate values when $\nu \approx 0$, the stimuli become a more difficult task. Individuals find themselves undecided whether to share the content or not, and this is due to its neutral nature. Jointly, solid lines are incorporated by varying the bias $z$ from less prone to share (top) to more prone to sharing (bottom).
        For the social patterns, we focused on synthetic networks spanning from more heterogeneous to more homogeneous, following similar characteristics as the empirical networks of X (former Twitter) \cite{de2013anatomy}, Mastodon \cite{la2022information} and Facebook \cite{facebook_copenhagen} and \cite{facebook_high_school}. Finally, for the case of X which is the most heterogeneous structure, we also explore an extreme intervention scenario with exponential cutoff for high-degree users proposed by \cite{newman2002spread}. See Supplementary Fig.~\ref{cutoff} for more details. Notice that in our case study, we are not considering the competition between news; there is only one type of news spreading, and its viralization only depends on the characteristics of the population, the network structure, and the nature of the content. To be able to apply analytically the estimated critical thresholds, we work under the simplifying assumption of considering tree-like networks, without clustering; and the presence of only one community in the system. 
        
        In general, from all panels in Fig.~\ref{theory}, we can observe that the network structure plays an important role in the dissemination of content; for more heterogeneous networks —Twitter and Mastodon—, we can see a greater viral phase where even strong unpersuasive and confusing content may easily become viral. As we move towards more homogeneous structures these phases tend to get reduced. Besides, we can observe two general trends: i) For {\it less cautious societies}, the bias and persuasiveness play a more crucial role in going viral. Meaning that, if individuals are a priori towards not sharing a certain type of content, the story should be more compelling to become viral. Conversely, if the individuals are biased towards sharing, almost any type of news will go viral. ii) For {\it more cautious societies}: since individuals tend to be more cautious and collect more evidence before answering, thus the bias and persuasiveness play a less important role. The nature of the news has to be at least neutral to be successful and there is a trade-off between the medium/low bias and homogeneous networks that discourage viralization.
        
        Modifying the structure of a network could be challenging, particularly in topologies as large and complex as Twitter, Facebook, or Mastodon. Yet, intervention strategies, such as limiting the number of accounts that can follow an individual may be possible to implement. Thus, we explore the impact of reducing the maximum degree of users. In Fig.~\ref{theory}, are displayed in dash lines the case of a Twitter network with an extreme case of degree cutoff $P(k) \sim k^{-\gamma} \; \exp(-k/\kappa)$ with $\gamma = 2.2$,  $\kappa=20$. As we can see, the reliability of the content becomes crucial, bringing up a significant reduction in the viral phase areas. Now, for the unbiased scenario (black dashed line) unreliable content cannot become viral anymore. The network structure still favors viralization, yet not enough for any type of content to spread. Interestingly, for the cases of biased populations toward not sharing ($z<0.5$), only persuasive content will become viral. And, for biased populations toward sharing, no major changes appear. 
        
        Furthermore, in the Supplementary Section in Fig.~\ref{theory-dunbar}, we explore a more realistic scenario with a cutoff $\kappa=150$, corresponding to Dunbar's number. In previous work \cite{gonccalves2011modeling}, authors analyzed Twitter conversations and found that there is a cognitive limit on the number of stable social relationships users can maintain, and this number ranges from 100 to 200 stable relationships. We find this scenario to be a good compromise where the impact of the cutoff is comparable with the cutoff $\kappa = 20$ discussed so far. Therefore, a potential containment strategy would be limiting users to reach a number of connections at least equally or less than the natural Dunbar's number and, in this way, reduce the leverage provided by social media that empowers few "influencers" to have almost unlimited reachability.
        
        Finally, we incorporate the set of values of the parameters obtained from the empirical data, where each grey dot corresponds to a different age range (increasing as the color gets darker), and the size increases with respect to the value of $z$. As we can see in Fig.~\ref{parameters-all}, for the case of misleading content the set of values of $a \in [7,10] $, $\nu \approx -0.3$, and $z \approx 0.55$ extracted from the data belongs to the viral phase. This means that in a homogeneous population, those characteristics and contact patterns virtualization may probably occur, however, with the extreme intervention scenario, it is possible to modify the topology enough so that it will not. Remarkably, in the case, $\kappa=150$ (Dunbar's number, see Supplementary Fig.~\ref{theory-dunbar}), seems to lie closely to the critical point where news might become viral (or not) randomly. Notice that in the conditions we are simulating here, there is only one piece of news monopolizing the media, and the 
        cognitive behavior of the population is homogeneous.
    
    \section*{Discussion}
        Not all information is trustworthy and safe to share. Occasionally, false information could be malicious and have the capability to change our perception of reality, creating confusion and compromising our welfare. In this information landscape, we wonder: Why does information successfully spread through an entire society? Which characteristics should the populations have to provide a fertile ground for this phenomenon to emerge? At what level do the neurocognitive mechanisms of each individual affect the spreading, and conversely, at what level do the social contact patterns impact?

        In this research, we study the decision-making process of sharing information —misleading and reliable— online. We take advantage of the universality of the Drift Diffusion Model to mathematically describe the response of individuals to share potentially (mis)informative content. By obtaining a solid accuracy (in the response times) between our model and the data, we analyzed the neurocognitive mechanisms of the process by tracking the free parameters of the DDM. The model is well-known for describing decision-making processes with two-option responses, different from our three-option response problem. Yet, here, we can reduce our problem and find a perfect accuracy between the DDM and the data. Then, we built a bridge between this individual behavior (micro-level) and the collective one (macro-level), using a social network approach. Finally, we use mathematical modeling to obtain the phase diagram of the viralization of content, where all the micro and macro information is merged. The diagram is a joint junction of the neurocognitive characteristics of the individuals, the features of the content, and the social context patterns, from where we may be able to predict if the content will go viral in a particular society. This approach also allowed us to study, as an intervention strategy, the impact of limiting the maximum number of followers a user can have.
        
        Regarding the neurocognitive parameters of the model, there is a difference in all our results, which is related to the veracity of the content. Firstly, individuals tend to be more cautious and hence take more time to decide whether to share misleading content. And this cautiousness is further accentuated with age. Secondly, individuals of all ages a priori tend to be slightly prone to instinctively share news headlines and more likely to if the content is misleading. This difference may be explained by the fact that misleading content creators are actively attempting to leverage emotions to generate engagement in their readers \cite{bakir2018fake}. Thirdly, in general, when individuals gather information concerning a news headline, the stimuli they receive tend to be directional in favor of not sharing content. And in the case of misleading content, the trend runs deeper.
        
        By differentiating societies in their level of cautiousness when making a decision, our results suggest different and interesting scenarios. For less cautious societies, the bias and persuasiveness play a more important role in the spread of information. Individuals here tend to respond faster, being more reckless and directing their answers toward their pre-existing concepts and even getting caught by uncompelling stories. Conversely, in more cautious societies, the effect of the bias fades off and the nature of the content should be at least neutral for individuals to consider sharing it. This behavior runs much deeper when there is a limit on the number of users an individual can be connected to. In the context of intervention strategies, such as in predicting the effects of pre-bunking \cite{lewandowsky2021countering} or automated debunking interventions \cite{russo2023countering}, our results suggest that it is of utmost importance to detect the features of the targeting population. Both the average age of individuals and the structural characteristics of the social networks are important in determining the spreading dynamics. Younger and less cautious societies tend to be a more fertile ground for misinformation spreading. At the same time, even strongly unpersuasive content may become viral if the network structures, where a smaller number of influencers dominate the spreading process, provide that individuals are biased toward sharing content. Unfortunately, our result suggests that the natural shape of online platforms —X, Mastodon, and Facebook— tendentially provide fertile ground for news to go viral rapidly. As modifying the network topology could be challenging, we propose to target users with large amounts of followers \cite{baribi2024supersharers}. We find that an extremely reduced maximum in-degree should be imposed in order to enhance the importance of the characteristics of the population (measured by the bias and the barrier) and the nature of the content (measured by the drift).
        
        Our work has limitations. On one hand, our aggregation in the data considers that all individuals are a replica of the same ``average subject'' of a specific age range, independently of gender, or political orientation, for instance. Our choice is motivated by the need to propose a method that can be scaled to what is commonly done in decision-making theory and cognitive neuroscience, where experiments are usually done in a controlled environment, and data is challenging to gather. Another aspect is related to cognitive mechanisms; when we scale the individual to the collective behavior, we consider a homogeneous population with the same values of the free parameters for all individuals. We know this is a limiting assumption, yet our research is motivated by how information spreads on online platforms, and we expect to have a strong presence of specific average age ranges as the most prominent one among all the users and different for each platform. Besides, in our case study, we are considering experimental conditions representing a monopolization of the media, where there is only one piece of information propagating in a society. Lastly, we recognize the need for a real-world network structure. In general, a real topology has a strong clustering presence along with the appearance of communities and homophily. All of them are important features for information diffusion. Motivated by these open questions, we would like to explore them more in detail in future studies, jointly with intervention strategies to diminish misinformation from going viral.
        
        Understanding the mechanisms behind sharing information online can be challenging but crucial to win the battle against misinformation. We hope that the methods and our findings are a step forward toward understanding the role of the characteristics of individuals in the propagation of information in a society from a theoretical and empirical perspective. The threat that the misinformation crisis imposes has socioeconomic and political consequences, and we advocate for continued progress in the mathematical modeling of this phenomenon.
        
    \section*{Methods}
        \subsection*{Decision-making individual data}

        The data used for this paper was collected, described, and analyzed by G. Pennycook \& D. G. Rand in \cite{pennycook2019lazy}. In this study, the authors aim to understand whether analytic thinking supports or undermines misleading political content susceptibility. For the data collection, authors reported a total of 5 replicas of the same experiment, with a total sample of $N=2644$ participants with an age range from $16$ to $88$ (average of $36.9$) and an almost equal balance between females and males genders. Each experiment consisted of presenting to participants 24 headlines, 12 of them reliable and 12 of them misleading, in random order. All misleading news headlines were originally taken from Snopes.com, a well-known fact-checking website. On the other hand, reliable news headlines were selected from mainstream news sources (e.g., NPR, The Washington Post) and were contemporary with the misleading news headlines. Participants were asked the following questions: 1) "To the best of your knowledge, how accurate is the claim in the above headline" with response options: not at all accurate, not very accurate, somewhat accurate, or very accurate; and 2) "Would you consider sharing this story online (for example, through Facebook or Twitter)?" with response option: {\it yes, maybe or not}.
        
        \subsection*{Drift Diffusion Model to describe human decision-making}
        
        A vast number of theoretical models describing the human decision-making process have been developed \cite{ratcliff2016diffusion, bogacz2006physics}. Among all these models, the statistical Drift Diffusion model (DDM) is one of the most prominent ones, especially due to its simplicity and accuracy. The model assumes a one-dimensional random walk behavior that represents the accumulation of noisy evidence -mimicking the accumulation of information- in our brain in favor of two alternative options \cite{ratcliff2008diffusion}. The process begins from a starting point $x(0) = z \; . \; a$, where $z$ is the bias (prior beliefs), and $a$ is the threshold (length of the barrier), see Fig.~\ref{model-network} in Introduction Section. At each time step, the individual gathers and processes information until one of the two boundaries is reached, $x = 0$ or $x = a$, and we say a decision is made. The continuous integration of evidence is described by $x(t)$ and is given by the equation $dx \; = \; \nu \; dt\;+ \; \sqrt{D} \; \xi(t)$; a one-dimensional Brownian motion with a drift term with parameter $\nu$, and diffusive term with parameter $\sqrt{D}$ as the diffusion coefficient plus white Gaussian noise $\xi$. The probability distribution of the response times (RTs), $i.e.$, the times at which the process reaches one of the boundaries, is given by
        \begin{equation} \label{RT-PDF}
            \begin{split}
                P(t;\nu,a,z) & = \frac{\pi}{a^2} \; exp \left(- \nu\;z\;a - \frac{\nu^2t}{2}\right)\\
                             & \times \sum^{\infty}_{m=1} \; exp \left(-\frac{m^2\pi^2t} {2a^2} \right)\; sin(m \; z \; \pi)
            \end{split}
        \end{equation}
        and is known as the F\"urth formula for first passages. Notice Eq.~\ref{RT-PDF} is a reduced expression obtained by taking $D=1$ \cite{bogacz2006physics,gallotti2019quantitative}. For simplicity, this is a common practice in order to use the remaining three parameters ($a$, $z$, and $\nu$) independently to fit the data with the curve. By imposing $D=1$ with $[D]=[t^{-1}]=sec^{-1}$ we have, for dimensional reasons that $[a]=[t^{1/2}]=sec^{1/2}$ and $[\nu]=[t^{-1/2}]=sec^{-1/2}$. 
        
        In our context, we say the decision not to share the content is made when the boundary $x = 0$ is reached; otherwise, if $x = a$ is reached, the decision is to share. The free parameters $z$, $a$, and $\nu$ played an important role in the process. The $z$ describes the biases that an individual has initially, the prior beliefs. In our context, if it is greater than $0.5$, this means that it is towards sharing the content, below towards not sharing, and an unbiased scenario is given when $z=0.5$. Then, the drift is related to the information gathering and can be analyzed in two parts: on one hand, the module of the drift $|\nu|$ is the signal-to-noise ratio representing the amount of evidence supporting the two alternative options, the lower the $|\nu|$, the more difficult the task; and on the other hand, the sign of $\nu$ is related to the direction supporting one of the two options, when $\nu>0$ (positive), the gathered of information is tendentiously in favor of sharing, while when $\nu<0$ (negative) the evidence gathered is mostly supporting not sharing. Hence, the probability distribution in Eq.(\ref{RT-PDF}) describes the decision-making process where the ultimate decision is to share. Conversely, the probability distribution associated with not sharing will be given under the conditions $P(t, \; -\nu, \; a, \; 1-z)$.
        
        A decision-making process with two options as possible answers can be well-modeled by the Drift Diffusion Model \cite{ratcliff2008diffusion} (see Methods). However, this is not the case for three or more responses. Typically, this kind of scenario can not be well described with a one-dimensional dynamic, and there are more suitable models, such as the race model \cite{bogacz2006physics}, in which accumulators for each alternative integrate evidence independently. To be able to use DDM to describe the dynamic, we reduce our problem to two dimensions. We assume that ``no'' responses are the only representative of not sharing (individuals here are determined not to share the content), and the rest of the answers, ``yes'' and ``maybe'', are in an undecided grey area, mostly prone to share the content. In this way, we test the accuracy of the DDM for more complex cases, and we reduce a complex problem and use a simple theoretical approach to describe it.
        
        
        \subsubsection*{Model fitting}
        
        The data was fitted using a Python-based toolbox called Hierarchical Drift Diffusion Model (HDDM) \cite{wiecki2013hddm} version=$0.6.0$. The library uses hierarchical Bayesian estimation for the free parameters of the DDM ($a$, $b$, $\nu$, and $t_0$), which are the ones that we use afterward in the analysis jointly with Eq.\ref{RT-PDF}. Response and decision times were disaggregated by age range ($16-24$, $25-31$, $38-47$, and $48-88$ years old), headlines ($24$ in total), and the veracity of the content ($12$ false headlines and $12$ reliable headlines). Then, we assume that all the participants are copies of an average subject by aggregating across participants.
        
        \subsection*{(Mis)information in a population}
        
        As explained before, the DDM model has been well studied on a micro-level, in which single individual decisions are taken and described. However, we know that interesting and unexpected outcomes can emerge from collective behavior. Furthermore, these outcomes can be shaped according to the social contact patterns when they are considered. Thus, it seems natural for us to explore what happens if we extend this individual decision-making process to a population in which individuals are socially connected. In other words, how the cognitive processes will affect the decision made by an individual, whom at the same time influences the decision of their neighbors. In this context, individuals are deciding whether to share or not misleading content after they see it online when shared by a neighbor. Hence, we have two processes taking place at the same time: on one side, there is the diffusion process of the content among the neighbors in the network structure; and on the other side, there is the final individual decision-making process of sharing the content driven by the cognitive mechanisms of each individual. 
        
        \subsubsection*{Network structure Data}
        
        The complexity of social-human interactions plays a fundamental role in studying this kind of phenomena. The network structure on top of which the process takes place could shape differently the resulting outcomes. Here, we analyzed four empirical networks and characterized the degree distribution to generate synthetic networks with similar characteristics to work within our theoretical framework. The datasets are presented in the following table,
        \begin{table}[h!]
            \centering
            \begin{tabular}{>{\centering\arraybackslash}p{4cm} >{\centering\arraybackslash}p{5cm} >{\centering\arraybackslash}p{5cm}} 
            \hline 
            \cellcolor{tableheadcolor!80}\textcolor{black}{\textbf{Empirical Network}} & \cellcolor{tableheadcolor!80}\textcolor{black}{\textbf{Degree Distribution}} & \cellcolor{tableheadcolor!80}\textcolor{black}{\textbf{Estimated Parameters}} 
            \\
            \hline 
            
            X (former Twitter) \cite{de2013anatomy}
            & \multirow{2}{*}{$P(k) = (1-\lambda) \; [k_{max}^{(1-\lambda)}- k_{min}^{(1-\lambda)}] \; k^{-\gamma}$} 
            & $\lambda = 2.2$, $k_{min} = 1$ and $k_{max} = 10^5$ 
            \\
            mastodon.social \cite{la2021understanding,la2022information}                              
            & Scale-Free 
            & $\lambda = 2.5$, $k_{min} = 10^2$ and $k_{max} = 10^4$ 
            \\
            \hline 
            
            \cellcolor{gray!30} 
            & \multirow{2}{*}{\cellcolor{gray!30}} 
            &  \cellcolor{gray!30}
            \\
            \cellcolor{gray!30} Facebook \cite{facebook_high_school} & $P(k)=\frac{1}{x}\;e^{-\frac{(log(x)-\mu)^2}{2\sigma^2}}$ \cellcolor{gray!30} & $\mu = 58$ and $\sigma = 24.4$ \cellcolor{gray!30}
            \\
            \cellcolor{gray!30} Facebook \cite{facebook_copenhagen}    
            & \cellcolor{gray!30} Log-Normal                               
            & \cellcolor{gray!30} $\mu = 16$ and $\sigma = 13.9$ \\
            \hline 
            \end{tabular}
            \caption{}
        \end{table}
        
        As intervention strategies for the case of X, we consider an exponential cutoff of: (i) $\kappa=20$ \cite{newman2002spread} and (ii) $\kappa=150$ \cite{gonccalves2011modeling} (Dunbar's number). For simplicity, we consider an undirected network. 
        
        \subsubsection*{Spreading information process}
        
        We took a modified version of the Susceptible-Infected-Recover(SIR) \cite{bailey1975mathematical,pastor2015epidemic} model to describe the advanced misleading information in the population. In the classic SIR model, individuals can be in one of three compartments: susceptible as an initial condition until they get in contact with an infected neighbor and become infected with probability $\beta$; and after $t_r$ time steps, infected individuals become recover and immune, $e.i$ cannot be re-infected. Here, we say an individual is ``infected'' once they see misleading content shared by a neighbor and decide to share it with probability $\beta$. Generally, in online platforms \cite{de2013anatomy} we observe that once a user shares a specific post, most probably in the near future, the user will re-share it; meaning $t_r=1$ for our dynamic. At each time step, individuals are divided into two compartments:
        \begin{itemize}
            \item {\bf Non-Sharers}: if they have not shared the content yet (either because they have not seen it yet or decided not to do it).
            \item {\bf Sharers}: if they have seen the content from a neighbor and, with probability $\beta$, decided to share it and will not do it again.
        \end{itemize}
        
        
        Every time an individual visualizes misleading content -shared by a connection online- deals with the decision to share the content or not. And this process is driven by the cognitive mechanisms inside the brain which can be modeled by DDM. From the DDM theoretical framework, we obtain in Eq.(\ref{RT-PDF}) the expression for the probability distribution associated with sharing $P(t, -\nu, a, 1-z)$, and the area under the curve corresponds to the fraction of individuals sharing content,
        \begin{equation}\label{c_share}
            \begin{split}
                C_{SHARE}(\nu,a,z) & = \int_{0}^{\infty} dt \; P(t;-\nu,a,1-z)\\
                 & =  \frac{\pi}{a^2} \int_{0}^{\infty} dt \; exp{\left[ \nu (1-z) a - \frac{\nu^2 t}{2} \right]} \times \sum_{m=1}^{\infty} \exp{\left[ \frac{-m^2 \pi^2 t}{2 a^2} \right]} sin(m \pi (1-z)),
            \end{split}
        \end{equation}
        Here, $C_{SHARE}$ represents the probability of sharing content, which varies depending on the free parameters' values. Hence, for non-sharer individuals, every time a neighbor in the network shares the content with probability $C_{SHARE}$, they will decide whether to share or not the news, $e.i.$, $C_{SHARE}=\beta$. From contact network epidemiology \cite{meyers2005network, newman2002spread}, we know how the SIR model behaves in the presence of contact patterns, how to characterize the disease in a population through the basic reproductive number $R_0 = \frac{\langle k \rangle}{\langle k^2 \rangle - \langle k \rangle} \times (1 - (1-\beta)^{t_r})$, and under which conditions will prevail in the population $R_0 \geq 1$ (Supplementary Fig.~\ref{standard-SIR}). See subsection Agent-based SIR model in Supplementary Information for details. By extrapolating this concept into our problem, considering that $C_{SHARE}=\beta$, $t_r=1$ and Eq.(\ref{c_share}), we obtain the condition under which the misleading content goes viral in a population,
        \begin{equation} \label{criticality}
                1 = \left[ \frac{\langle k \rangle}{\langle k^2 \rangle - \langle k \rangle} \right] \times \frac{\pi}{a_c^2} \int_{0}^{\infty} dt \; exp{\left[ \nu (1-z_c) a_c - \frac{\nu_c^2 t}{2} \right]} \times \sum_{m=1}^{\infty} \exp{\left[ \frac{-m^2 \pi^2 t}{2 a_c^2} \right]} sin(m \pi (1-z_c)),
        \end{equation}
        where $a_c, \nu_c, z_c$ are the critical values for the free parameters. Notice that the first term considers the network structure of the population; different social contact patterns will result in different sets of critical values that satisfy the condition.
        
        \subsubsection*{Scenario without bias}
        
        As a final step, we study the particular scenario of an unbiased society. This means that individuals have no a priori inclinations or opinions; thus, they are not towards sharing or not sharing. This manifests in DDM when $z = 0.5$, and the fraction of individuals sharing content takes the shape $C_{SHARE}(\nu, a, z=0.5) = \frac{1}{1+exp(-a \nu)}$ \cite{gallotti2019quantitative}. Then, the condition under which the misleading content goes viral is reduced to,
        \begin{equation}
            \centering
            a_c \times \nu_c = ln\left(  \frac{\langle k \rangle}{\langle k^2 \rangle - 2 \langle k \rangle}  \right).
        \end{equation}
        On the left side, we have the contribution from the cognitive mechanisms, the social characteristics of the individuals in terms of cautiousness ($a$), and the ability to gather information ($\nu$). While on the right side, we have the contribution from the network structure, related to how the social contact patterns are distributed. Altogether, this expression gives us the critical minimum condition the population should have in order to favor the virtualization of the content. In the Supplementary Information, the phase diagrams for the cases of ER and SF networks are displayed. As we can see, a similar trend as Fig.~\ref{theory} emerges.
        
    \section*{Acknowledgements}
        Research reported in this publication was supported by the AI-TN project funded by the Autonomous Province of Trento. LLC was supported by project SERICS (PE00000014) under the MUR National Recovery and Resilience Plan funded by the European Union - NextGenerationEU. RG acknowledges the financial support received from the European Union’s Horizon Europe research and innovation program under grant agreement 101070190. LGAZ acknowledges financial support from the project `Understanding Misinformation and Science in Societal Debates' (UnMiSSeD), funded by the European Media and Information Fund.
     
    \bibliographystyle{unsrt}

    \clearpage

    \section*{Supplementary Information}
        \setcounter{figure}{0} 
        \setcounter{table}{0} 
        \subsection*{Sensitivity analysis of the Response Times}
        
        Here, we analyzed how the free parameters of the DDM vary as a function of the response time (RTs). For each threshold value, we consider those empirical responses corresponding to RTs lower than the threshold to then obtain the free parameters that better represent that pool of responses. As we can see, in both cases, misleading and reliable content, the values stabilized after $RTs = 100$. Each curve corresponds to a different age range, and for the sake of simplicity, we average among headlines.
        \begin{figure}[H]
            \renewcommand{\figurename}{Supplementary Figure}
            \centering
            \includegraphics[width=1\textwidth]{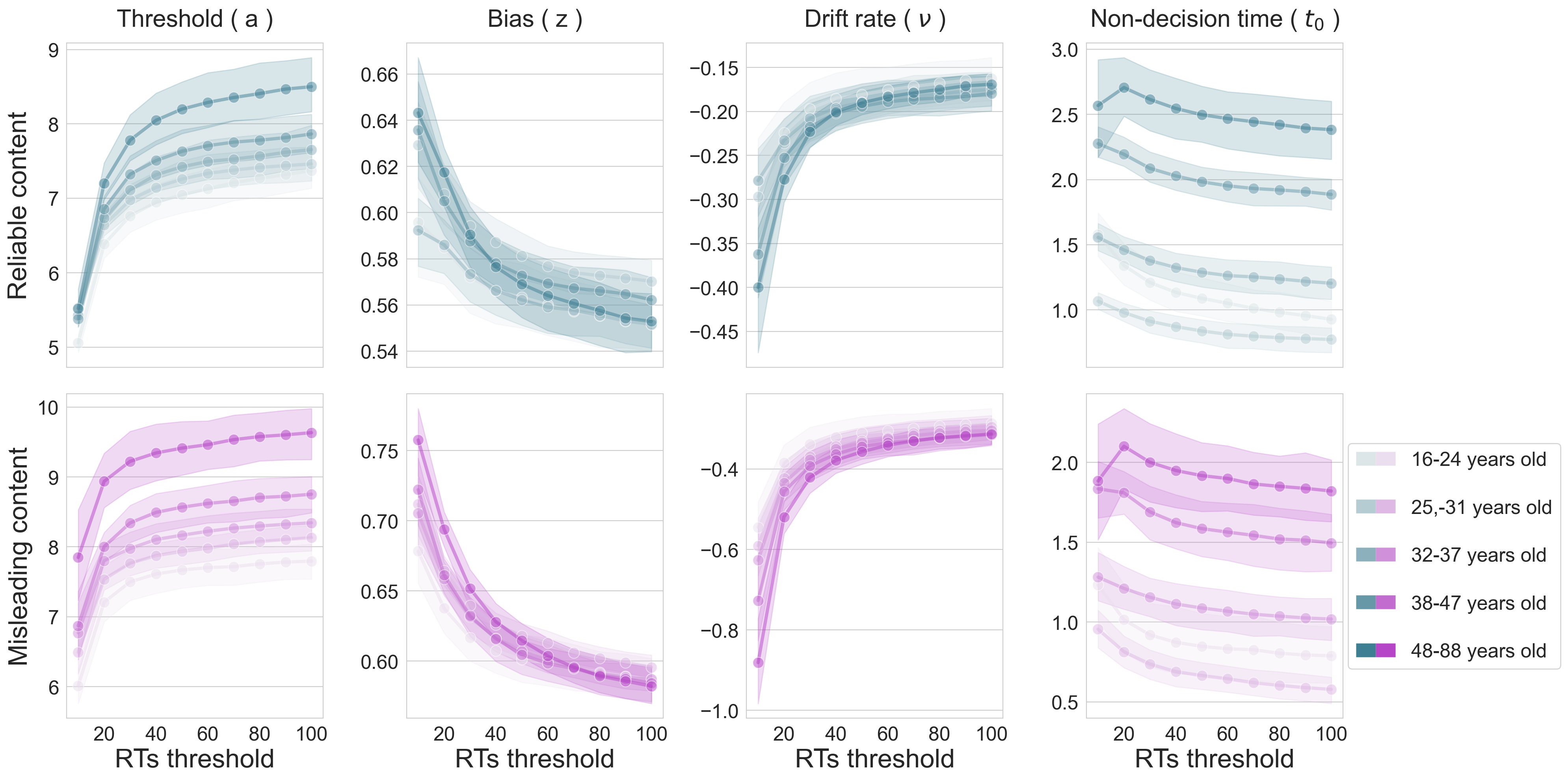}
            \caption{{\bf Sensitivity analysis of the free parameters of DDM as a function of RTs}. Each curve corresponds to a different age range; rows correspond to misleading (bottom) and reliable (top) content. We average among headlines ($12$ per row).}
            \label{RTs-analysis}
        \end{figure}
        %
        
        \subsection*{Unbiased case of study: theoretical analysis}
        
\begin{figure}[H]
            \renewcommand{\figurename}{Supplementary Figure}
            \centering
            \includegraphics[width=0.6\textwidth]{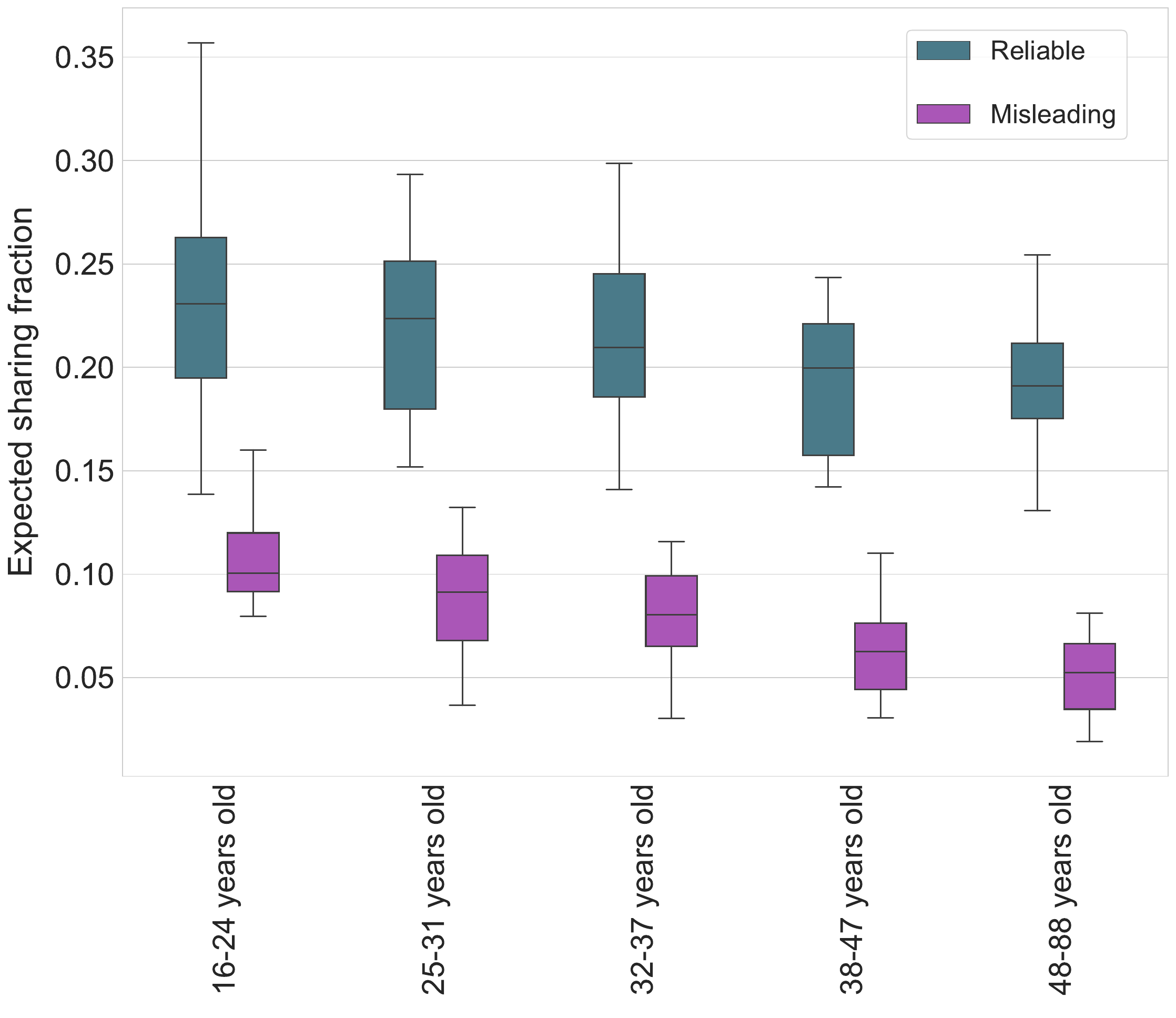} 
            \caption{{\bf Expected fraction of individuals towards sharing}. Each box plot shows the distribution of each age range averaging over $12$ headlines for the case of misleading (blue) and reliable (green) headlines.}
            \label{expected_share_fraction}
        \end{figure}

        In this subsection, we explore the results of spreading (mis)information in a simple unbiased case. Within this context, individuals do not have a priori inclination. First, we obtain the expected fraction of individuals towards sharing information online $C_{SHARE}$ (See Methods) obtained with the estimated parameters (Fig.~\ref{parameters-all}) for each age range and differentiating in the veracity of the content. As we can see from Fig.~\ref{expected_share_fraction}, people because more accurate with age, particularly for misleading content.
        
        \subsection*{Sensitivity analysis of the exponential cutoff on degree}
        
\begin{figure}[H]
            \renewcommand{\figurename}{Supplementary Figure}
            \centering
            \begin{subfigure}[b]{0.45\textwidth}
                 \centering
                 \includegraphics[width=1\textwidth]{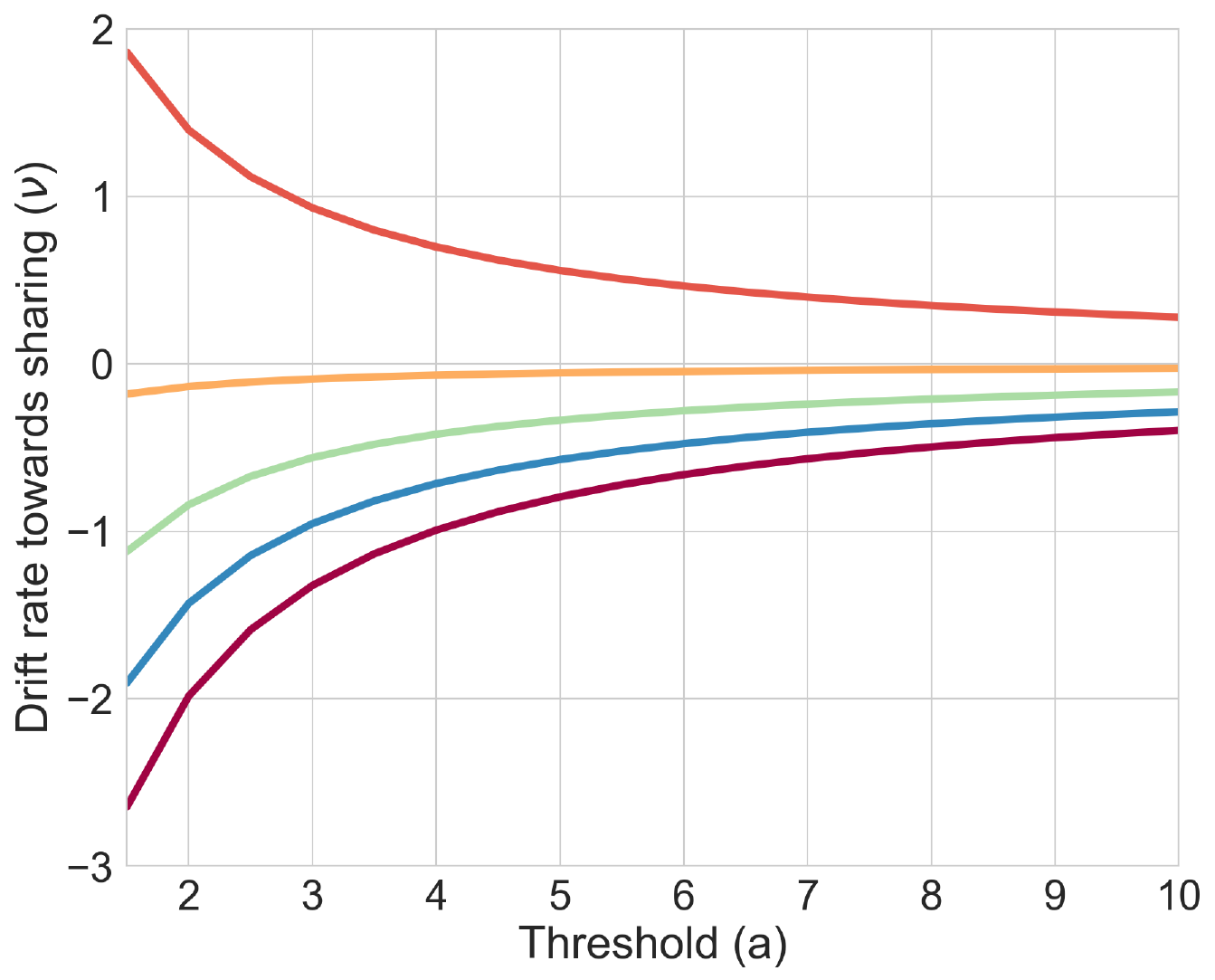}
                 \caption{}
            \end{subfigure}
            \begin{subfigure}[b]{0.45\textwidth}
                 \centering
                 \includegraphics[width=1\textwidth]{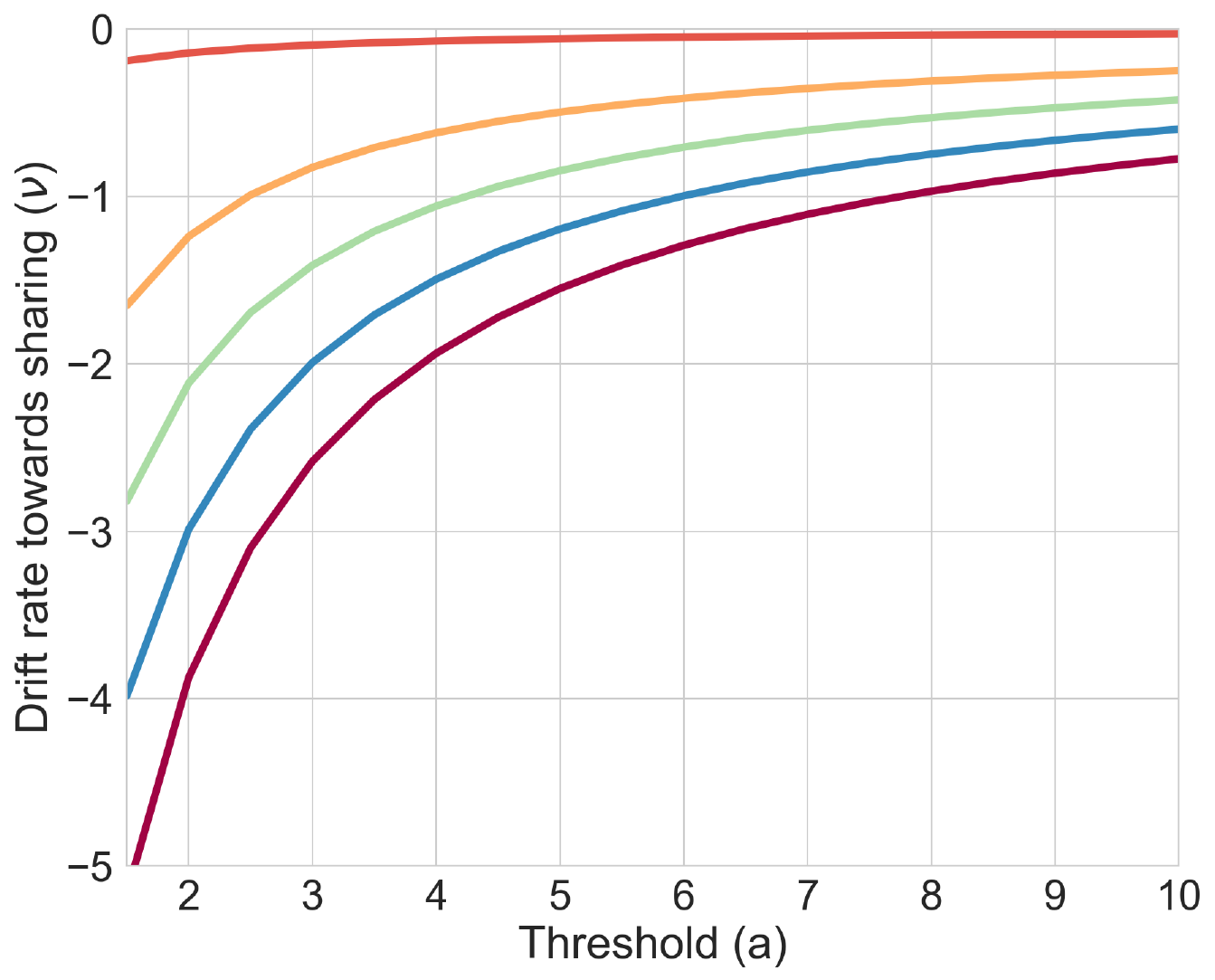}
                 \caption{}
            \end{subfigure}
            \begin{subfigure}[b]{0.45\textwidth}
                 \centering
                 \includegraphics[width=1\textwidth]{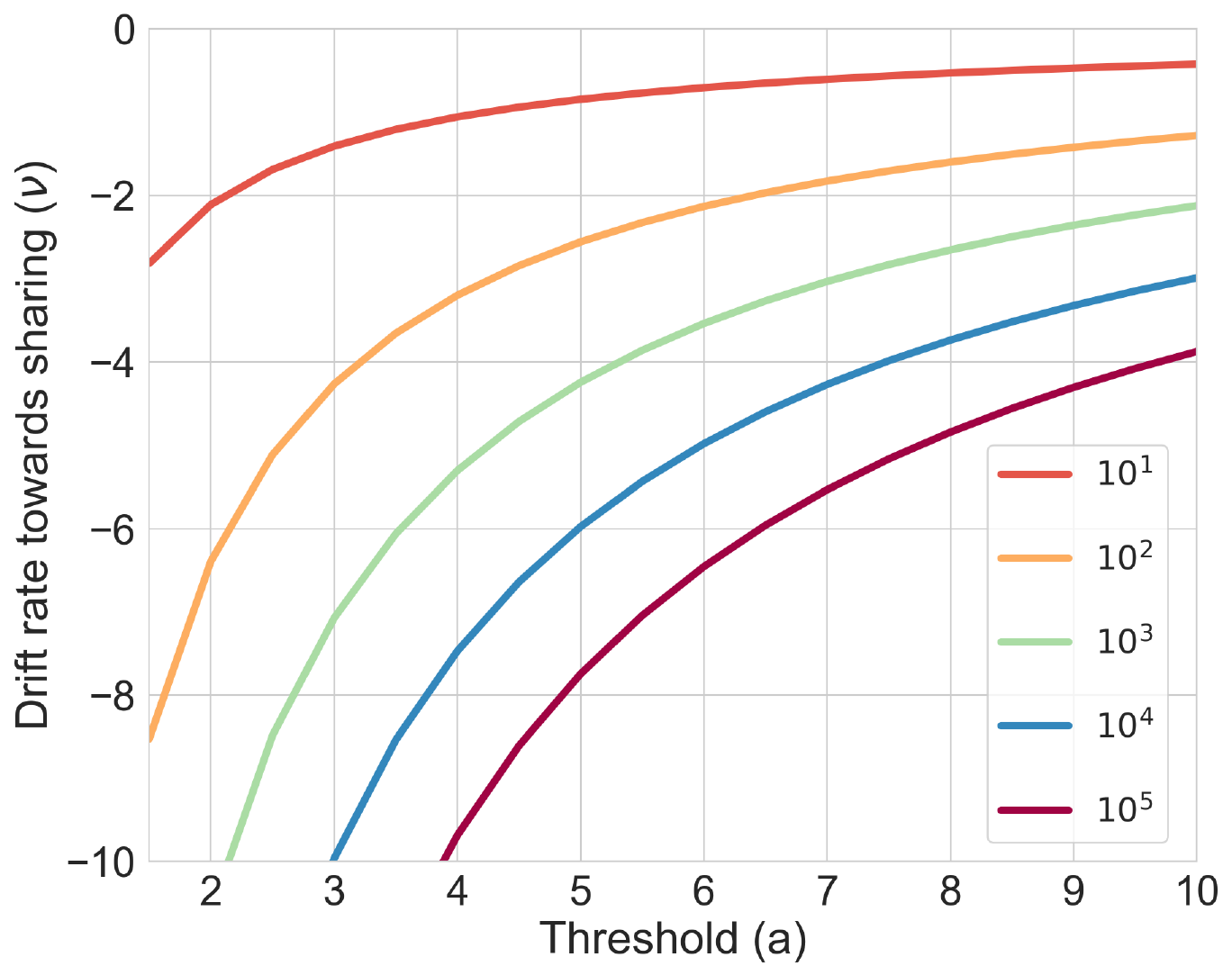}
                 \caption{}
            \end{subfigure}
            \caption{{\bf Sensitivity analysis of the exponential cutoff on the degree}. The network structure corresponds to a power-law with $P(k) \sim k^{-\gamma} \; exp(-k/ \kappa)$ with $\gamma = 2.2$ and different values of cutoff $\kappa$. The plots correspond to the phase diagram for content to go viral according to the free parameters of DDM. On the axis, we have the drift rate ($\nu$) as a function of the threshold ($a$); each curve corresponds to a different value of the degree cutoff while each panel to a different value of the bias (a) $z=0.1$, (b) $z=0.5$ and (c) $z=0.9$.}
            \label{cutoff}
        \end{figure}

        Previous research \cite{newman2002spread,boguna2003absence} has shown the absence of an epidemic threshold when pure power-law degree distributions with $2 < \lambda < 3$ are considered. For our purposes, this is a trivial case of study. Hence, we incorporate an exponential cutoff around the maximum degree $\kappa = 10^5$ obtained in \cite{de2013anatomy}. Additionally, here, we do a sensitivity analysis of the cutoff. As we can see in Fig.~\ref{cutoff}, low values of cutoff have a bigger impact on the free-parameters critical values, and as the cutoff increases, a steady behavior remains. This is due to the fact that the greater the cutoff, the closer the distribution is to the pure power-law behavior.
        As a final step, we explore the case of Dunbar's number $\kappa=150$, suggested in \cite{gonccalves2011modeling} by analyzing Twitter data. We display the phase diagram in Supplementary Fig.~\ref{theory-dunbar}, where the topology corresponds to Power Law network $P(k) \sim k^{-\gamma}  \; exp(-k/\kappa)$ with $\gamma = 2.2$, cutoff on the degree of $\kappa=10^5$ (solid lines) and $\kappa=150$ (dashed lines).
        \begin{figure}[H]
            \renewcommand{\figurename}{Supplementary Figure}
            \centering
            \includegraphics[width=1\textwidth]{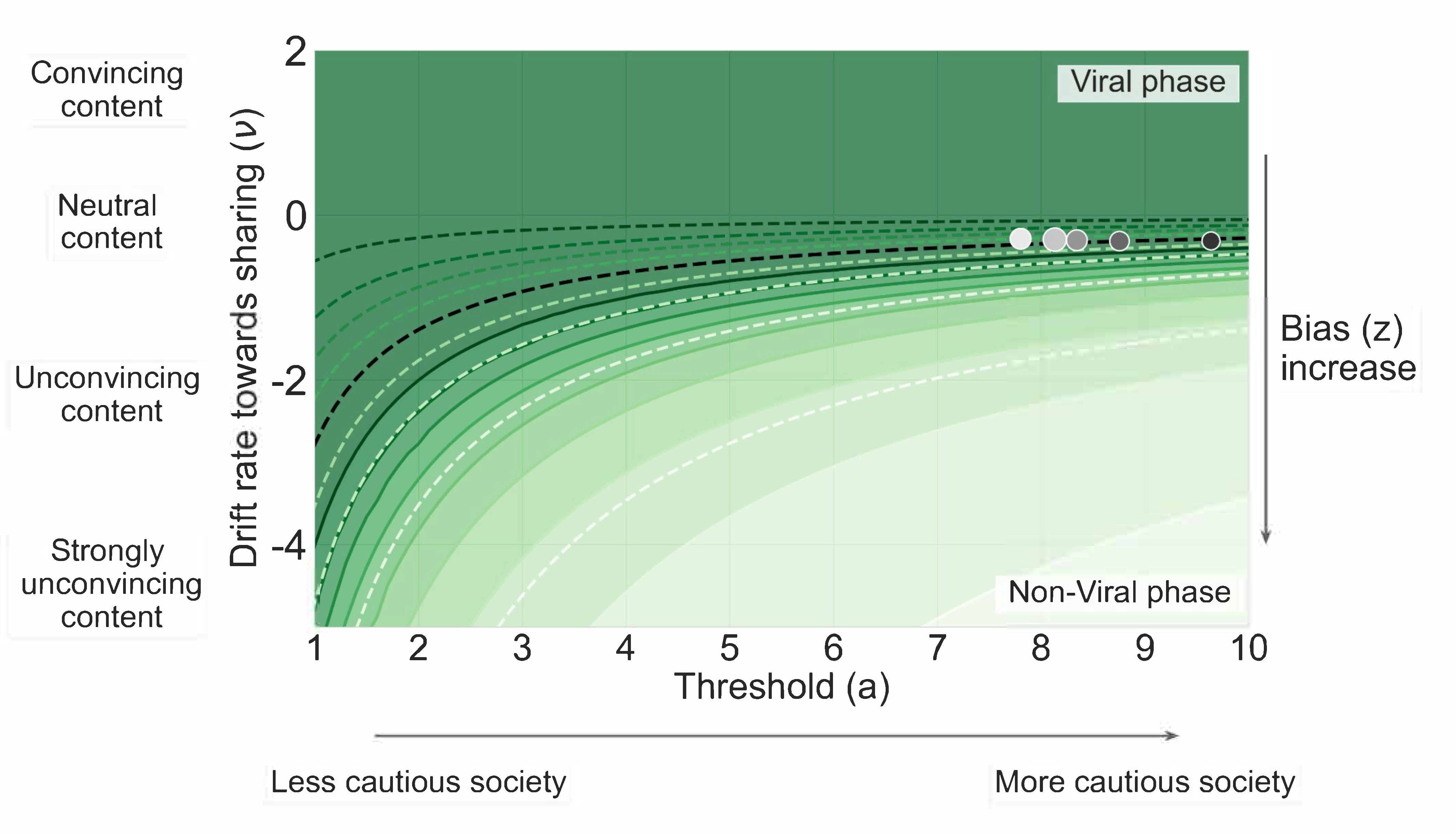}
            \caption{{\bf Phase diagram for content to go viral according to the free parameters of DDM}. The drift rate ($\nu$) as a function of the threshold ($a$), while each curve corresponds to a different value of bias $z$ increasing from $0.1$ (top) to $z=0.9$ (bottom) with $\Delta z = 0.1$. Black lines correspond to the unbiased scenario with $z=0.5$. In this case, we consider the topology of a Power Law network $P(k) \sim k^{-\gamma}  \; exp(-k/\kappa)$. Solid lines correspond to an exponent $\gamma = 2.2$ and cutoff on the degree $\kappa=10^5$, while dashed lines are the intervention scenario with the same exponent $\gamma = 2.2$ and a reduced value in the number of maximum degree an individual can have, $\kappa=150$ (Dunbar's number). Each dot corresponds to empirical data. For more details, see Fig.~\ref{parameters-all} in the main text.}
            \label{theory-dunbar}
        \end{figure}
        
        \subsection*{Agent-based SIR model}
        
\begin{figure}[H]
            \renewcommand{\figurename}{Supplementary Figure}
            \centering
            \includegraphics[width=0.6\textwidth]{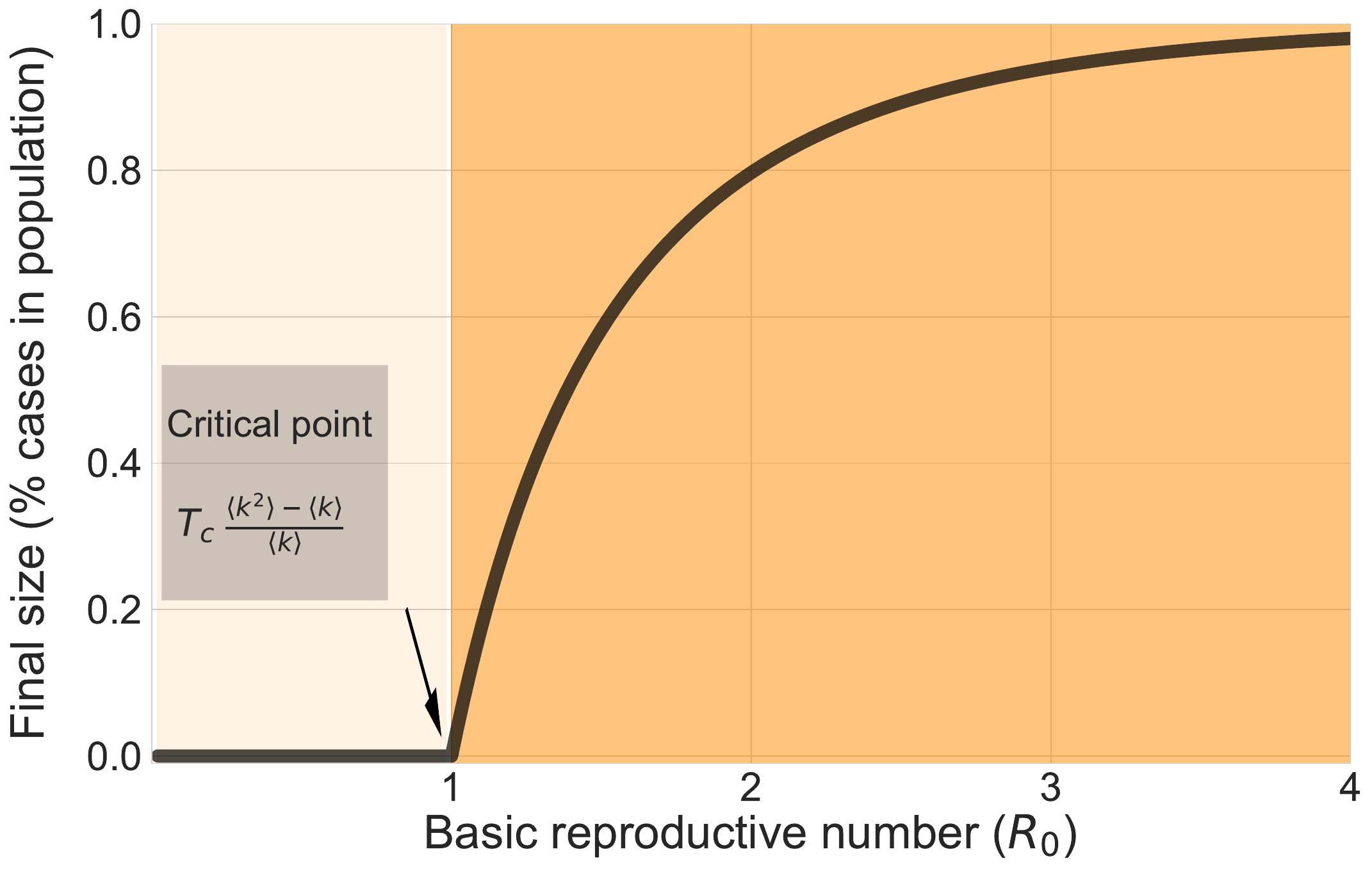}
            \caption{{\bf Final size of recovered individuals as a function of $R_0$}. The network structure corresponds to an Erdős–Rényi with $P(k)=\langle k \rangle^k \; e^{-\langle k \rangle}/ \; k!$ and $\langle k \rangle = 4$, hence $T_c = 0.25$. Results are obtained theoretically from \cite{meyers2005network}}
            \label{standard-SIR}
        \end{figure}

        In this subsection, we display the final size of the population reached by the diseases as a function of $R_0$ in an epidemiological context; if we consider a misinformation context, it is the percentage of individuals that shared the content. The basic reproductive number $R_0$ is the expected number of cases originated by the first case (zero patient) in a population where all individuals are susceptible. Notice this assumes that no other individuals are infected or immunized. As we can see, for $R_0 < 1$, the diseases vanishes. However, for $R_0>1$, a macroscopic size of the population is reached by the diseases causing an epidemic scenario. From previous work \cite{newman2002spread,meyers2005network}, we know the mathematical relationship in the critical point $R_0=1$ (Eq. within Fig.~\ref{standard-SIR}) between the contact patterns of a society (right side) and the characteristics of a disease (left side), $i.e.$, we can determine if a specific emerging infectious disease will become an epidemic in a specific population.
        
        \subsection*{Sensitivity analysis desegregating by headlines}
        
        Here, we display the probability distribution of response times for the empirical, theoretical, and simulated data. Each set of plots corresponds to a different headline and each plot to a different range of age.
        
        \begin{figure}[H]
            \renewcommand{\figurename}{Supplementary Figure}
            \centering
            \includegraphics[width=.45\textwidth]{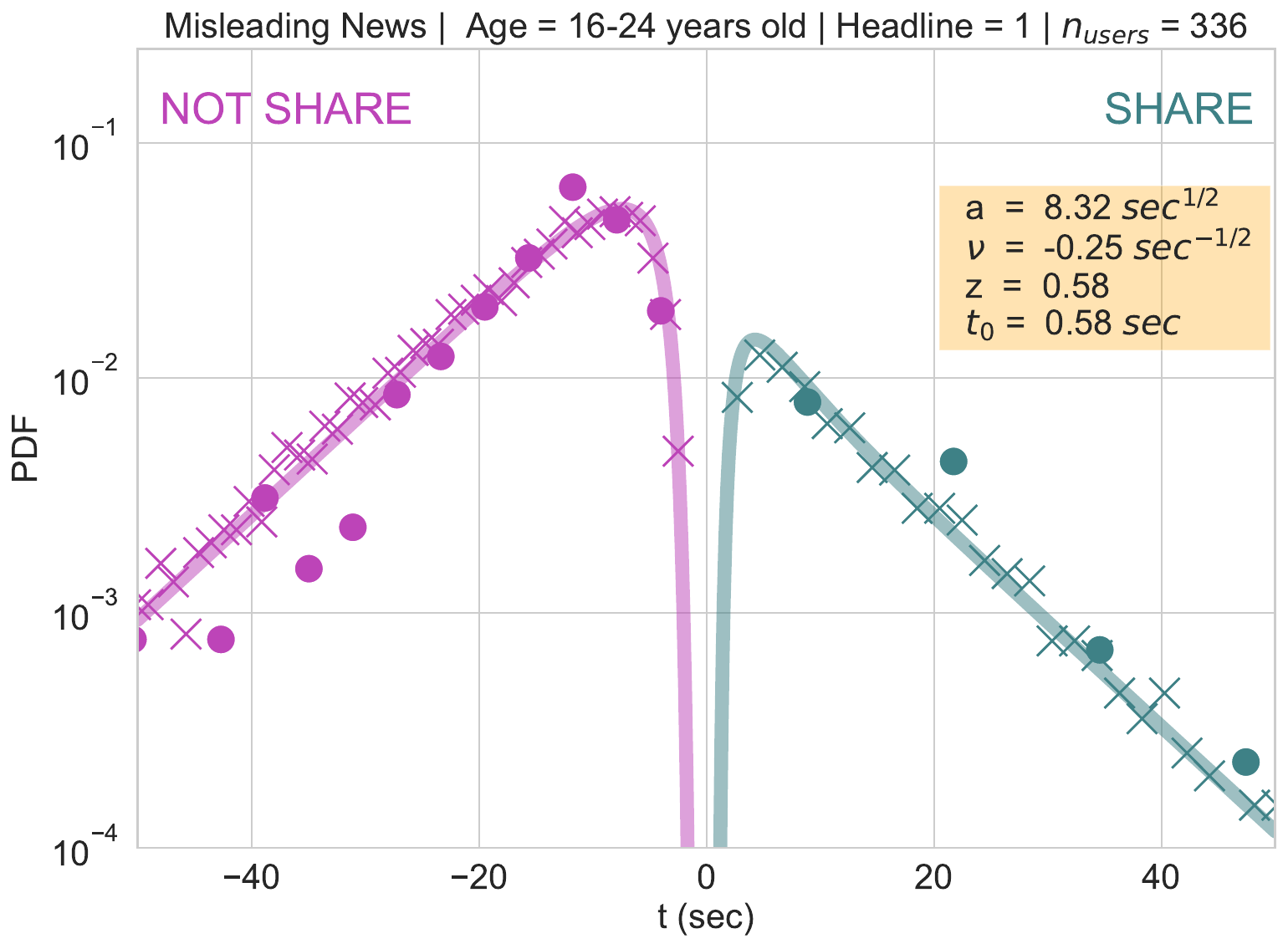}\quad
            \includegraphics[width=.45\textwidth]{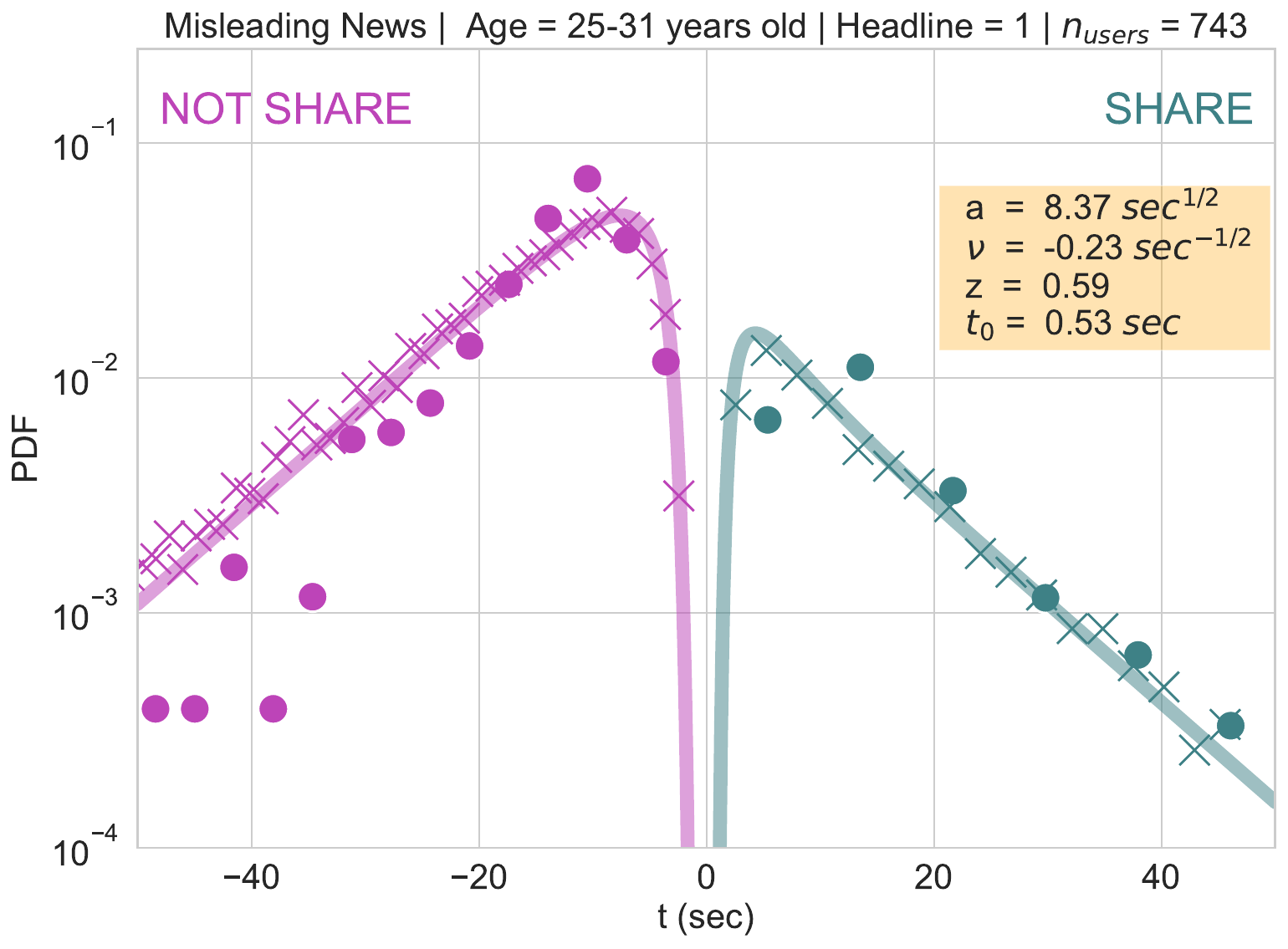}\quad
            \includegraphics[width=.45\textwidth]{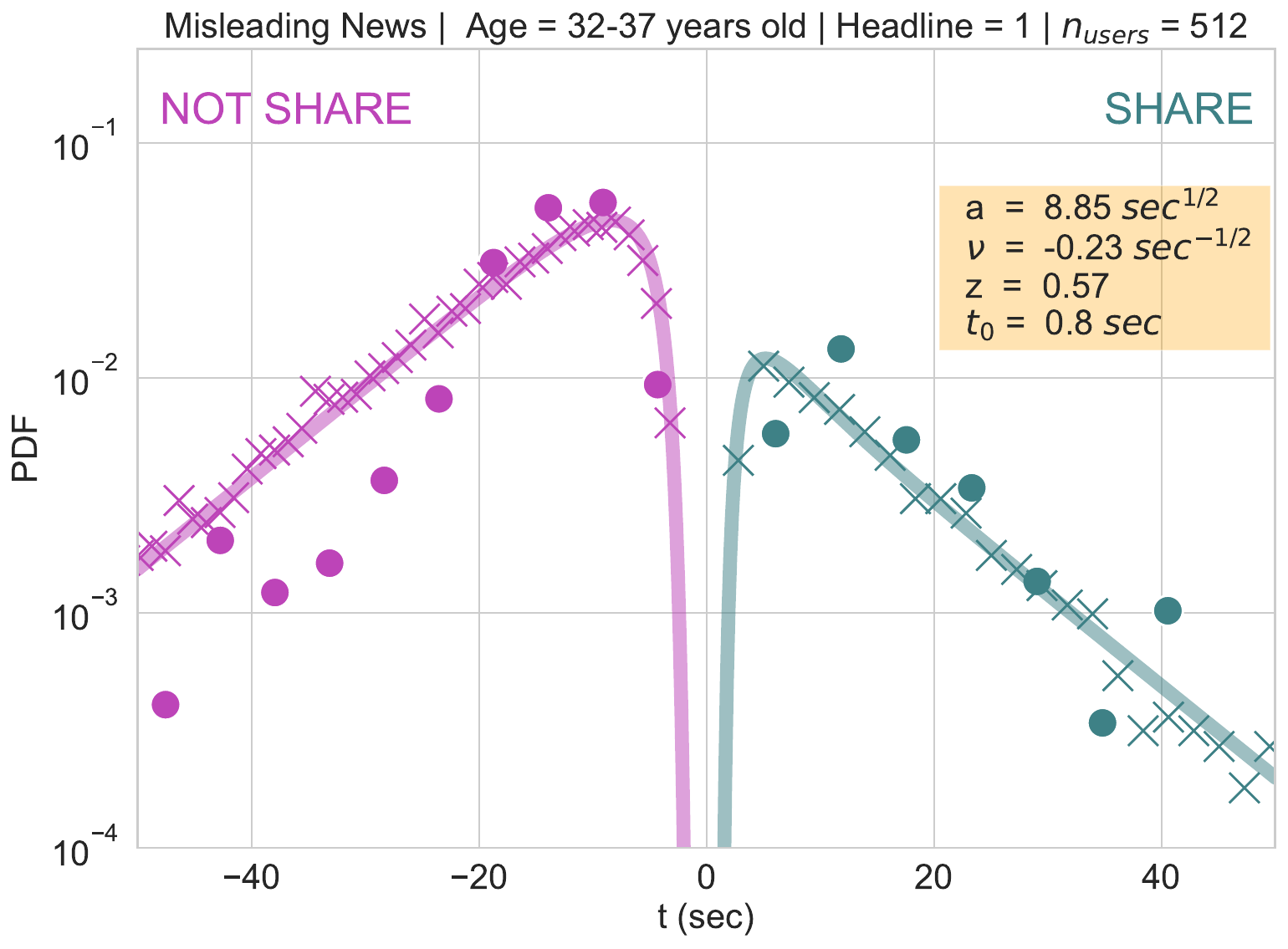}
            \medskip
            \includegraphics[width=.45\textwidth]{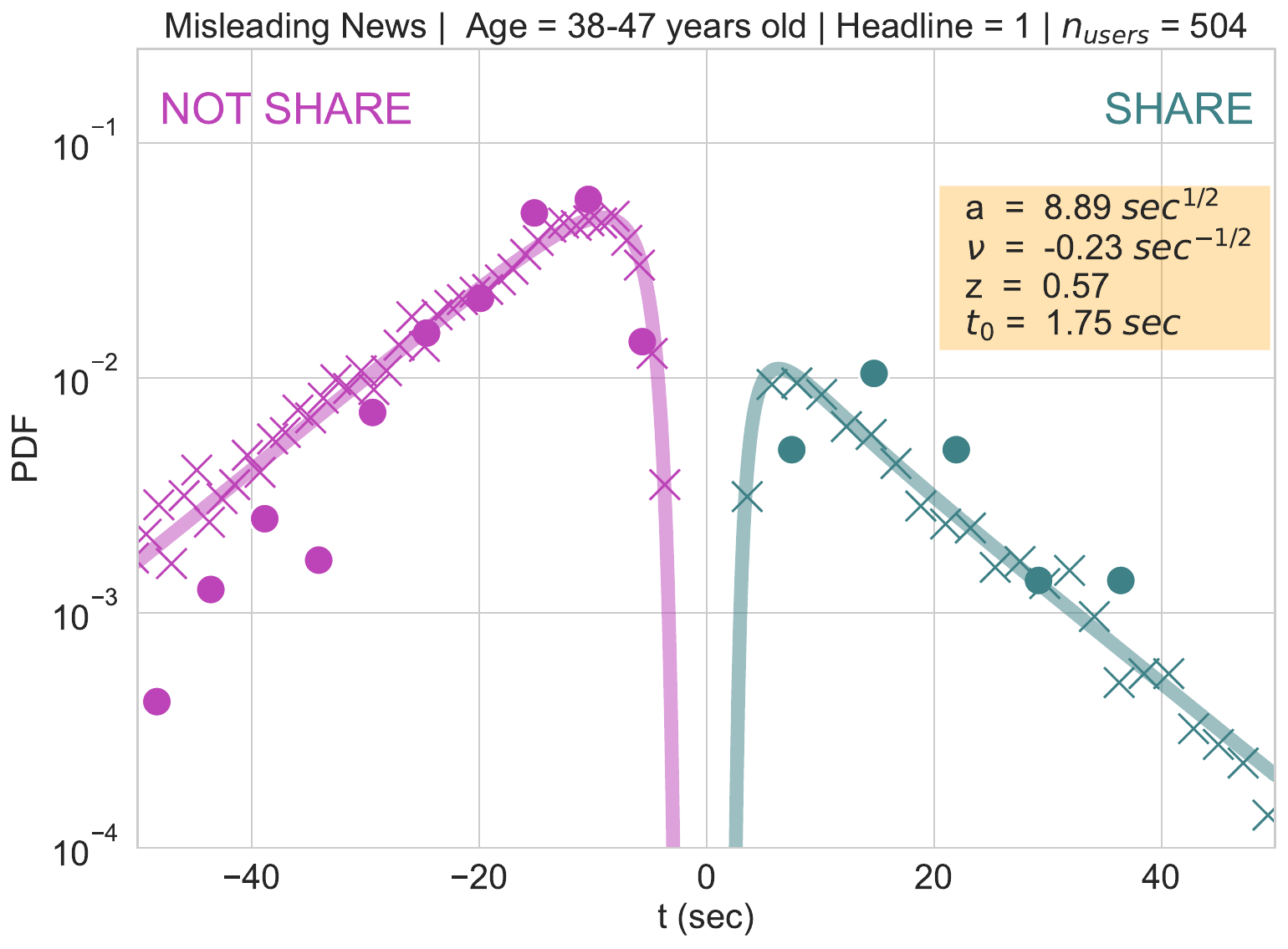}\quad
            \includegraphics[width=.45\textwidth]{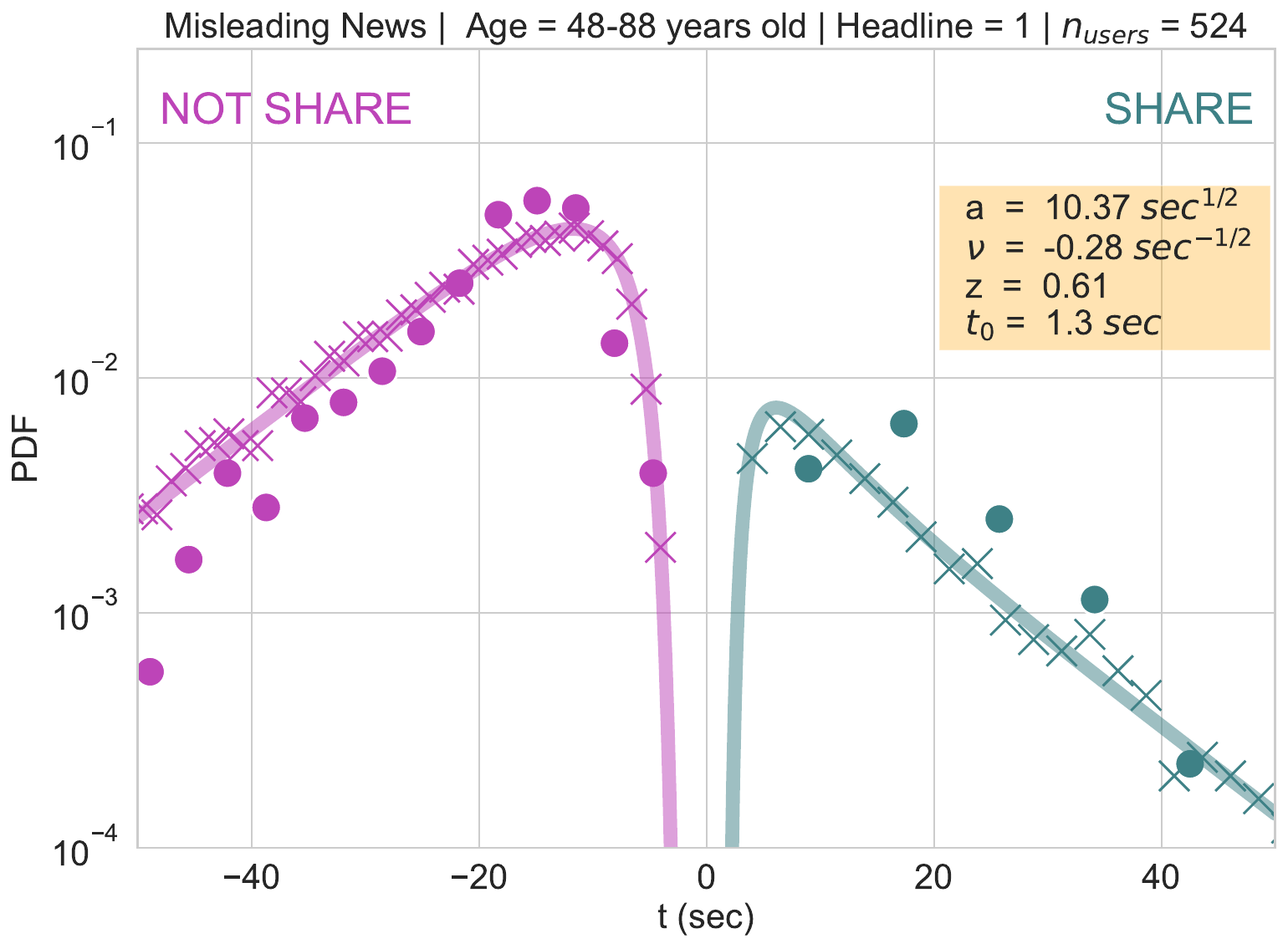}
            \caption{{\bf Headline 1}: Probability distribution of the response time for sharing and not sharing misleading information. Each figure corresponds to different age ranges. The solid line corresponds to theoretical results, dots correspond to empirical data and crosses to stochastic simulations.}
            \label{headline1Fake}
        \end{figure}
        
        \begin{figure}[H]
            \renewcommand{\figurename}{Supplementary Figure}
            \centering
            \includegraphics[width=.45\textwidth]{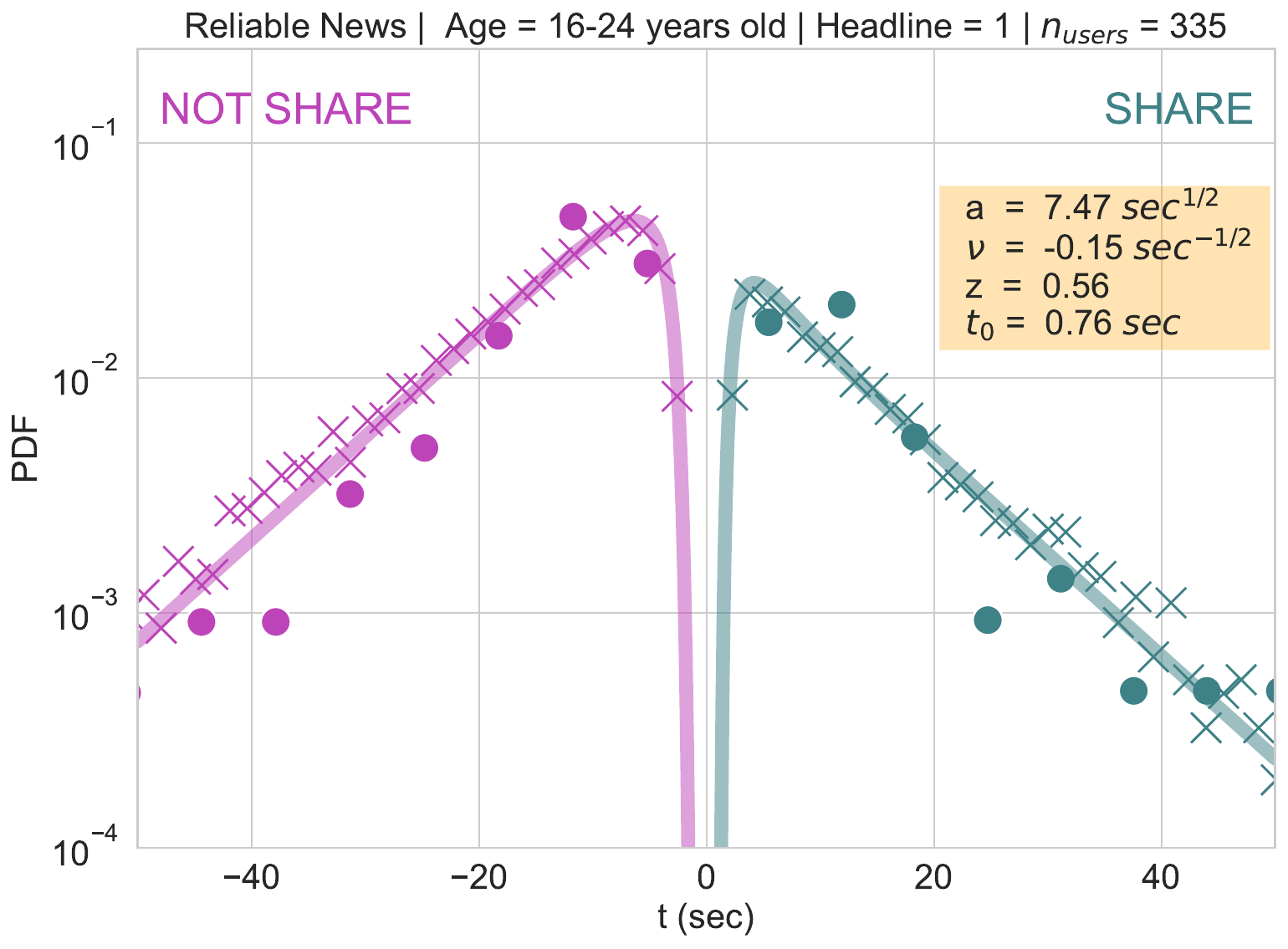}\quad
            \includegraphics[width=.45\textwidth]{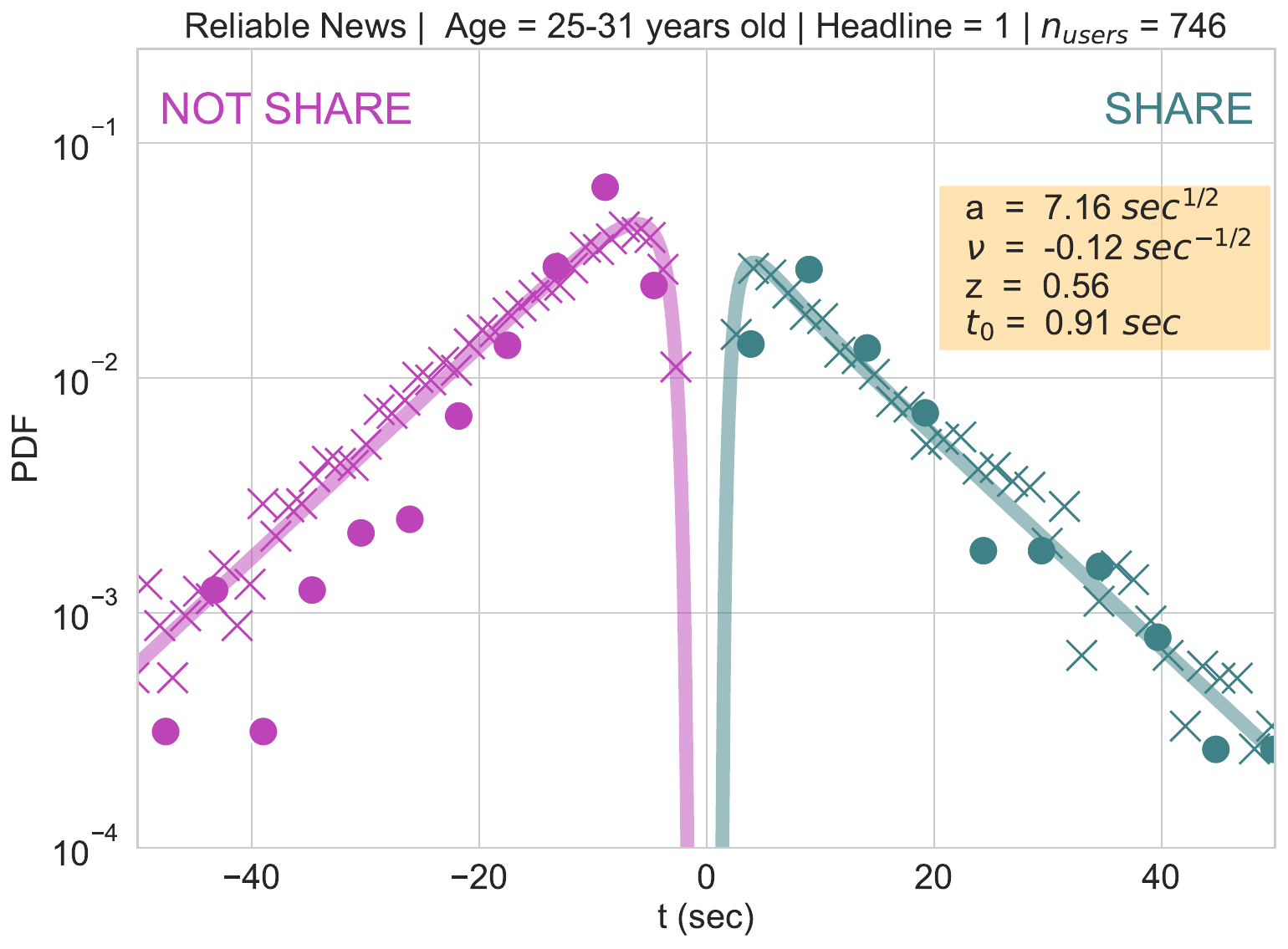}\quad
            \includegraphics[width=.45\textwidth]{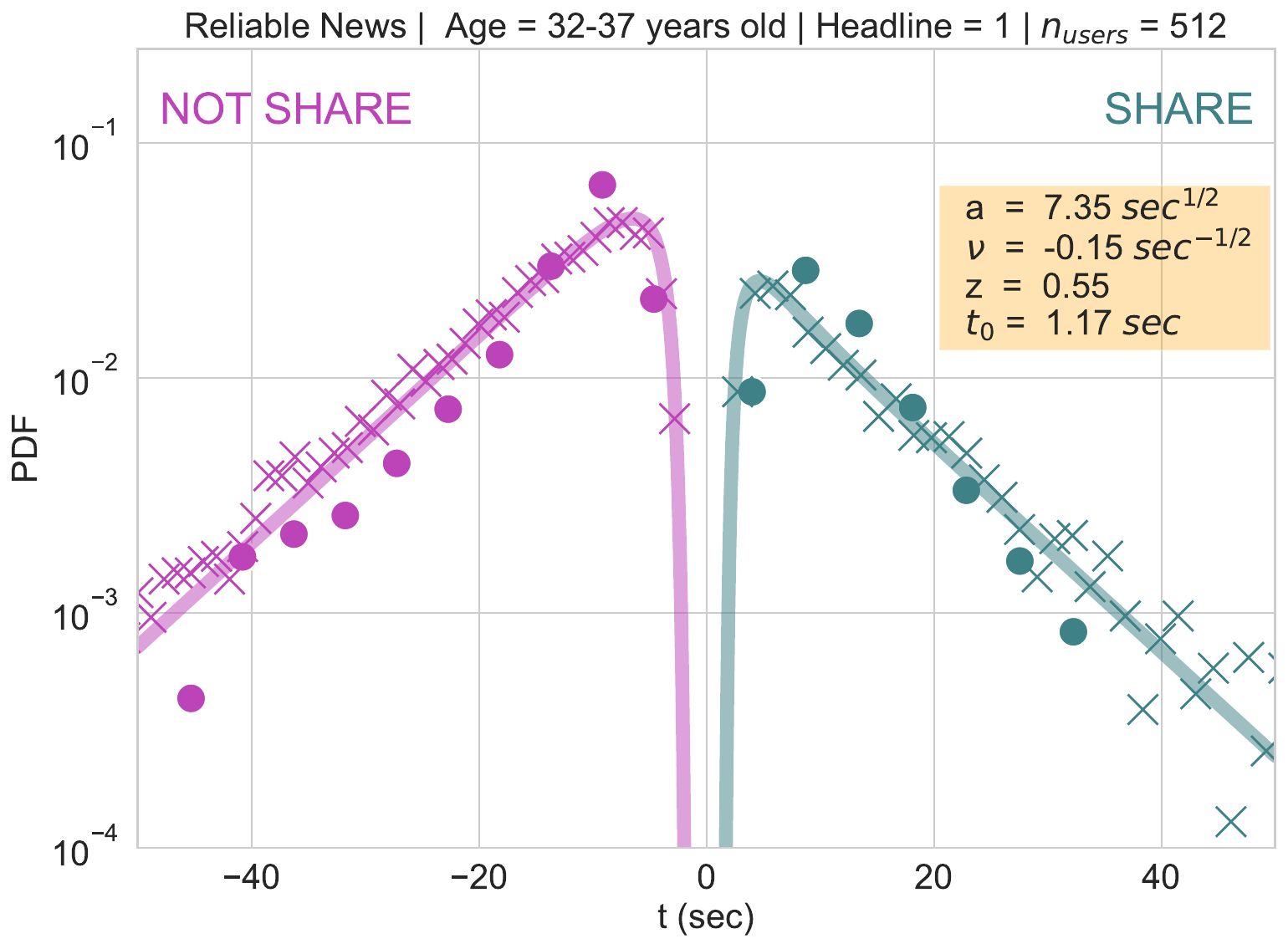}
            \medskip
            \includegraphics[width=.45\textwidth]{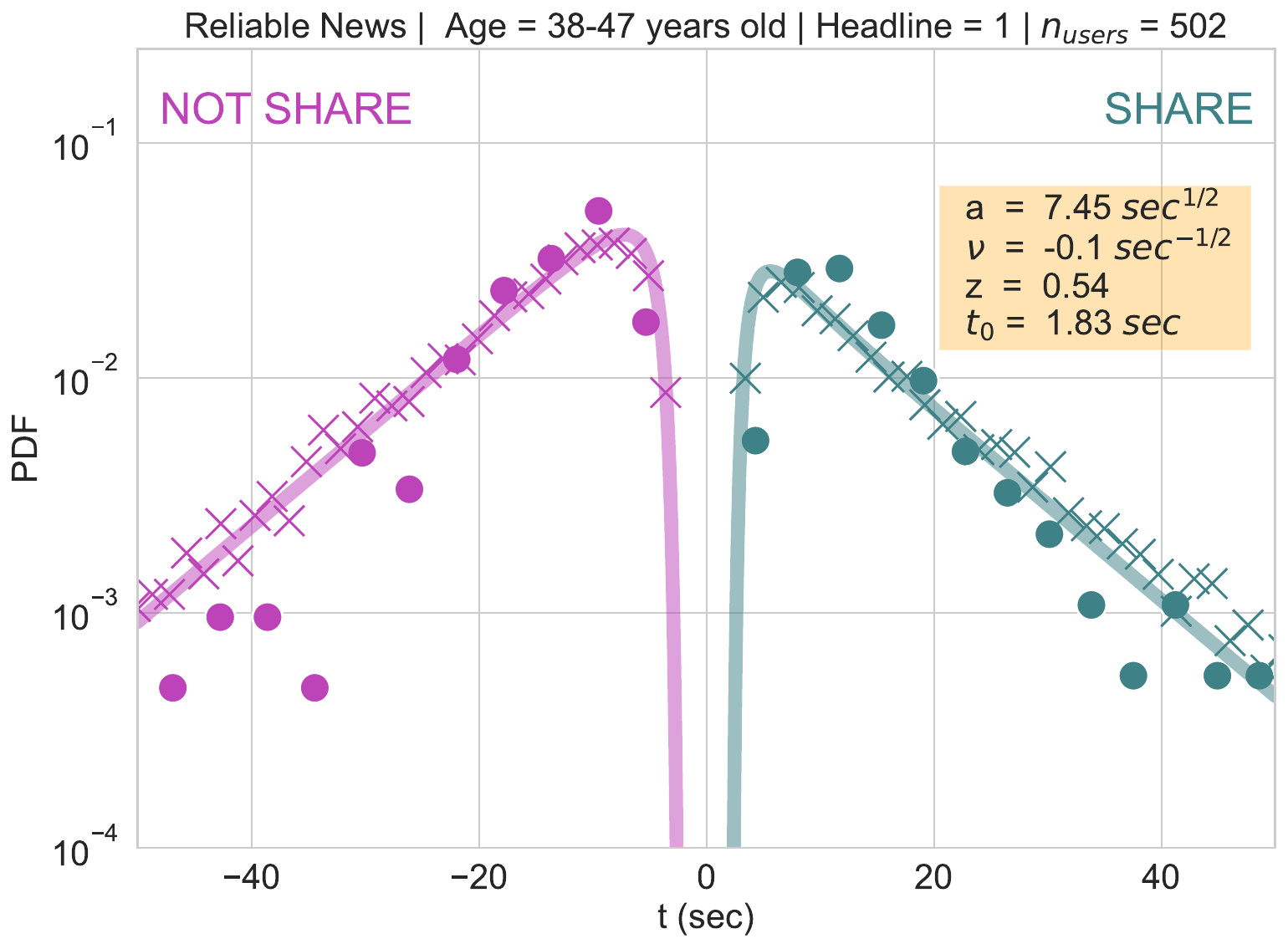}\quad
            \includegraphics[width=.45\textwidth]{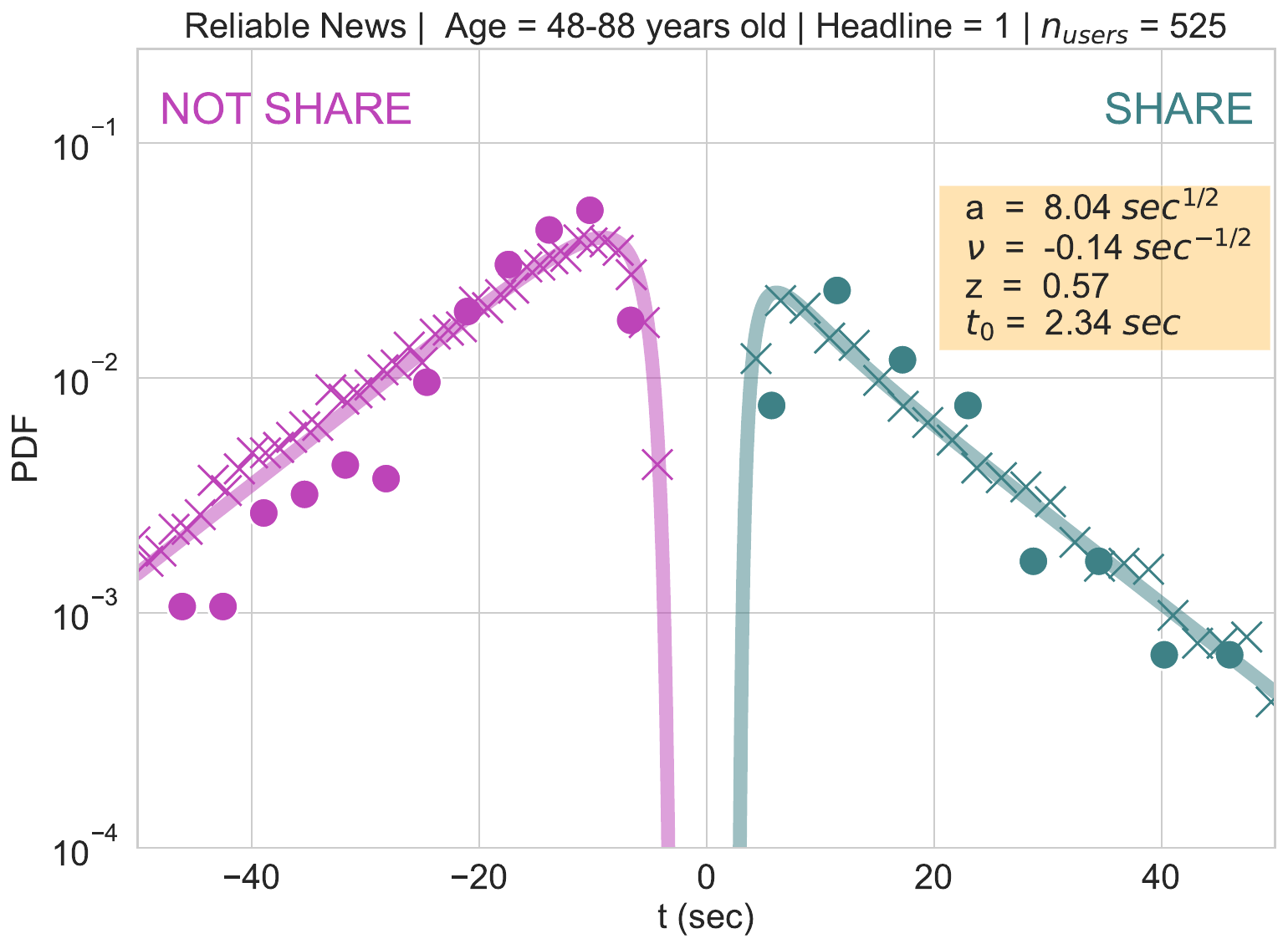}
            \caption{{\bf Headline 1}: Probability distribution of the response time for sharing and not sharing reliable information. Each figure corresponds to different age ranges. The solid line corresponds to theoretical results, dots correspond to empirical data and crosses to stochastic simulations.}
            \label{headline1Real}
        \end{figure}
        
        \begin{figure}[H]
            \renewcommand{\figurename}{Supplementary Figure}
            \centering
            \includegraphics[width=.45\textwidth]{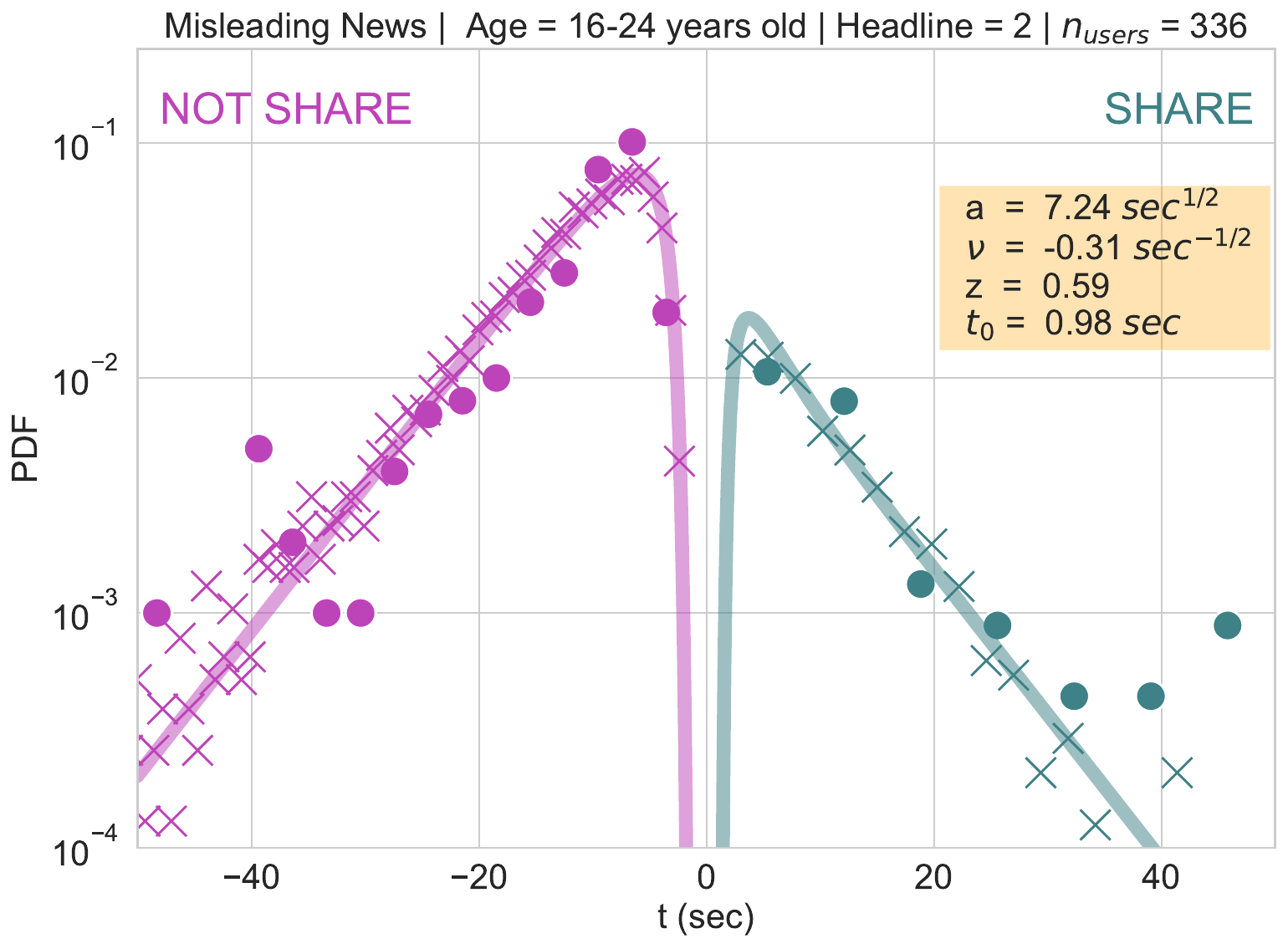}\quad
            \includegraphics[width=.45\textwidth]{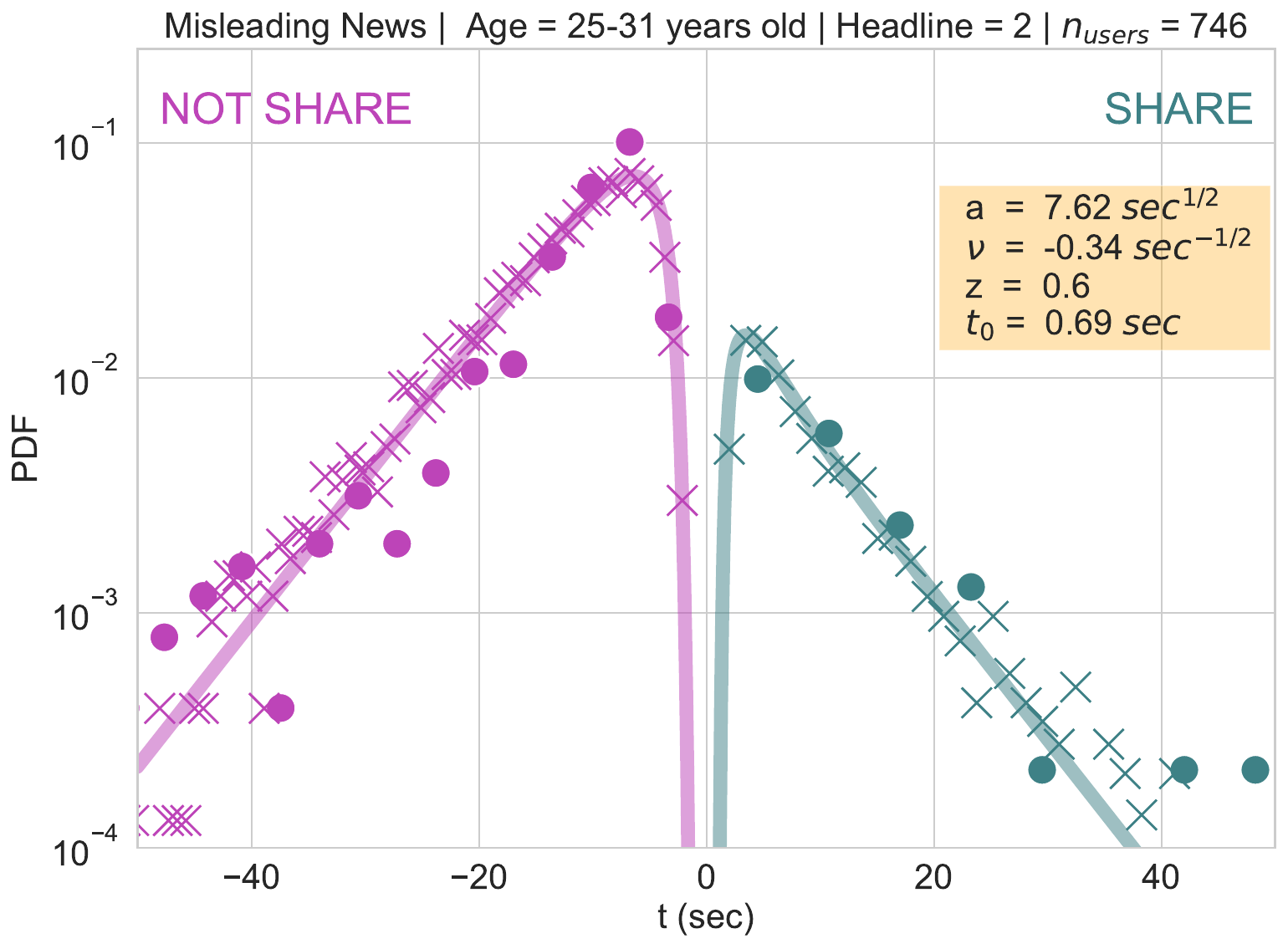}\quad
            \includegraphics[width=.45\textwidth]{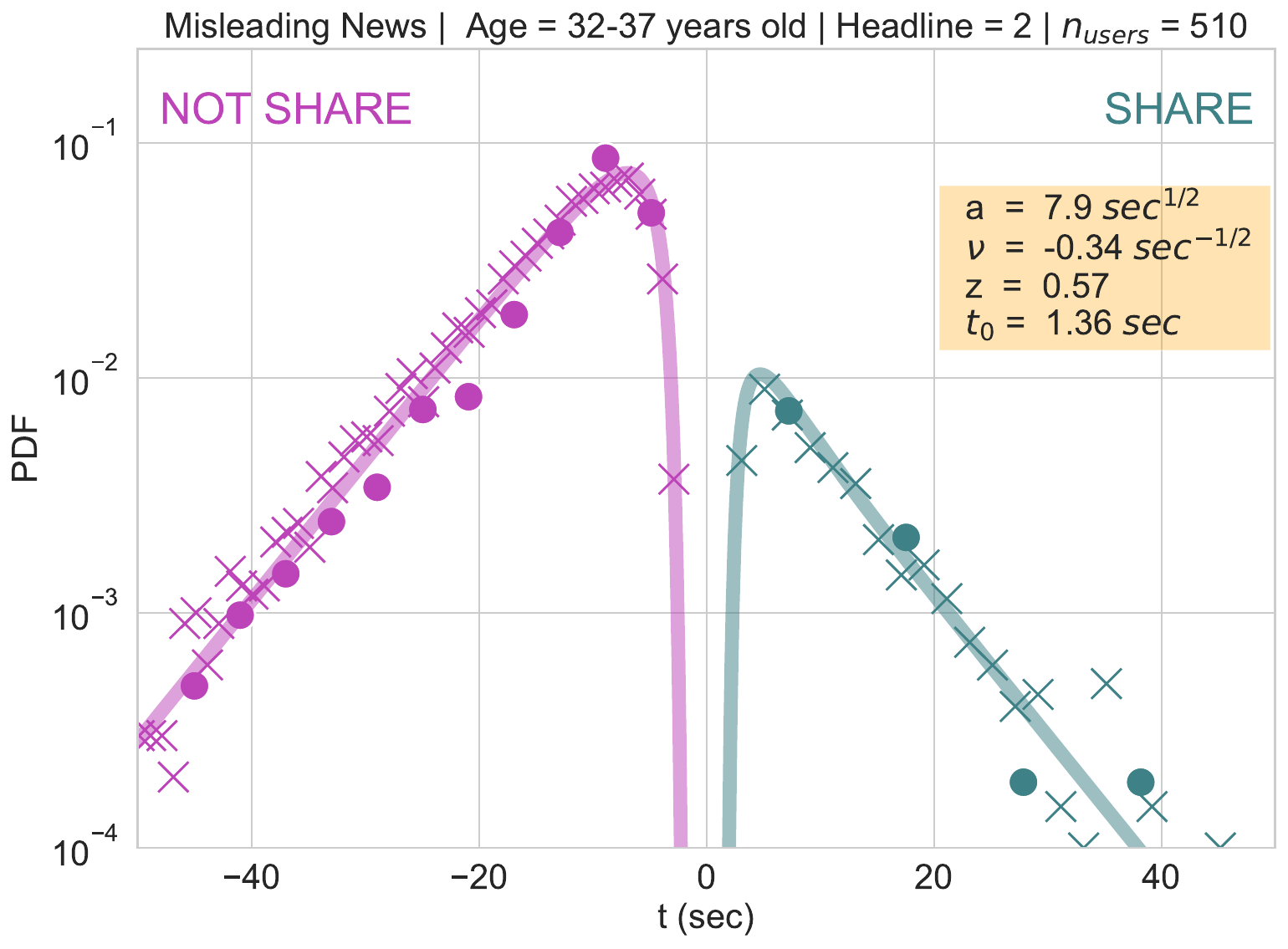}
            \medskip
            \includegraphics[width=.45\textwidth]{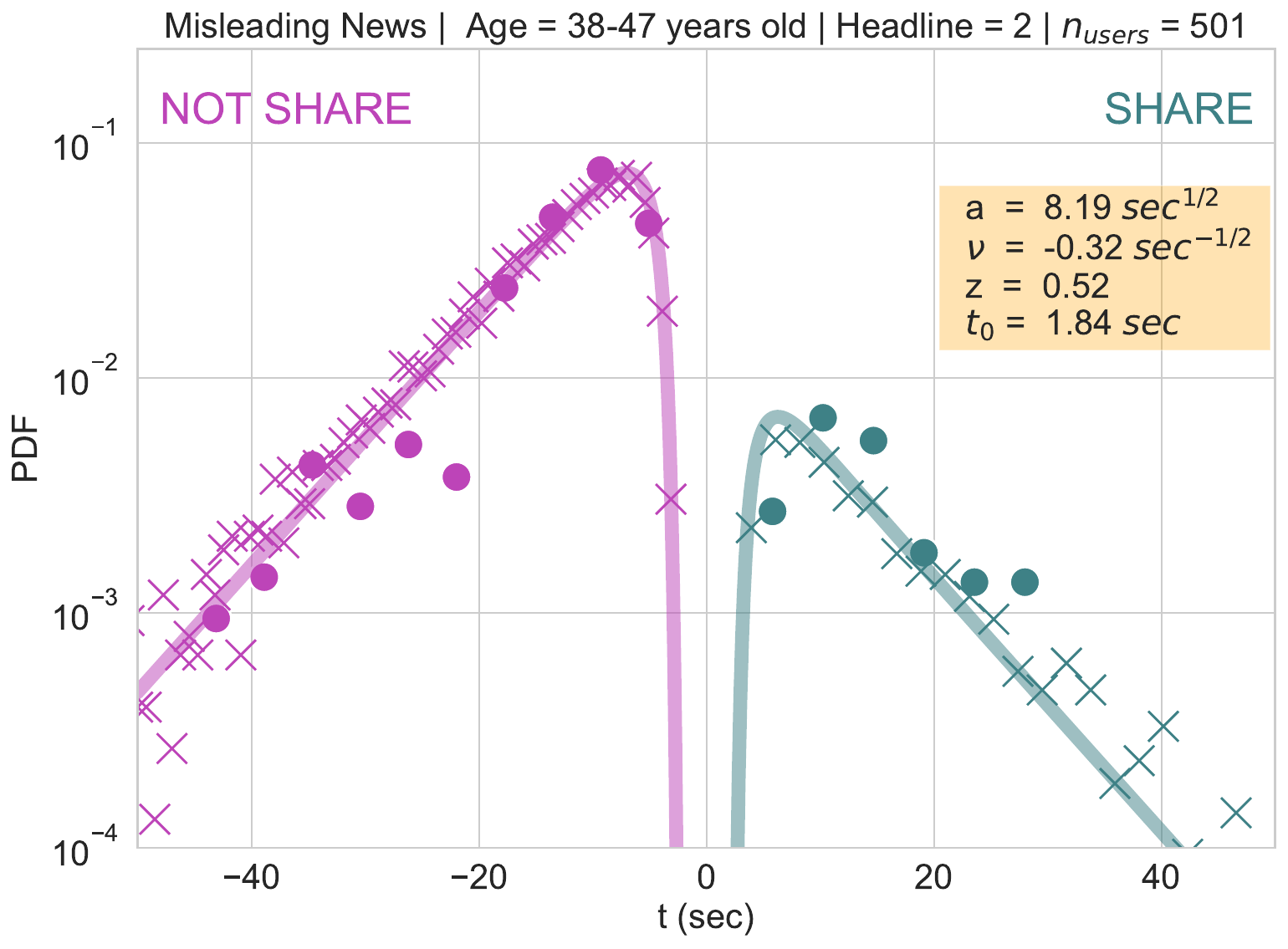}\quad
            \includegraphics[width=.45\textwidth]{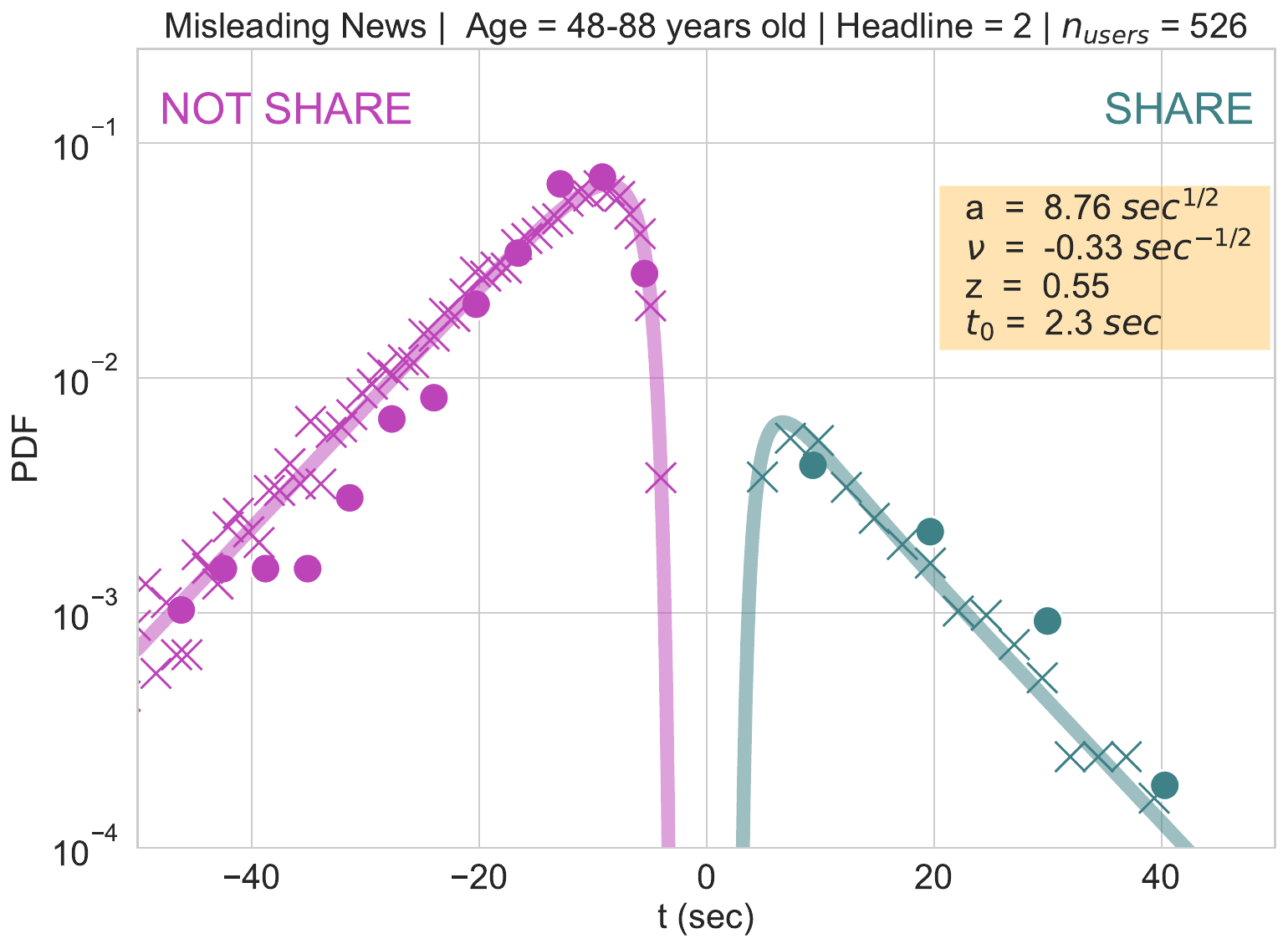}
            \caption{{\bf Headline 2}: Probability distribution of the response time for sharing and not sharing misleading information. Each figure corresponds to different age ranges. The solid line corresponds to theoretical results, dots correspond to empirical data and crosses to stochastic simulations.}
            \label{headline2Fake}
        \end{figure}
        
        \begin{figure}[H]
            \renewcommand{\figurename}{Supplementary Figure}
            \centering
            \includegraphics[width=.45\textwidth]{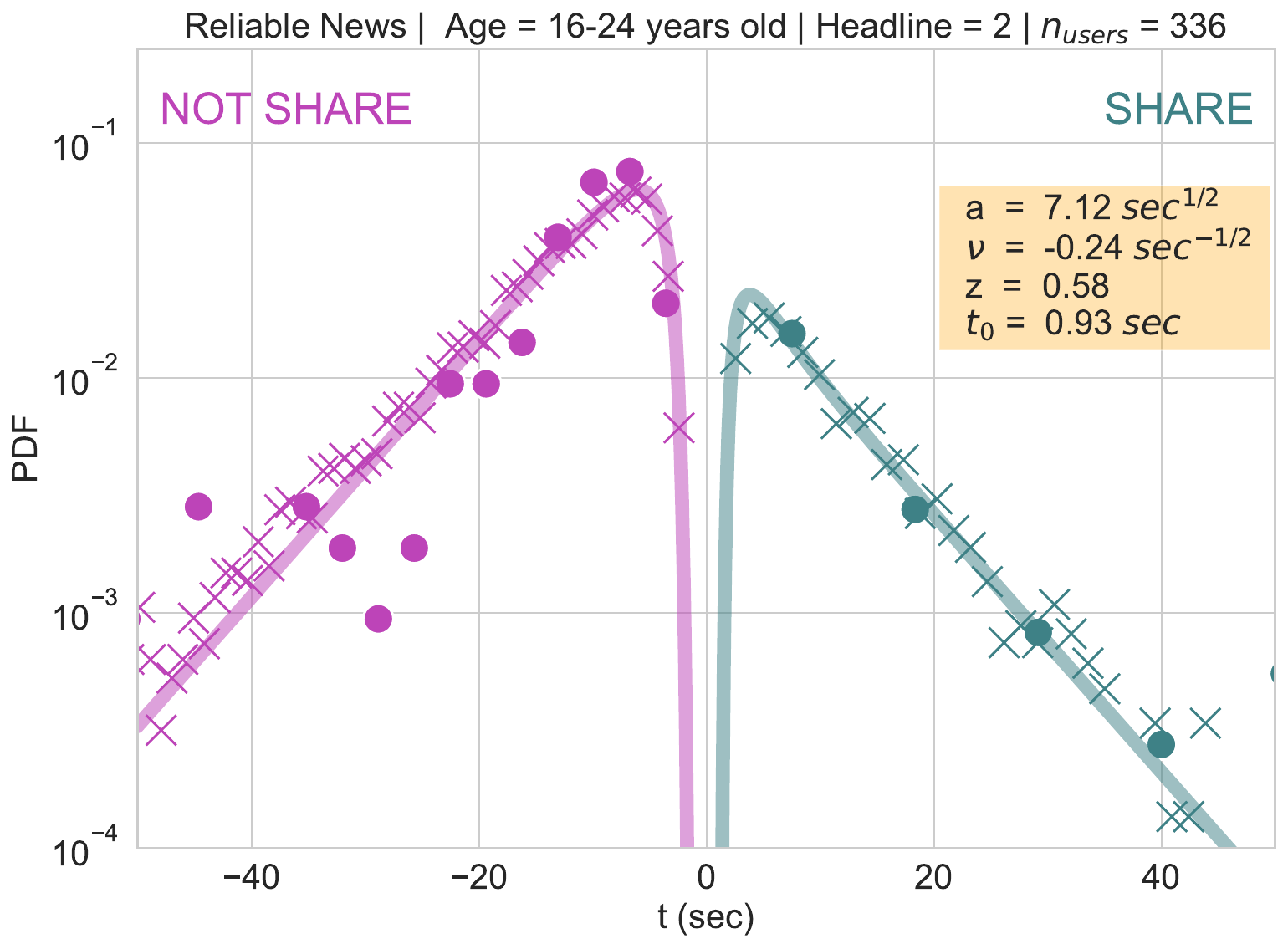}\quad
            \includegraphics[width=.45\textwidth]{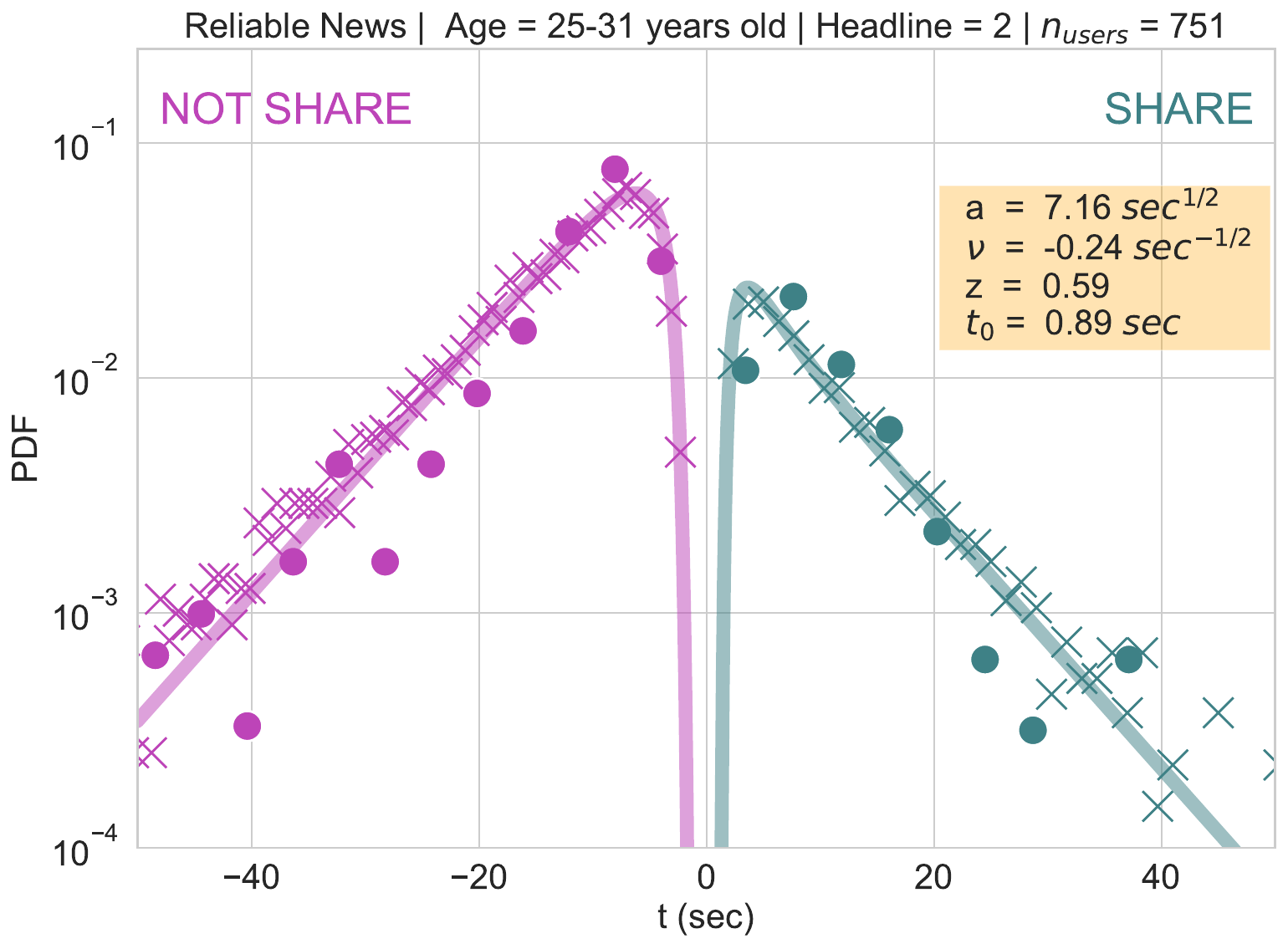}\quad
            \includegraphics[width=.45\textwidth]{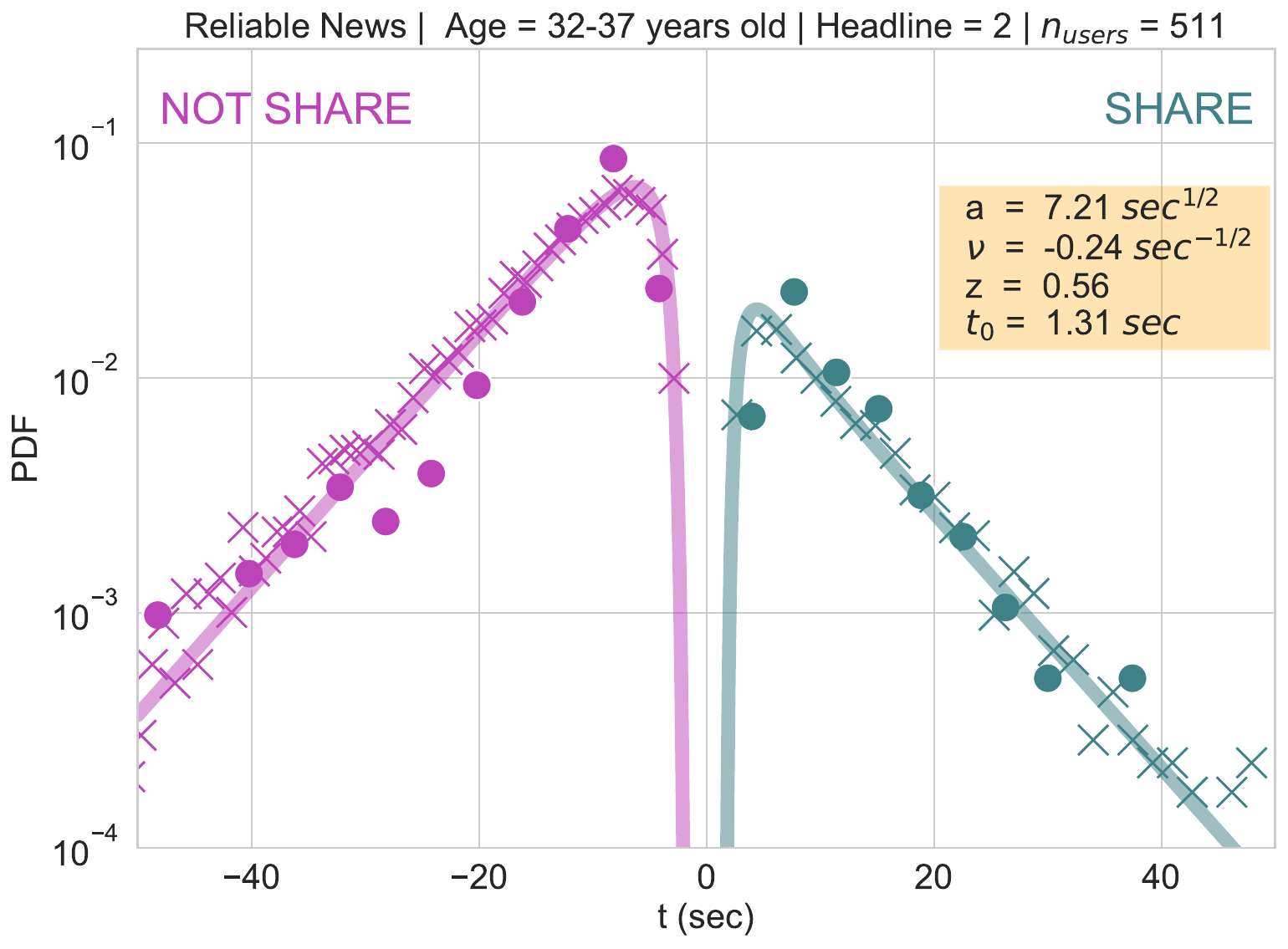}
            \medskip
            \includegraphics[width=.45\textwidth]{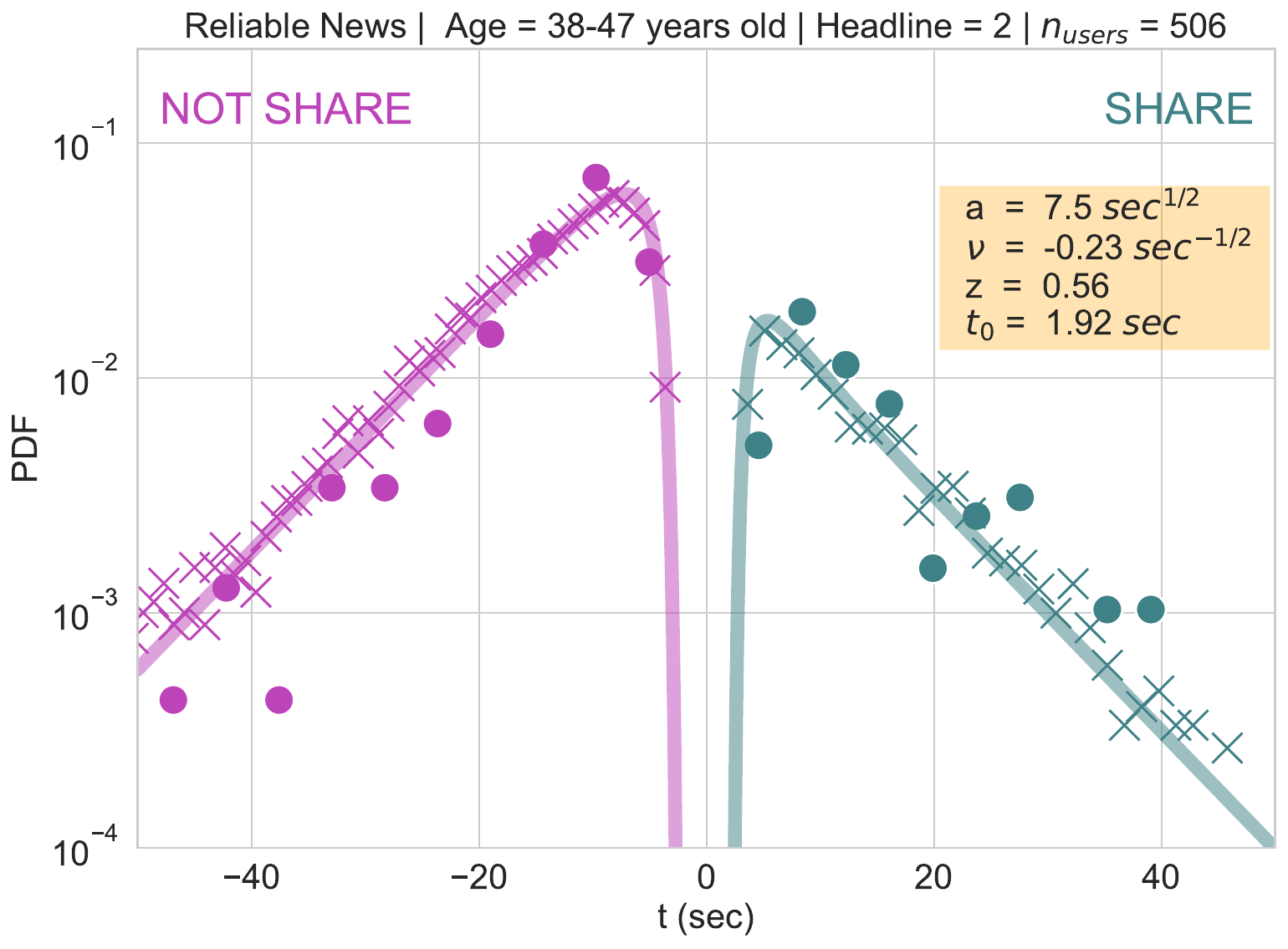}\quad
            \includegraphics[width=.45\textwidth]{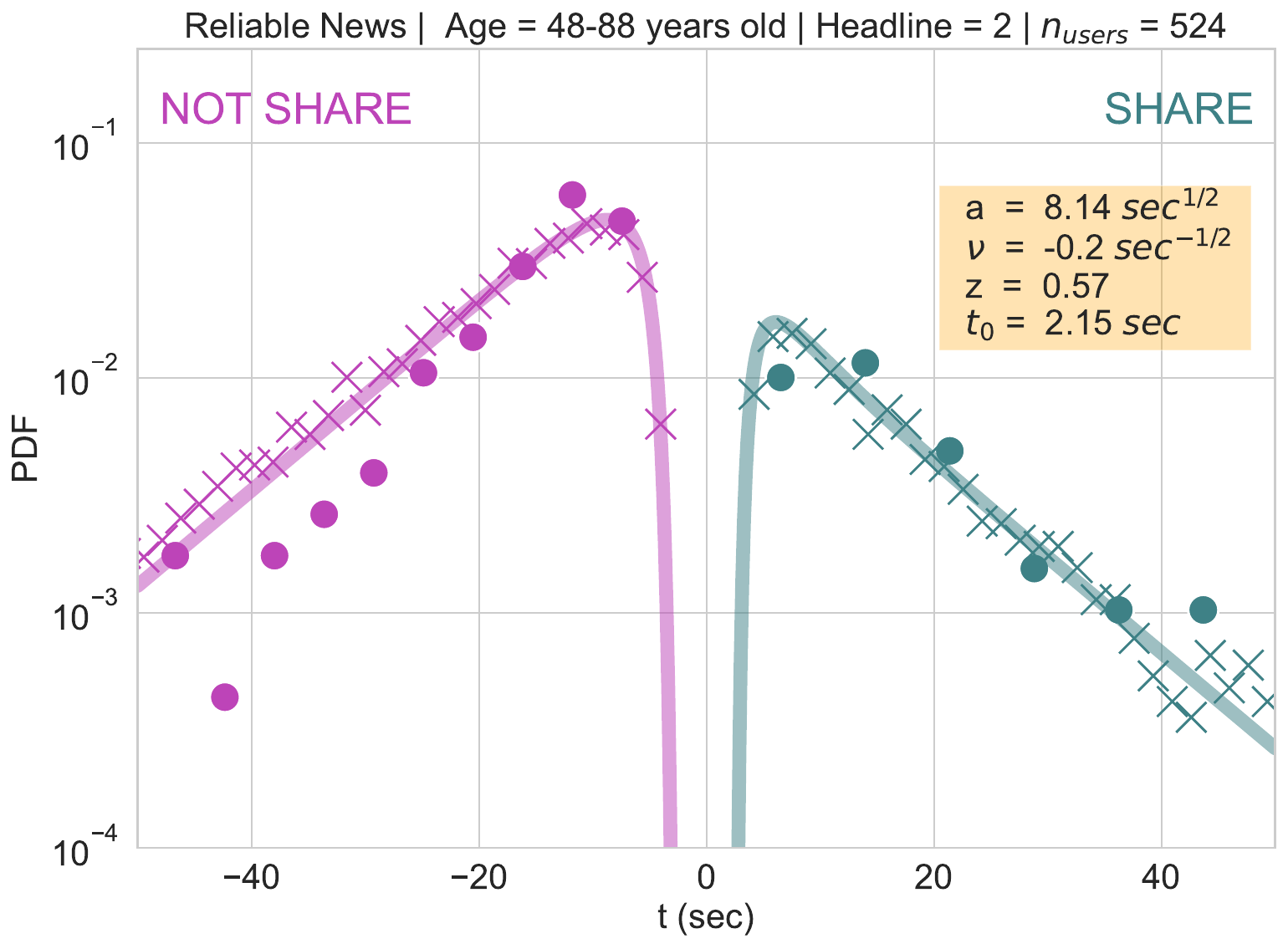}
            \caption{{\bf Headline 2}: Probability distribution of the response time for sharing and not sharing reliable information. Each figure corresponds to different age ranges. The solid line corresponds to theoretical results, dots correspond to empirical data and crosses to stochastic simulations.}
            \label{headline2Real}
        \end{figure}
        
        \begin{figure}[H]
            \renewcommand{\figurename}{Supplementary Figure}
            \centering
            \includegraphics[width=.45\textwidth]{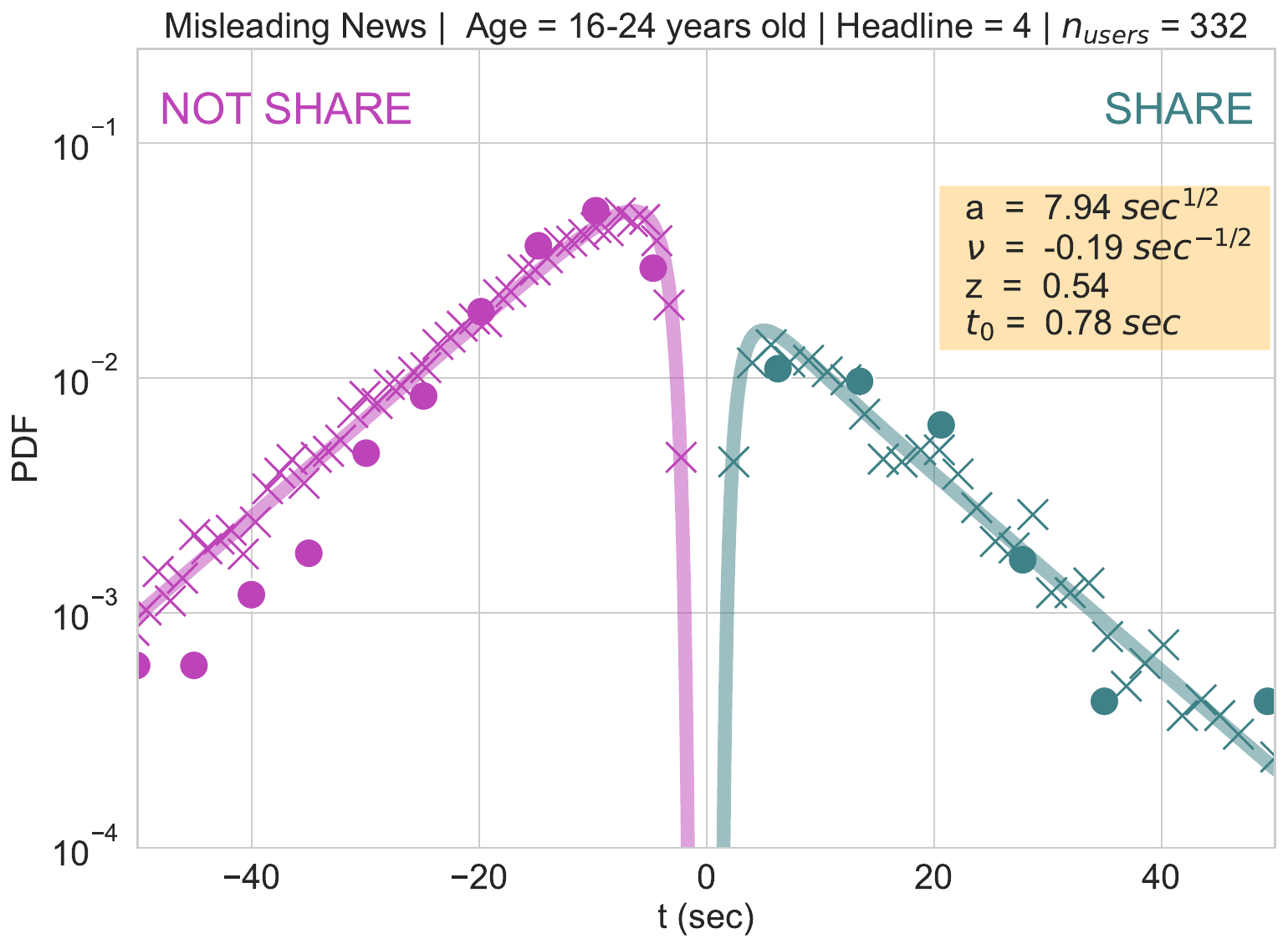}\quad
            \includegraphics[width=.45\textwidth]{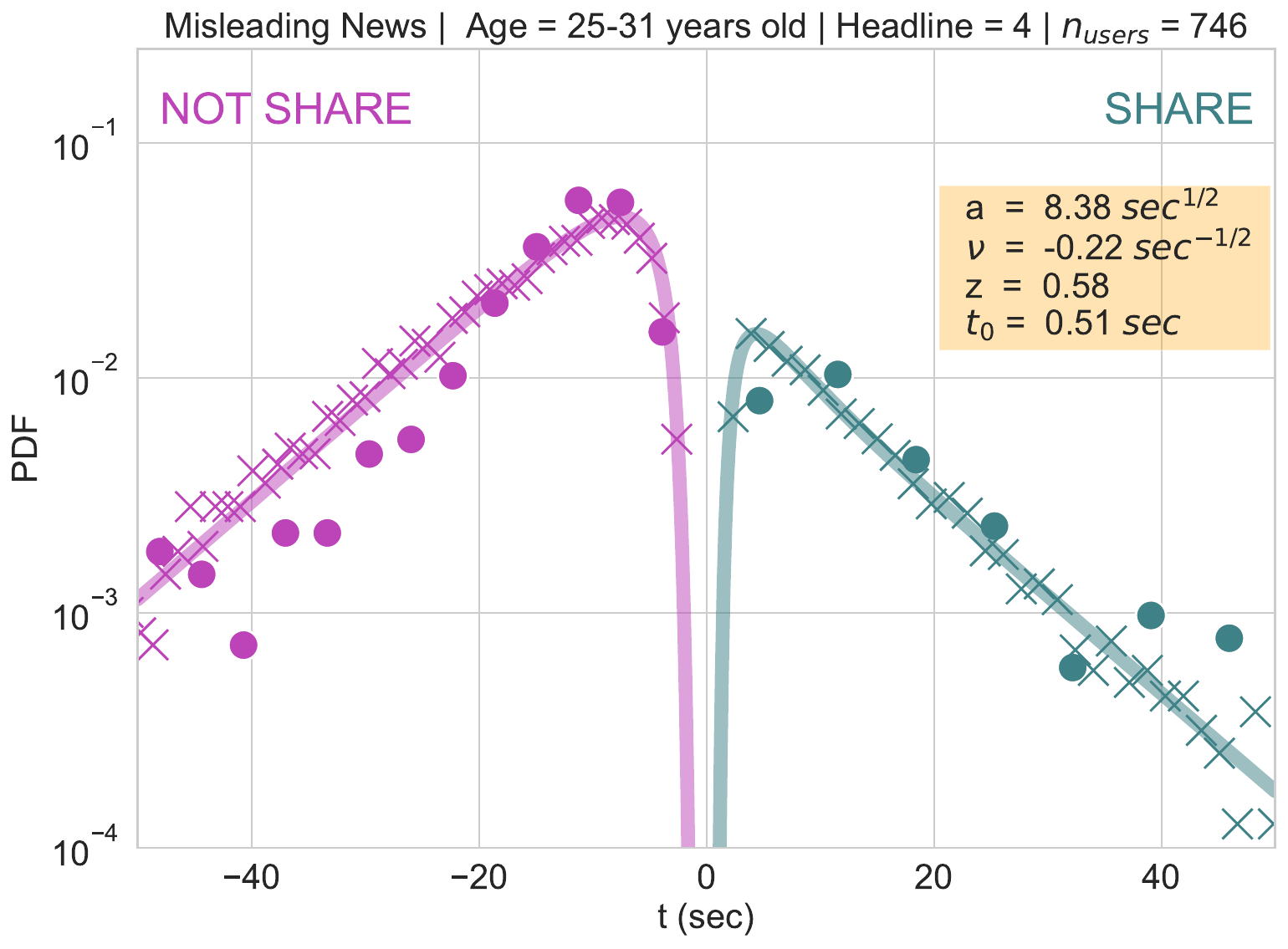}\quad
            \includegraphics[width=.45\textwidth]{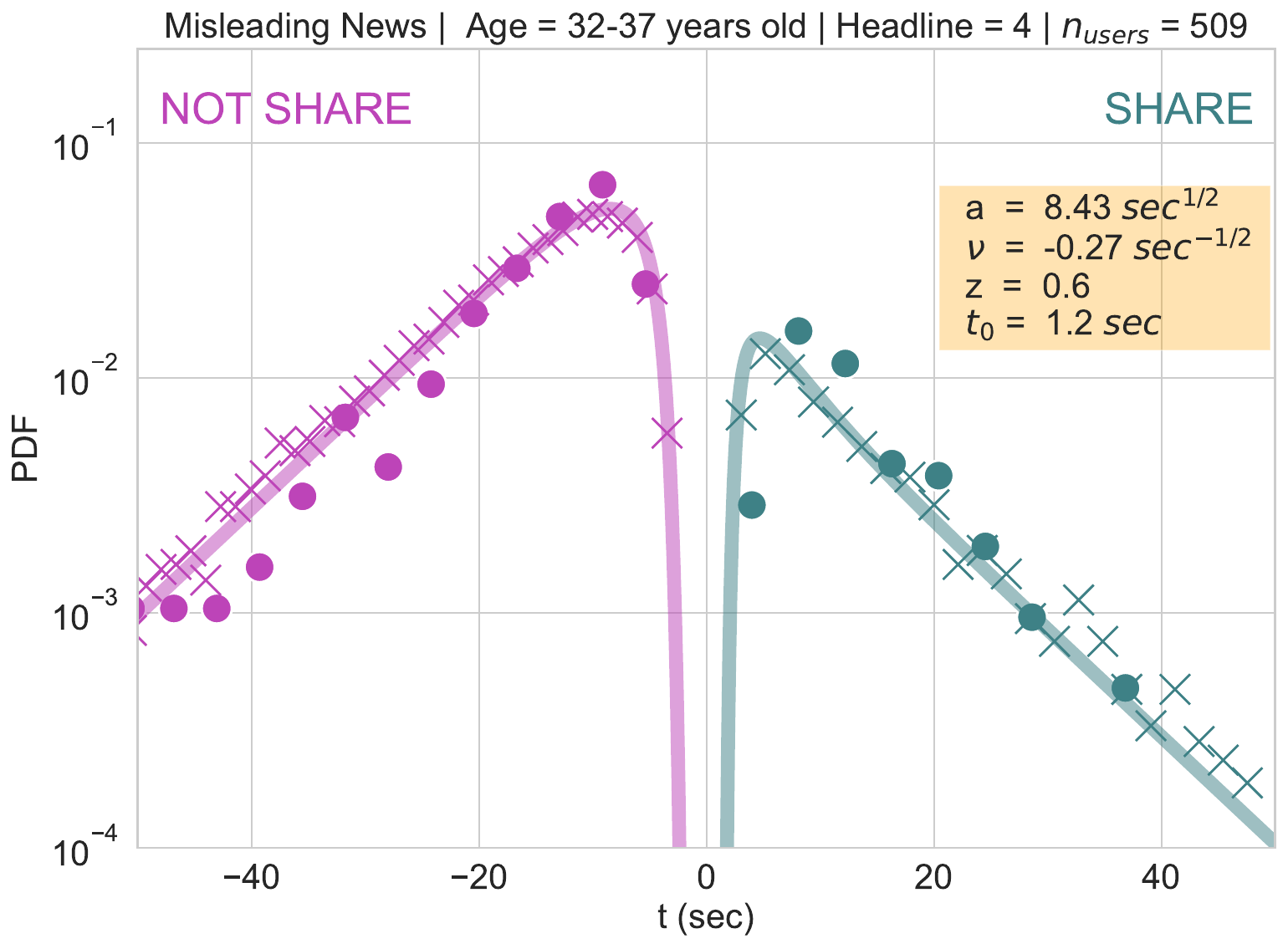}
            \medskip
            \includegraphics[width=.45\textwidth]{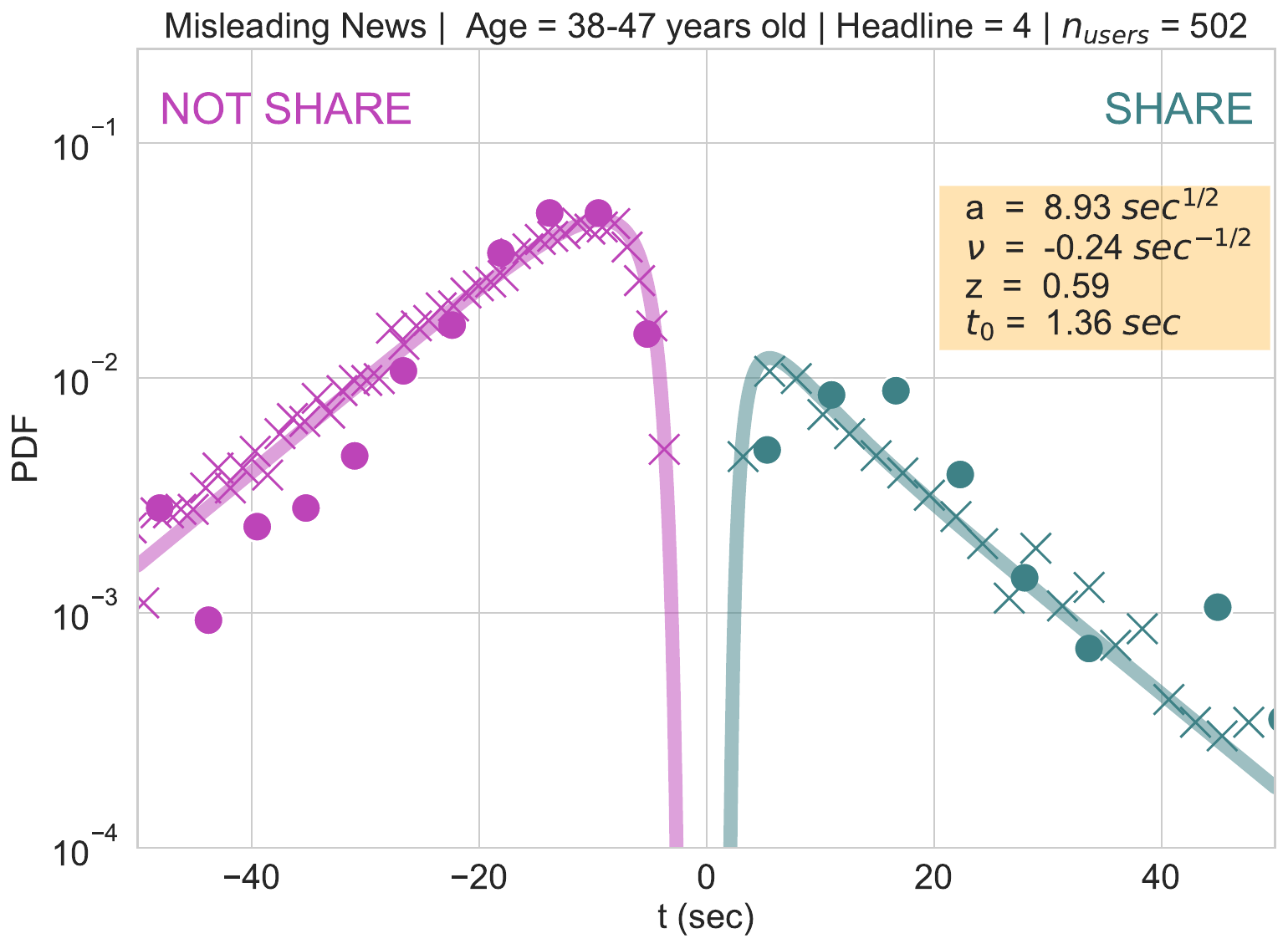}\quad
            \includegraphics[width=.45\textwidth]{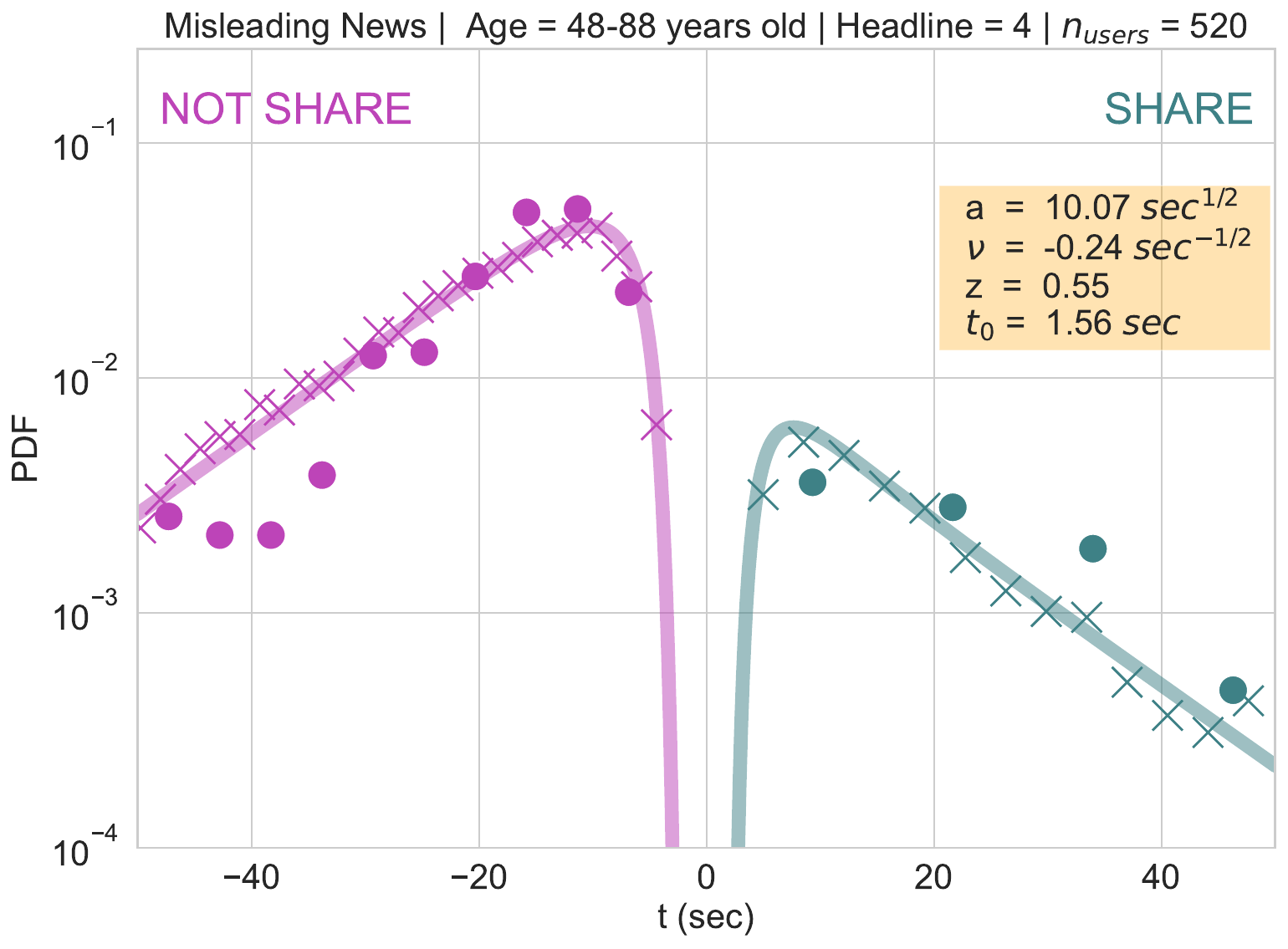}
            \caption{{\bf Headline 4}: Probability distribution of the response time for sharing and not sharing misleading information. Each figure corresponds to different age ranges. The solid line corresponds to theoretical results, dots correspond to empirical data and crosses to stochastic simulations.}
            \label{headline4Fake}
        \end{figure}
        
        \begin{figure}[H]
            \renewcommand{\figurename}{Supplementary Figure}
            \centering
            \includegraphics[width=.45\textwidth]{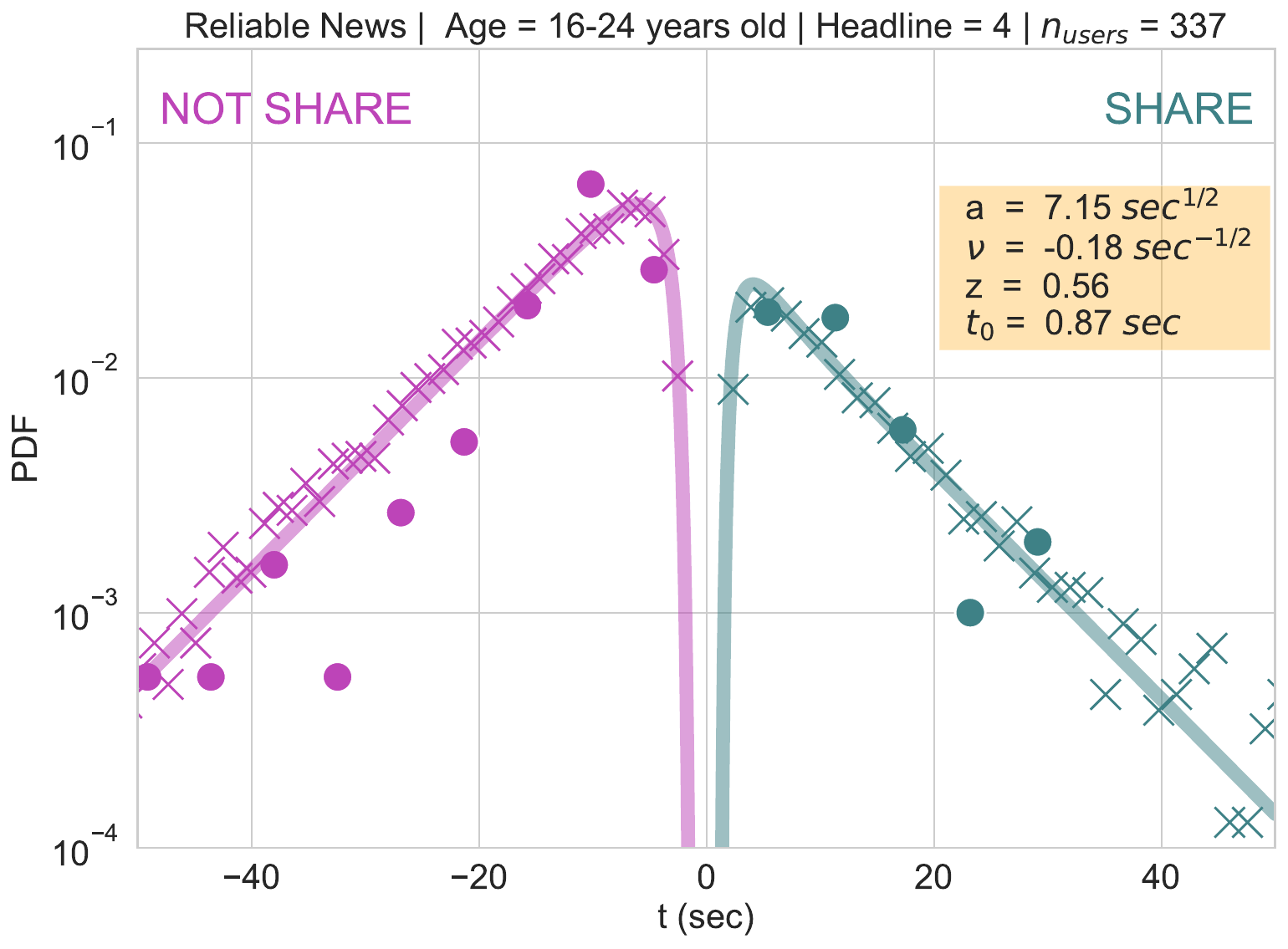}\quad
            \includegraphics[width=.45\textwidth]{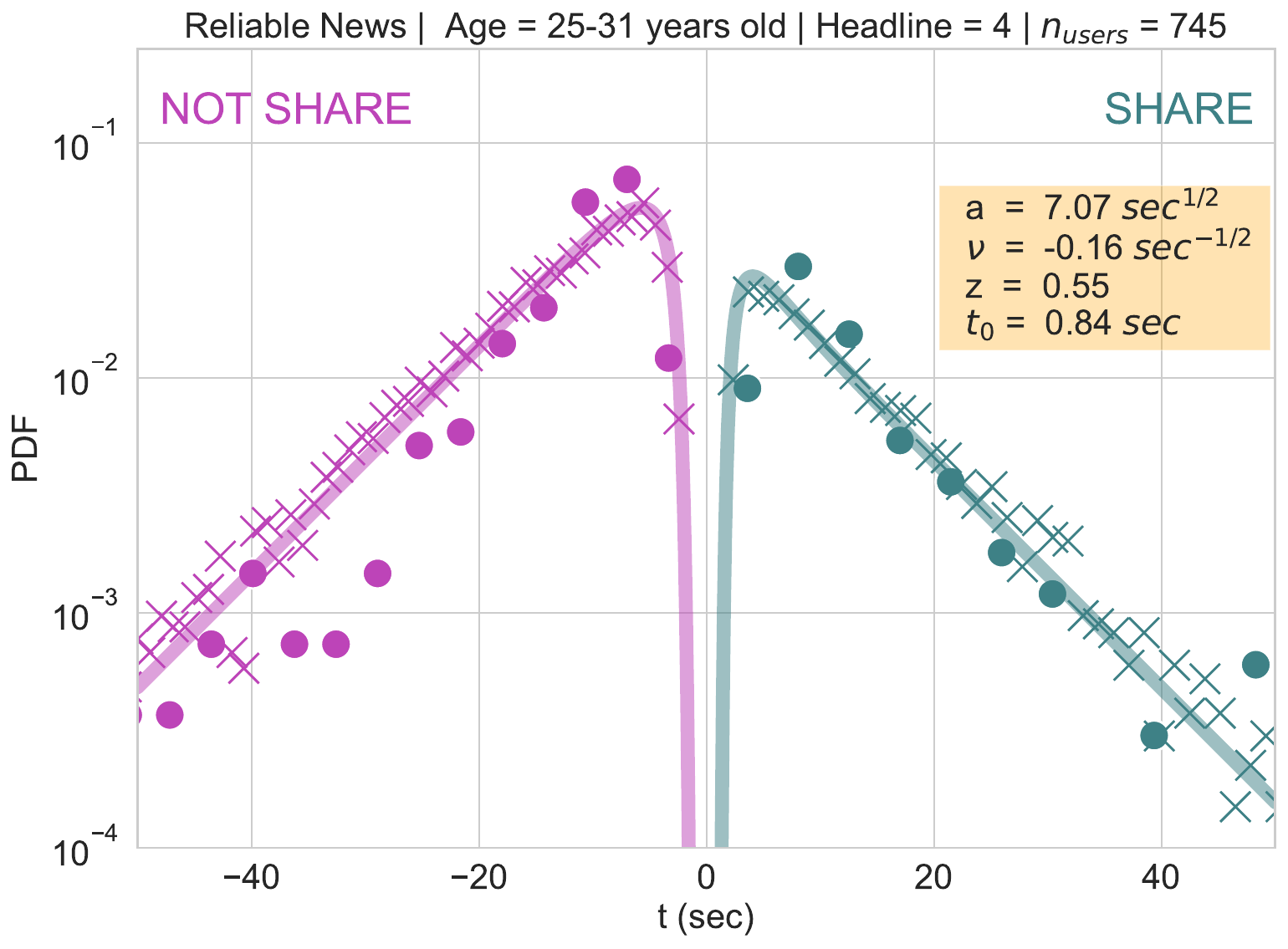}\quad
            \includegraphics[width=.45\textwidth]{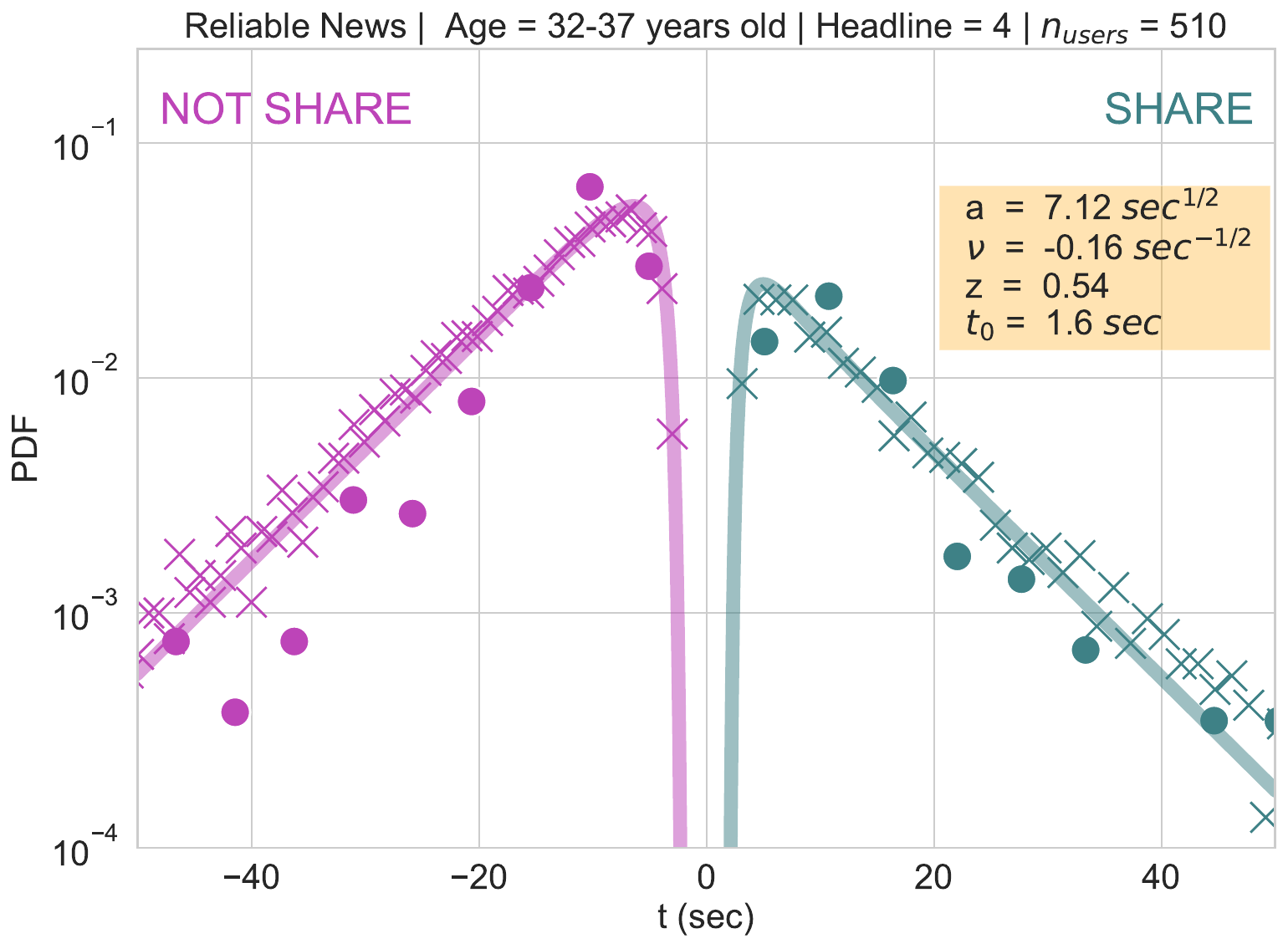}
            \medskip
            \includegraphics[width=.45\textwidth]{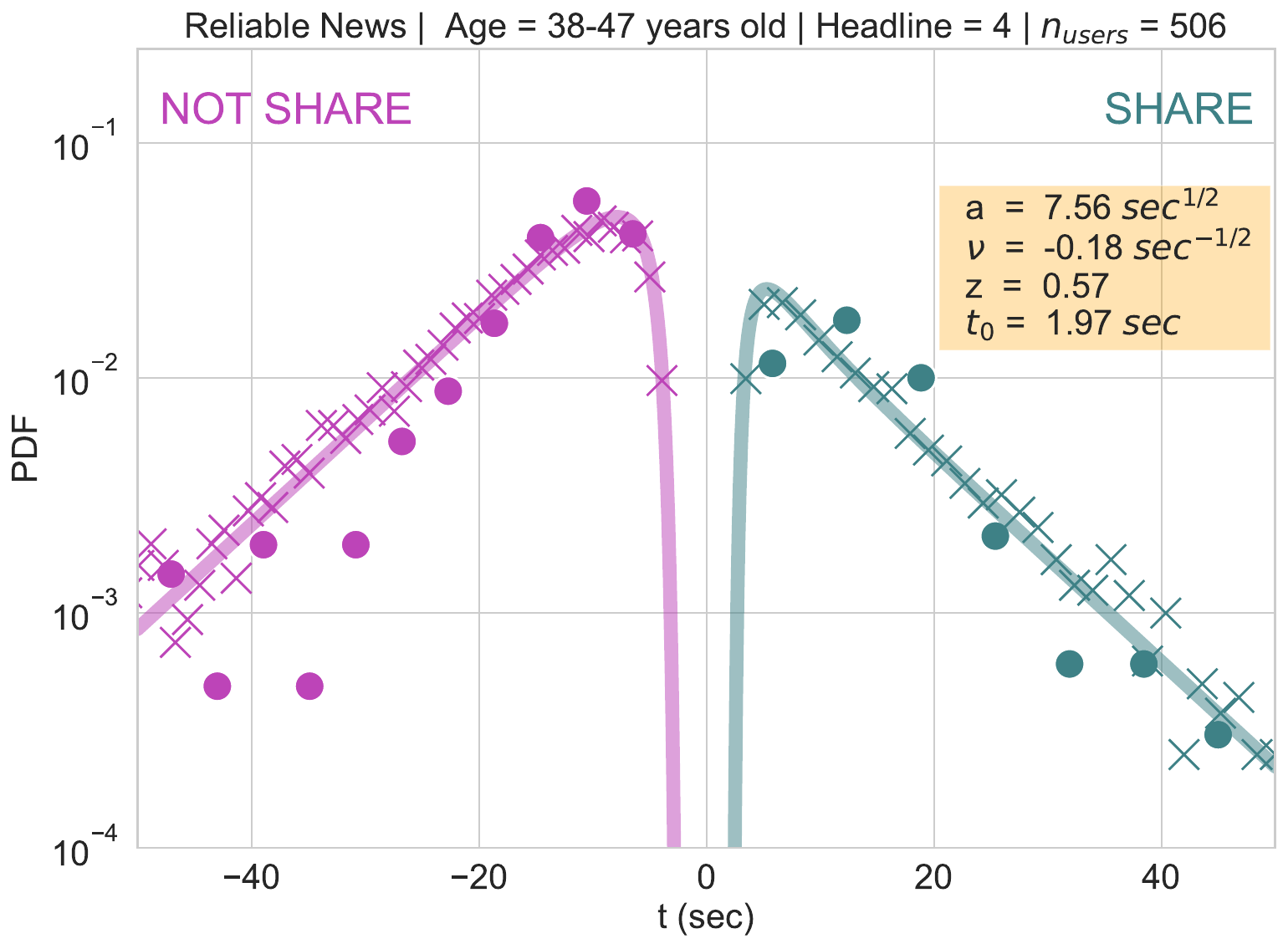}\quad
            \includegraphics[width=.45\textwidth]{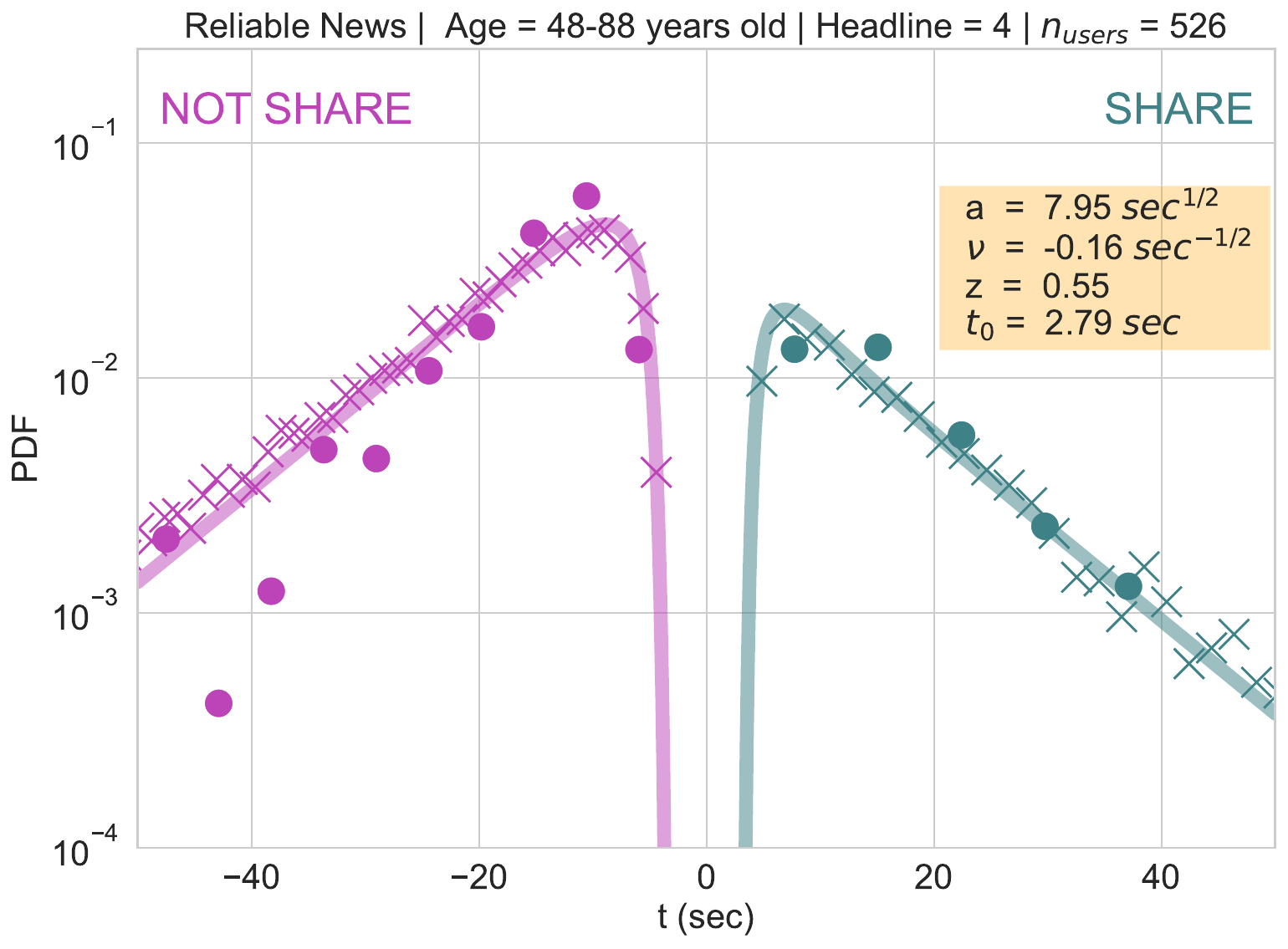}
            \caption{{\bf Headline 4}: Probability distribution of the response time for sharing and not sharing reliable information. Each figure corresponds to different age ranges. The solid line corresponds to theoretical results, dots correspond to empirical data and crosses to stochastic simulations.}
            \label{headline4Real}
        \end{figure}
        
        \begin{figure}[H]
            \renewcommand{\figurename}{Supplementary Figure}
            \centering
            \includegraphics[width=.45\textwidth]{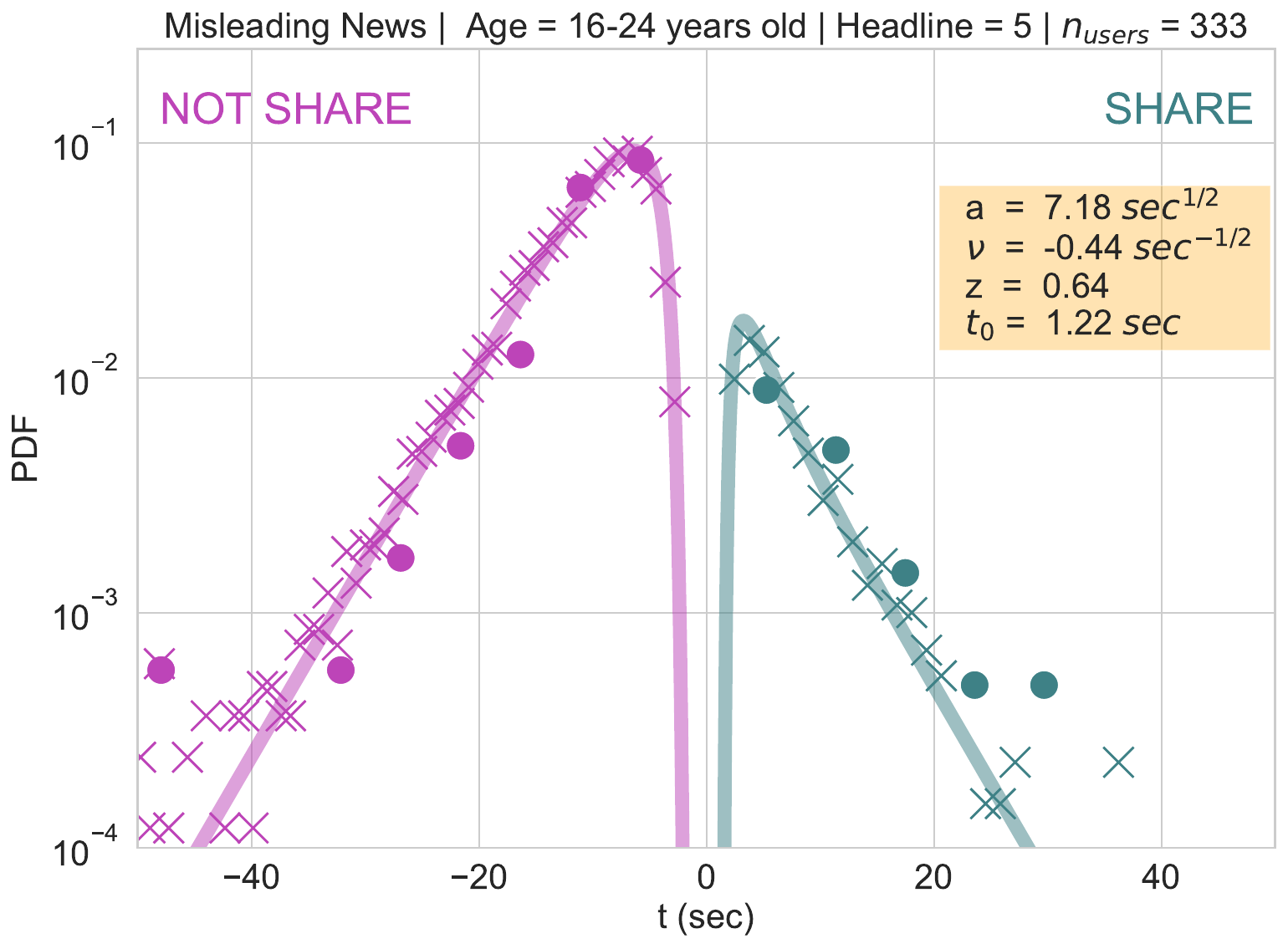}\quad
            \includegraphics[width=.45\textwidth]{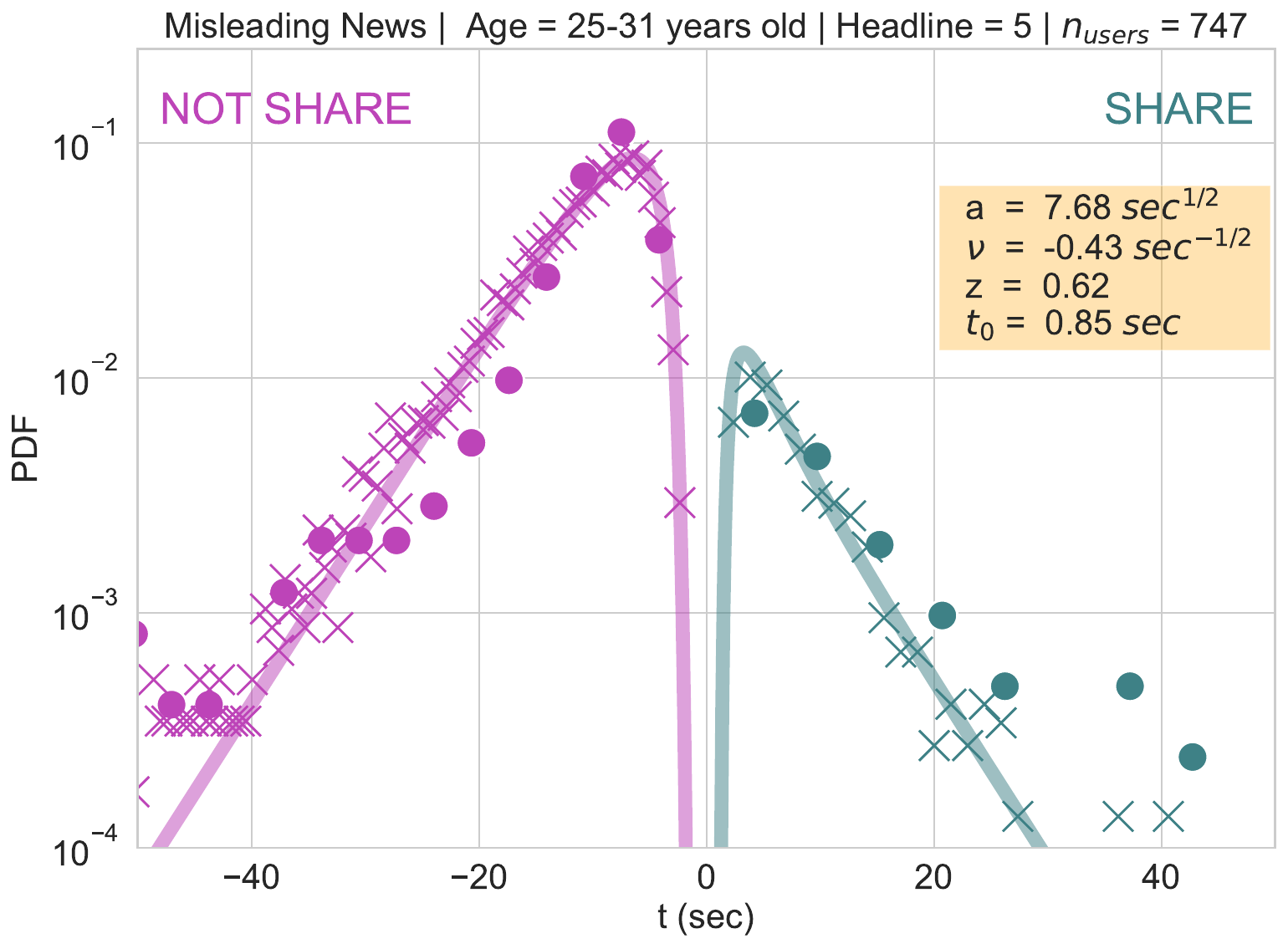}\quad
            \includegraphics[width=.45\textwidth]{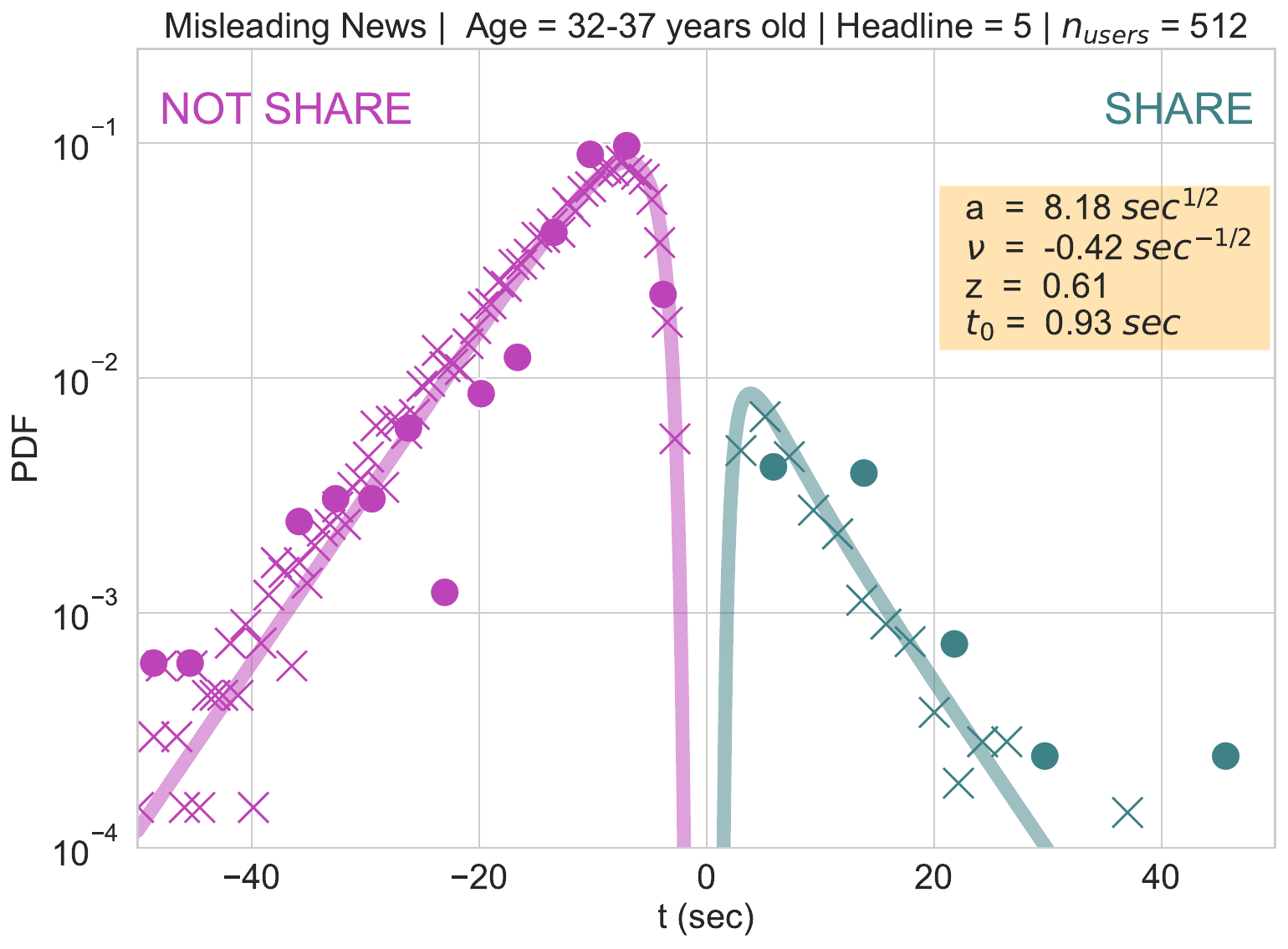}
            \medskip
            \includegraphics[width=.45\textwidth]{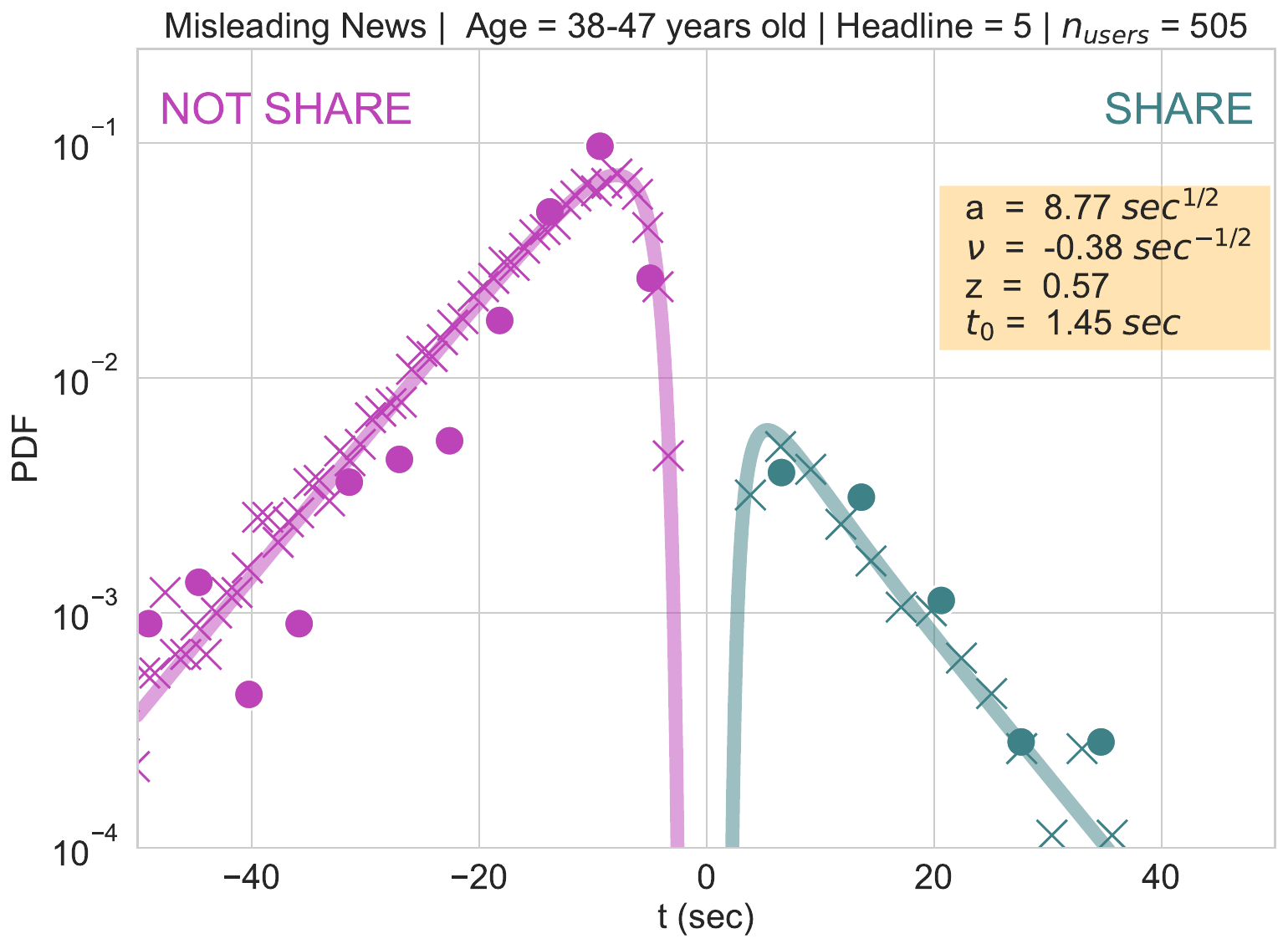}\quad
            \includegraphics[width=.45\textwidth]{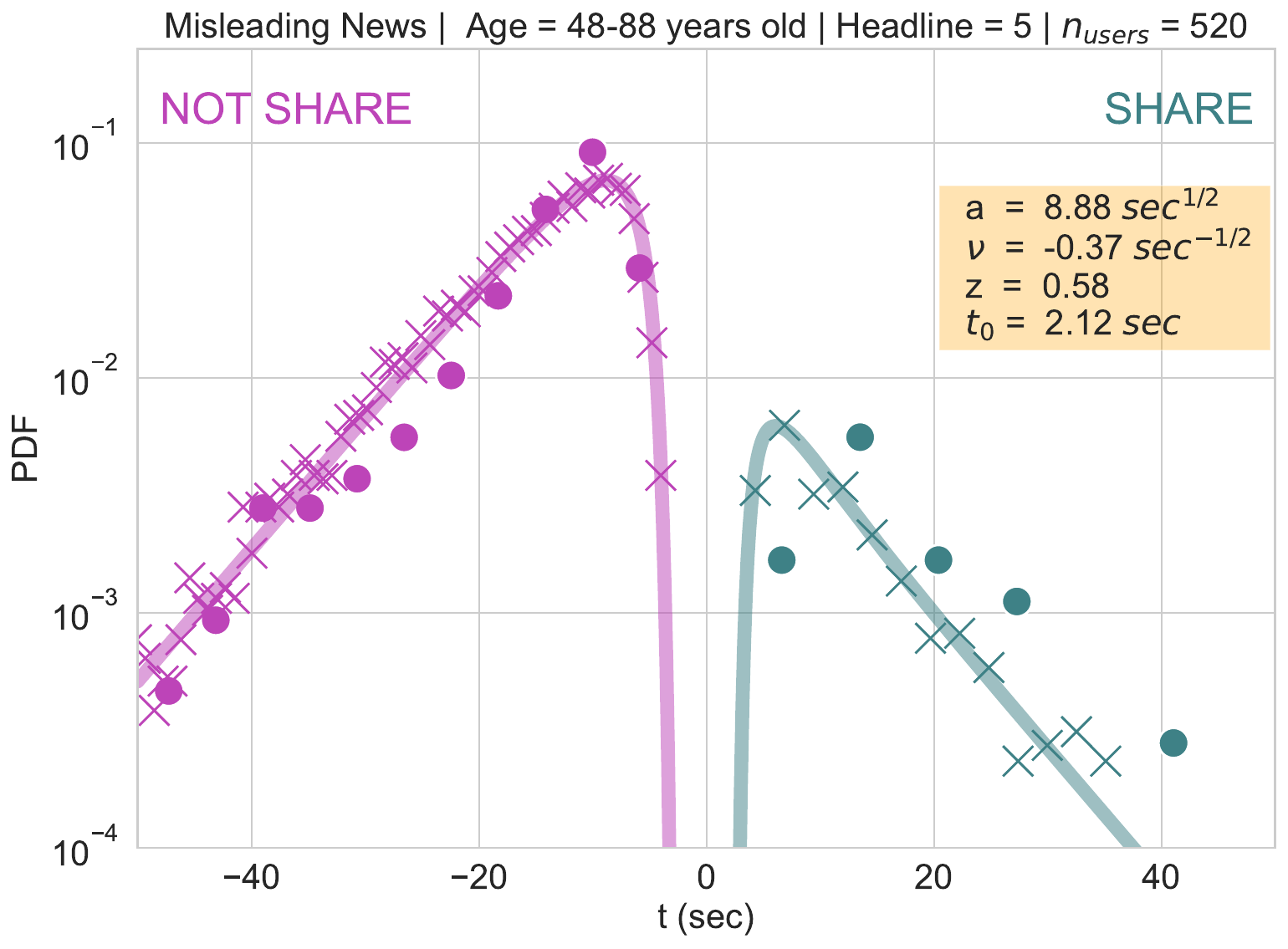}
            \caption{{\bf Headline 5}: Probability distribution of the response time for sharing and not sharing misleading information. Each figure corresponds to different age ranges. The solid line corresponds to theoretical results, dots correspond to empirical data and crosses to stochastic simulations.}
            \label{headline5Fake}
        \end{figure}
        
        \begin{figure}[H]
            \renewcommand{\figurename}{Supplementary Figure}
            \centering
            \includegraphics[width=.45\textwidth]{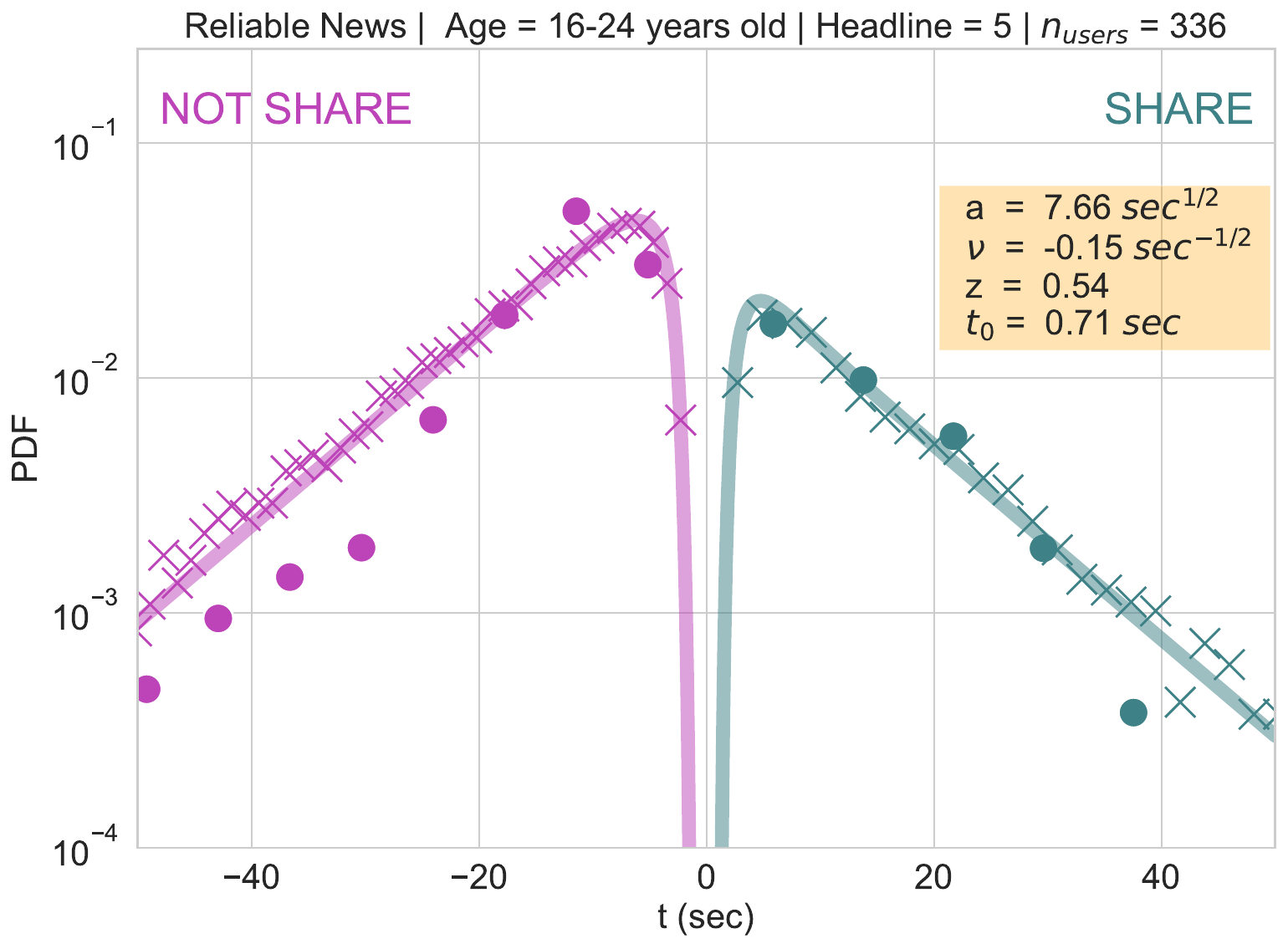}\quad
            \includegraphics[width=.45\textwidth]{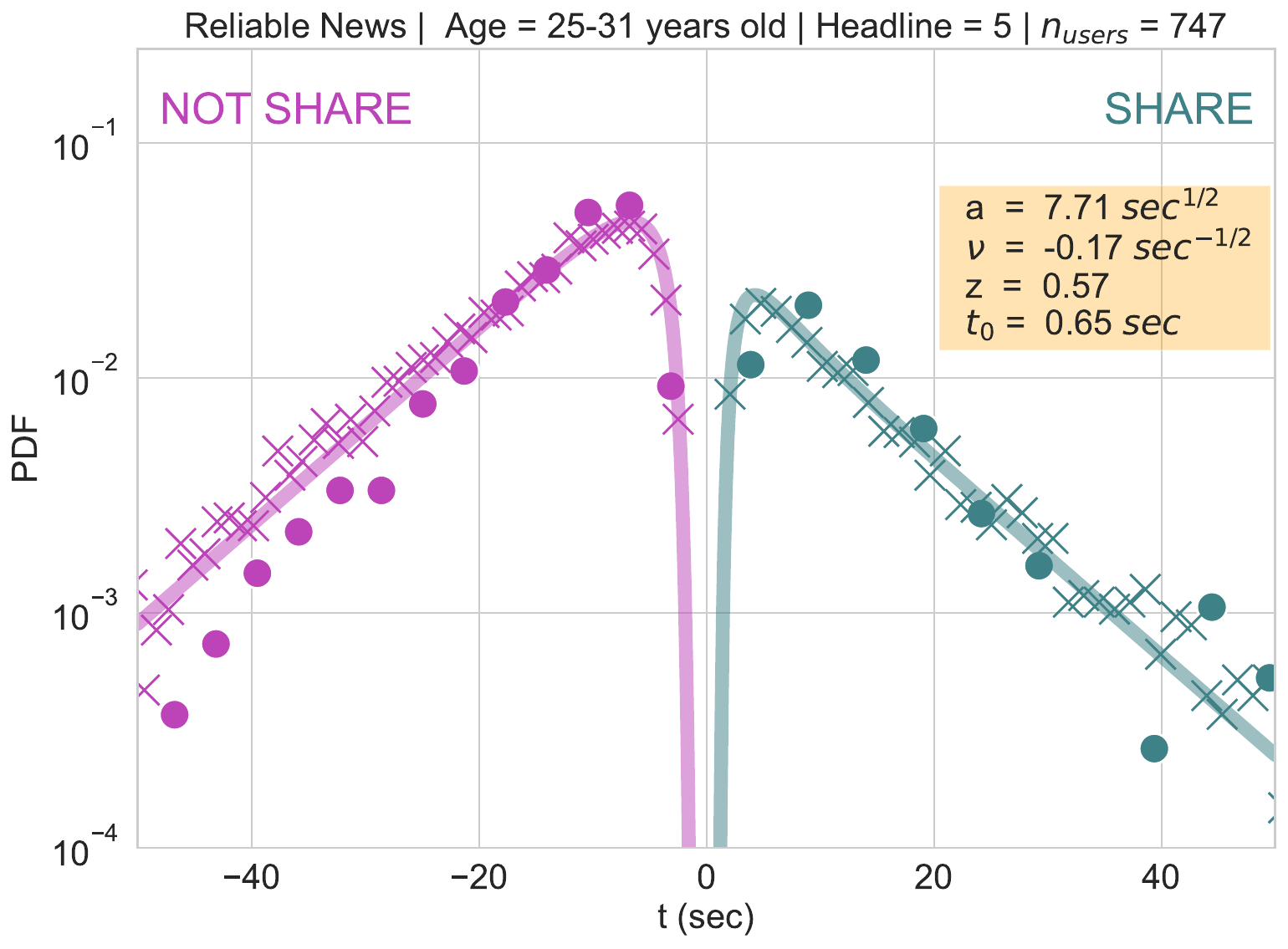}\quad
            \includegraphics[width=.45\textwidth]{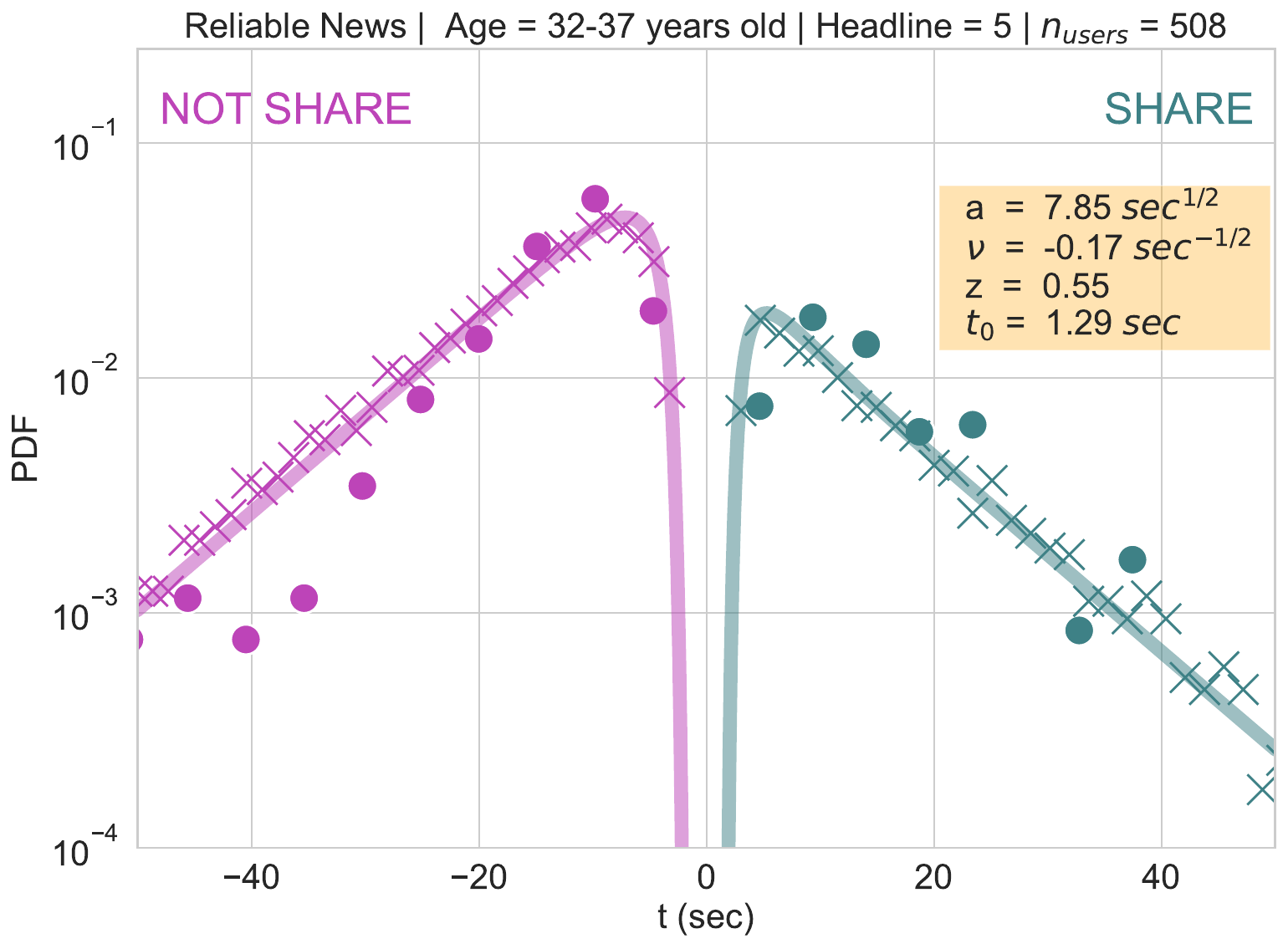}
            \medskip
            \includegraphics[width=.45\textwidth]{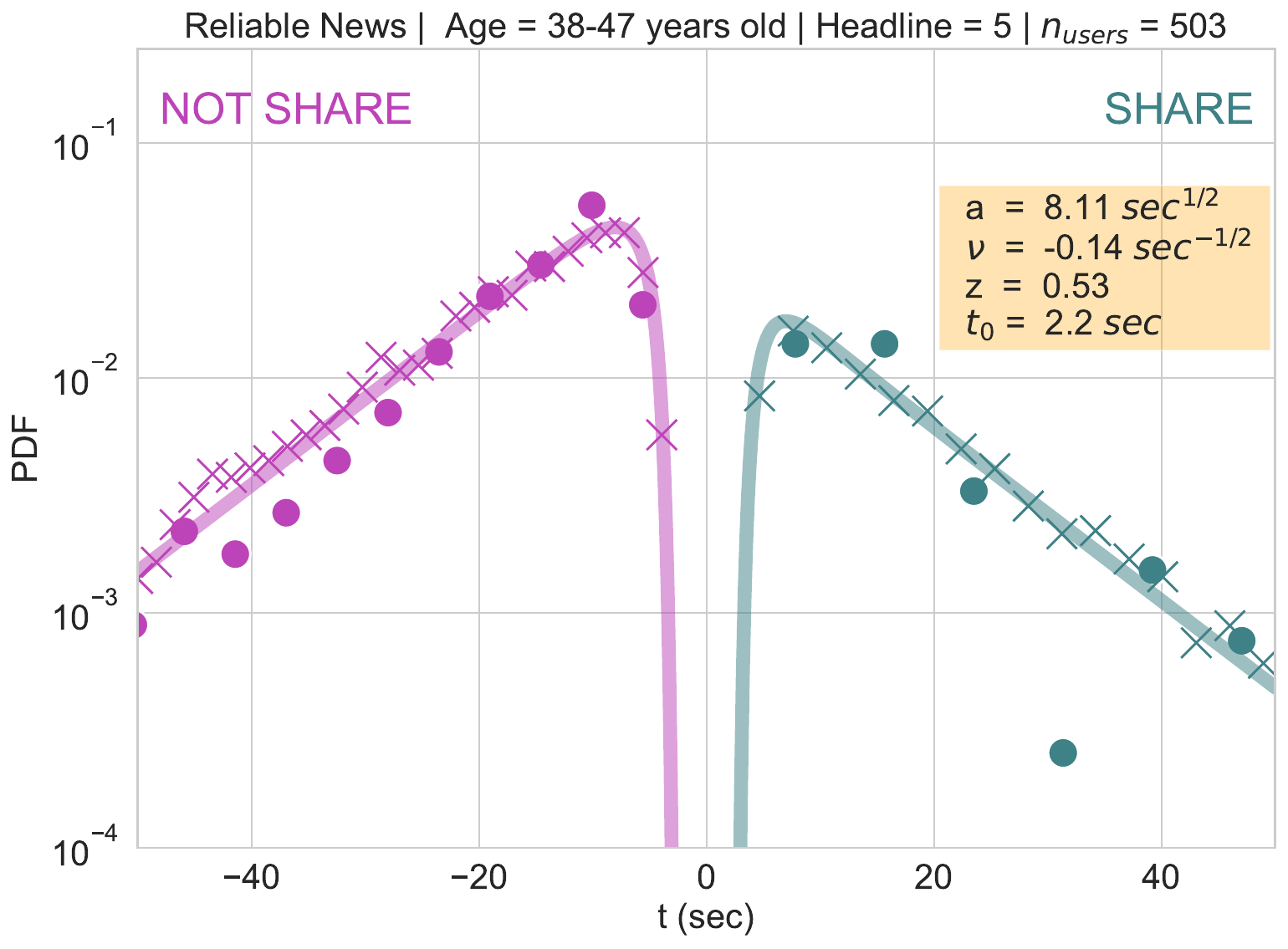}\quad
            \includegraphics[width=.45\textwidth]{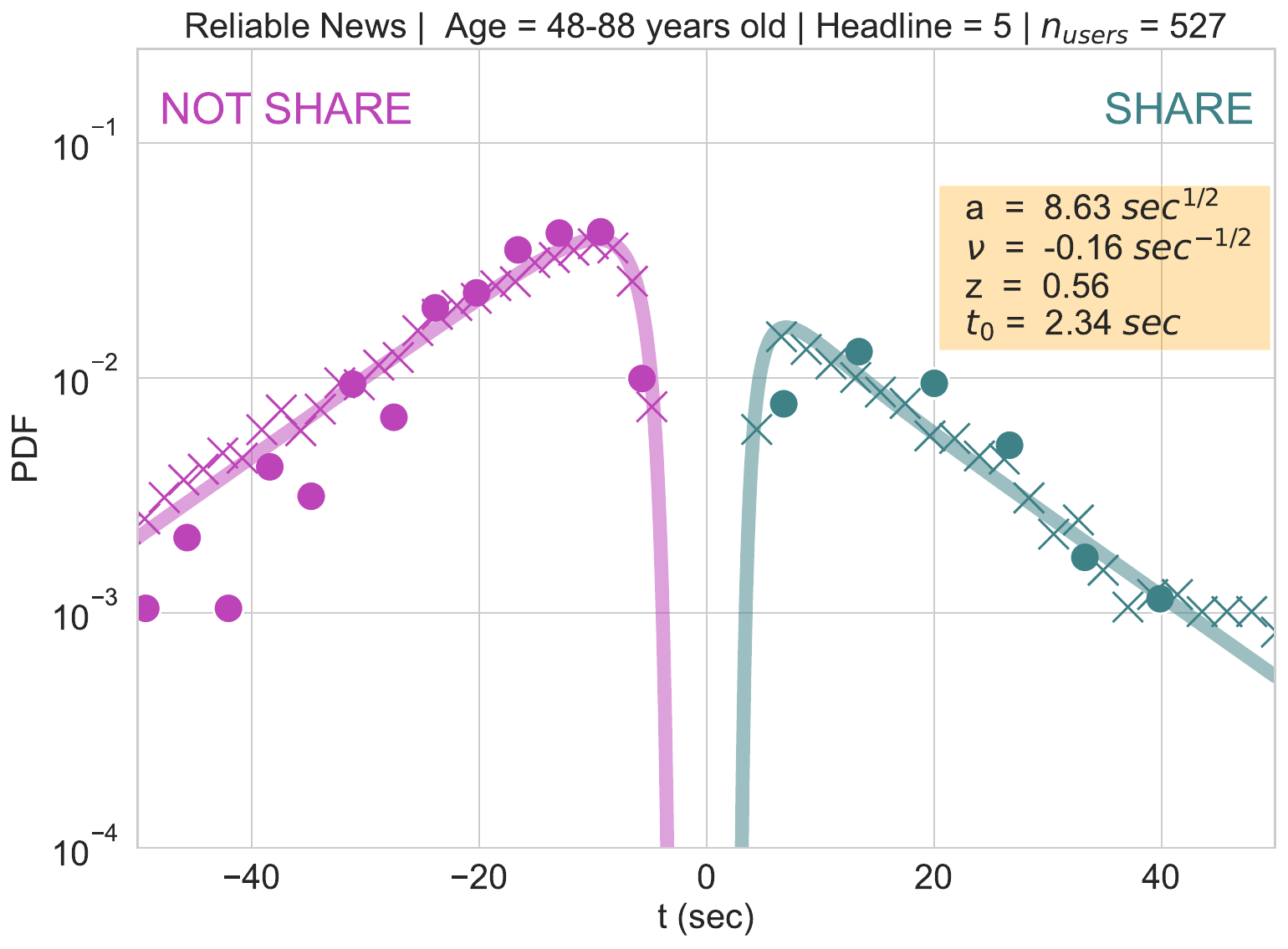}
            \caption{{\bf Headline 5}: Probability distribution of the response time for sharing and not sharing reliable information. Each figure corresponds to different age ranges. The solid line corresponds to theoretical results, dots correspond to empirical data and crosses to stochastic simulations.}
            \label{headline5Real}
        \end{figure}
        
        \begin{figure}[H]
            \renewcommand{\figurename}{Supplementary Figure}
            \centering
            \includegraphics[width=.45\textwidth]{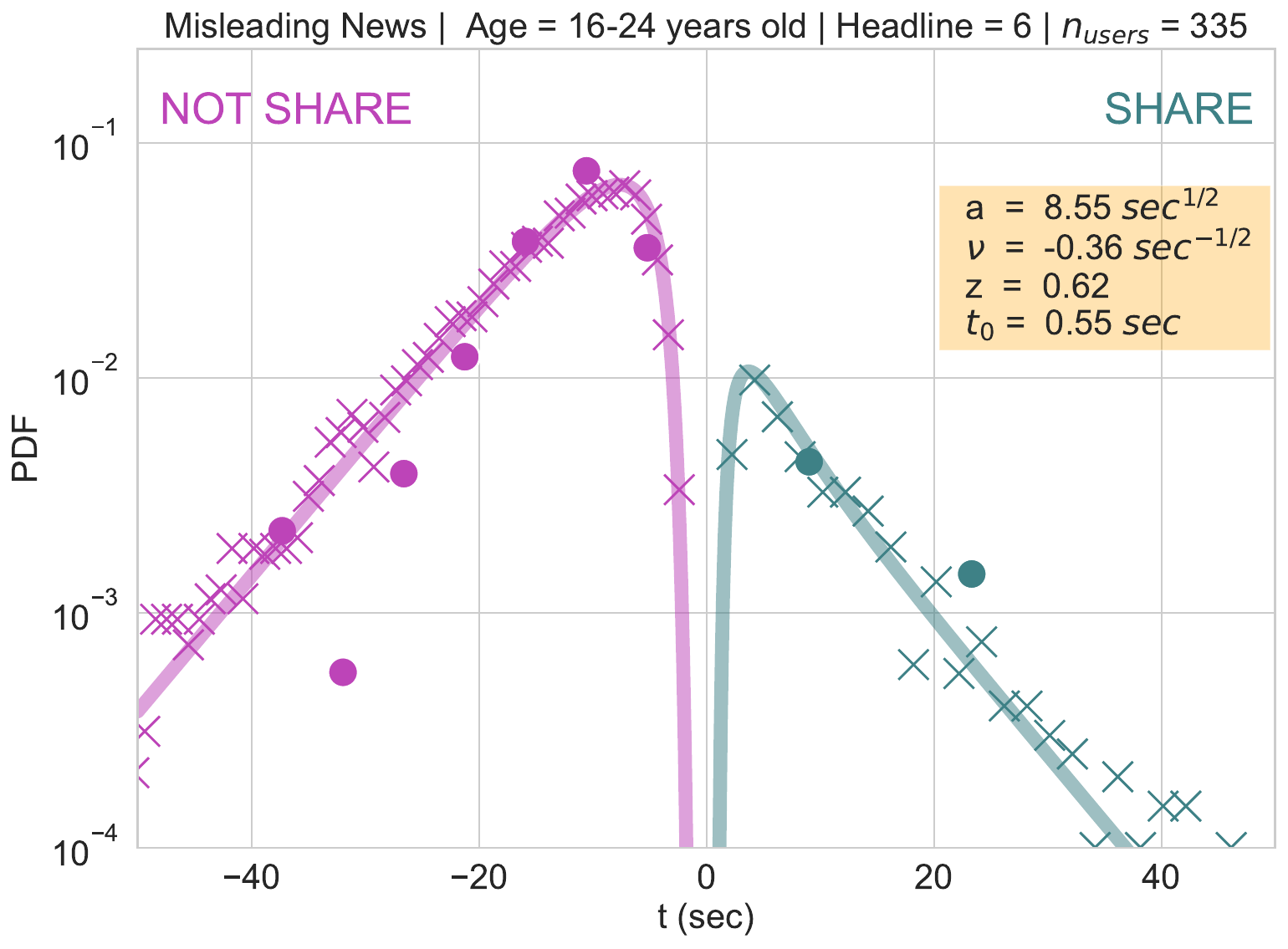}\quad
            \includegraphics[width=.45\textwidth]{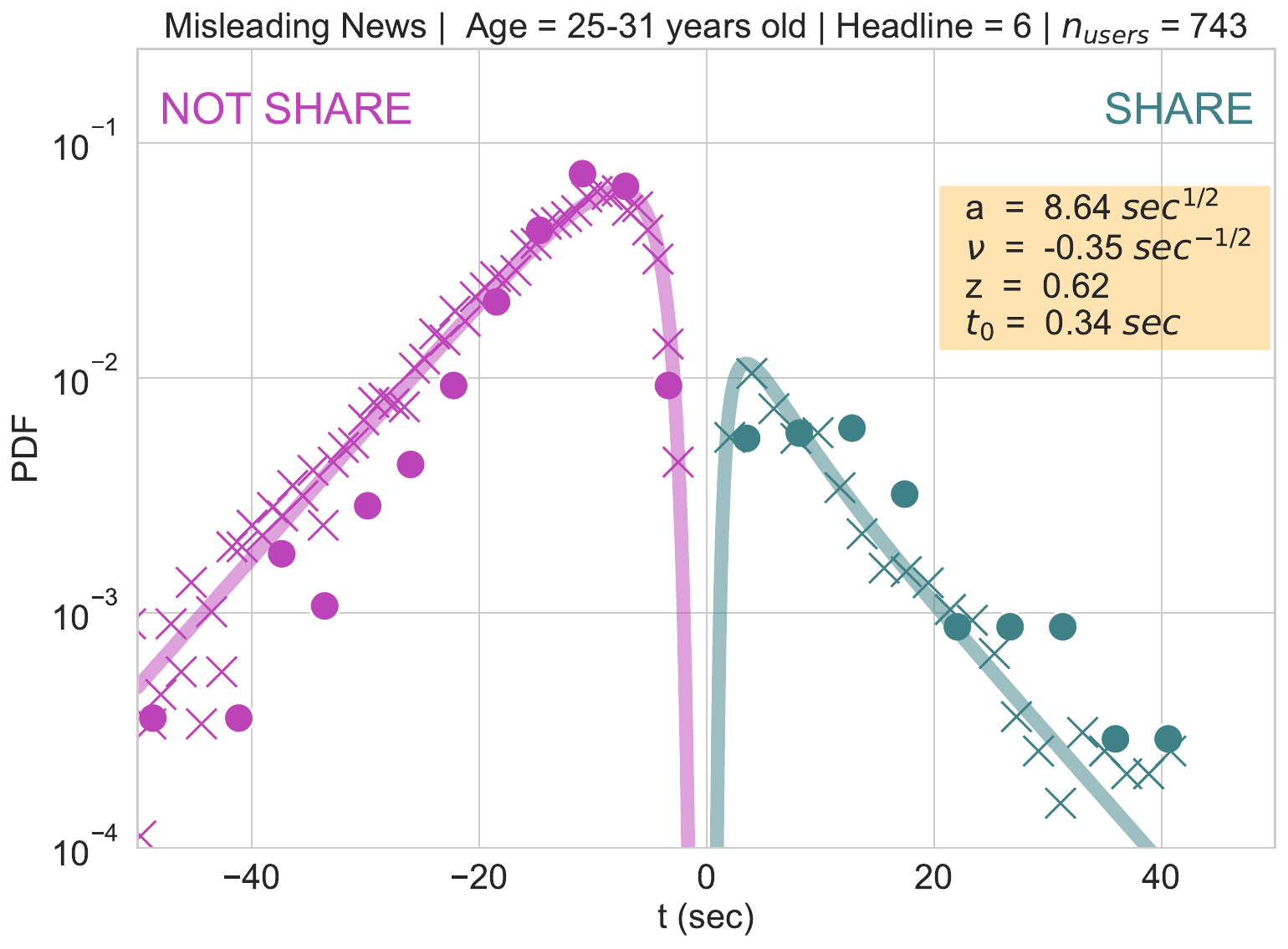}\quad
            \includegraphics[width=.45\textwidth]{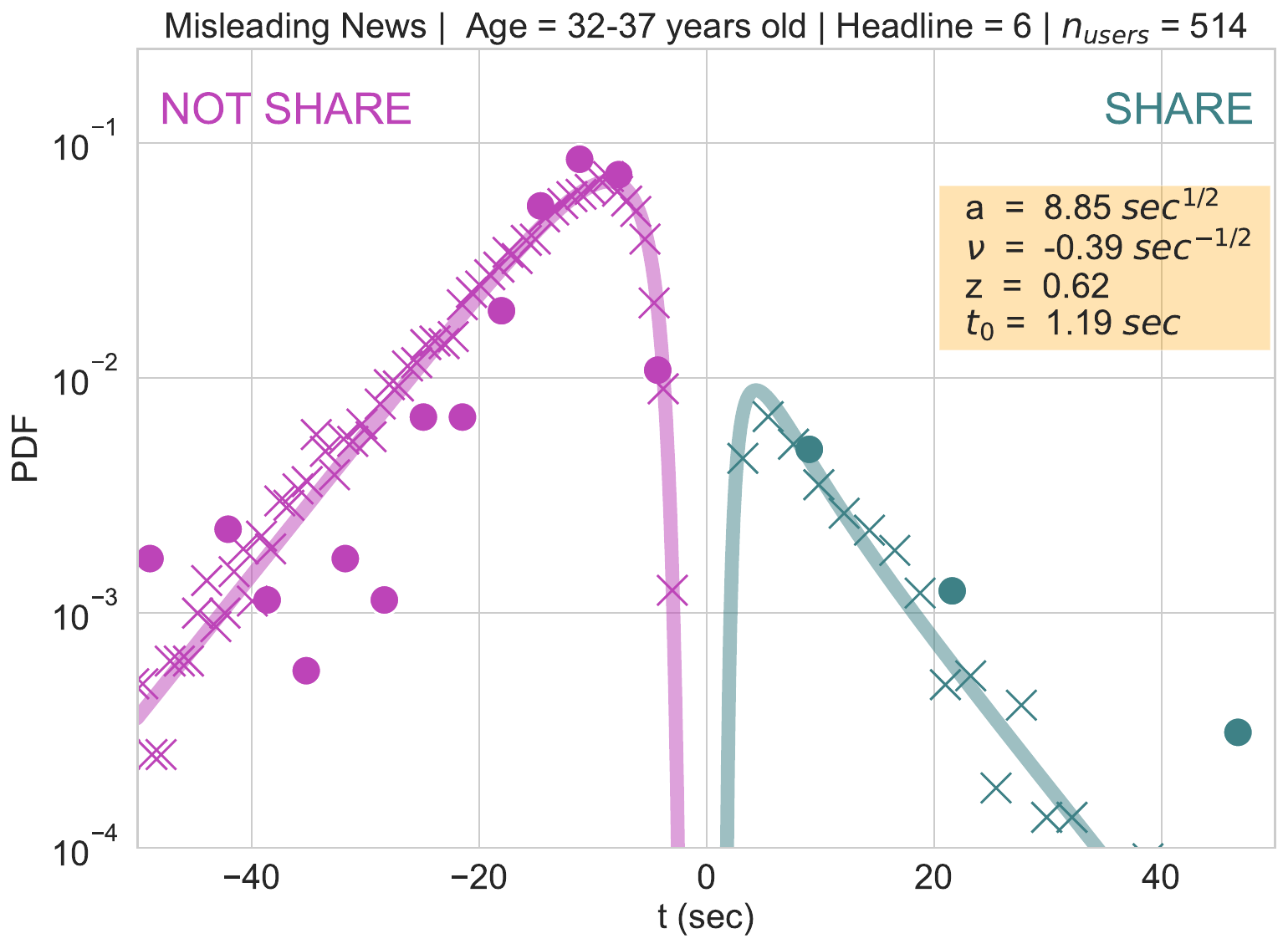}
            \medskip
            \includegraphics[width=.45\textwidth]{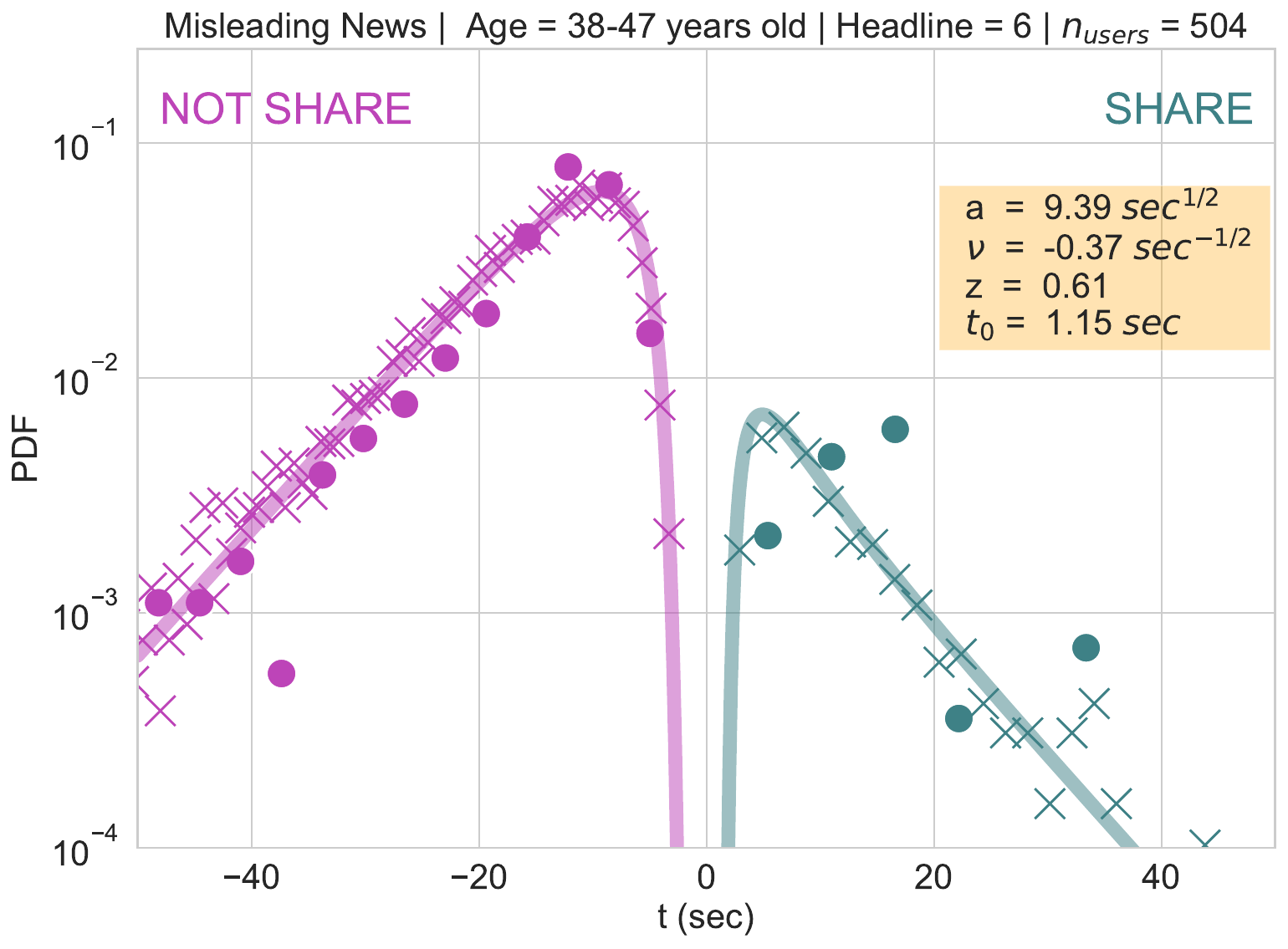}\quad
            \includegraphics[width=.45\textwidth]{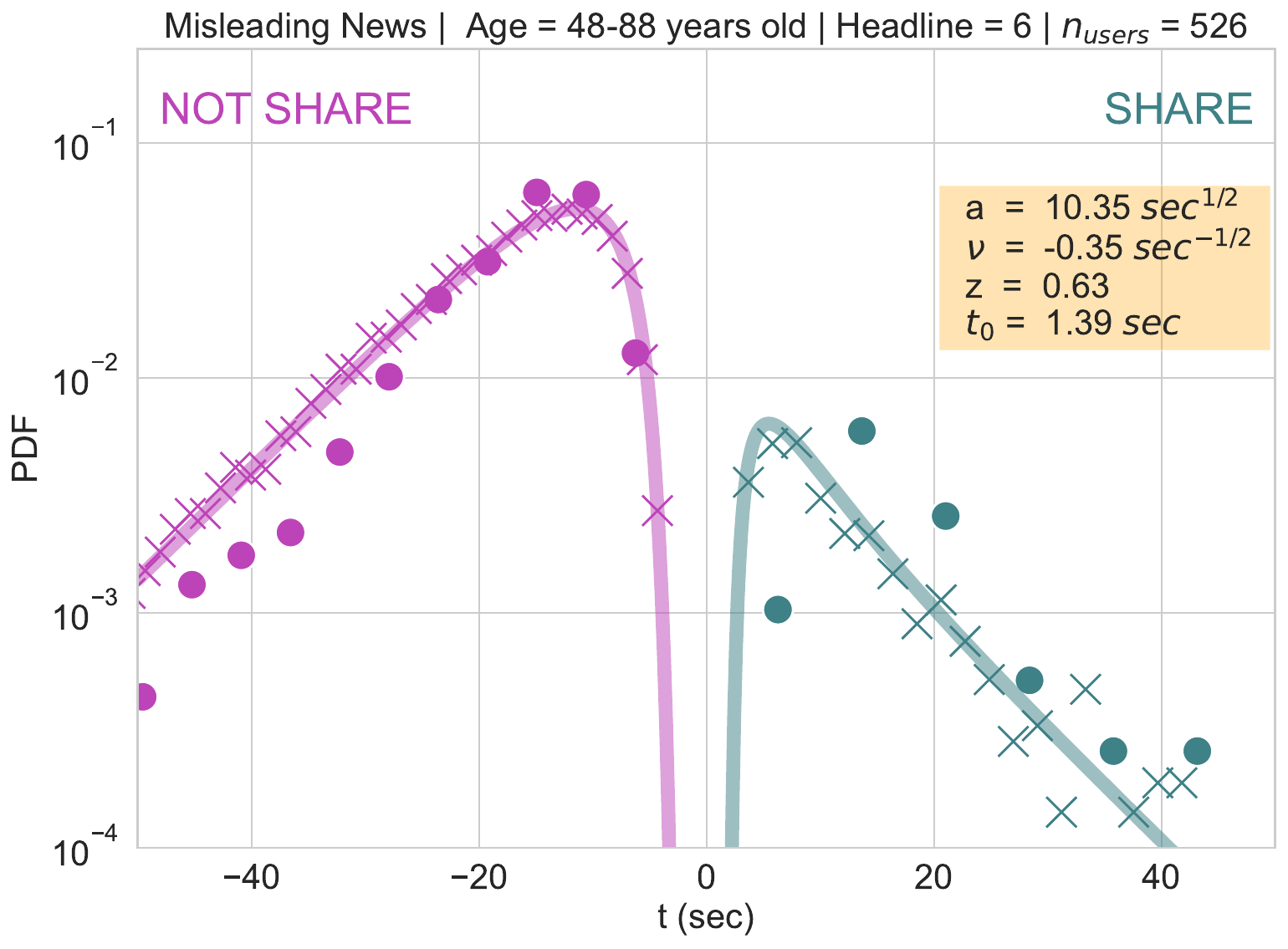}
            \caption{{\bf Headline 6}: Probability distribution of the response time for sharing and not sharing misleading information. Each figure corresponds to different age ranges. The solid line corresponds to theoretical results, dots correspond to empirical data and crosses to stochastic simulations.}
            \label{headline6Fake}
        \end{figure}
        
        \begin{figure}[H]
            \renewcommand{\figurename}{Supplementary Figure}
            \centering
            \includegraphics[width=.45\textwidth]{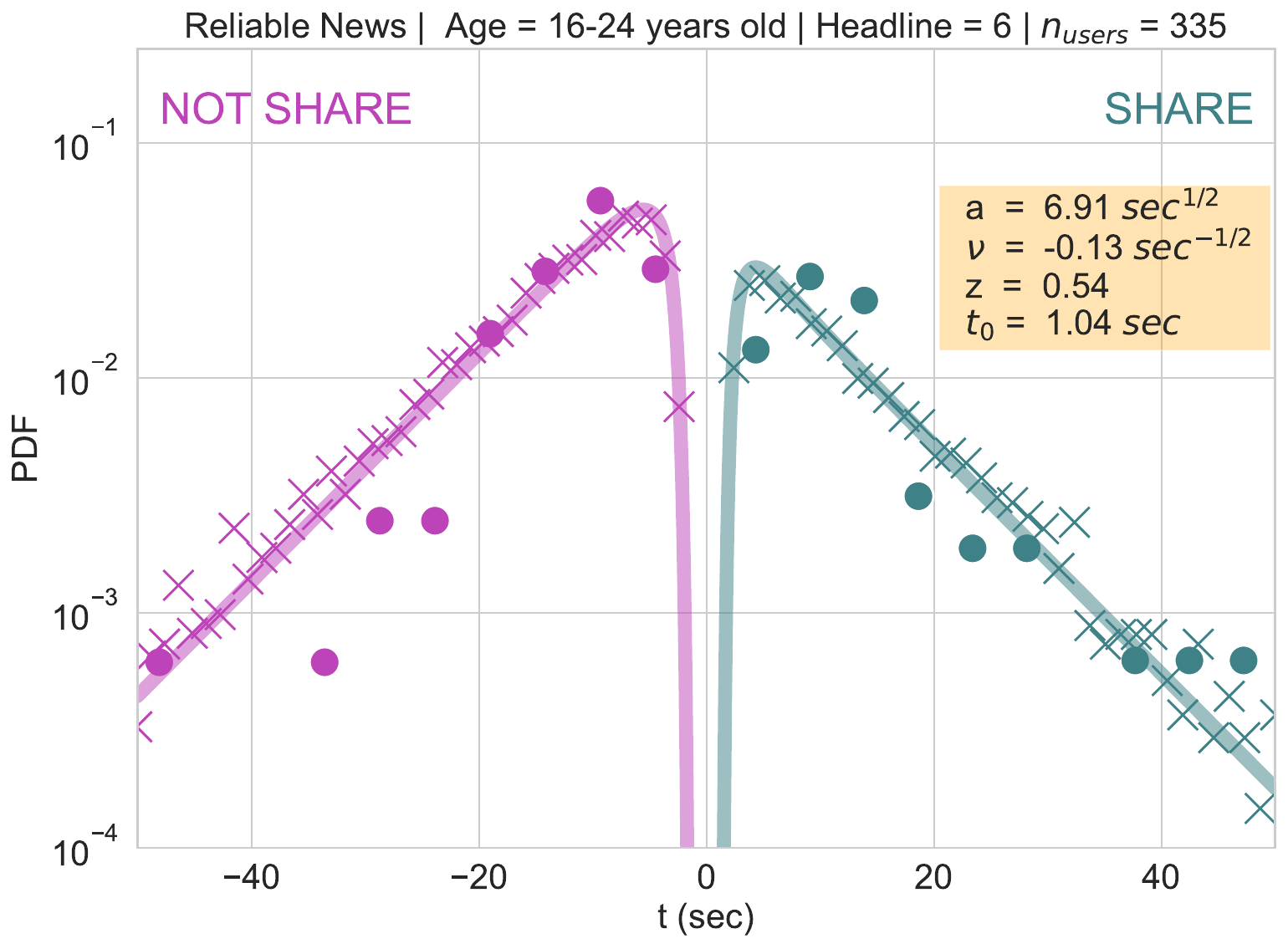}\quad
            \includegraphics[width=.45\textwidth]{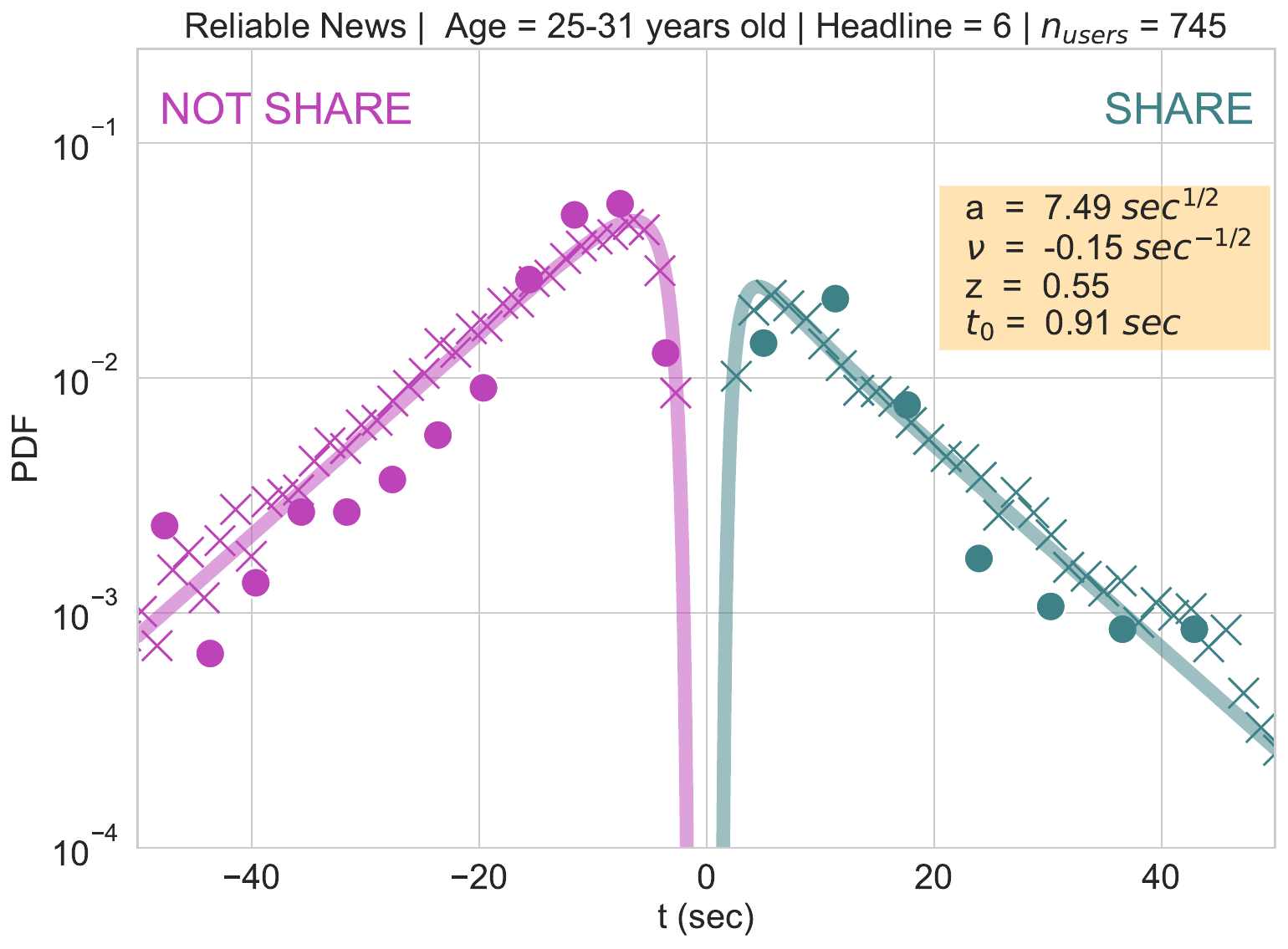}\quad
            \includegraphics[width=.45\textwidth]{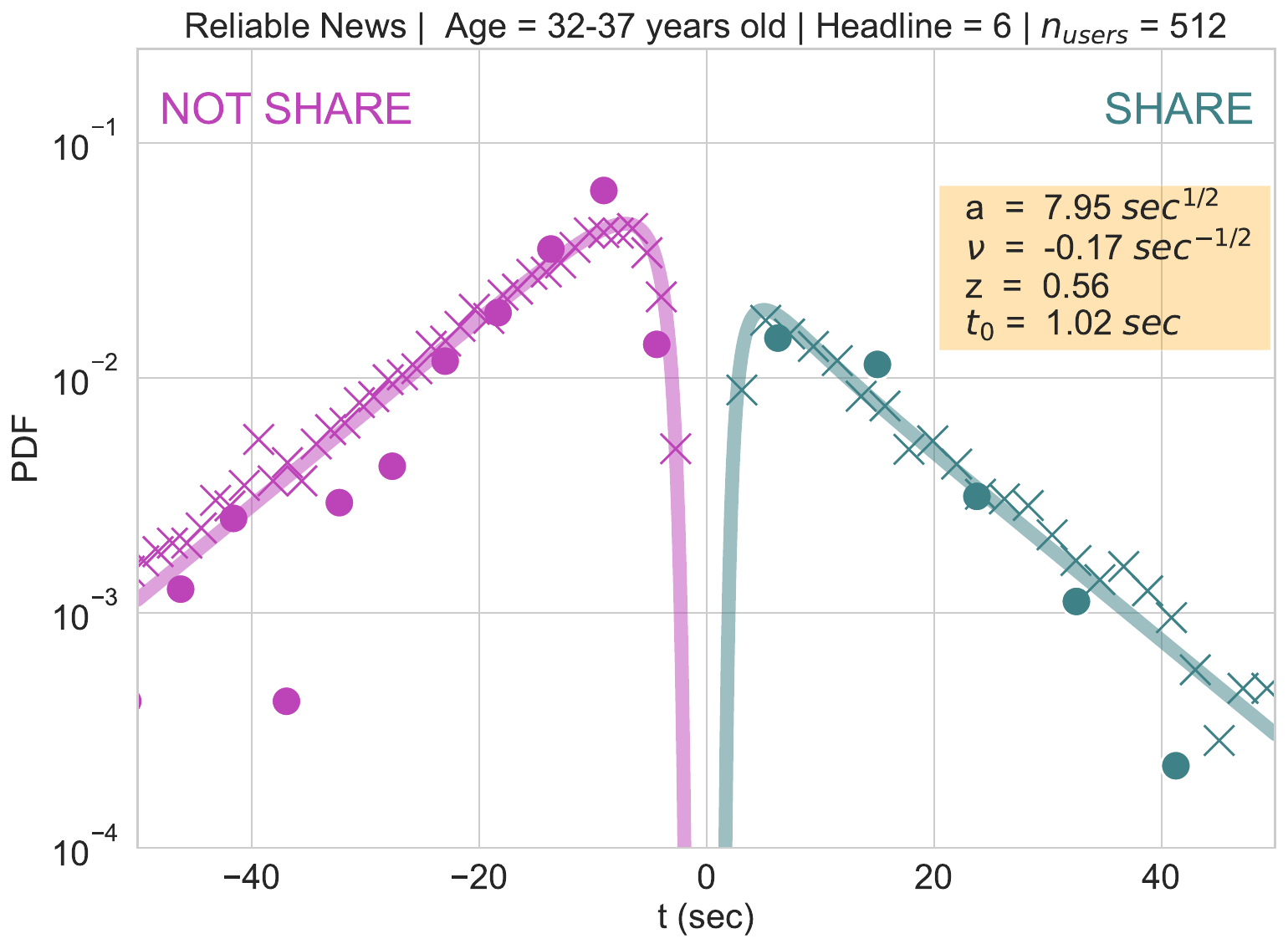}
            \medskip
            \includegraphics[width=.45\textwidth]{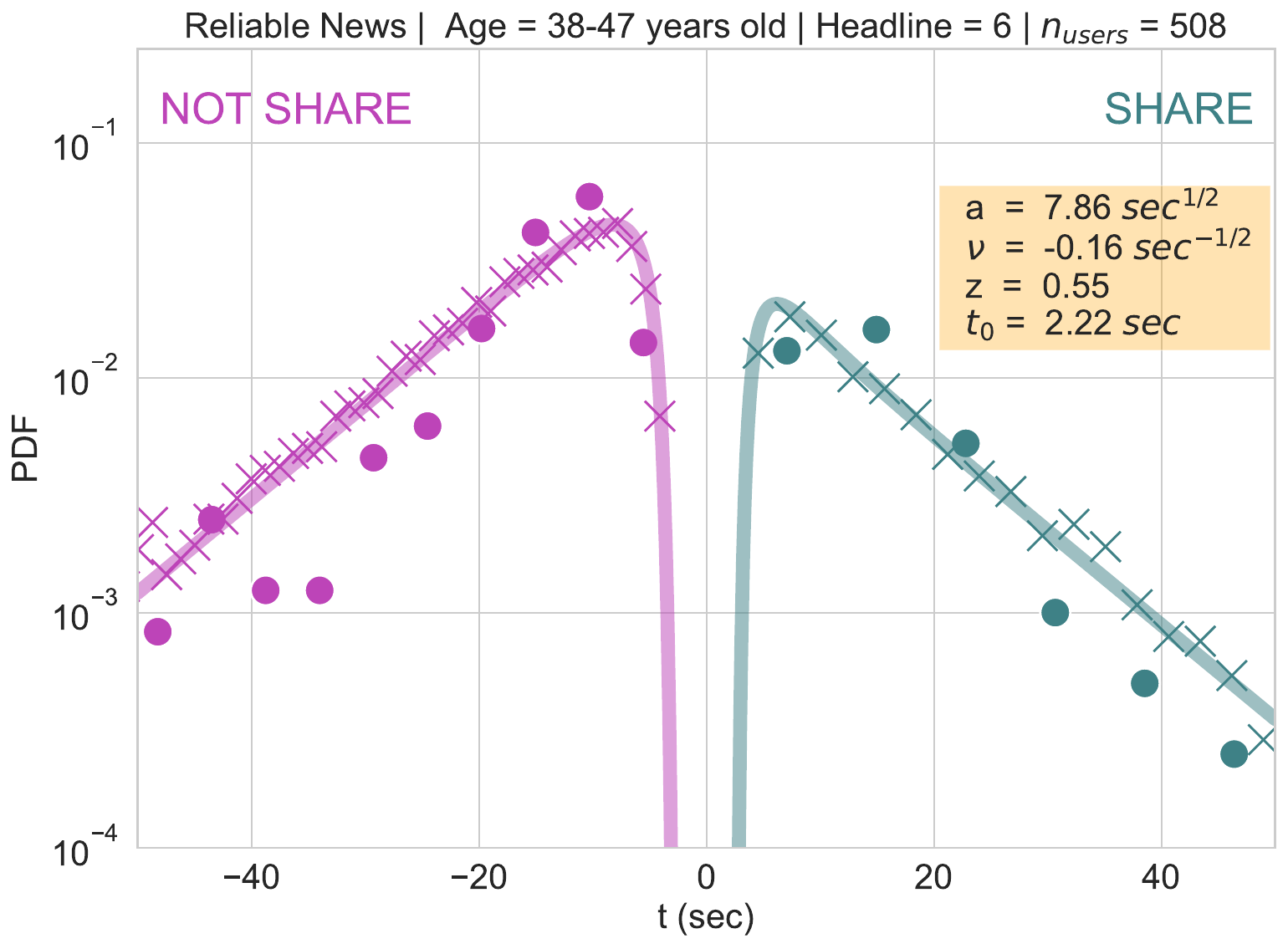}\quad
            \includegraphics[width=.45\textwidth]{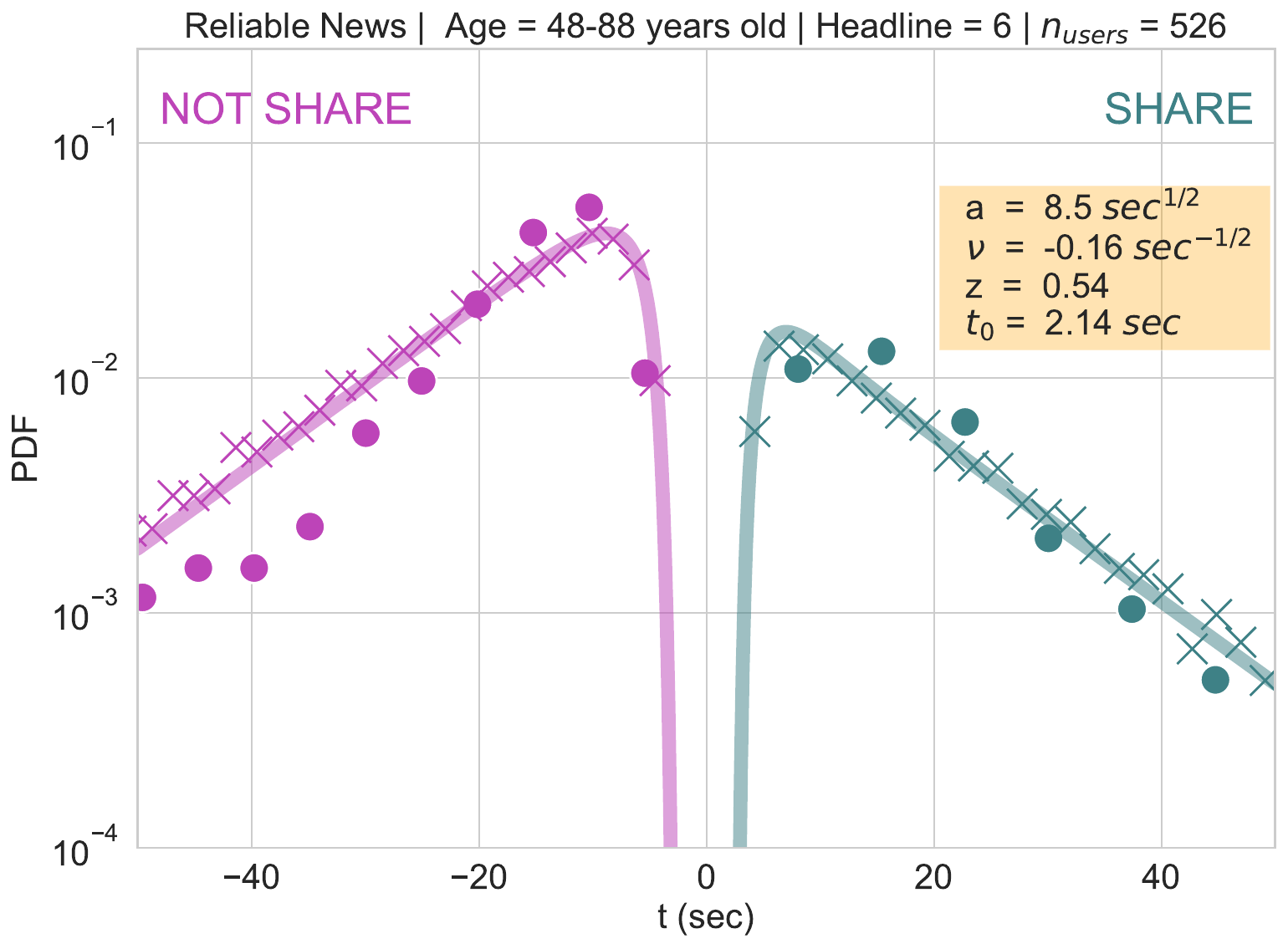}
            \caption{{\bf Headline 6}: Probability distribution of the response time for sharing and not sharing reliable information. Each figure corresponds to different age ranges. The solid line corresponds to theoretical results, dots correspond to empirical data and crosses to stochastic simulations.}
            \label{headline6Real}
        \end{figure}
        
        \begin{figure}[H]
            \renewcommand{\figurename}{Supplementary Figure}
            \centering
            \includegraphics[width=.45\textwidth]{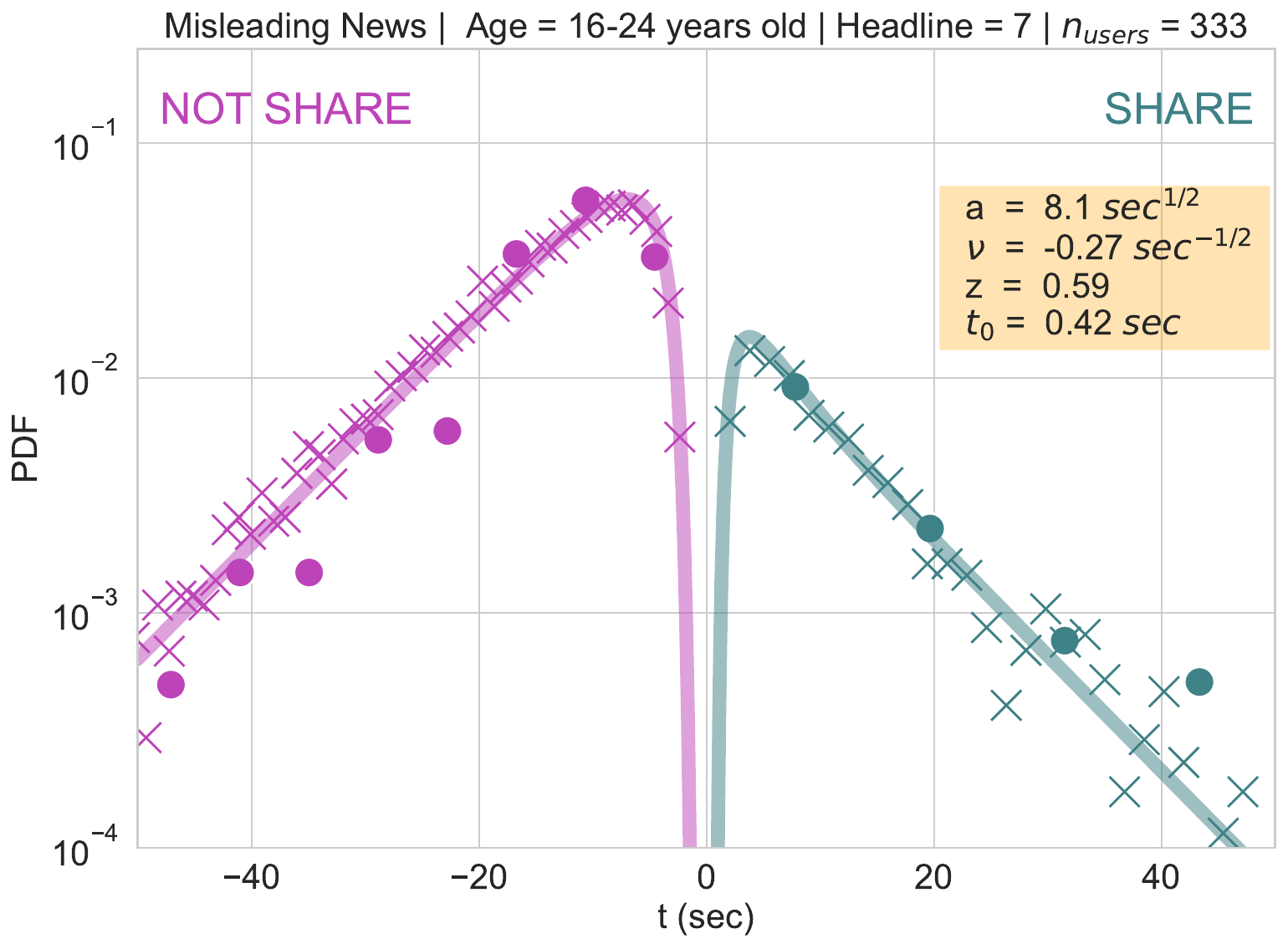}\quad
            \includegraphics[width=.45\textwidth]{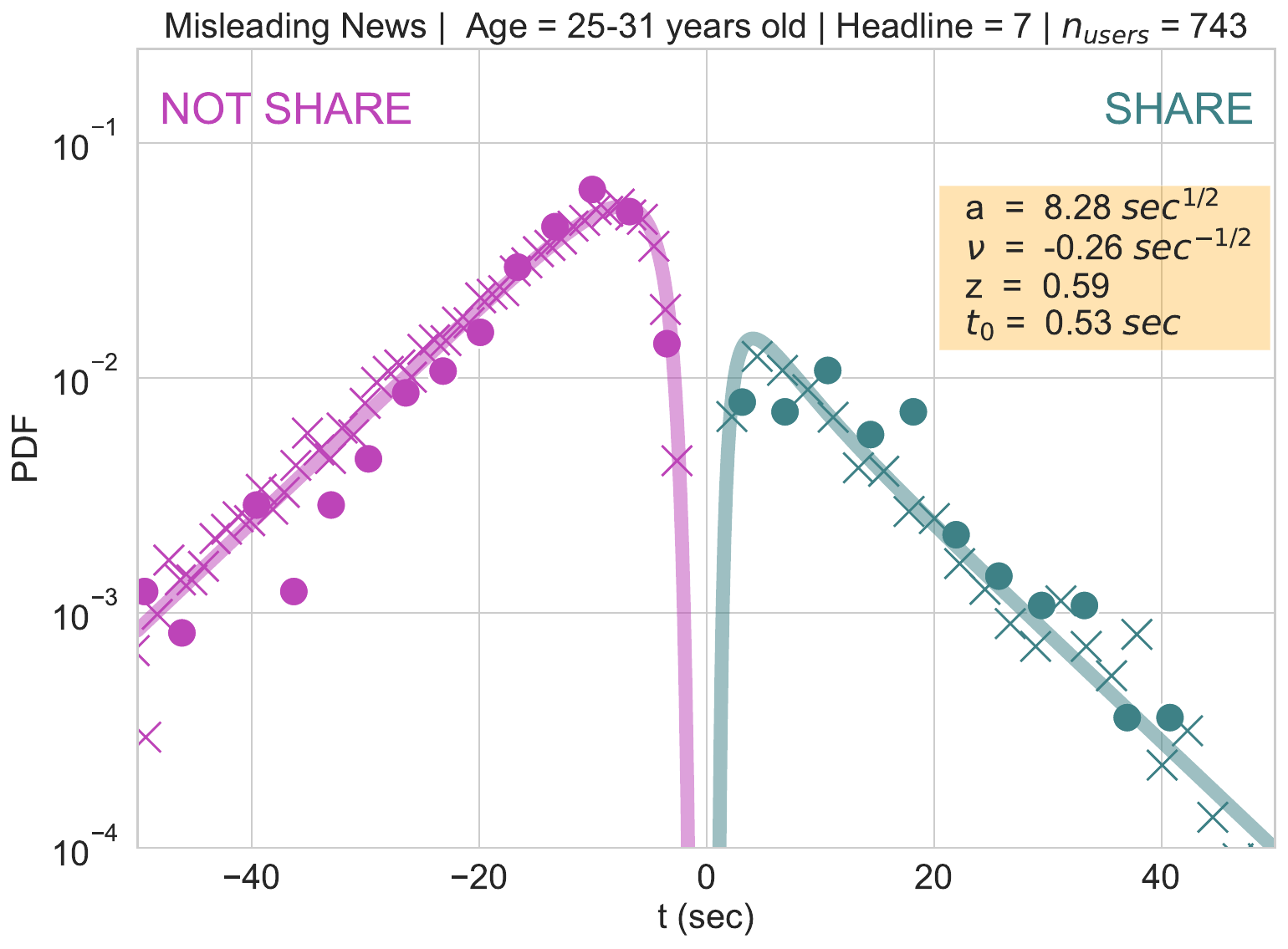}\quad
            \includegraphics[width=.45\textwidth]{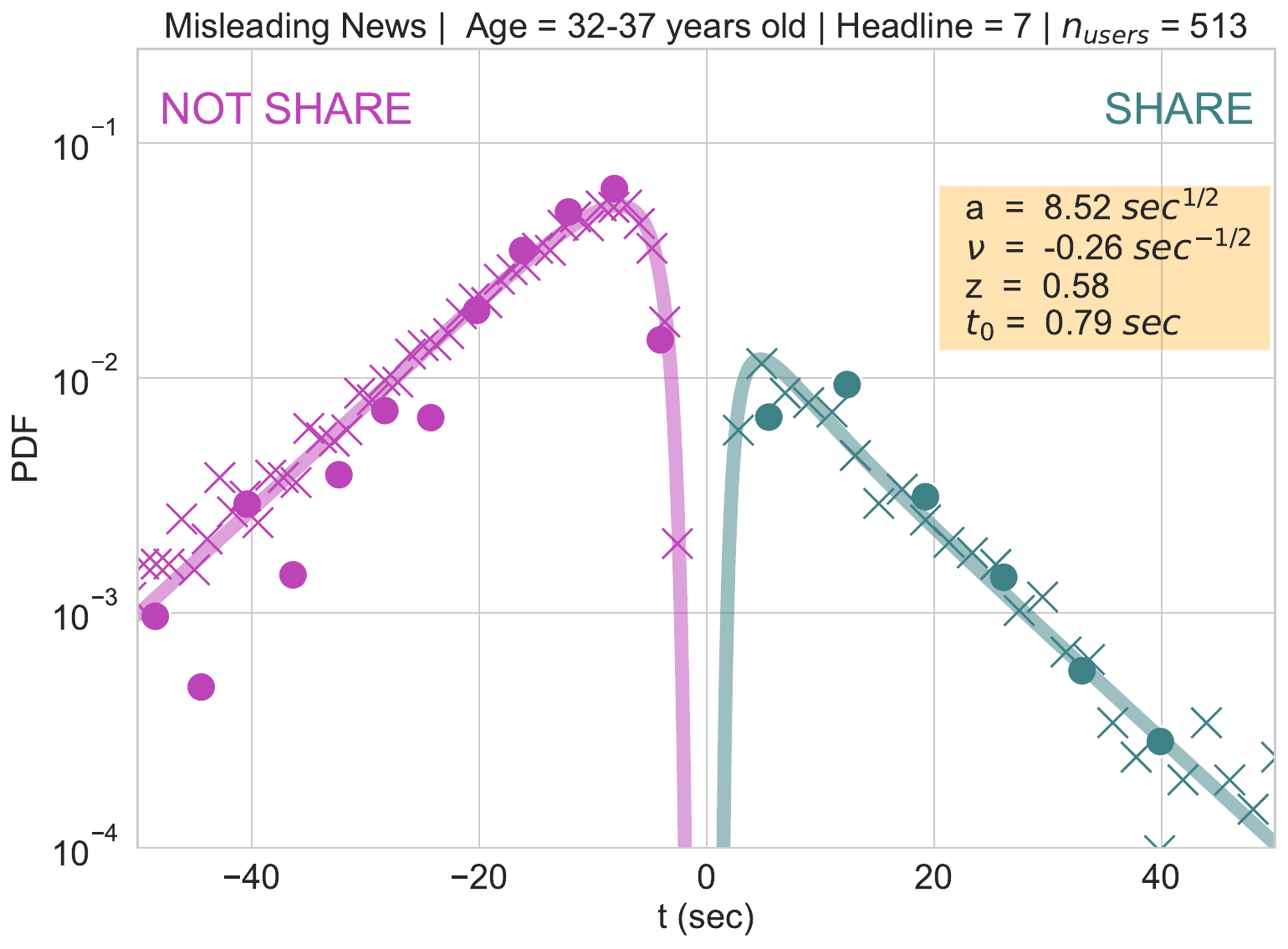}
            \medskip
            \includegraphics[width=.45\textwidth]{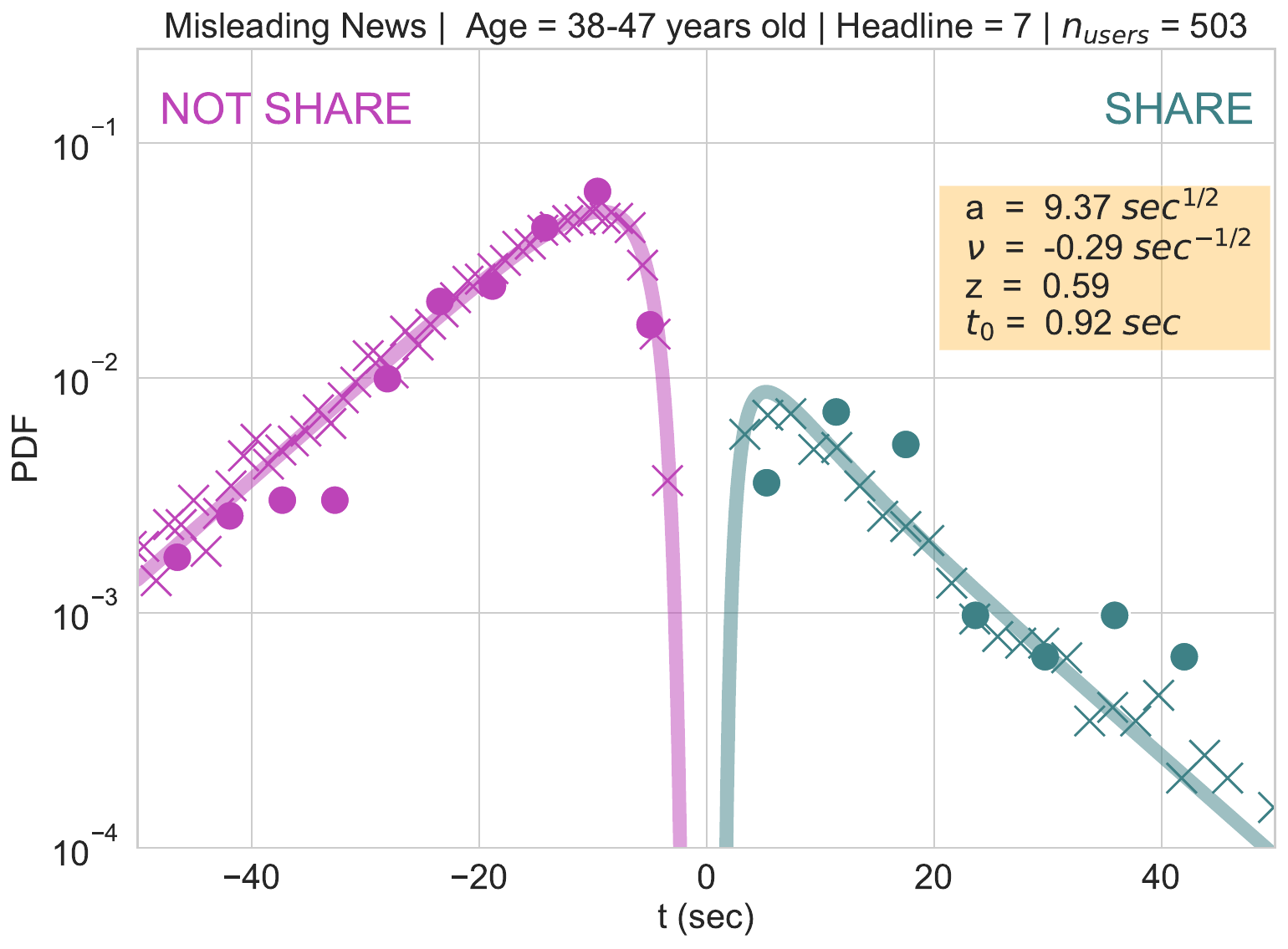}\quad
            \includegraphics[width=.45\textwidth]{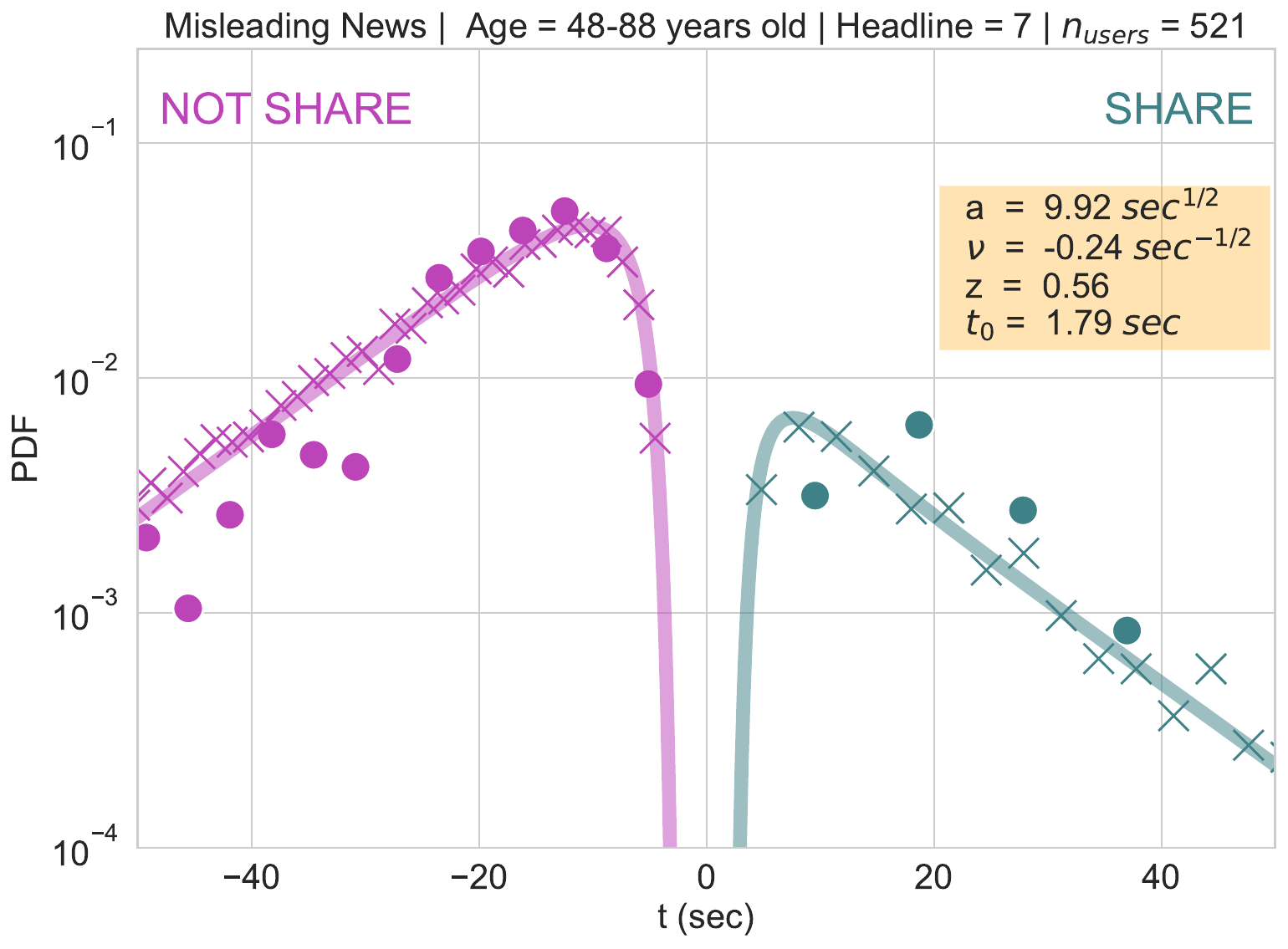}
            \caption{{\bf Headline 7}: Probability distribution of the response time for sharing and not sharing misleading information. Each figure corresponds to different age ranges. The solid line corresponds to theoretical results, dots correspond to empirical data and crosses to stochastic simulations.}
            \label{headline7Fake}
        \end{figure}
        
        \begin{figure}[H]
            \renewcommand{\figurename}{Supplementary Figure}
            \centering
            \includegraphics[width=.45\textwidth]{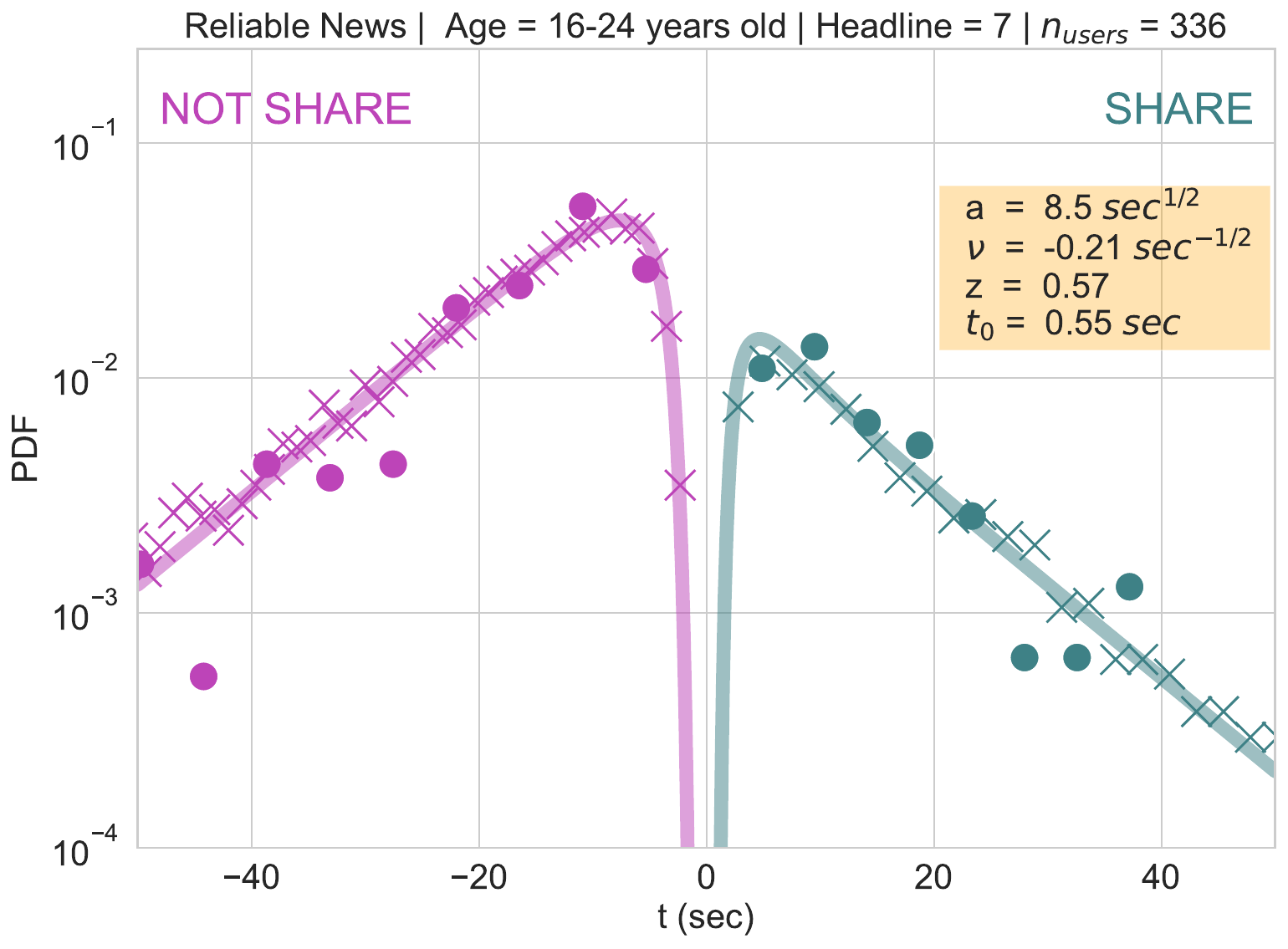}\quad
            \includegraphics[width=.45\textwidth]{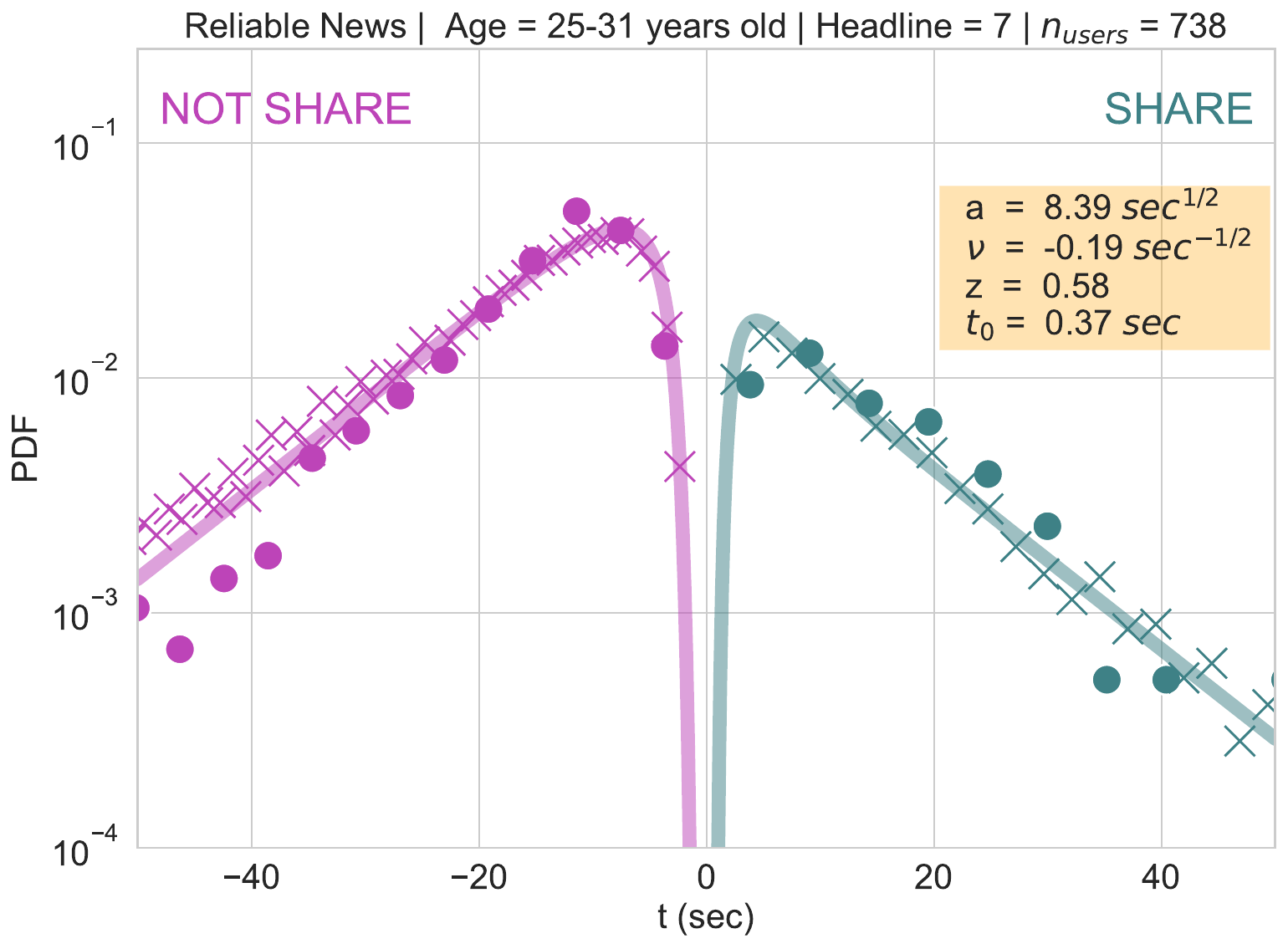}\quad
            \includegraphics[width=.45\textwidth]{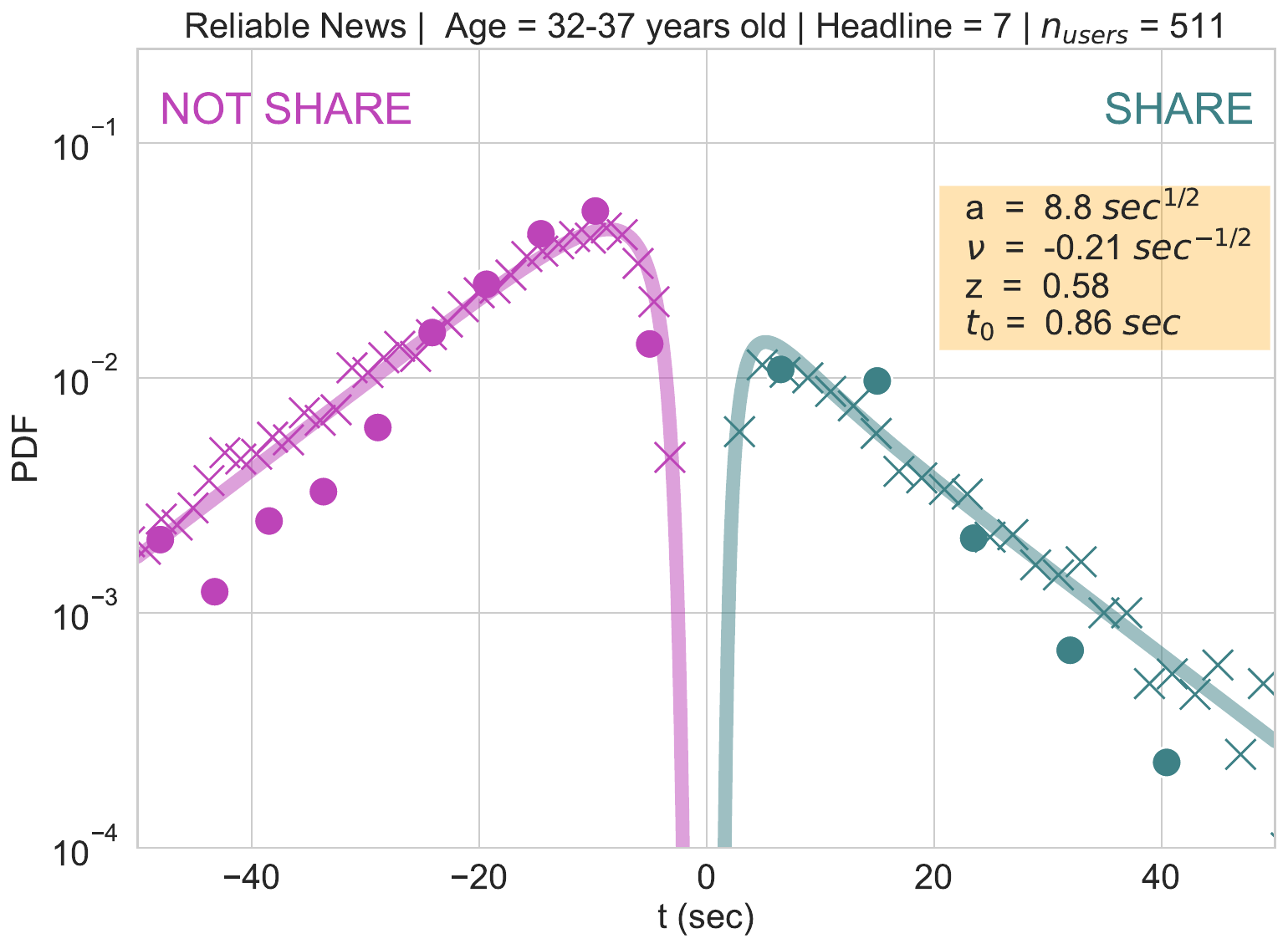}
            \medskip
            \includegraphics[width=.45\textwidth]{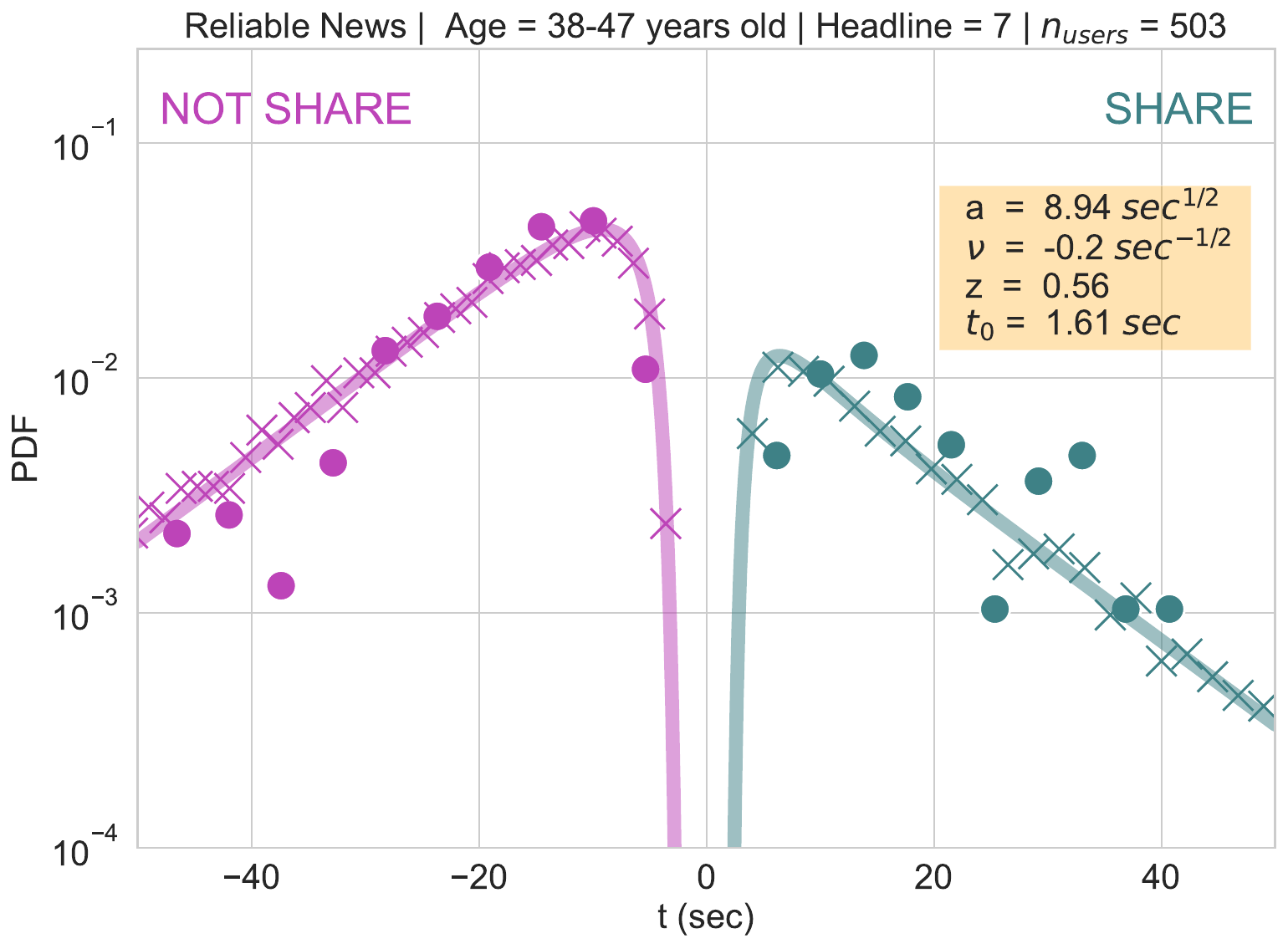}\quad
            \includegraphics[width=.45\textwidth]{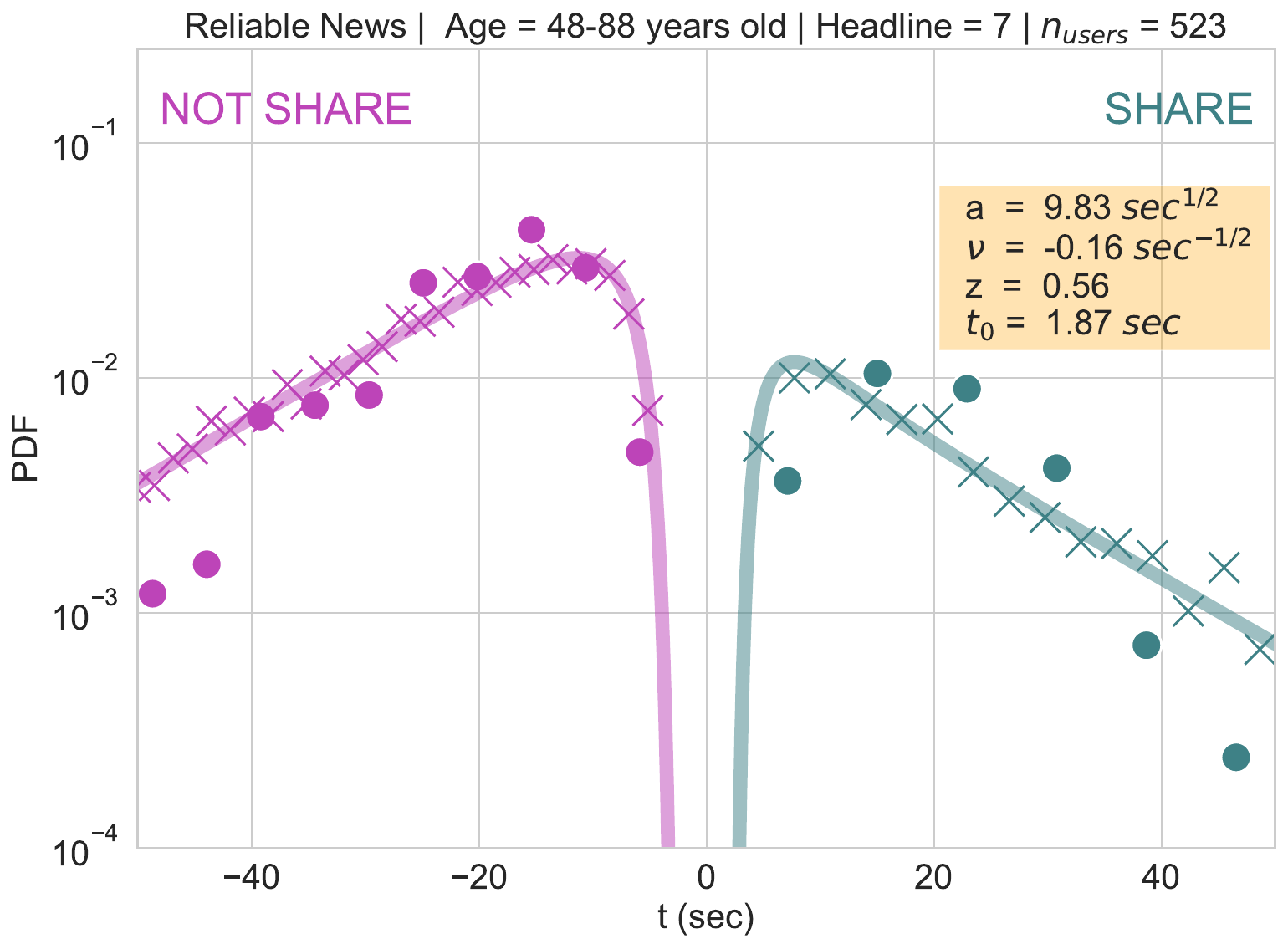}
            \caption{{\bf Headline 7}: Probability distribution of the response time for sharing and not sharing reliable information. Each figure corresponds to different age ranges. The solid line corresponds to theoretical results, dots correspond to empirical data and crosses to stochastic simulations.}
            \label{headline7Real}
        \end{figure}
        
        \begin{figure}[H]
            \renewcommand{\figurename}{Supplementary Figure}
            \centering
            \includegraphics[width=.45\textwidth]{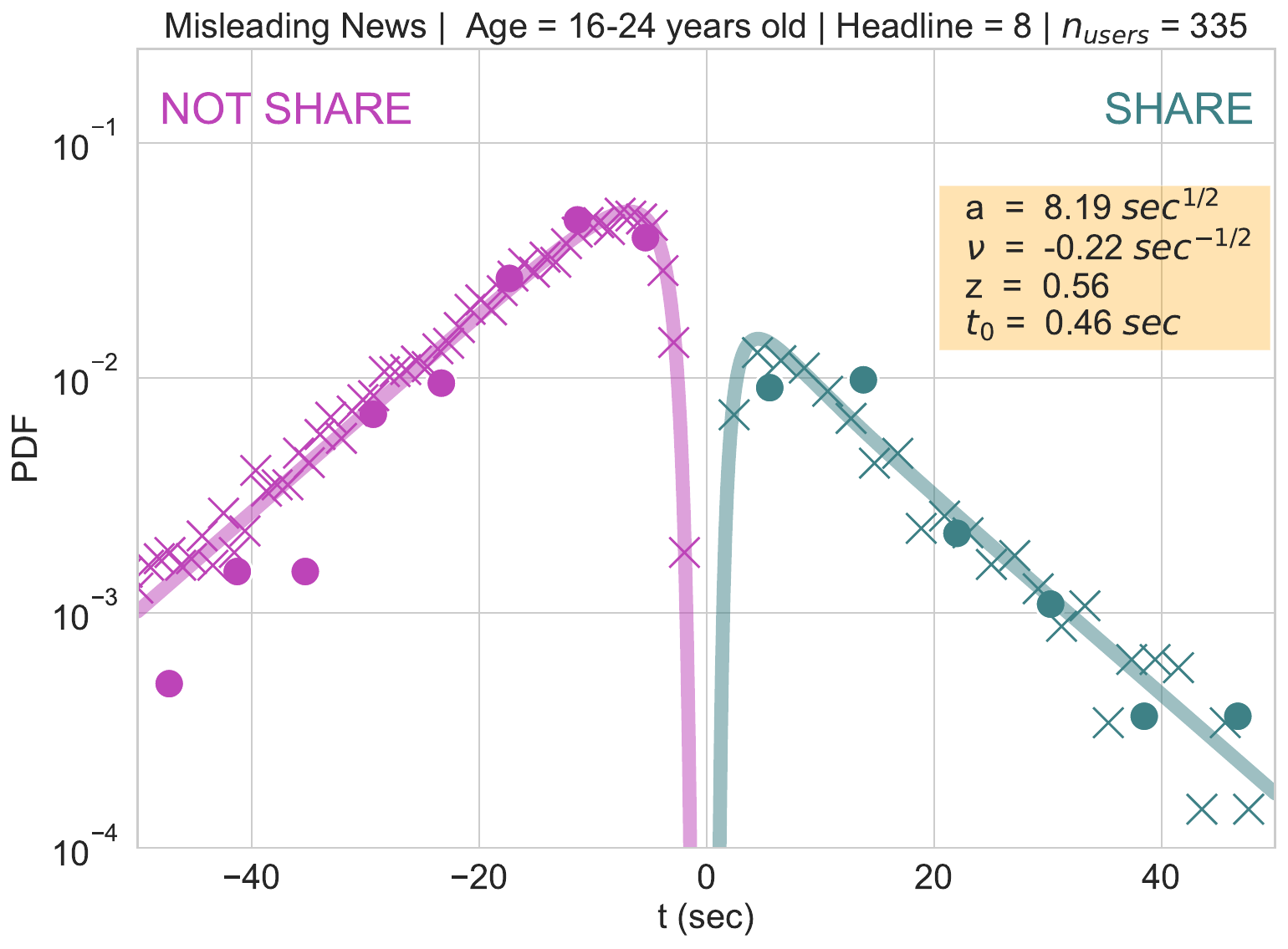}\quad
            \includegraphics[width=.45\textwidth]{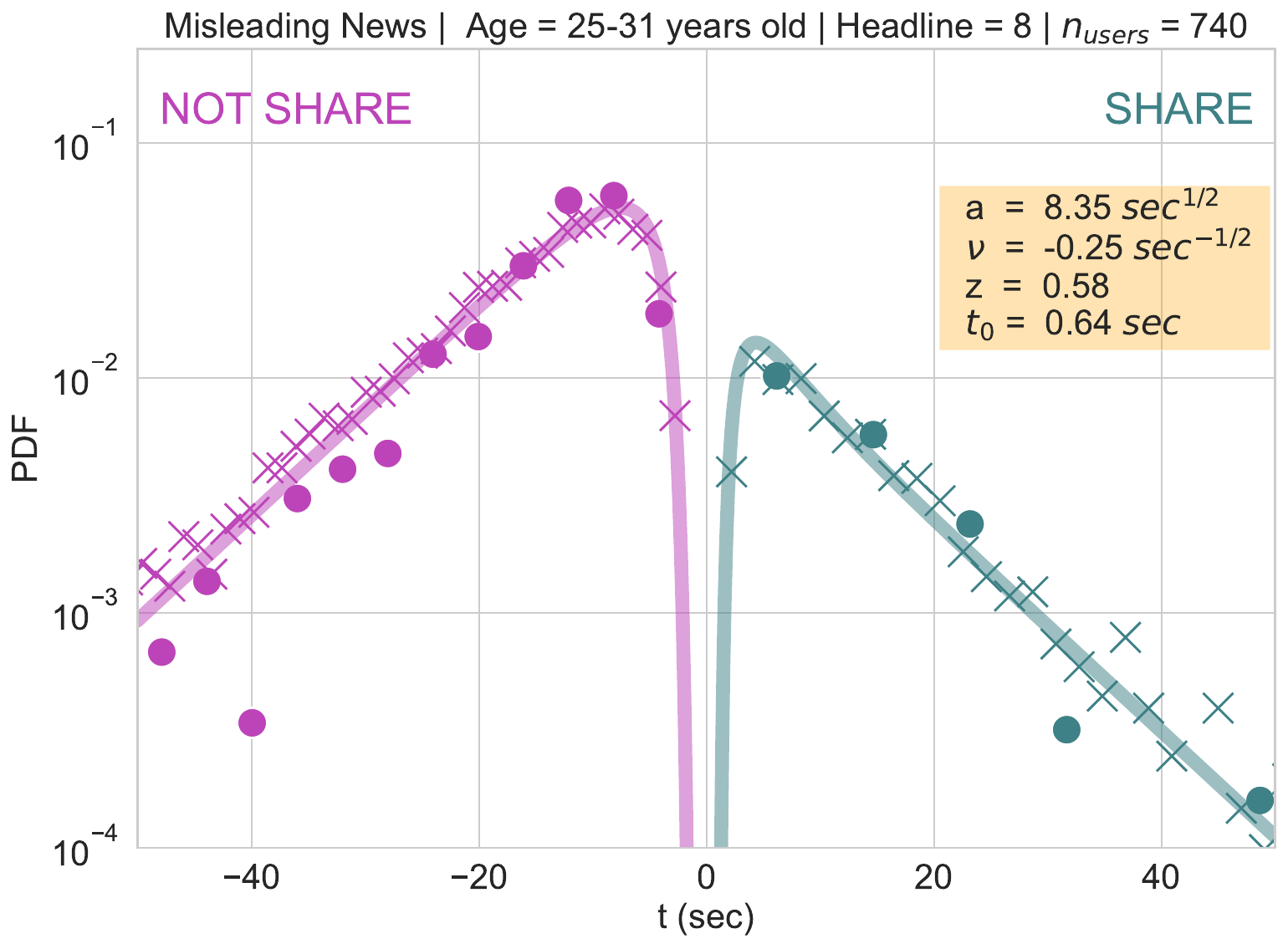}\quad
            \includegraphics[width=.45\textwidth]{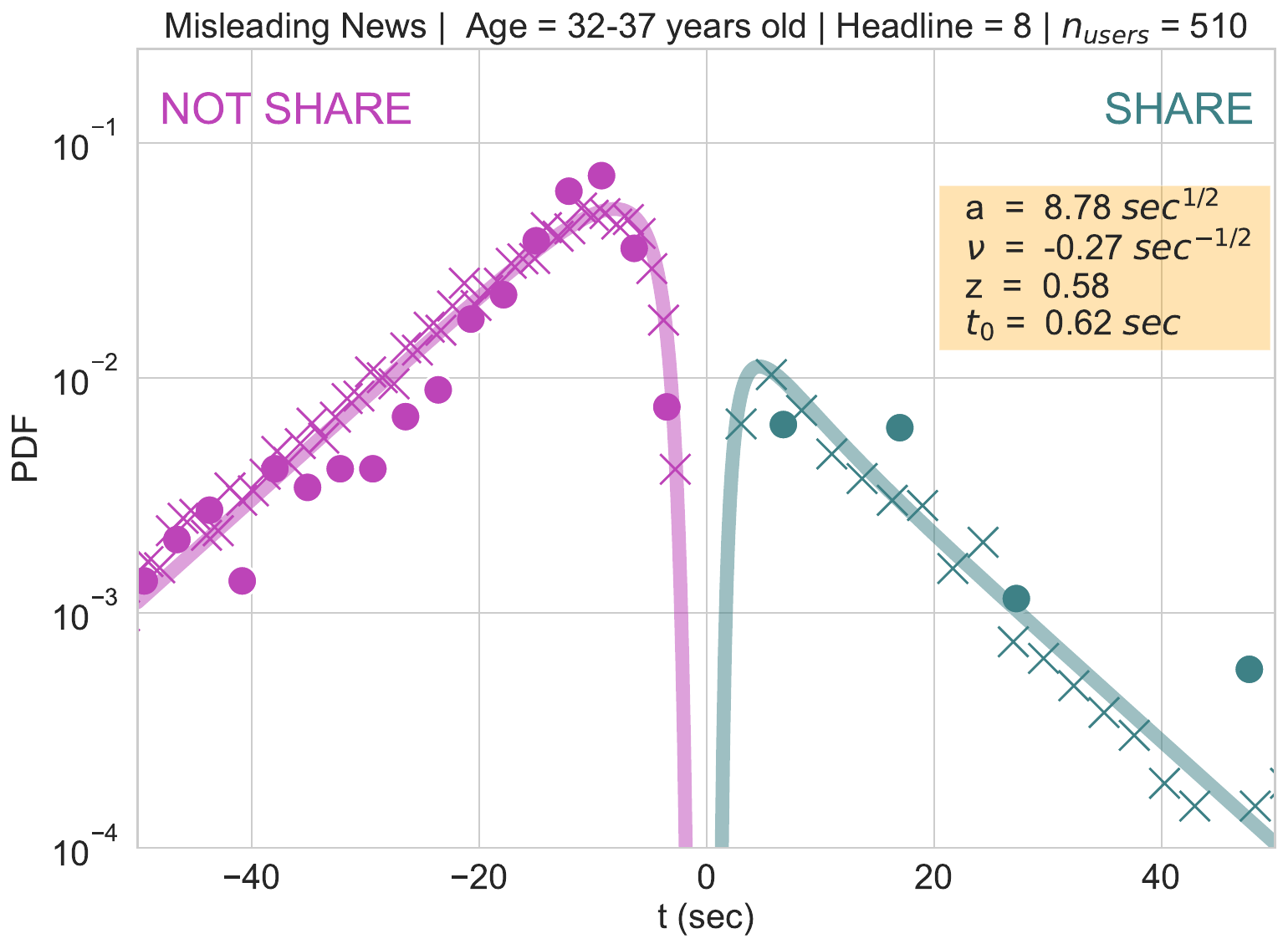}
            \medskip
            \includegraphics[width=.45\textwidth]{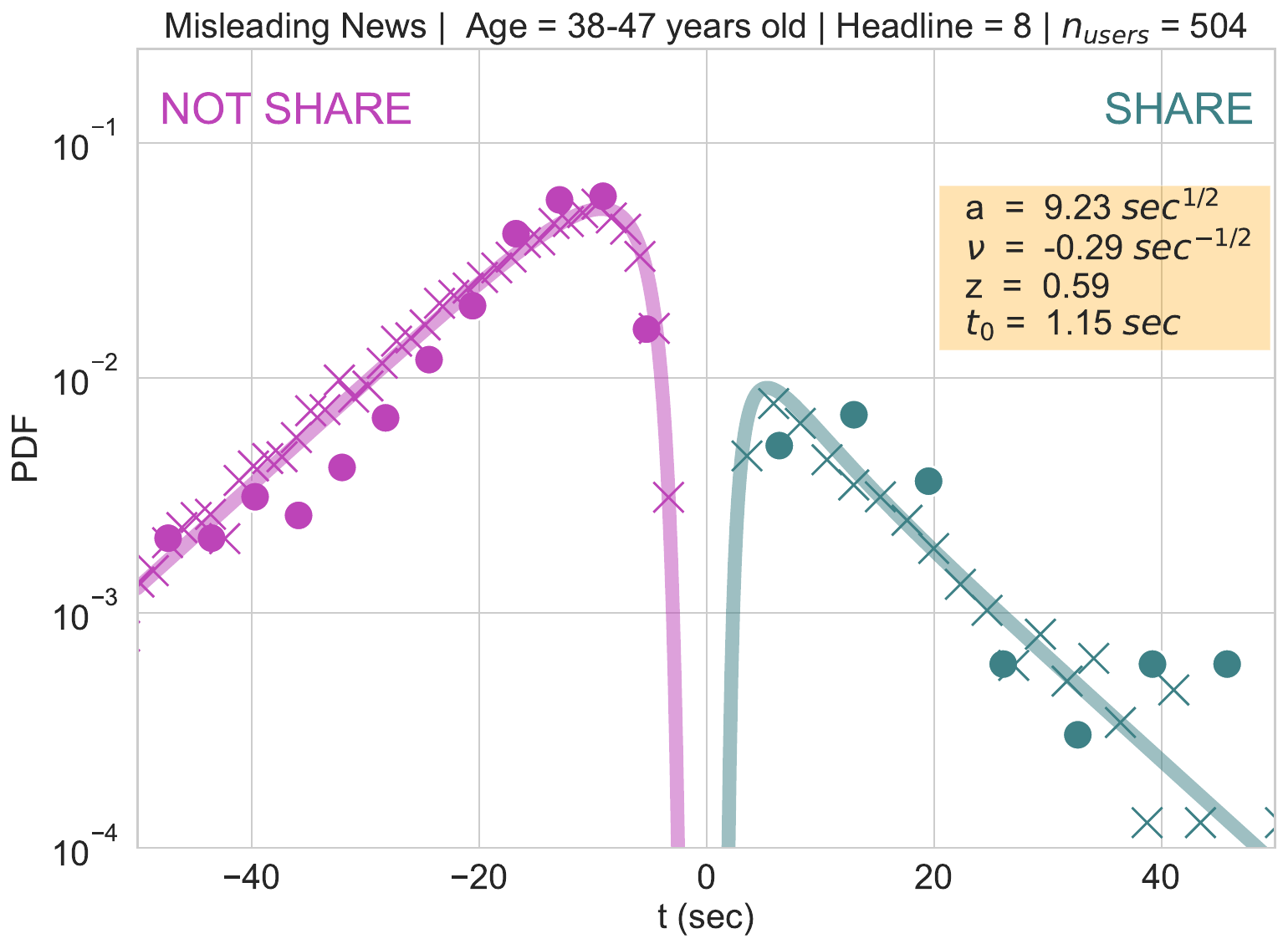}\quad
            \includegraphics[width=.45\textwidth]{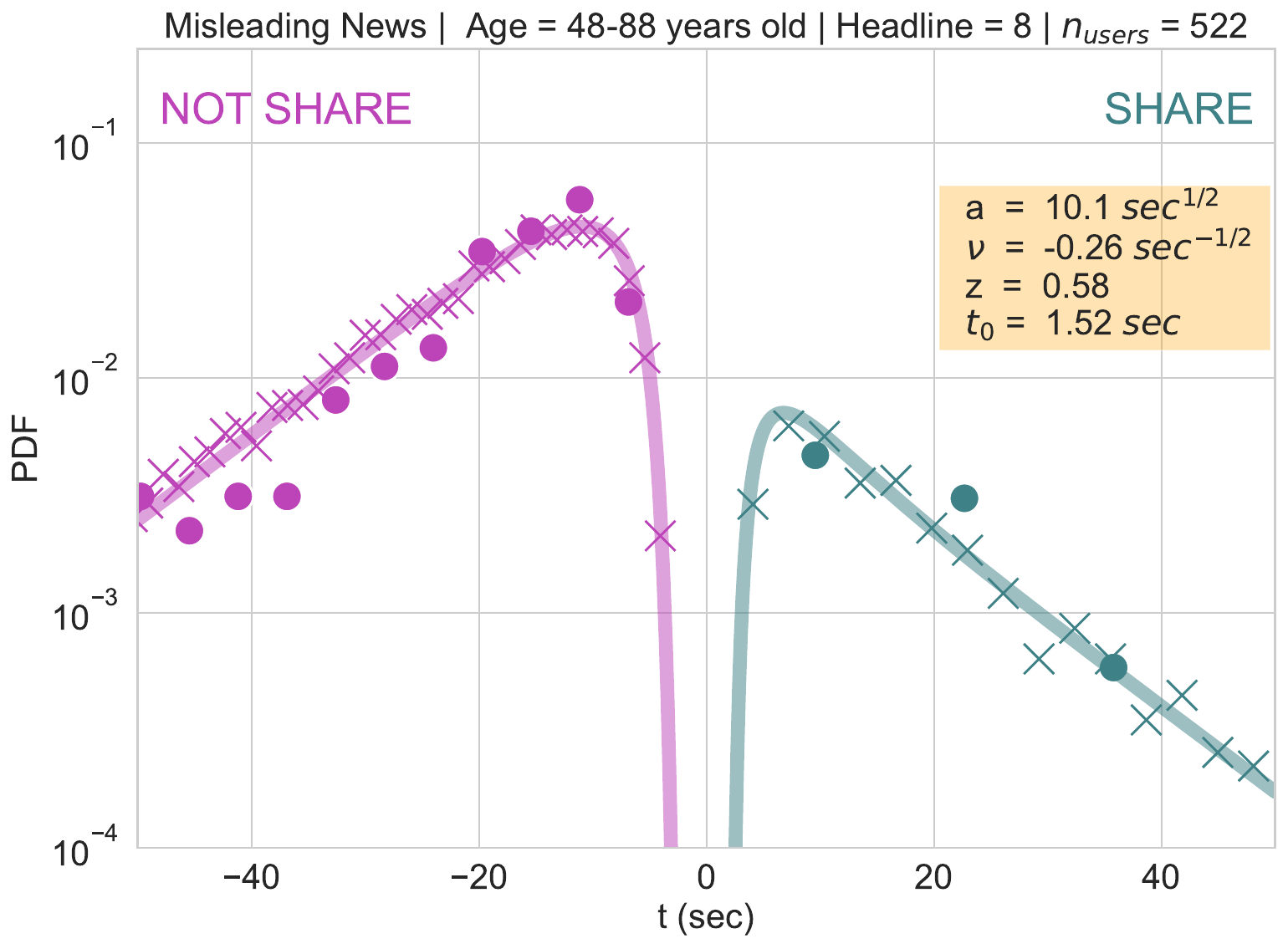}
            \caption{{\bf Headline 8}: Probability distribution of the response time for sharing and not sharing misleading information. Each figure corresponds to different age ranges. The solid line corresponds to theoretical results, dots correspond to empirical data and crosses to stochastic simulations.}
            \label{headline8Fake}
        \end{figure}
        
        \begin{figure}[H]
            \renewcommand{\figurename}{Supplementary Figure}
            \centering
            \includegraphics[width=.45\textwidth]{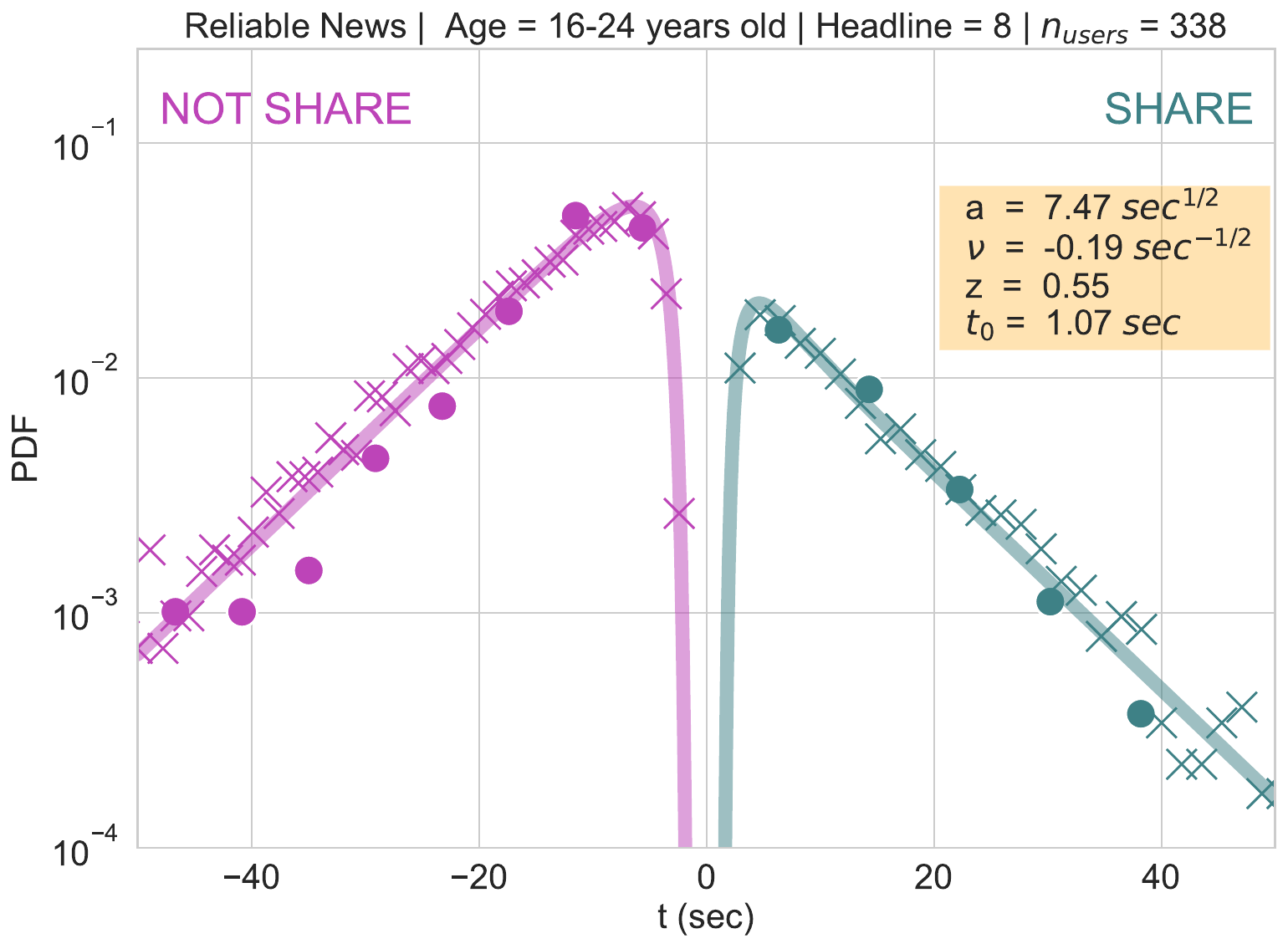}\quad
            \includegraphics[width=.45\textwidth]{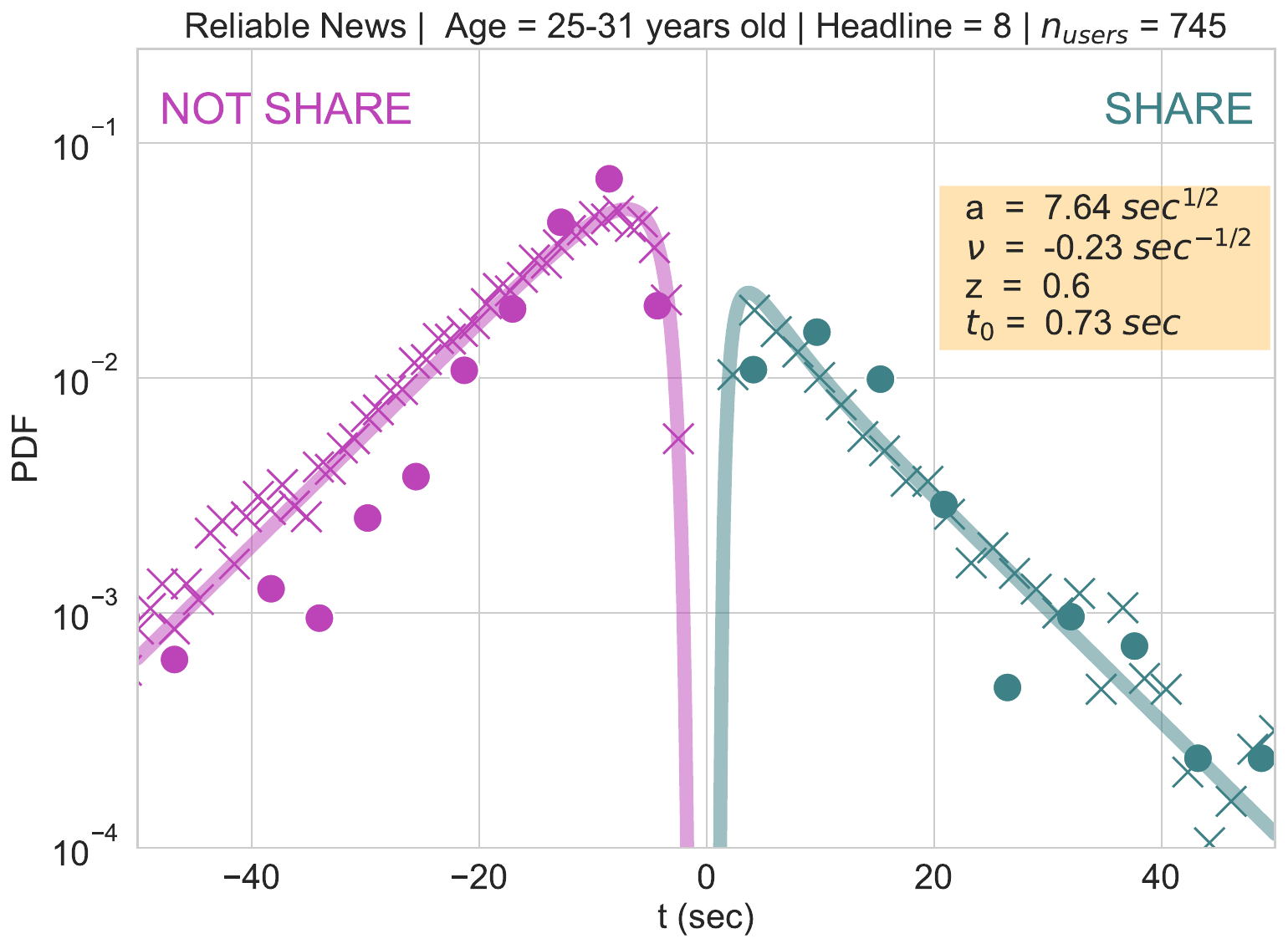}\quad
            \includegraphics[width=.45\textwidth]{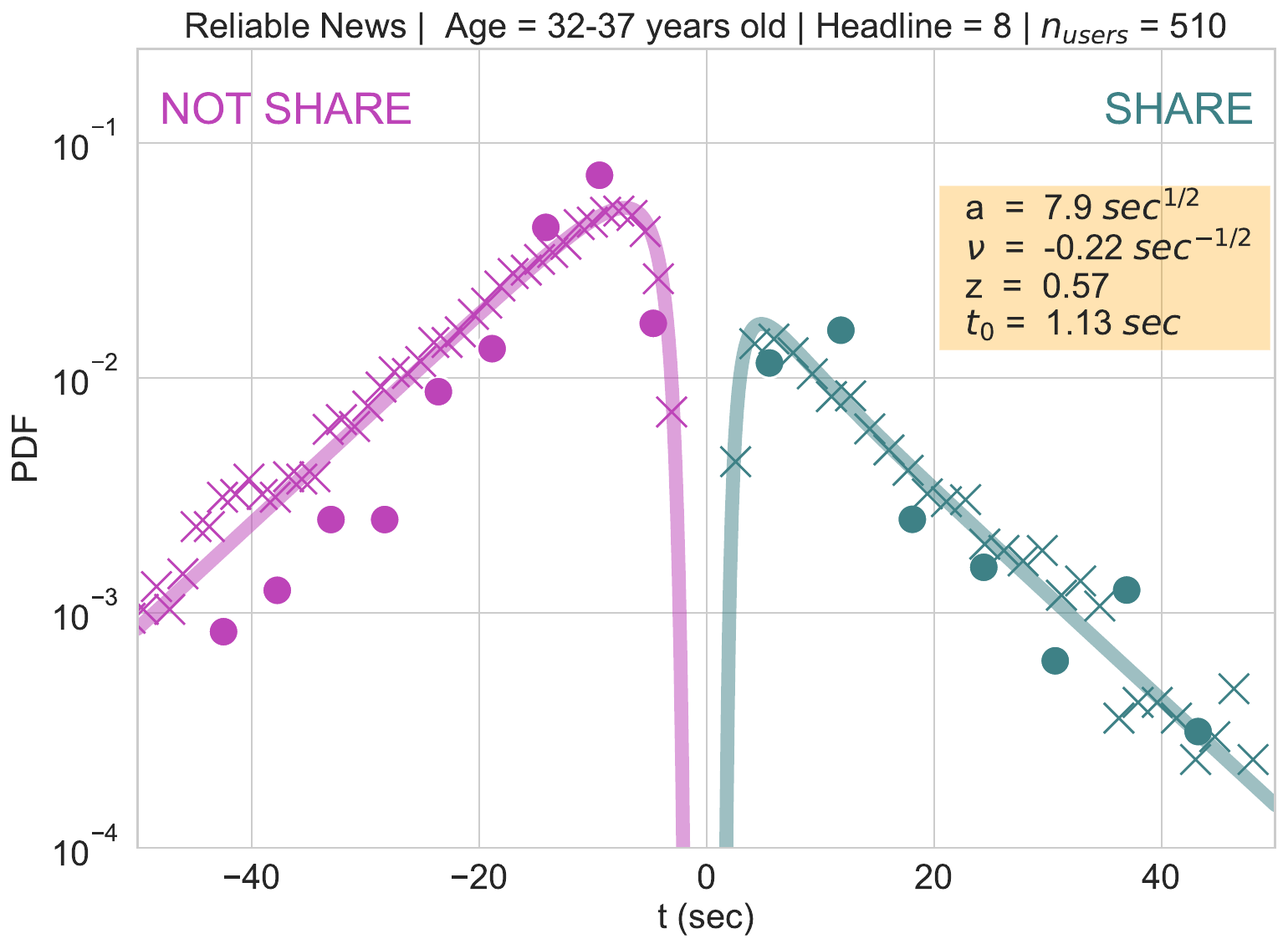}
            \medskip
            \includegraphics[width=.45\textwidth]{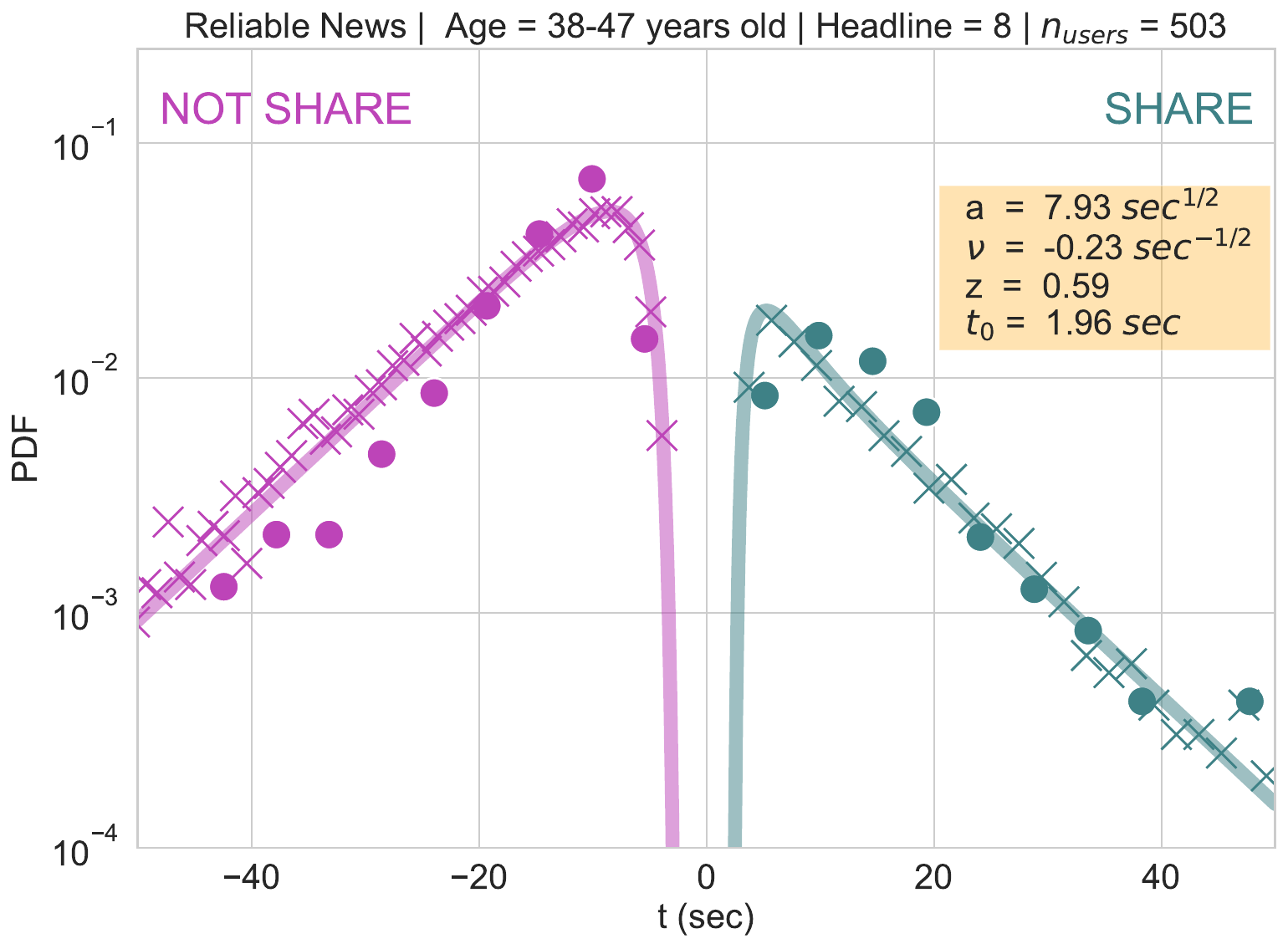}\quad
            \includegraphics[width=.45\textwidth]{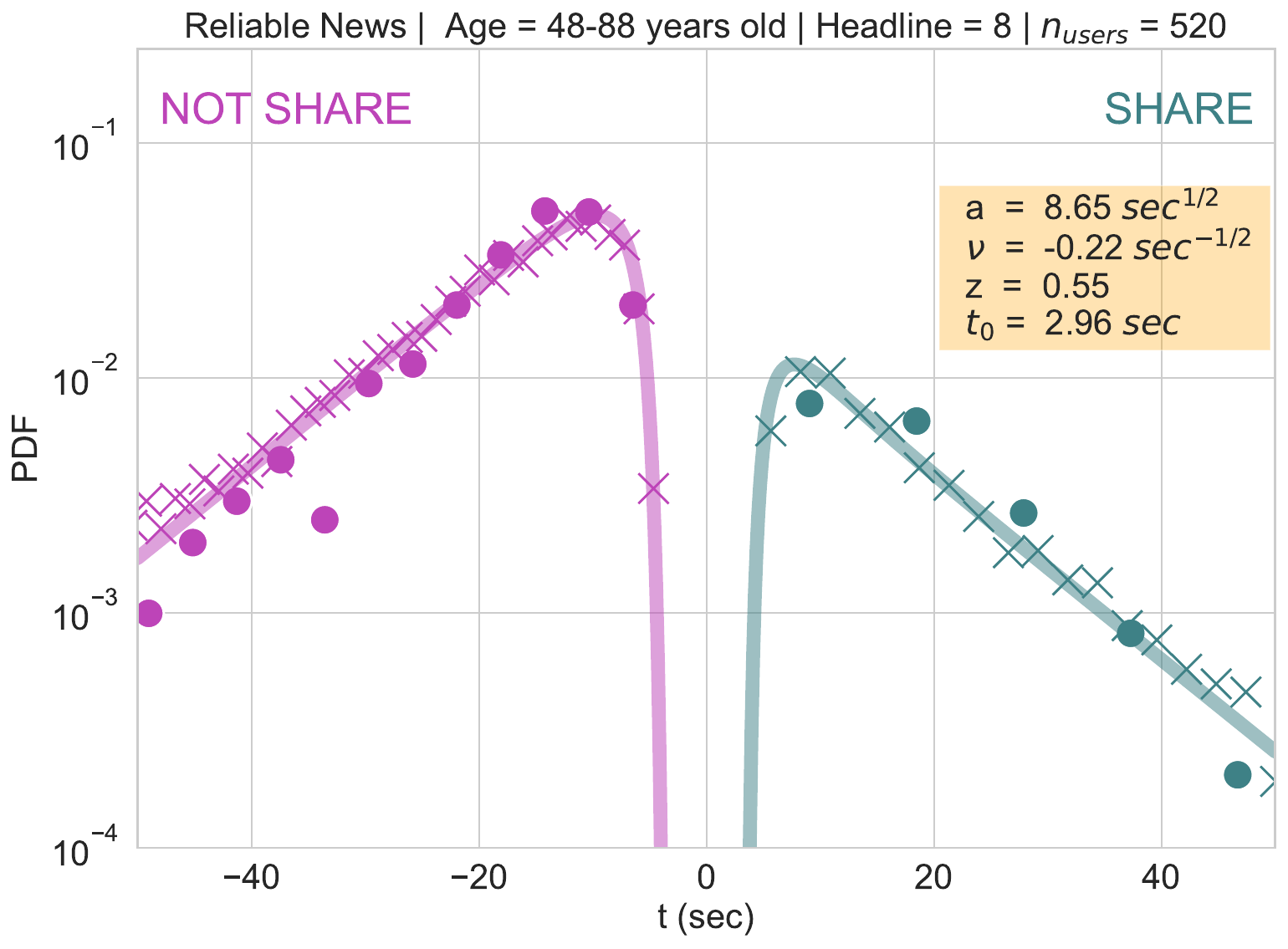}
            \caption{{\bf Headline 8}: Probability distribution of the response time for sharing and not sharing reliable information. Each figure corresponds to different age ranges. The solid line corresponds to theoretical results, dots correspond to empirical data and crosses to stochastic simulations.}
            \label{headline8Real}
        \end{figure}
        
        \begin{figure}[H]
            \renewcommand{\figurename}{Supplementary Figure}
            \centering
            \includegraphics[width=.45\textwidth]{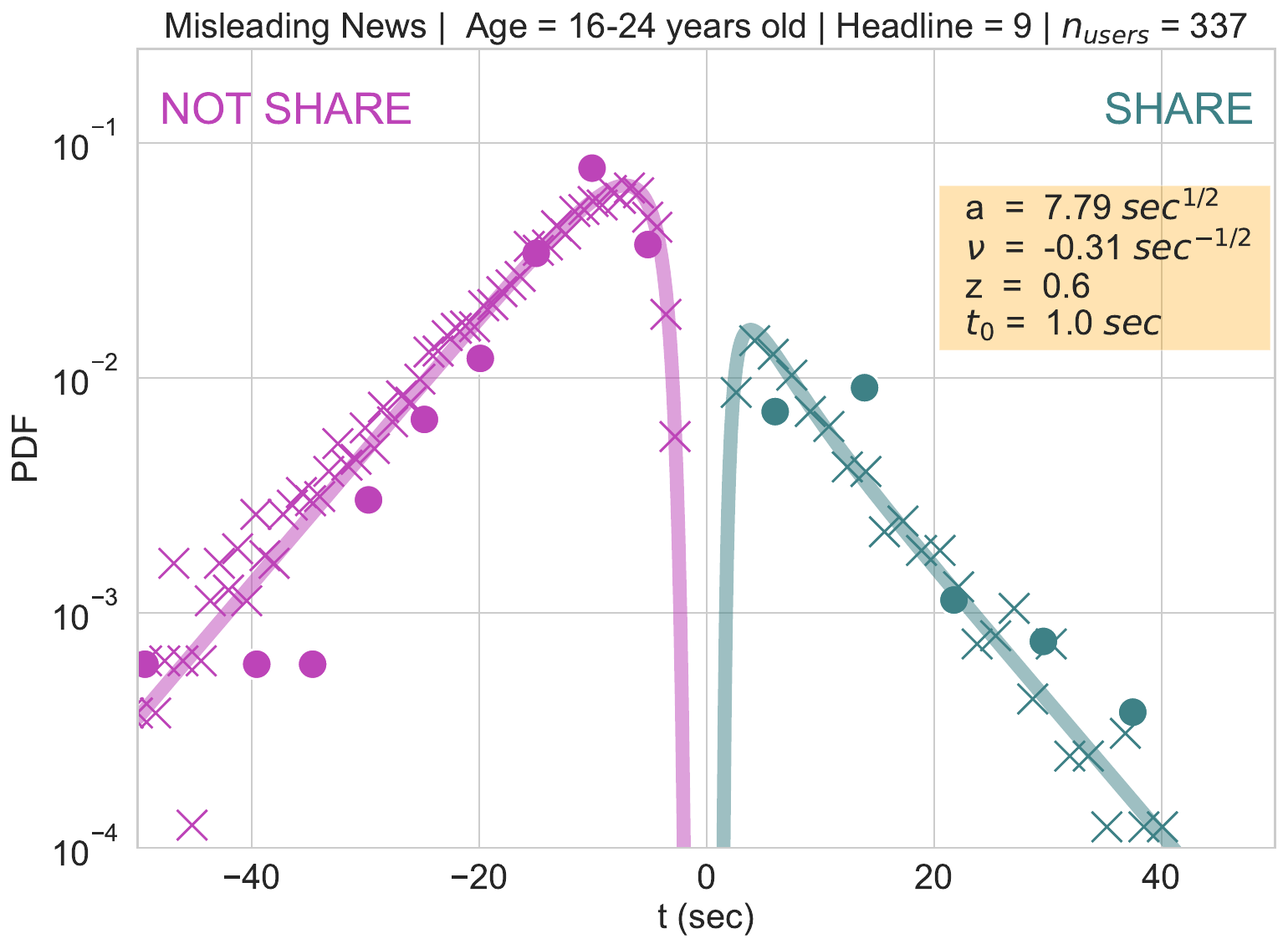}\quad
            \includegraphics[width=.45\textwidth]{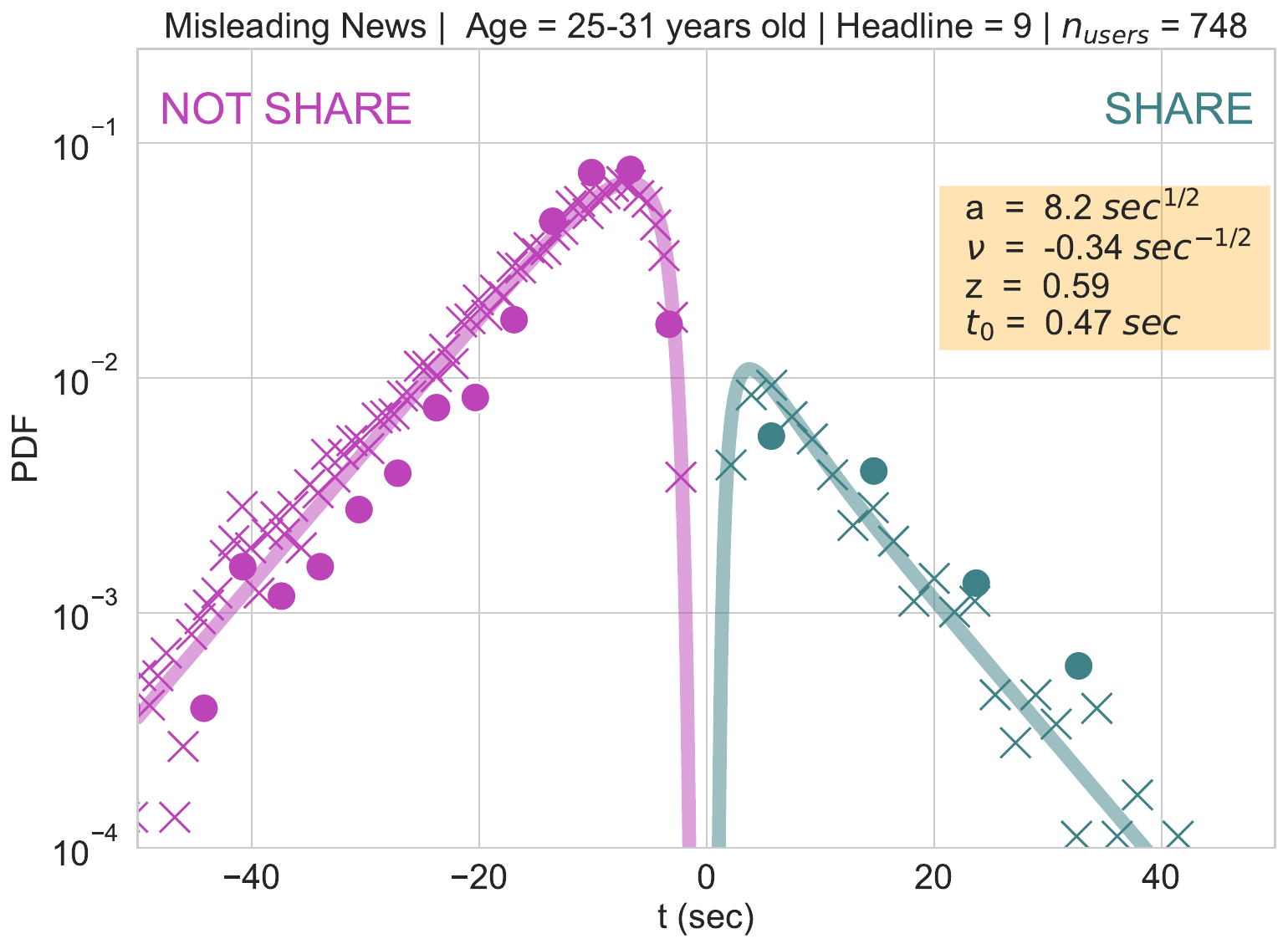}\quad
            \includegraphics[width=.45\textwidth]{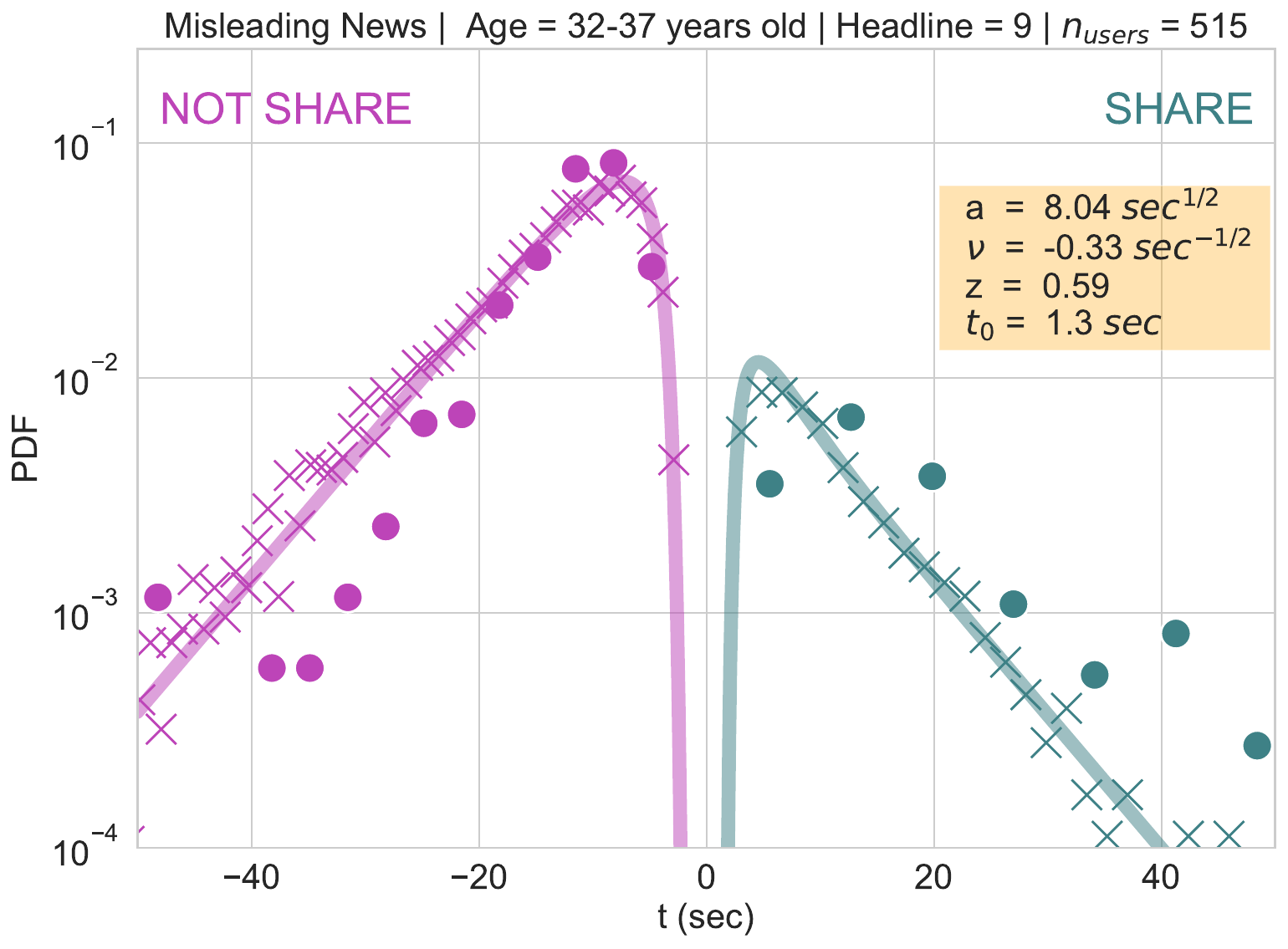}
            \medskip
            \includegraphics[width=.45\textwidth]{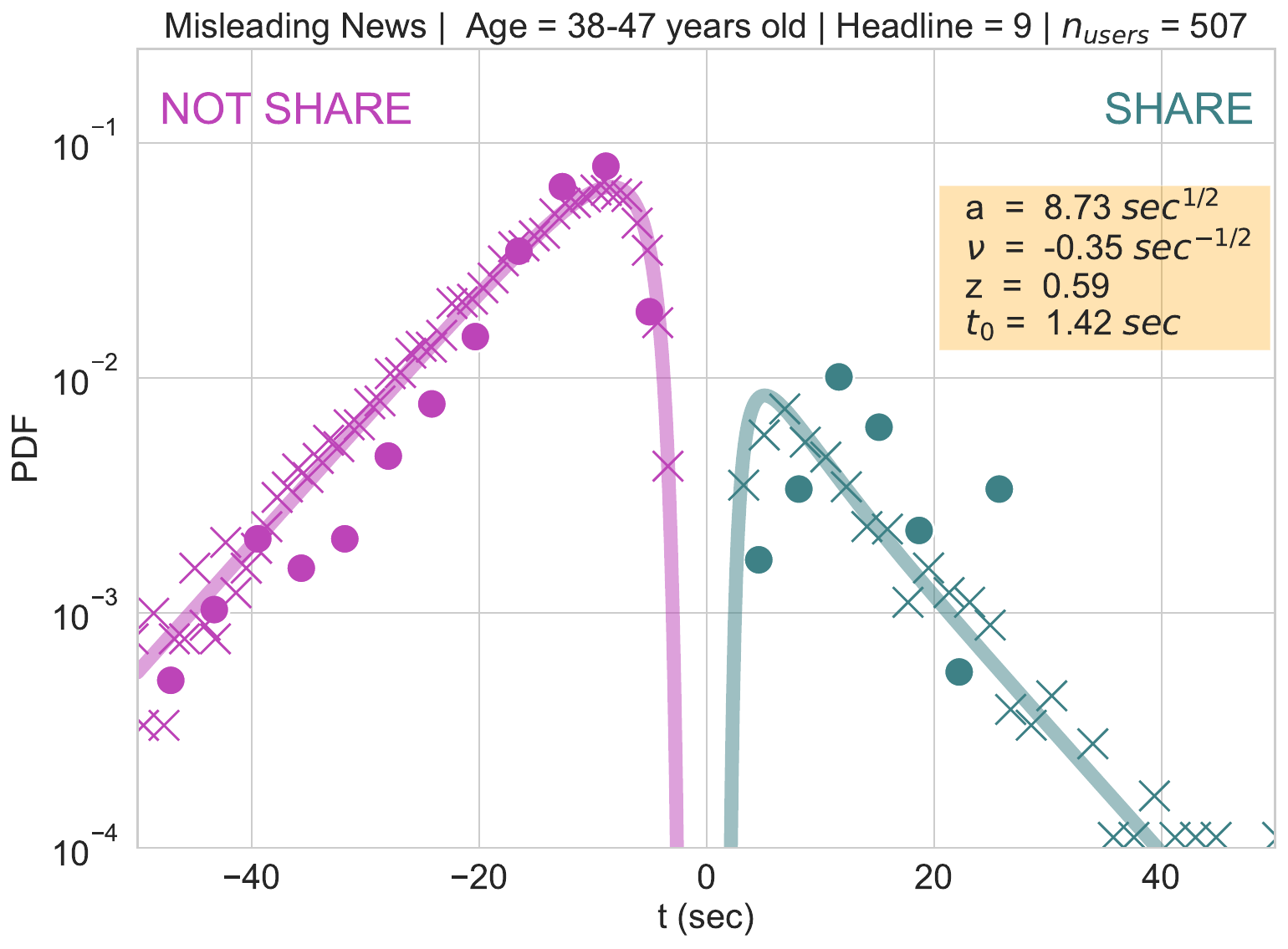}\quad
            \includegraphics[width=.45\textwidth]{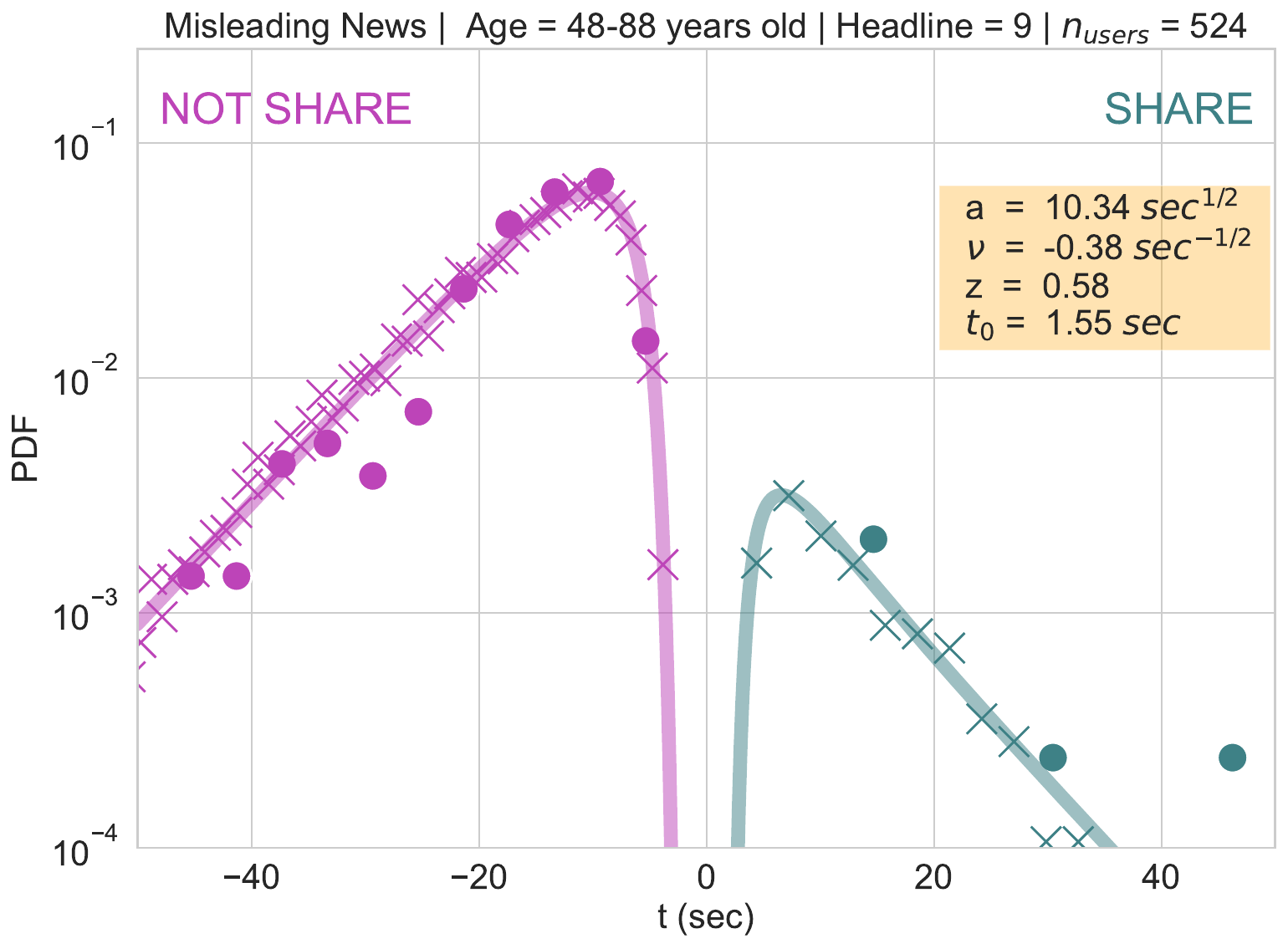}
            \caption{{\bf Headline 9}: Probability distribution of the response time for sharing and not sharing misleading information. Each figure corresponds to different age ranges. The solid line corresponds to theoretical results, dots correspond to empirical data and crosses to stochastic simulations.}
            \label{headline9Fake}
        \end{figure}
        
        \begin{figure}[H]
            \renewcommand{\figurename}{Supplementary Figure}
            \centering
            \includegraphics[width=.45\textwidth]{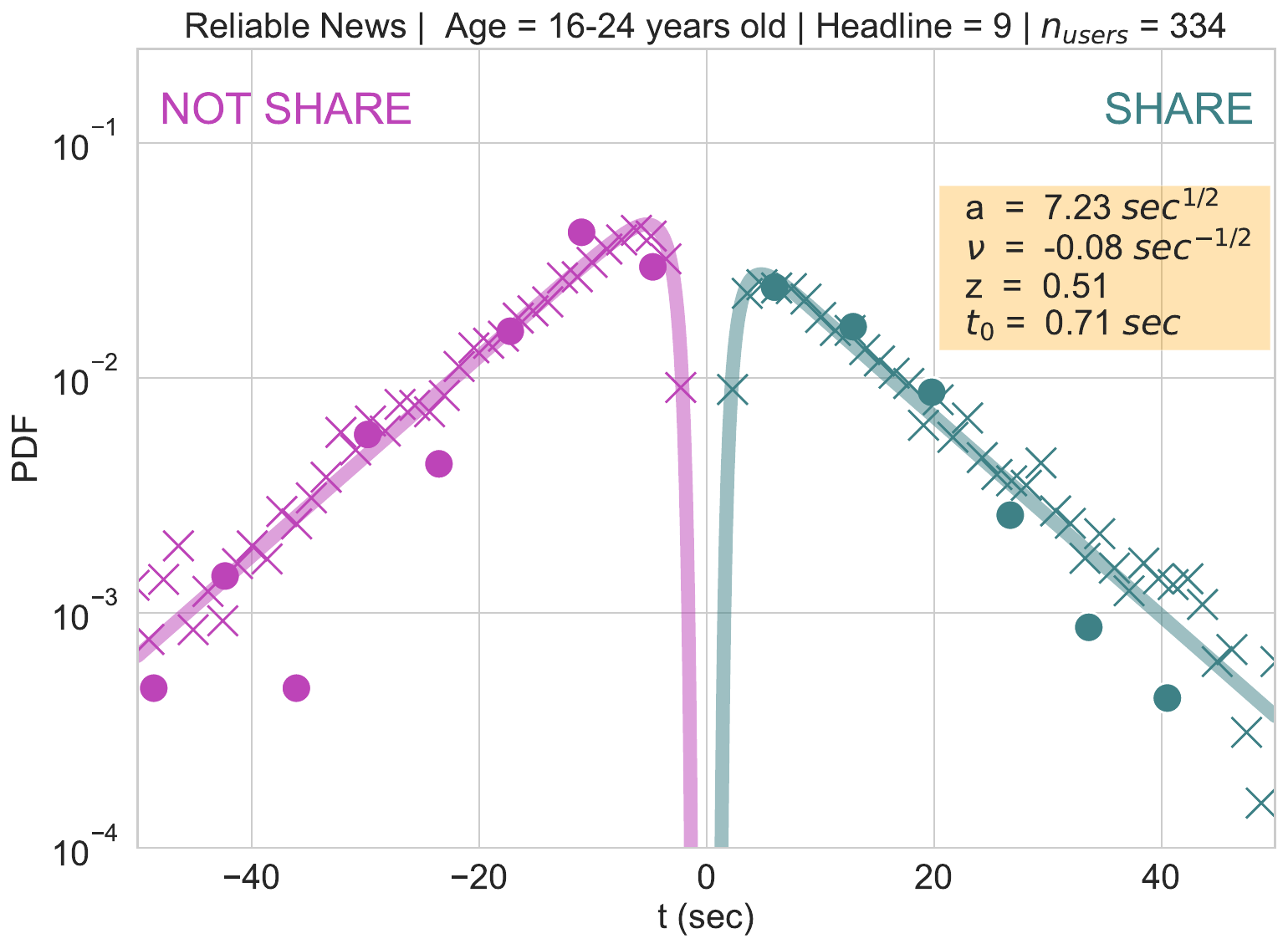}\quad
            \includegraphics[width=.45\textwidth]{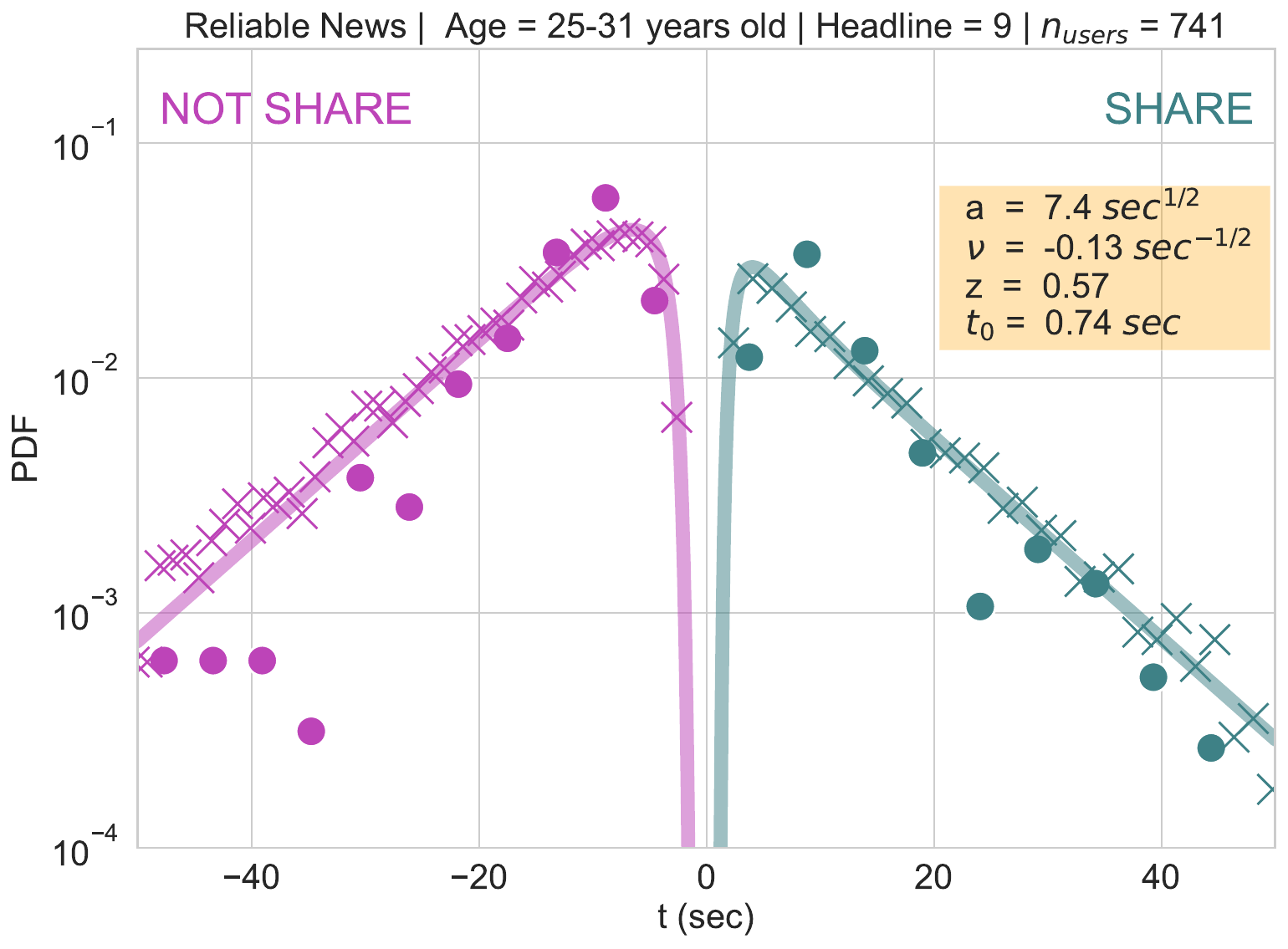}\quad
            \includegraphics[width=.45\textwidth]{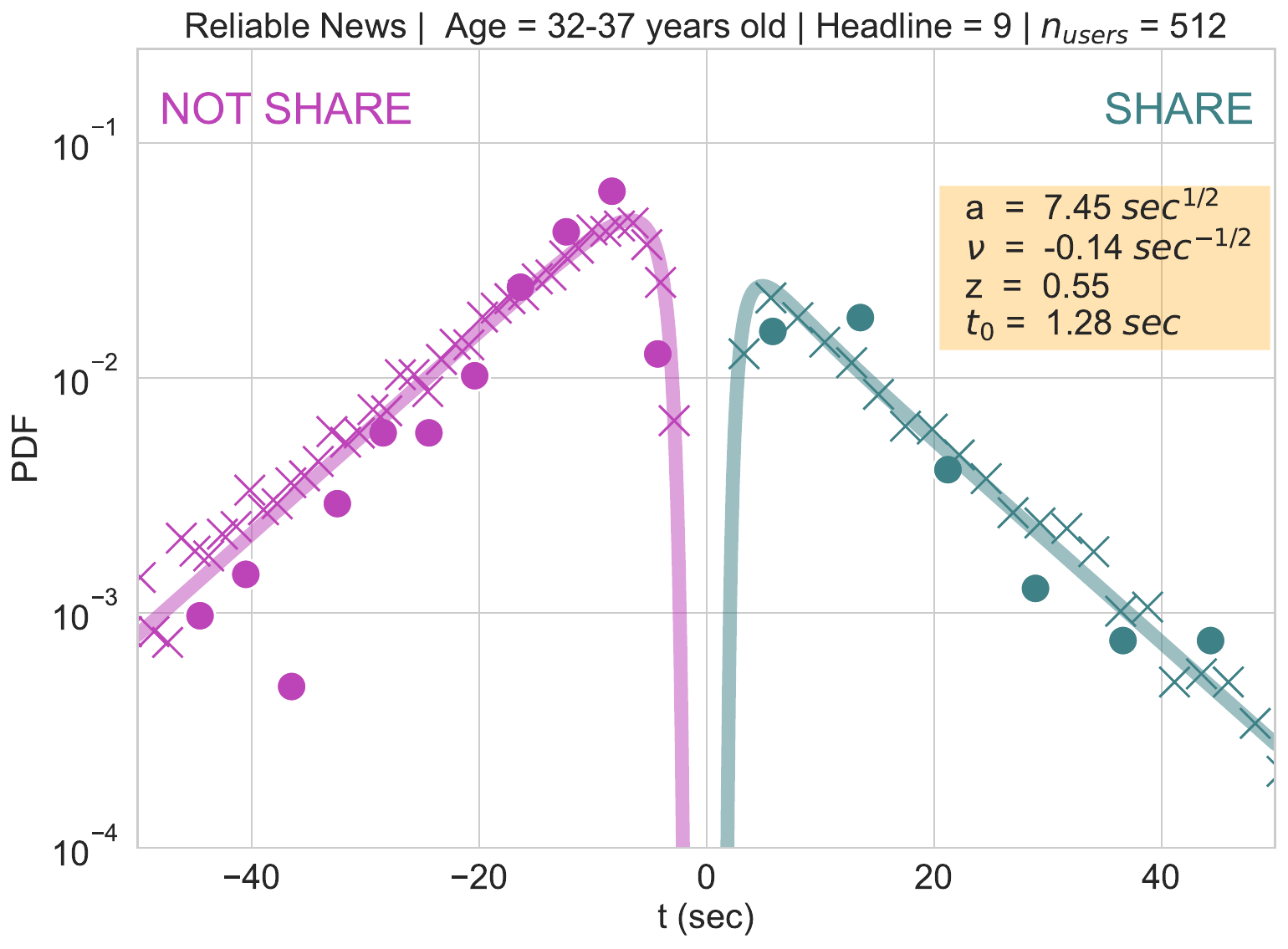}
            \medskip
            \includegraphics[width=.45\textwidth]{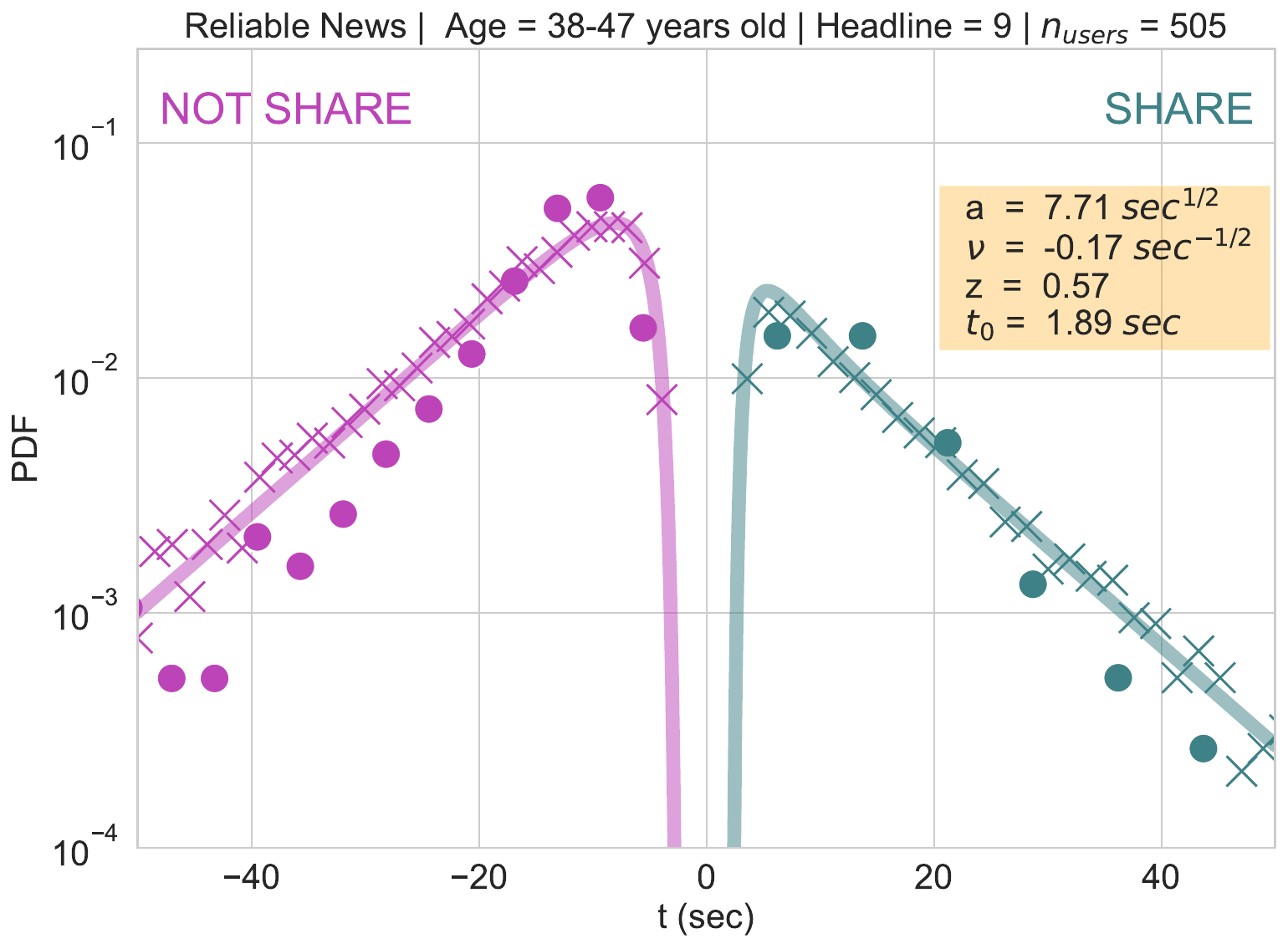}\quad
            \includegraphics[width=.45\textwidth]{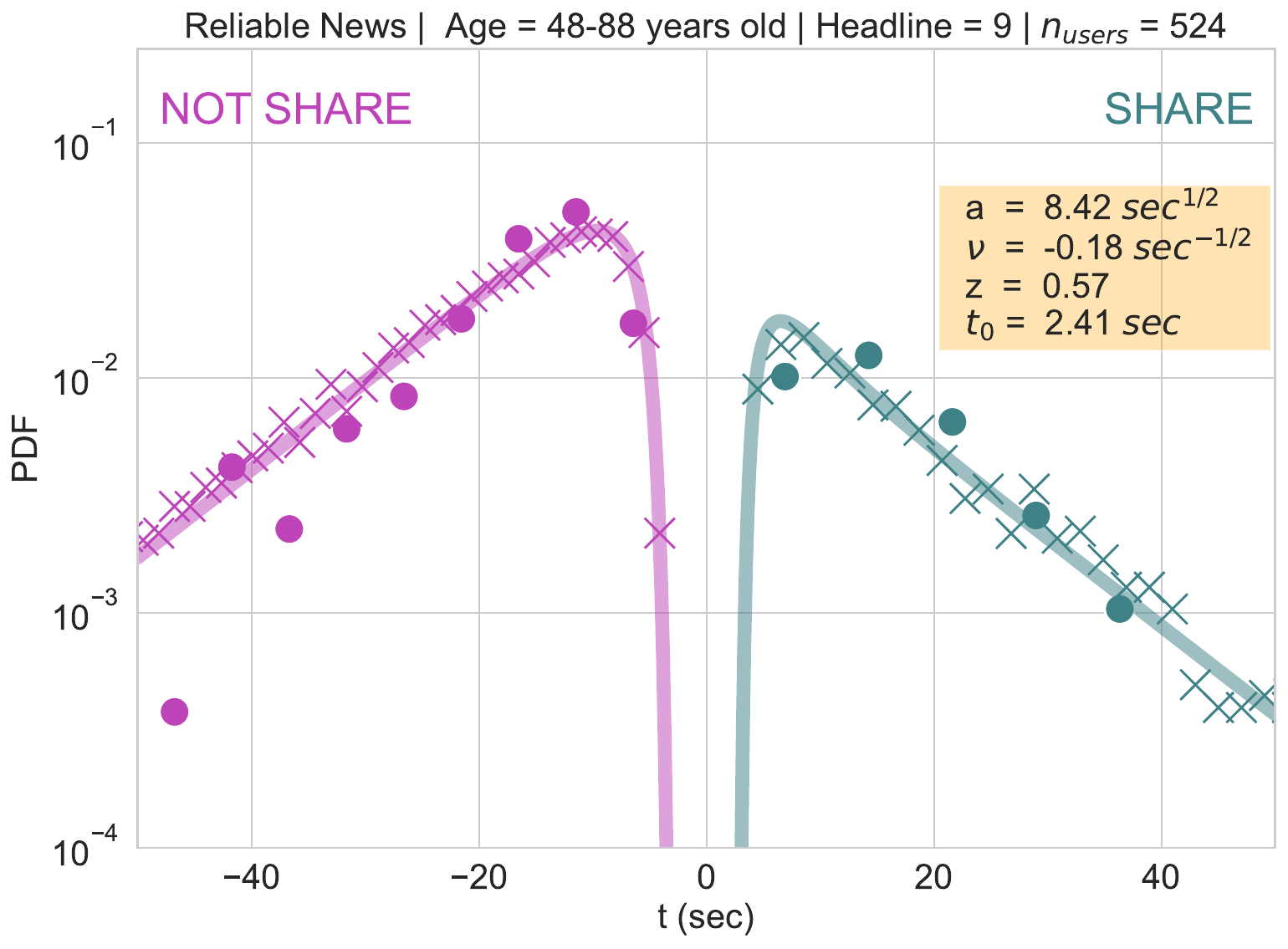}
            \caption{{\bf Headline 9}: Probability distribution of the response time for sharing and not sharing reliable information. Each figure corresponds to different age ranges. The solid line corresponds to theoretical results, dots correspond to empirical data and crosses to stochastic simulations.}
            \label{headline9Real}
        \end{figure}
        
        \begin{figure}[H]
            \renewcommand{\figurename}{Supplementary Figure}
            \centering
            \includegraphics[width=.45\textwidth]{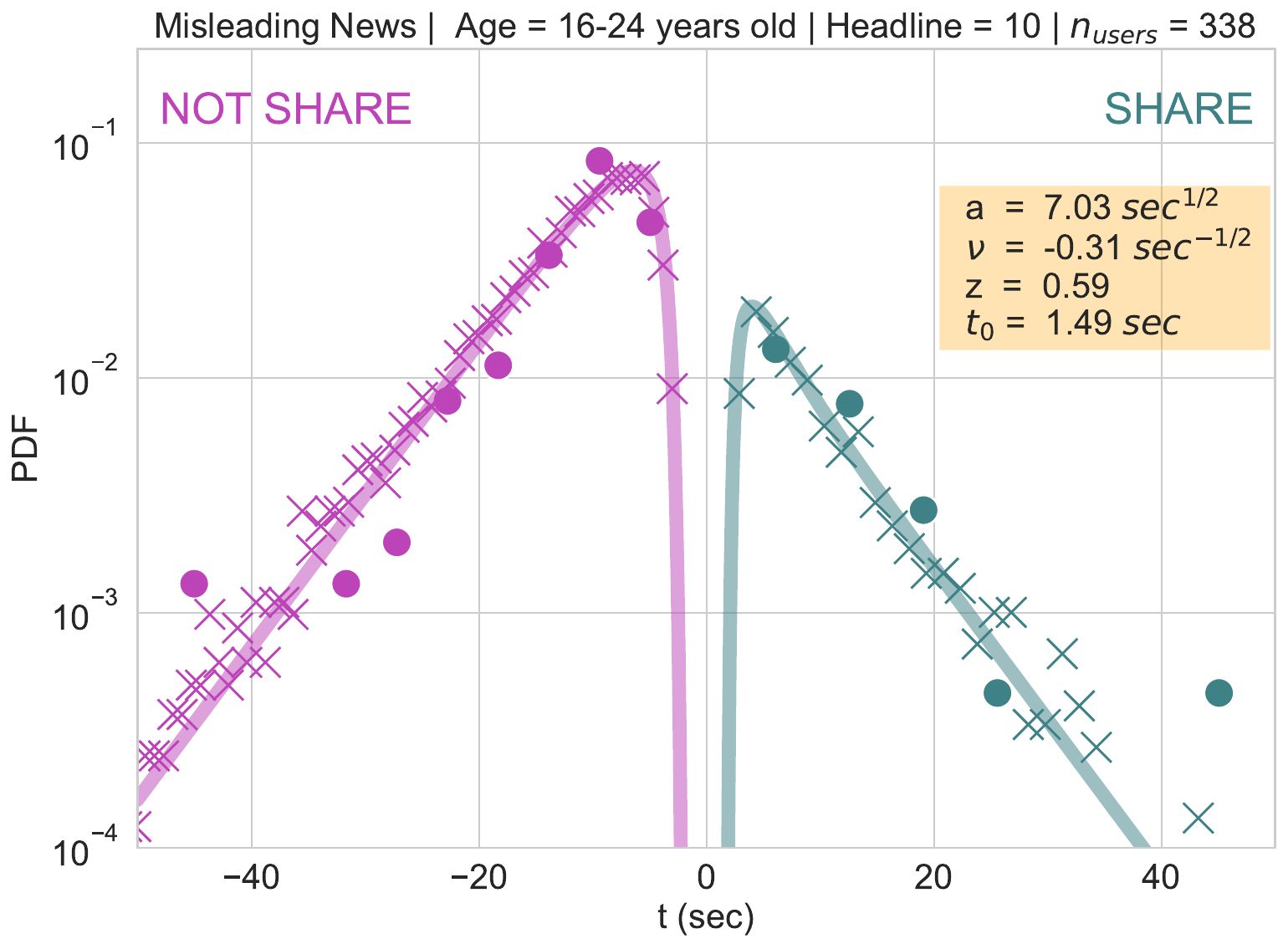}\quad
            \includegraphics[width=.45\textwidth]{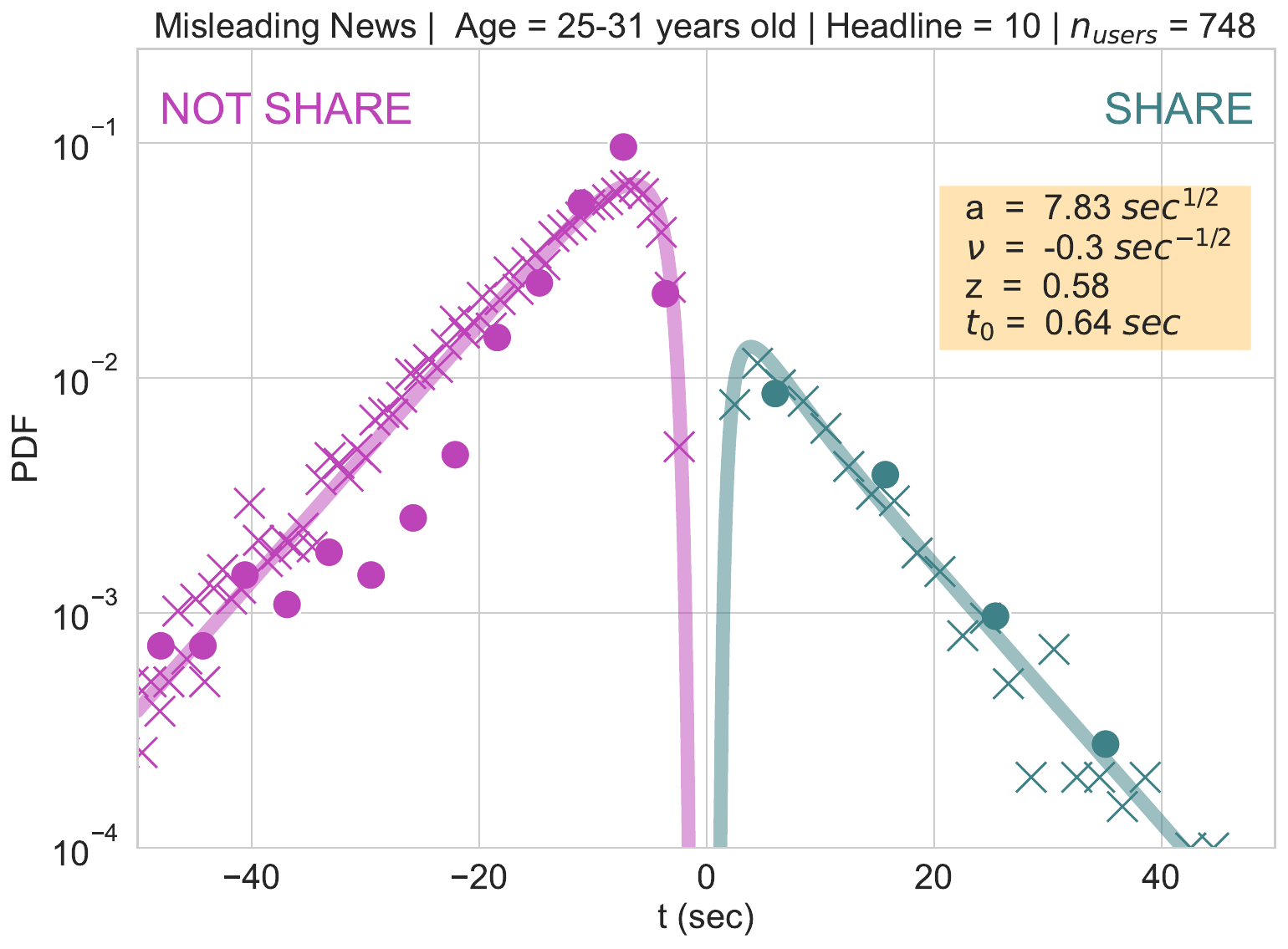}\quad
            \includegraphics[width=.45\textwidth]{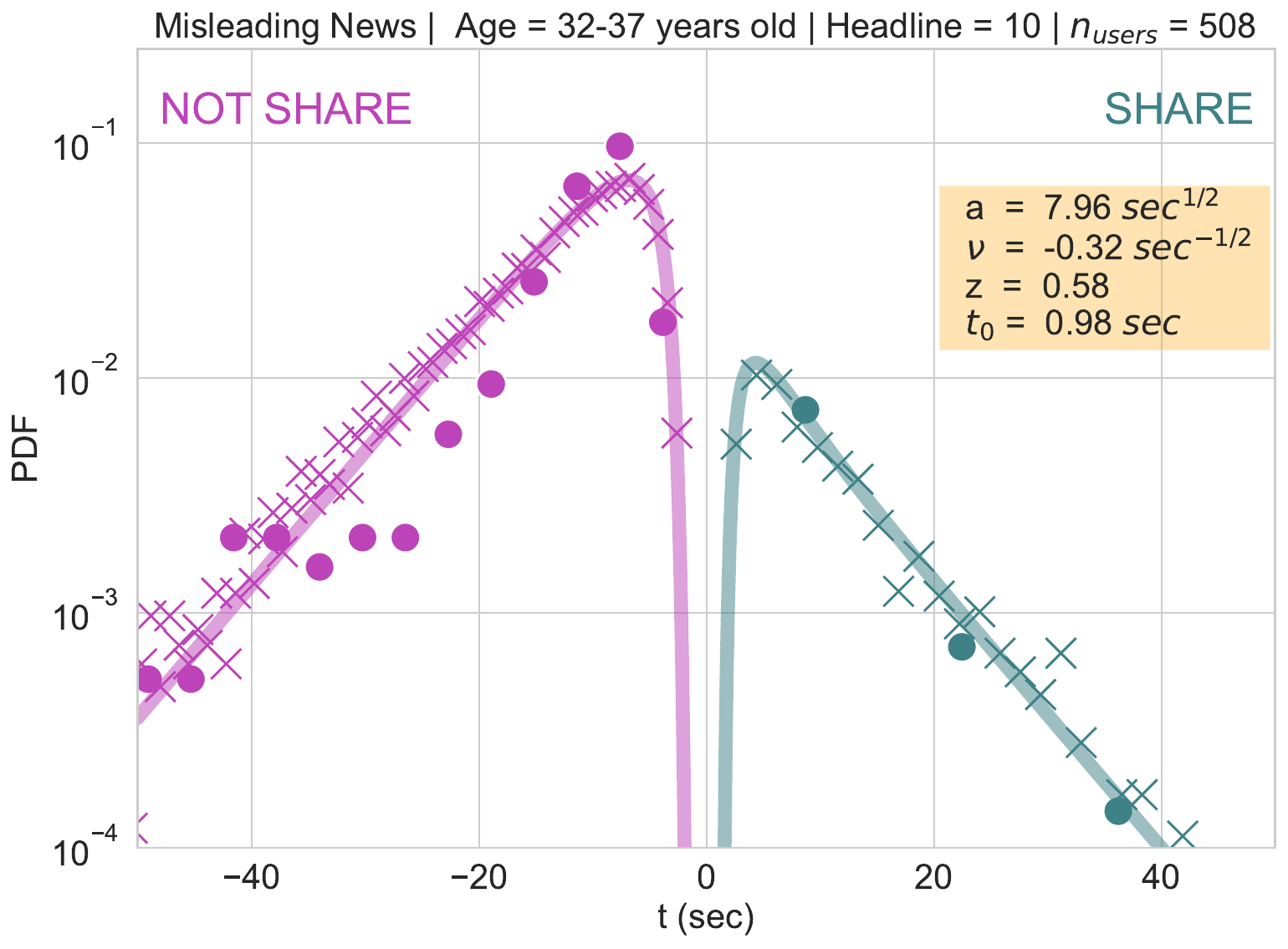}
            \medskip
            \includegraphics[width=.45\textwidth]{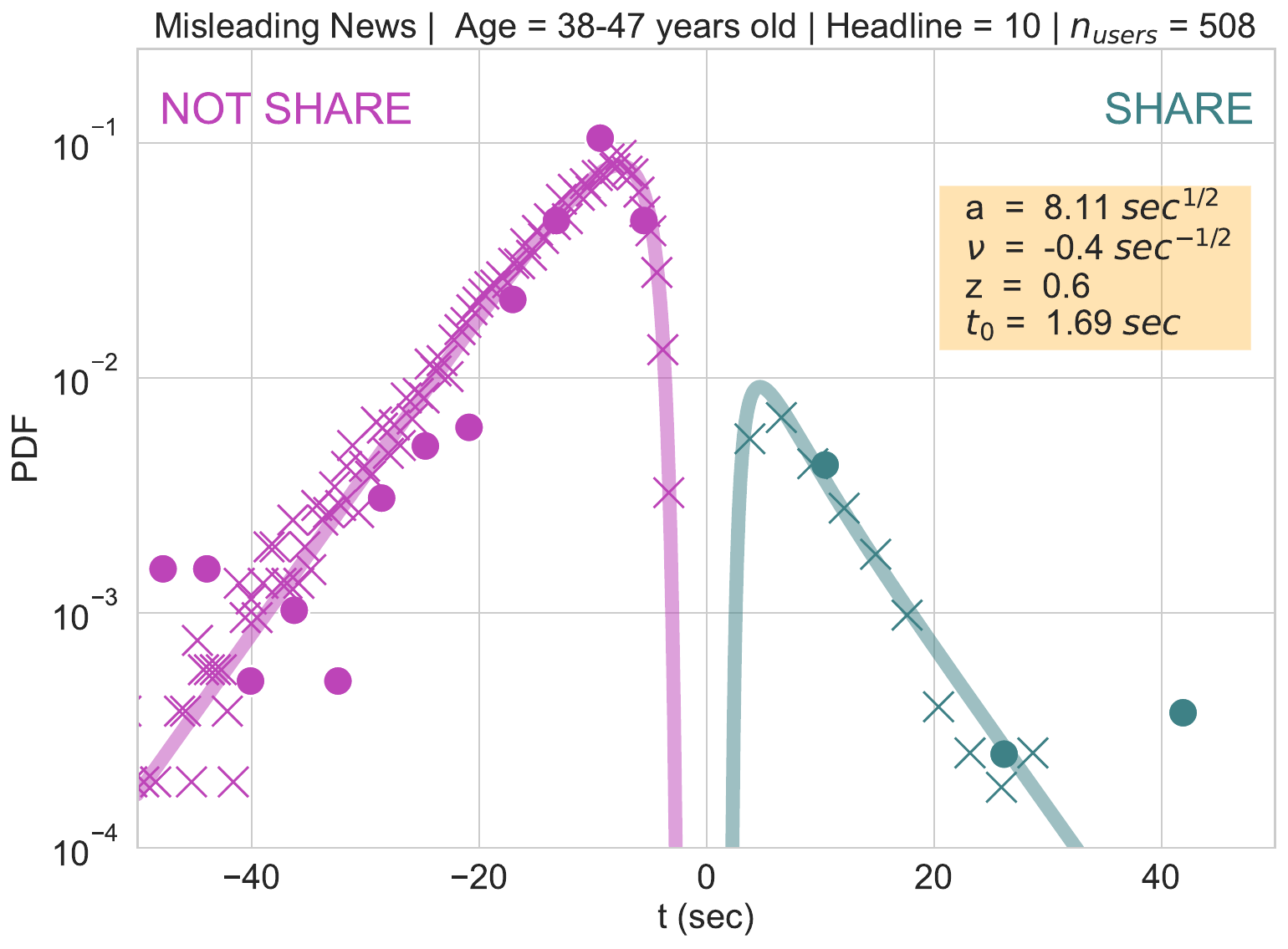}\quad
            \includegraphics[width=.45\textwidth]{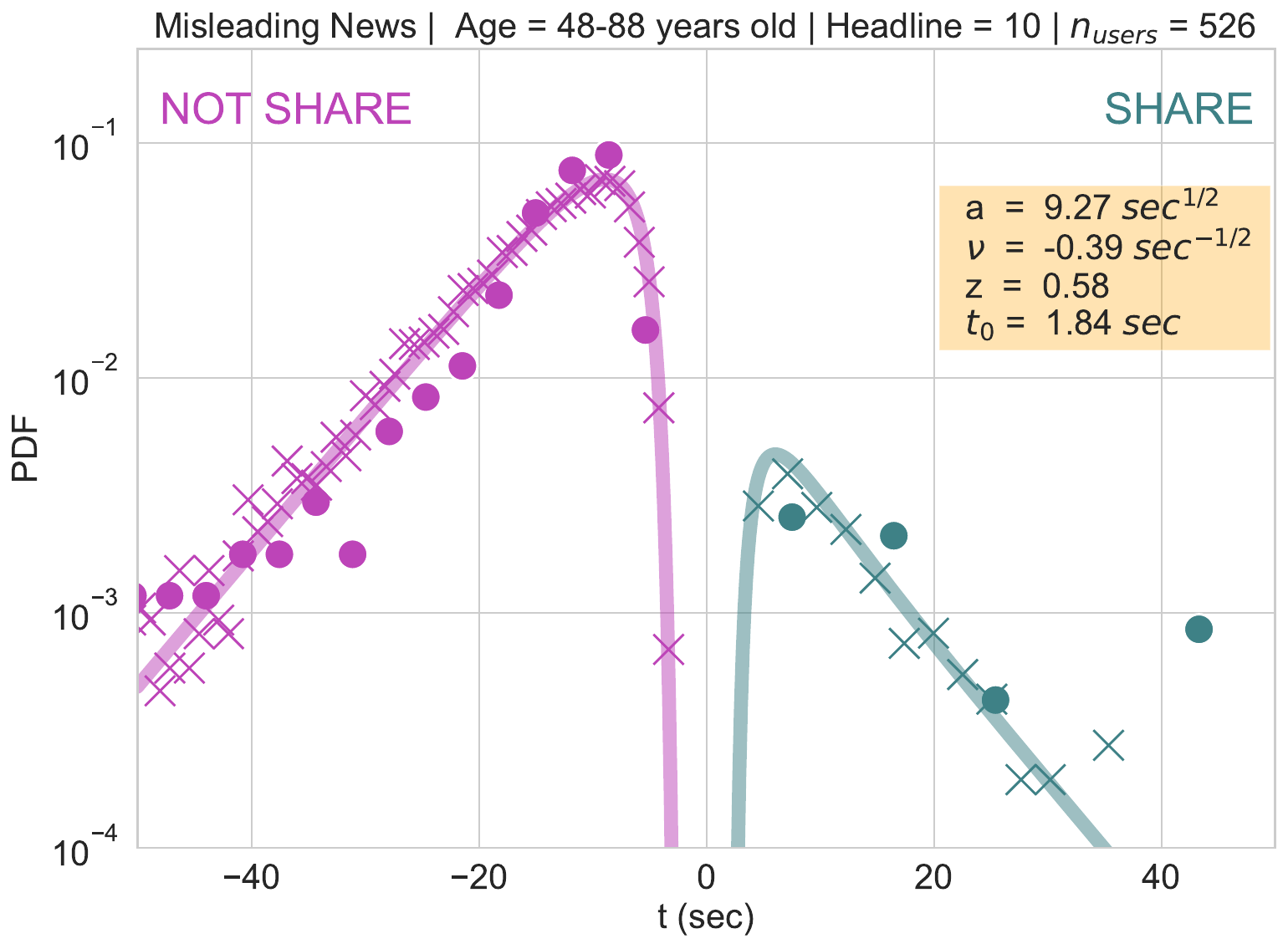}
            \caption{{\bf Headline 10}: Probability distribution of the response time for sharing and not sharing misleading information. Each figure corresponds to different age ranges. The solid line corresponds to theoretical results, dots correspond to empirical data and crosses to stochastic simulations.}
            \label{headline10Fake}
        \end{figure}
        
        \begin{figure}[H]
            \renewcommand{\figurename}{Supplementary Figure}
            \centering
            \includegraphics[width=.45\textwidth]{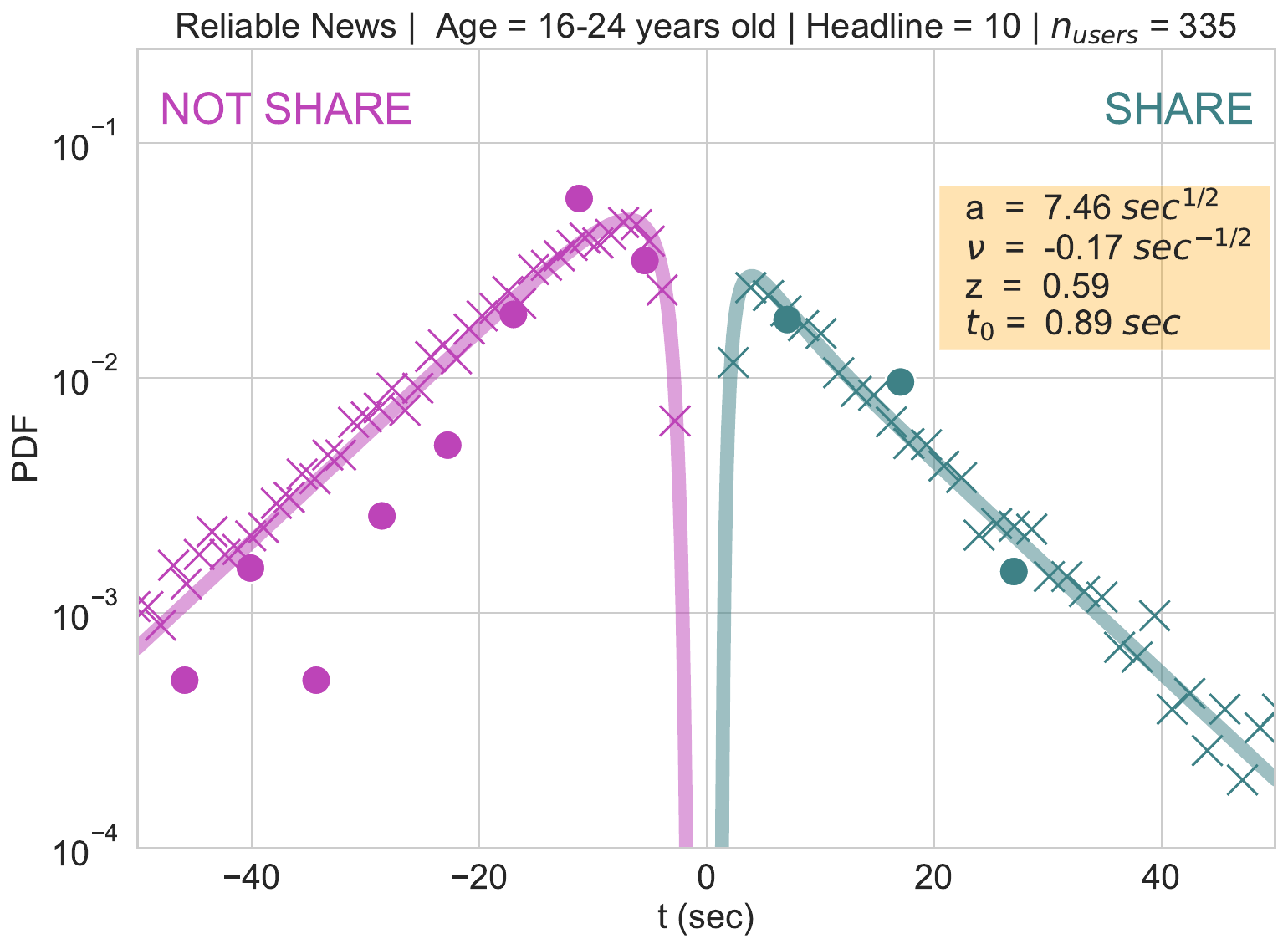}\quad
            \includegraphics[width=.45\textwidth]{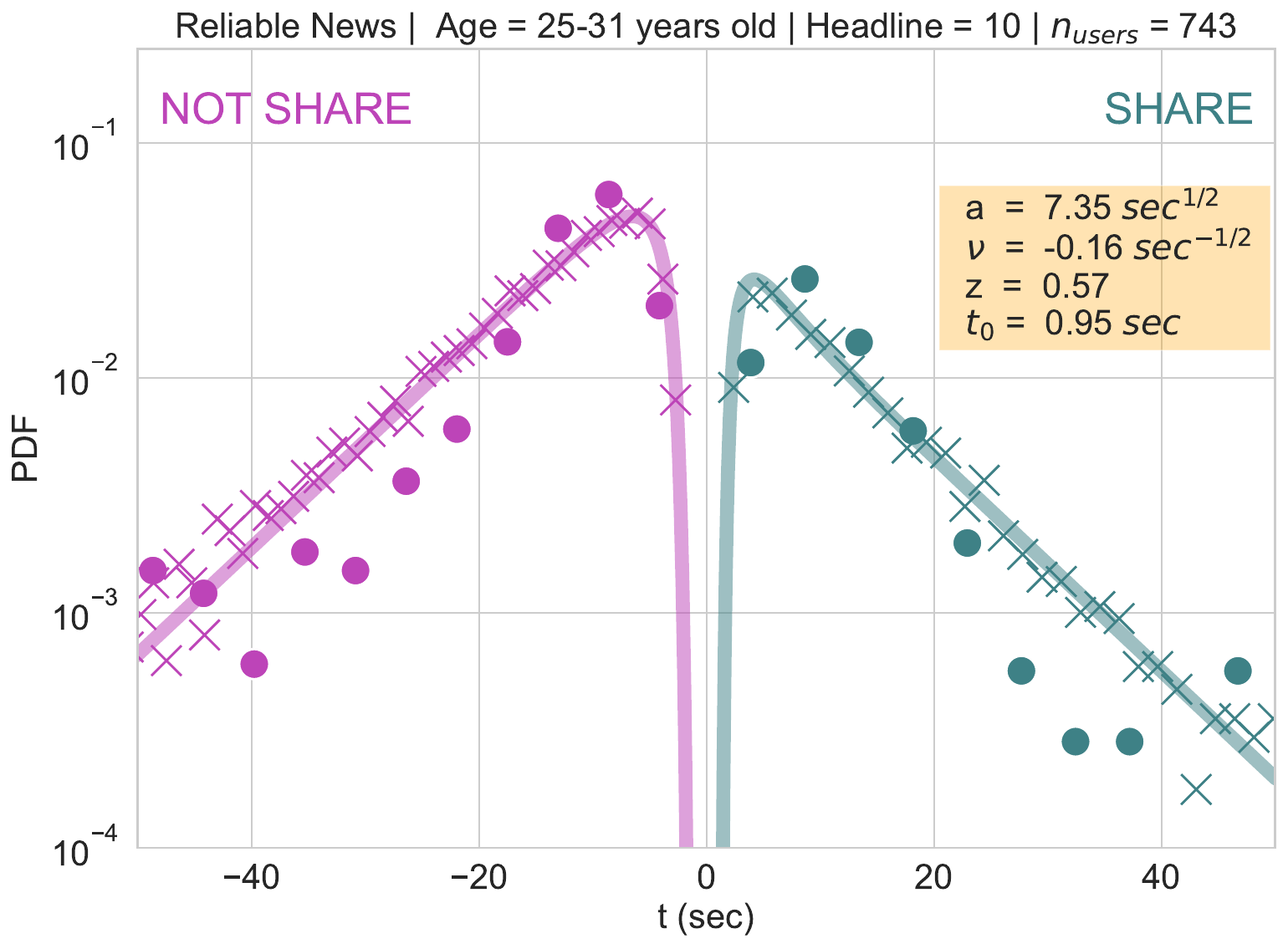}\quad
            \includegraphics[width=.45\textwidth]{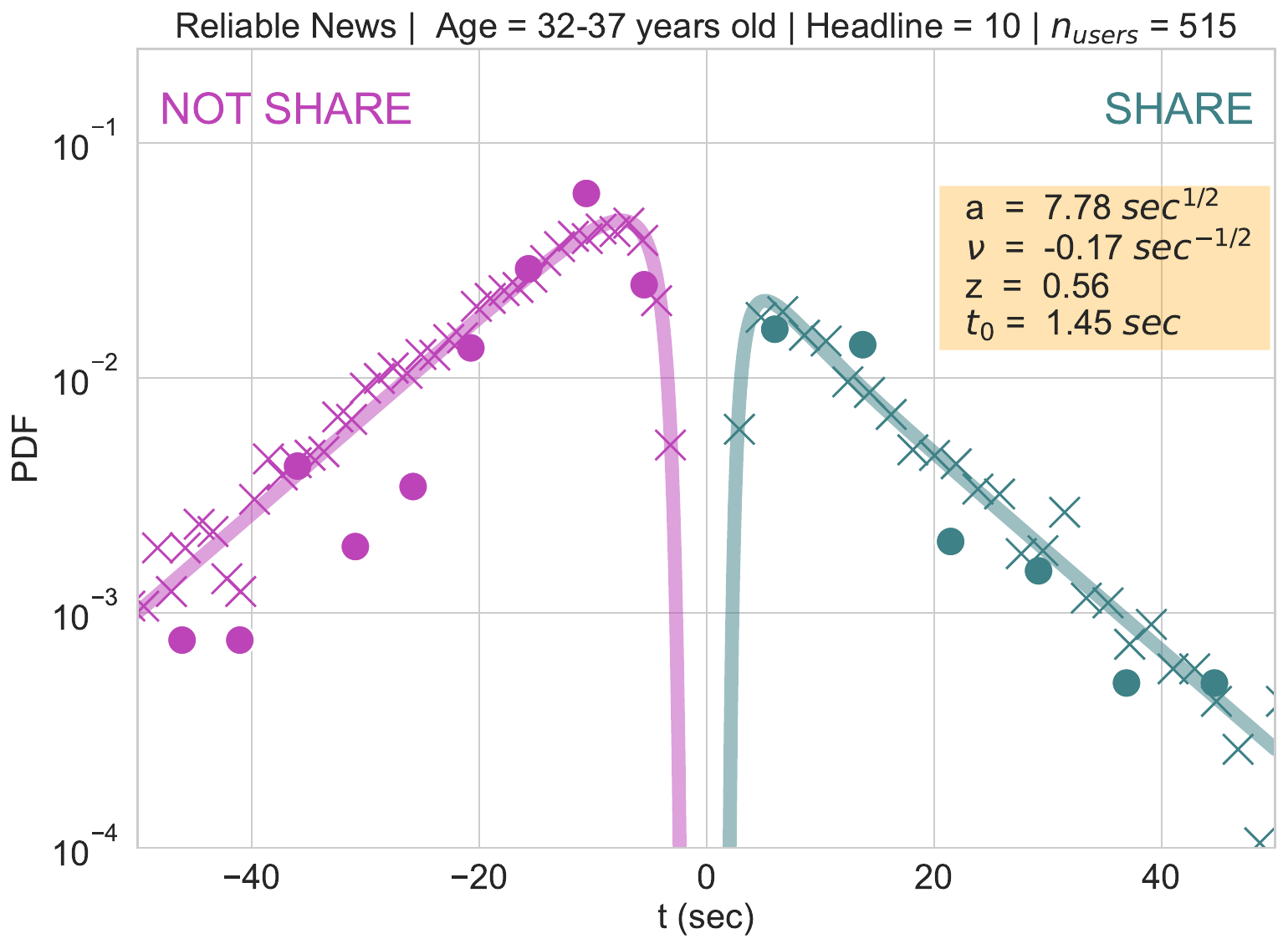}
            \medskip
            \includegraphics[width=.45\textwidth]{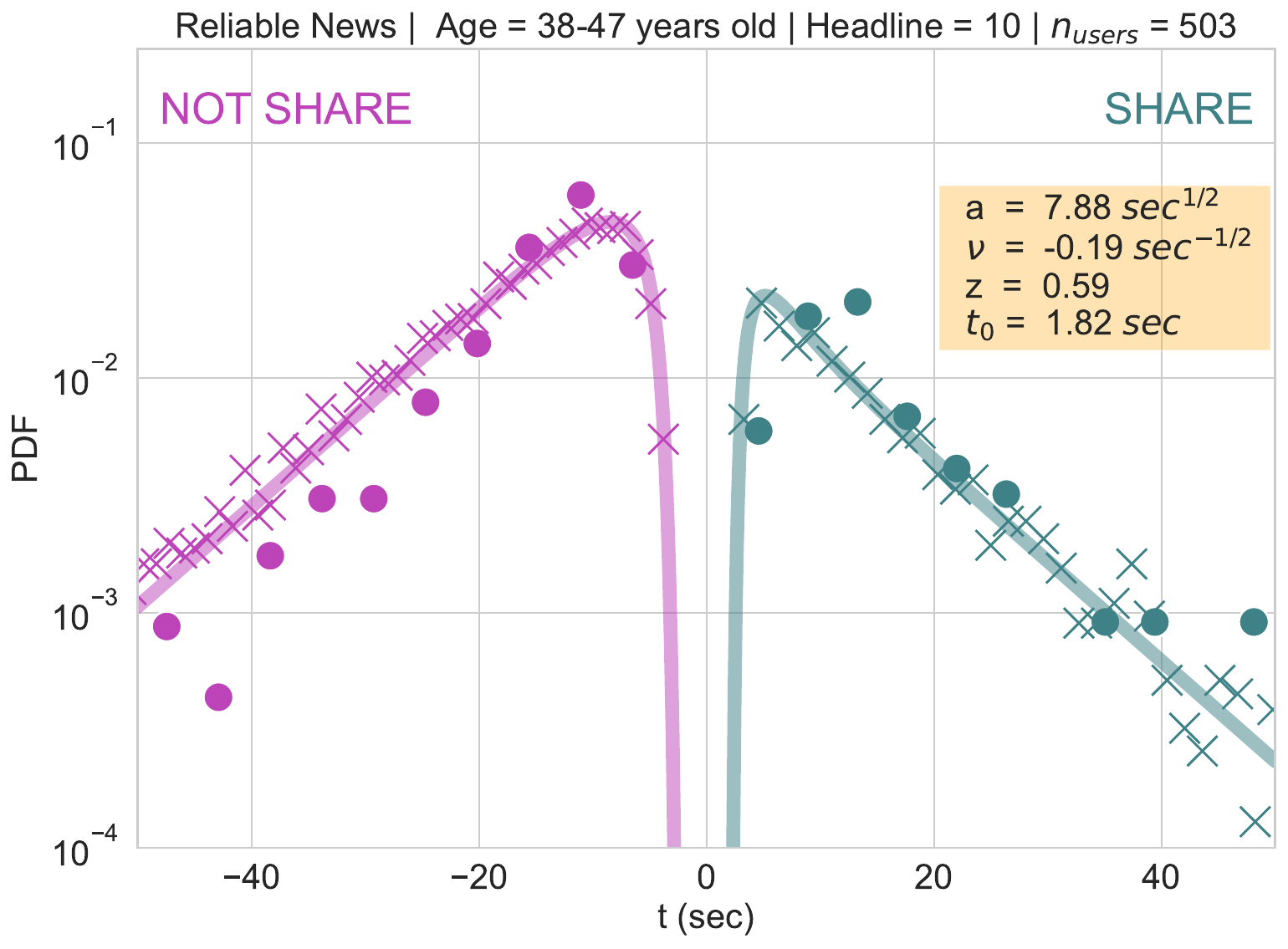}\quad
            \includegraphics[width=.45\textwidth]{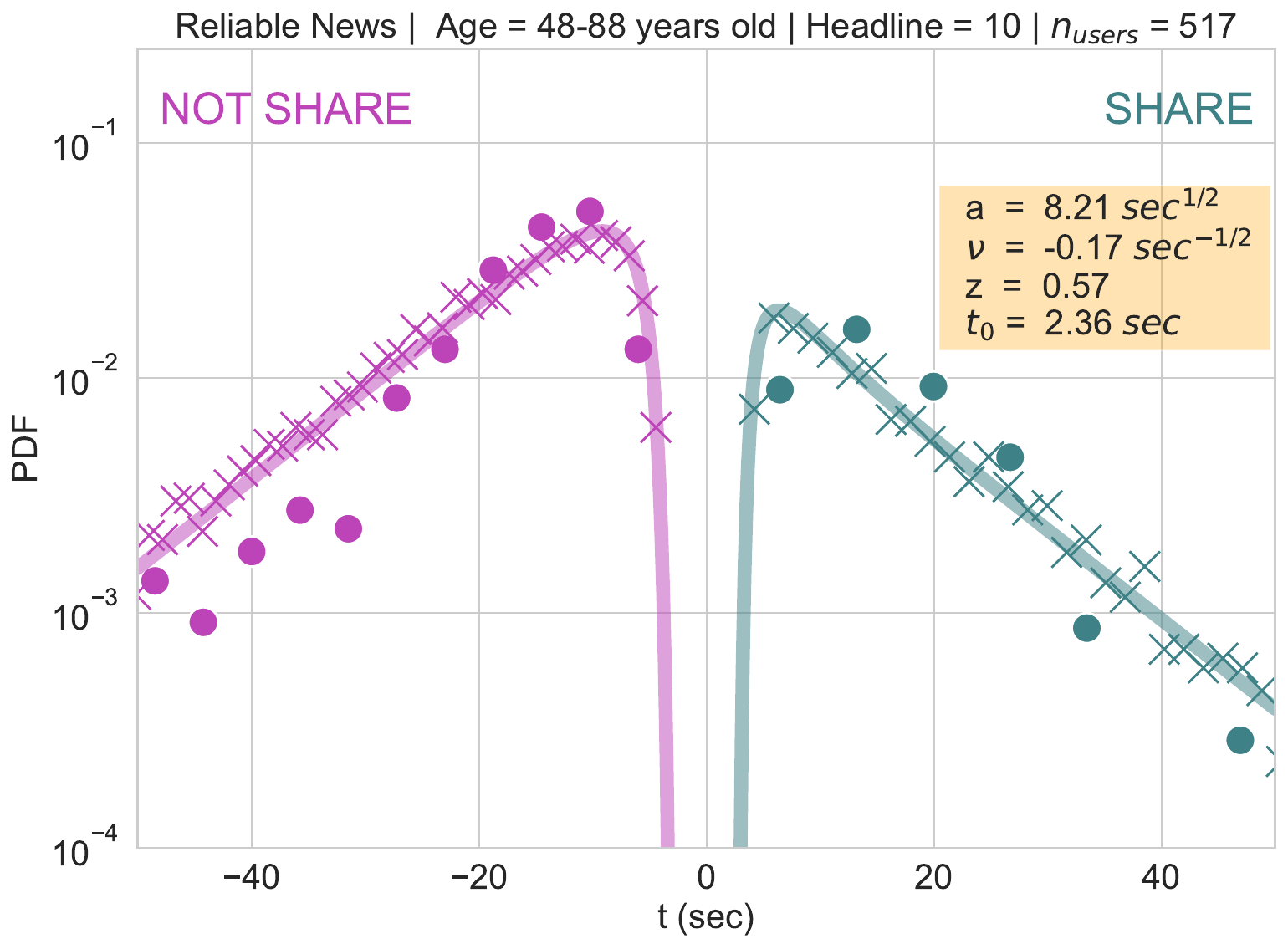}
            \caption{{\bf Headline 10}: Probability distribution of the response time for sharing and not sharing reliable information. Each figure corresponds to different age ranges. The solid line corresponds to theoretical results, dots correspond to empirical data and crosses to stochastic simulations.}
            \label{headline10Real}
        \end{figure}
        
        \begin{figure}[H]
            \renewcommand{\figurename}{Supplementary Figure}
            \centering
            \includegraphics[width=.45\textwidth]{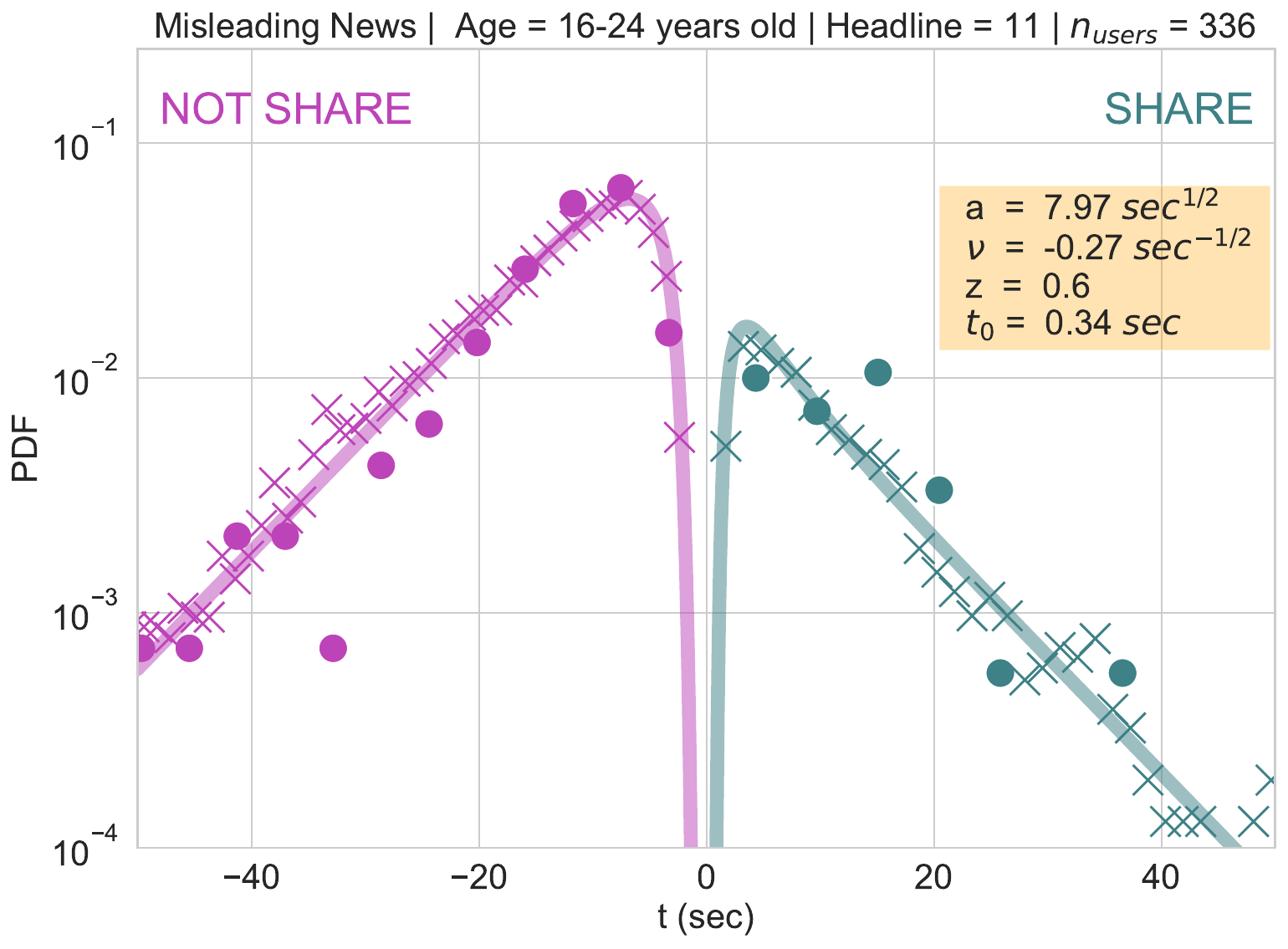}\quad
            \includegraphics[width=.45\textwidth]{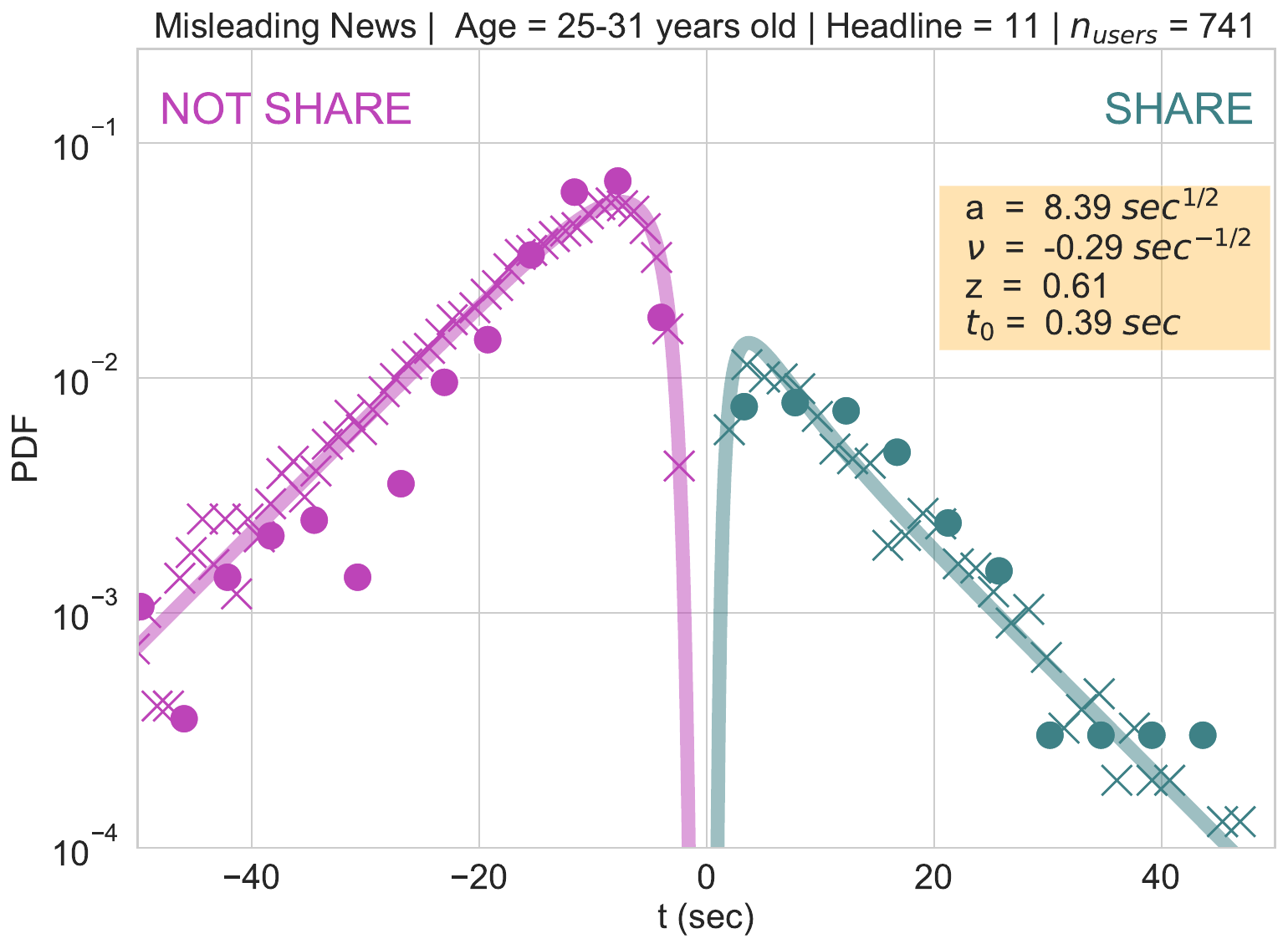}\quad
            \includegraphics[width=.45\textwidth]{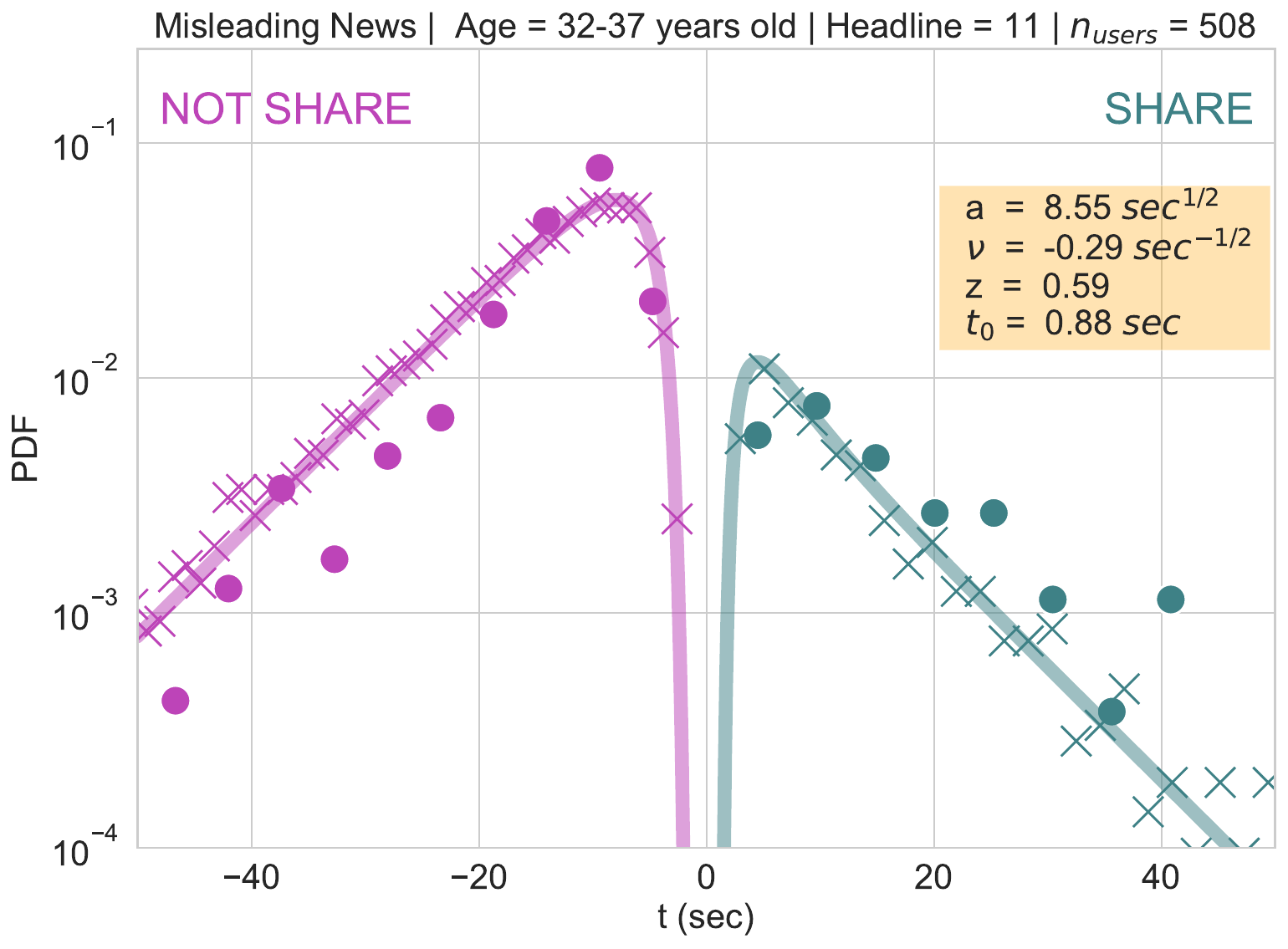}
            \medskip
            \includegraphics[width=.45\textwidth]{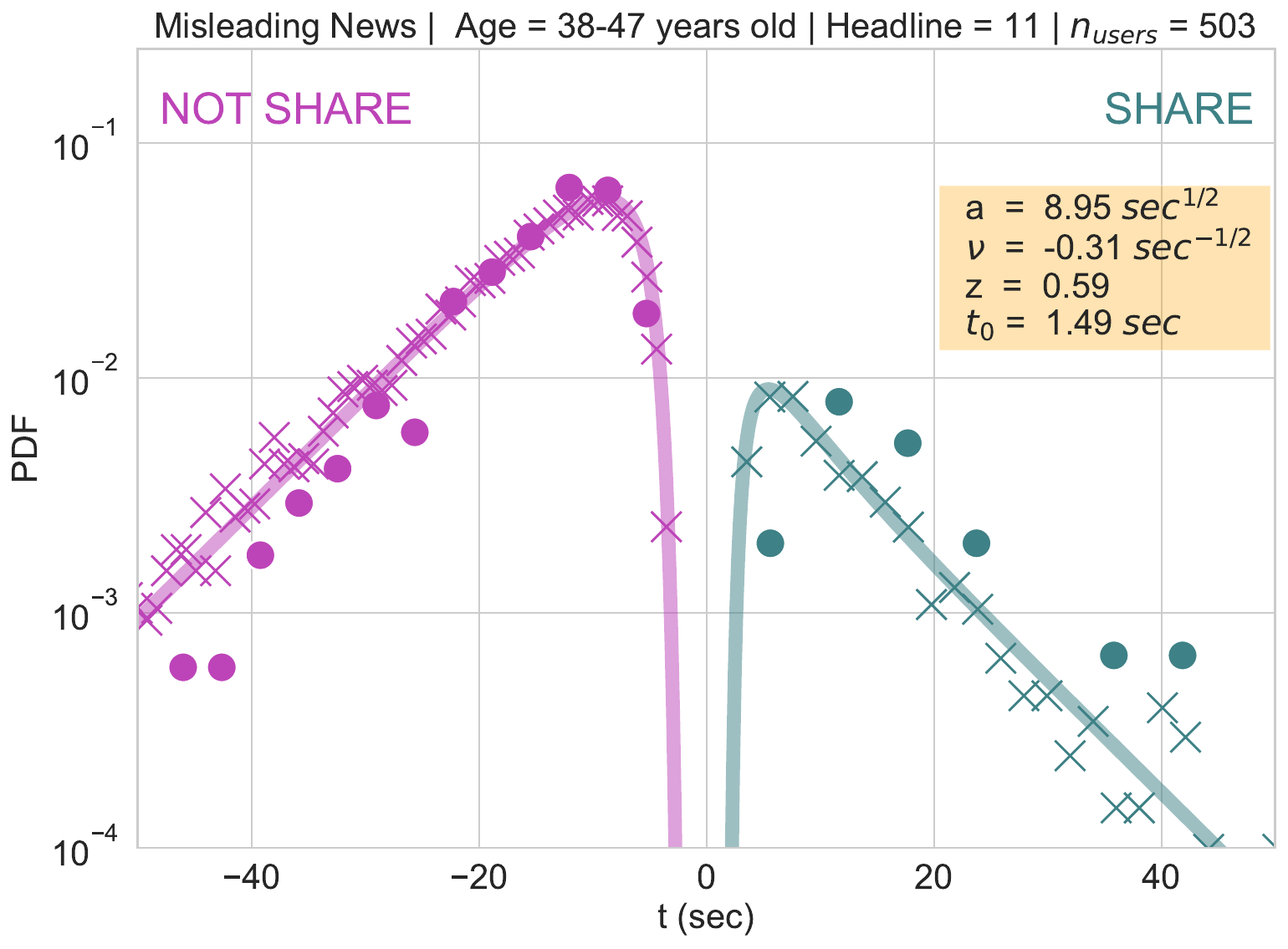}\quad
            \includegraphics[width=.45\textwidth]{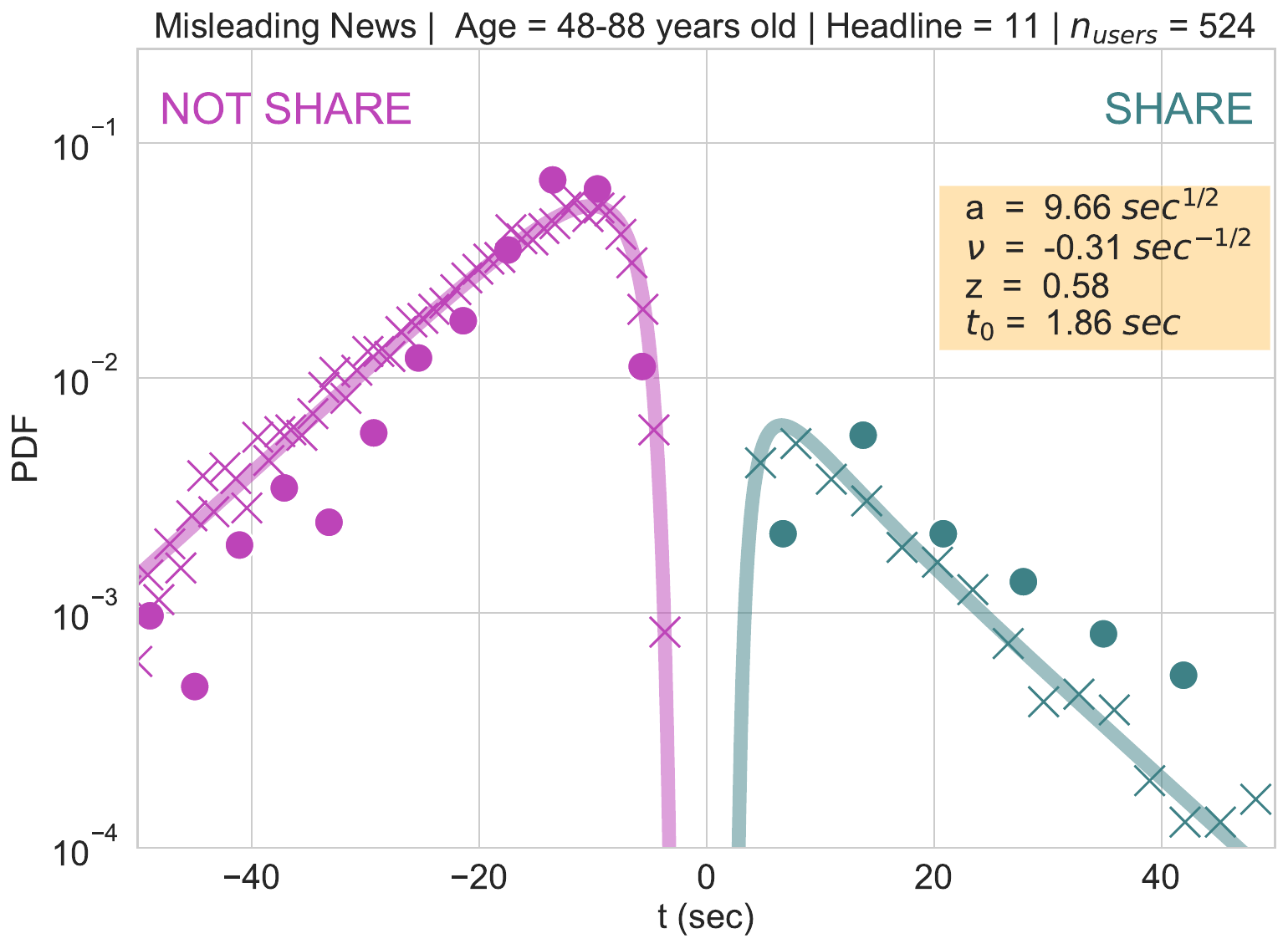}
            \caption{{\bf Headline 11}: Probability distribution of the response time for sharing and not sharing misleading information. Each figure corresponds to different age ranges. The solid line corresponds to theoretical results, dots correspond to empirical data and crosses to stochastic simulations.}
            \label{headline11Fake}
        \end{figure}
        
        \begin{figure}[H]
            \renewcommand{\figurename}{Supplementary Figure}
            \centering
            \includegraphics[width=.45\textwidth]{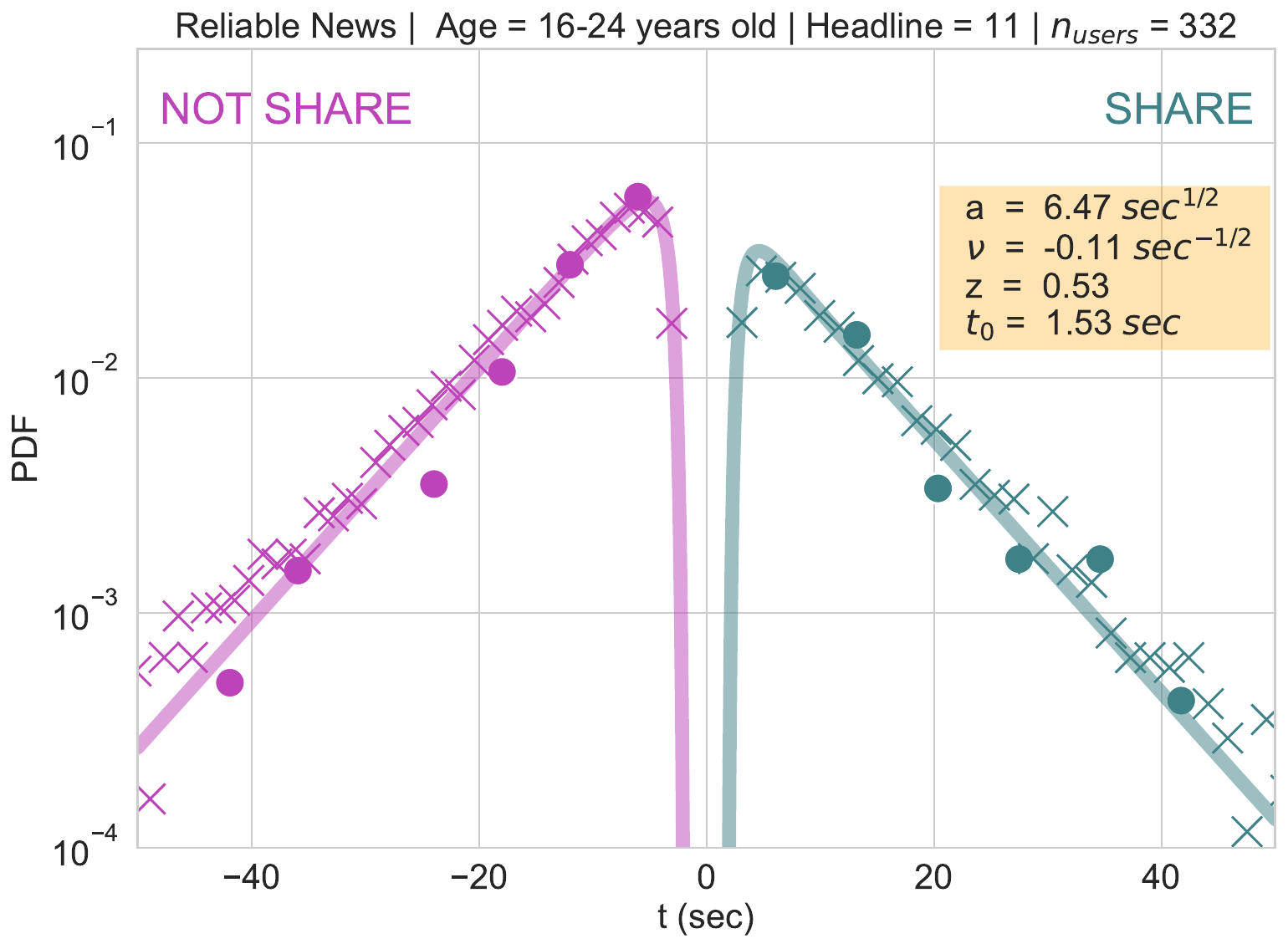}\quad
            \includegraphics[width=.45\textwidth]{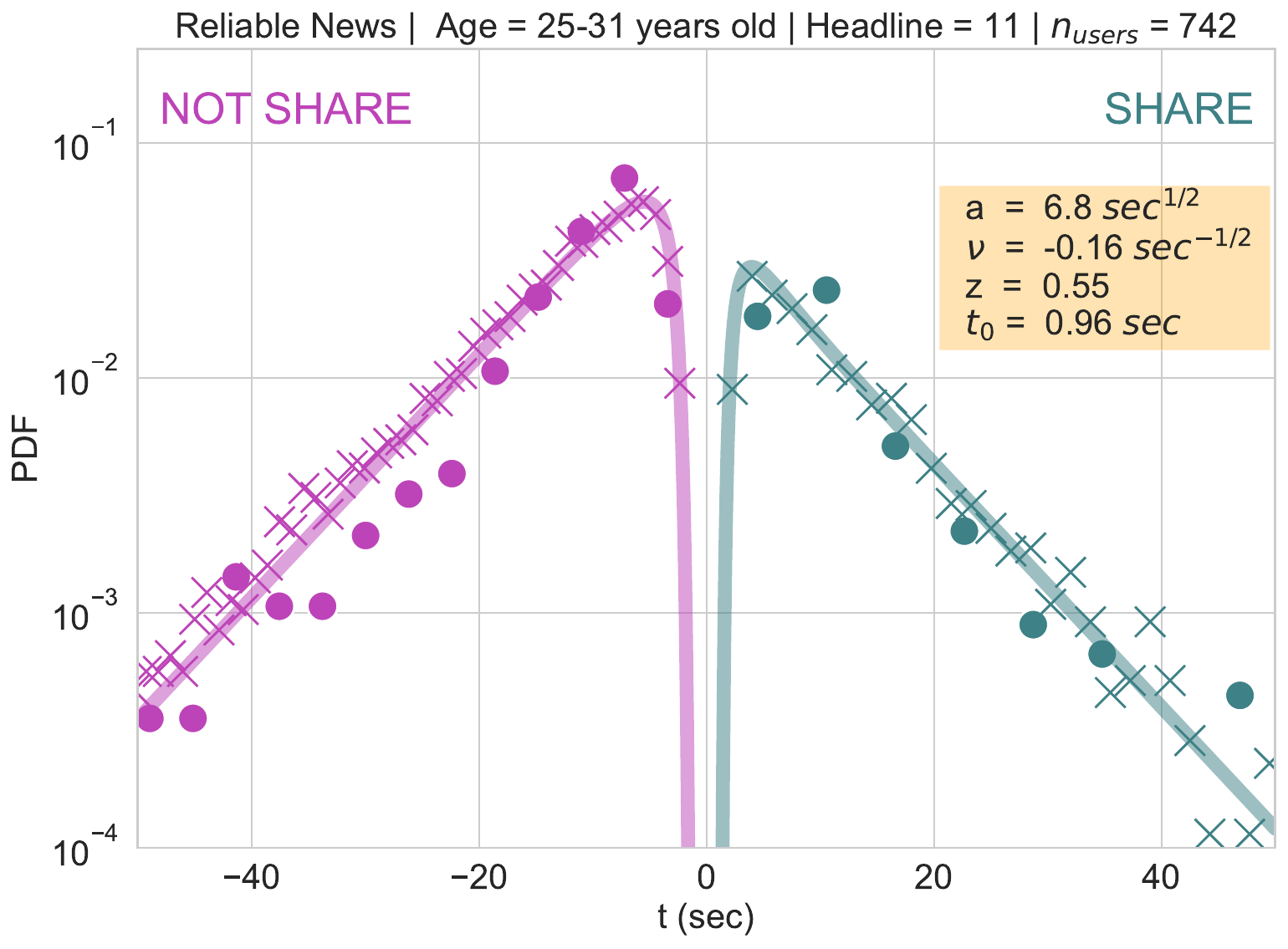}\quad
            \includegraphics[width=.45\textwidth]{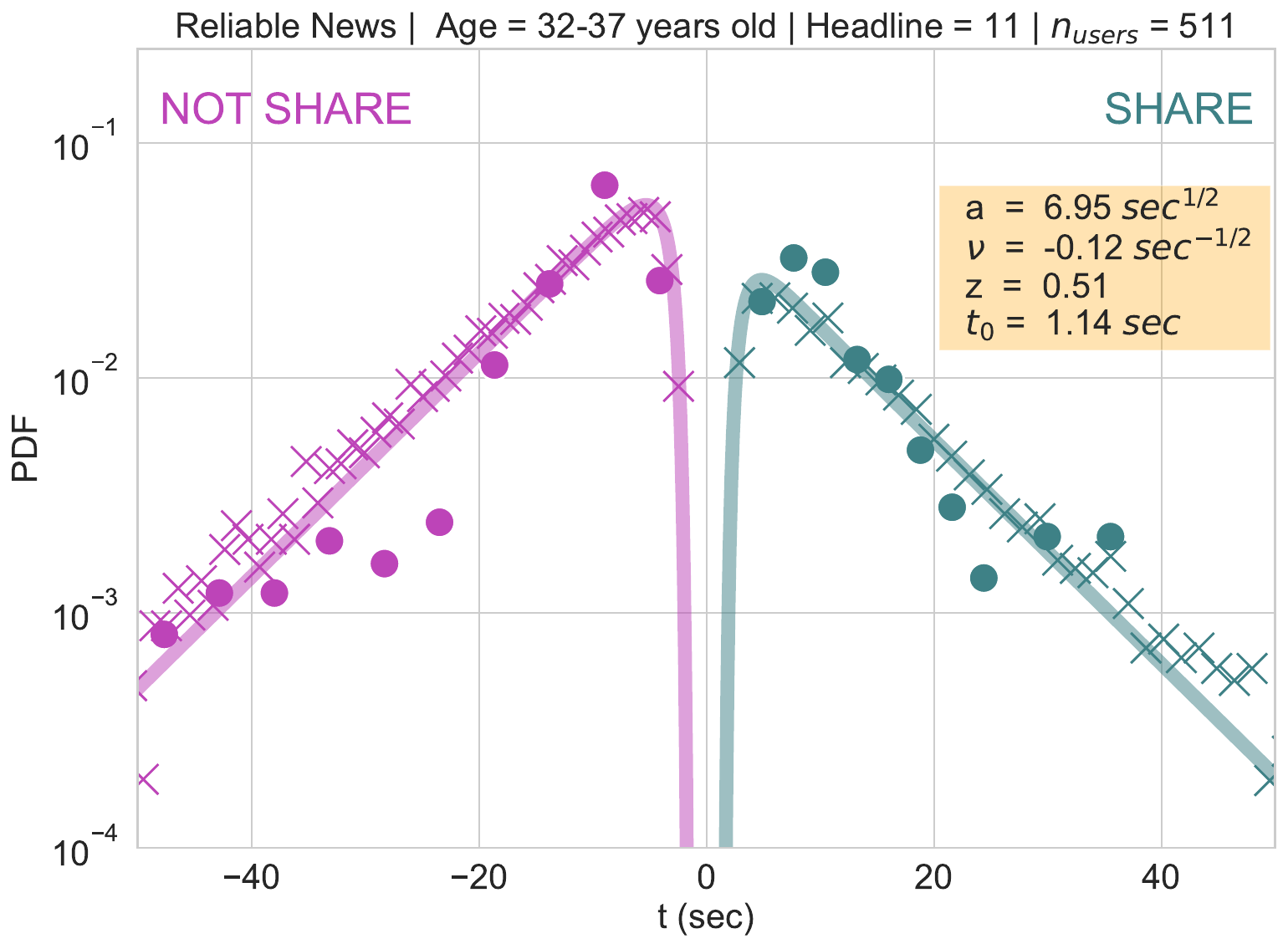}
            \medskip
            \includegraphics[width=.45\textwidth]{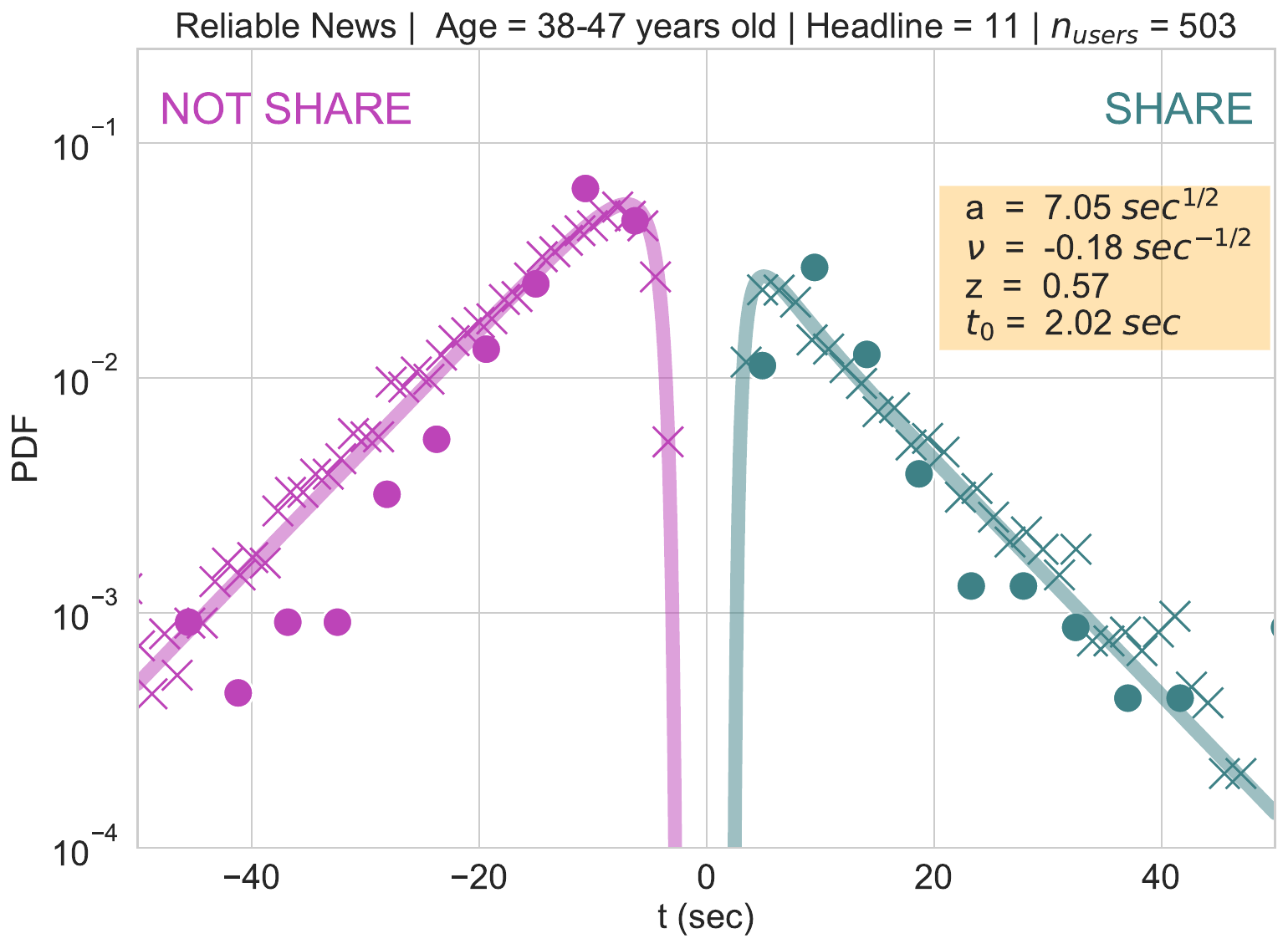}\quad
            \includegraphics[width=.45\textwidth]{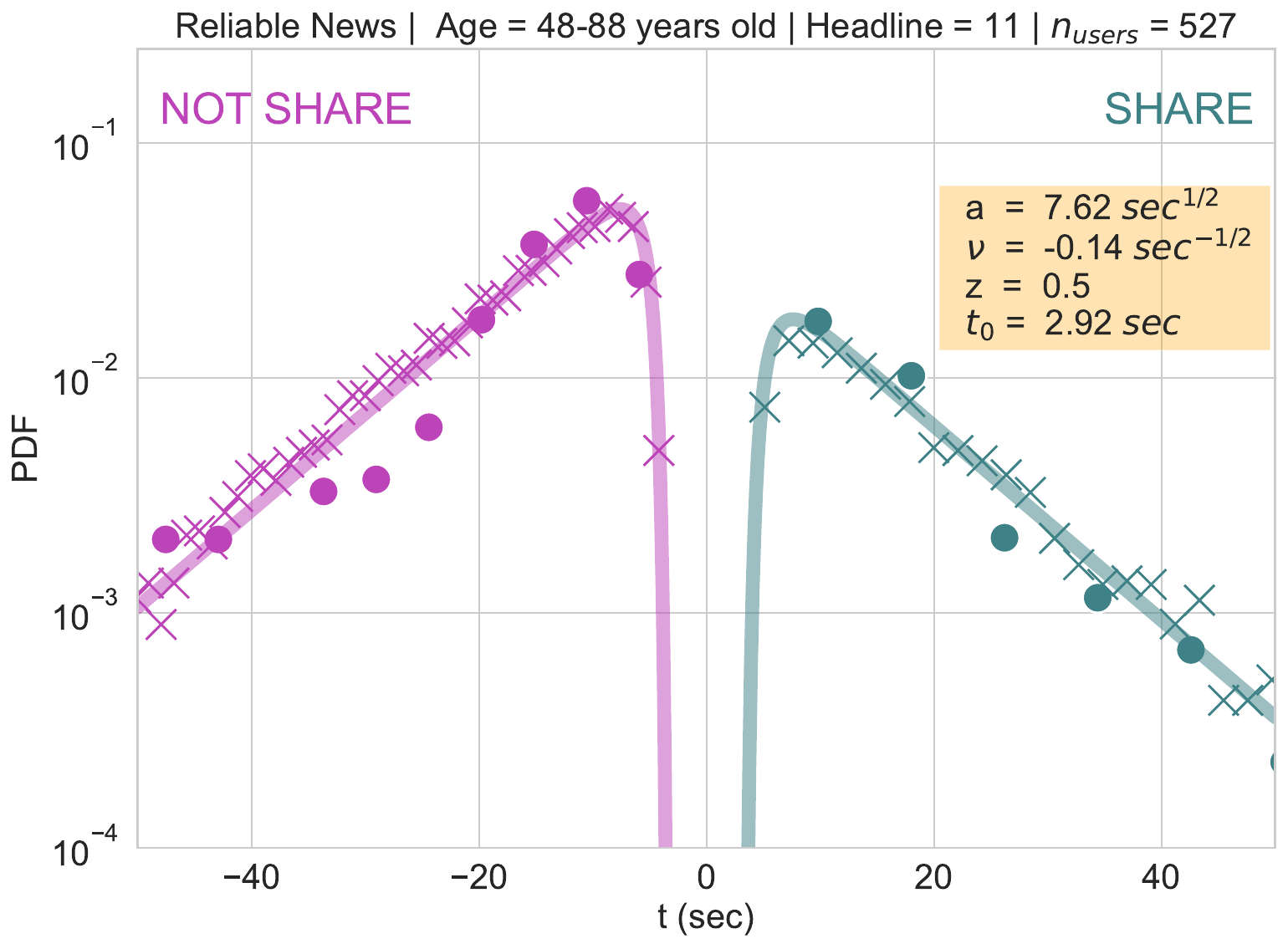}
            \caption{{\bf Headline 11}: Probability distribution of the response time for sharing and not sharing reliable information. Each figure corresponds to different age ranges. The solid line corresponds to theoretical results, dots correspond to empirical data and crosses to stochastic simulations.}
            \label{headline11Real}
        \end{figure}
        
        \begin{figure}[H]
            \renewcommand{\figurename}{Supplementary Figure}
            \centering
            \includegraphics[width=.45\textwidth]{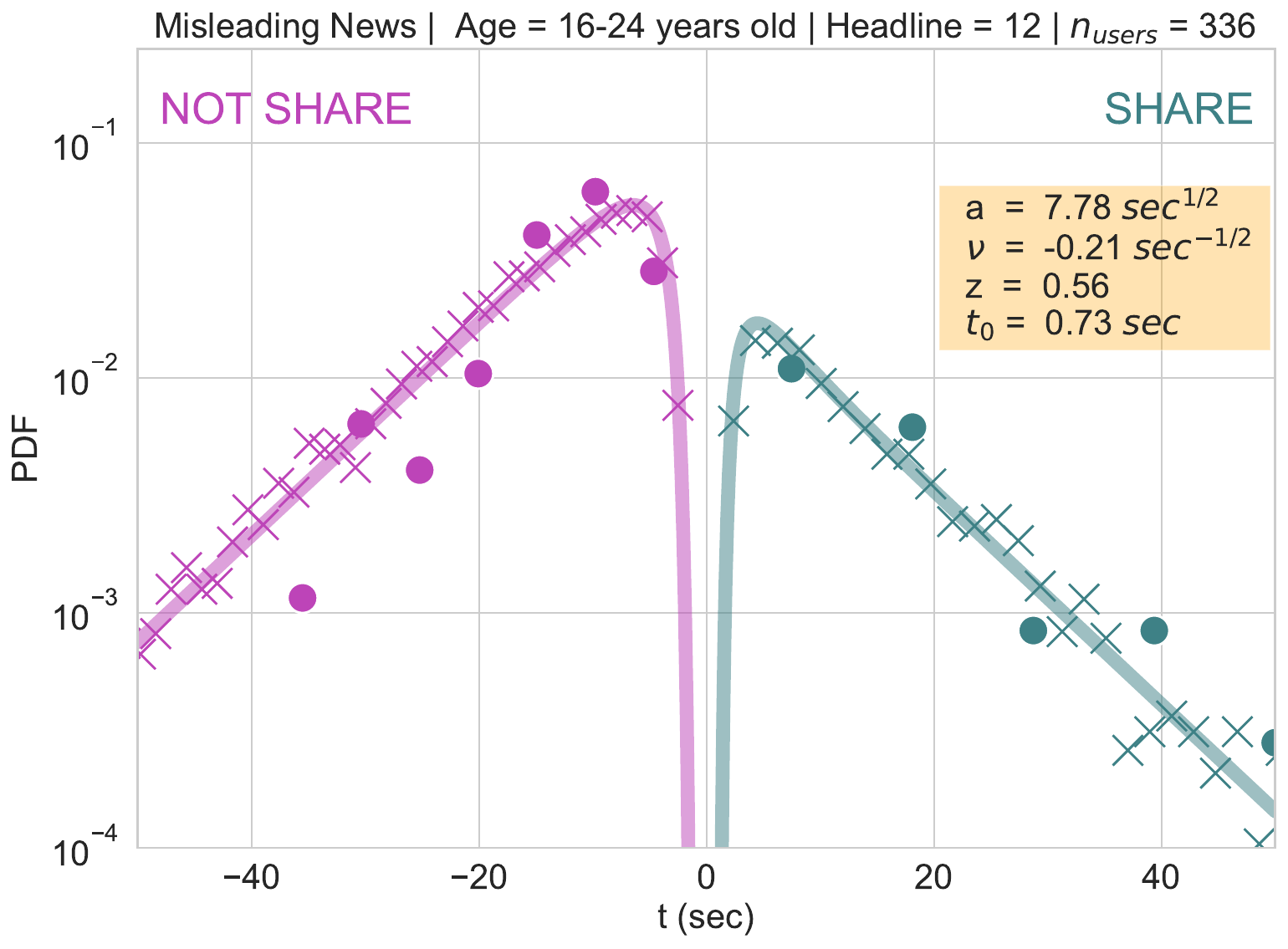}\quad
            \includegraphics[width=.45\textwidth]{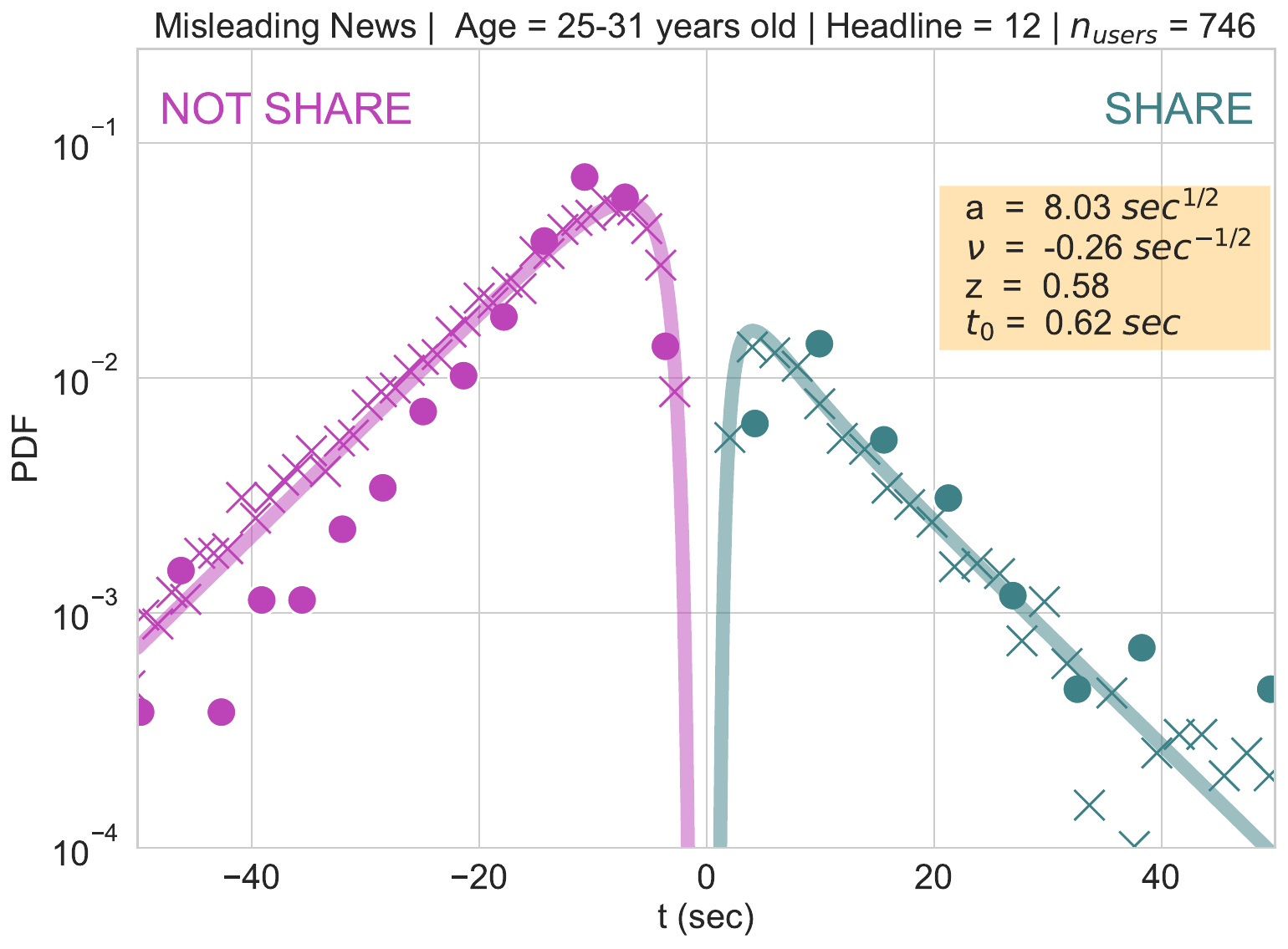}\quad
            \includegraphics[width=.45\textwidth]{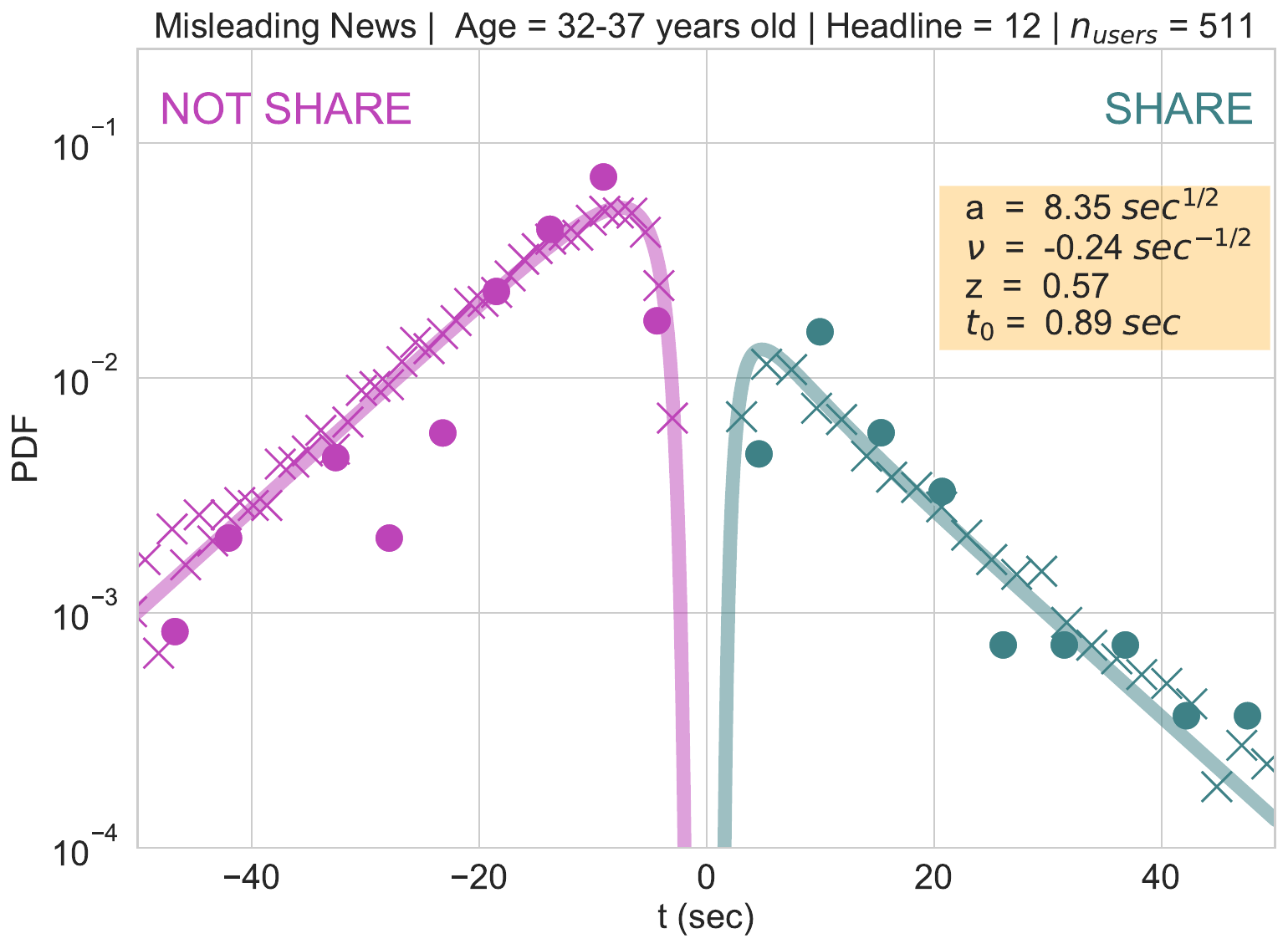}
            \medskip
            \includegraphics[width=.45\textwidth]{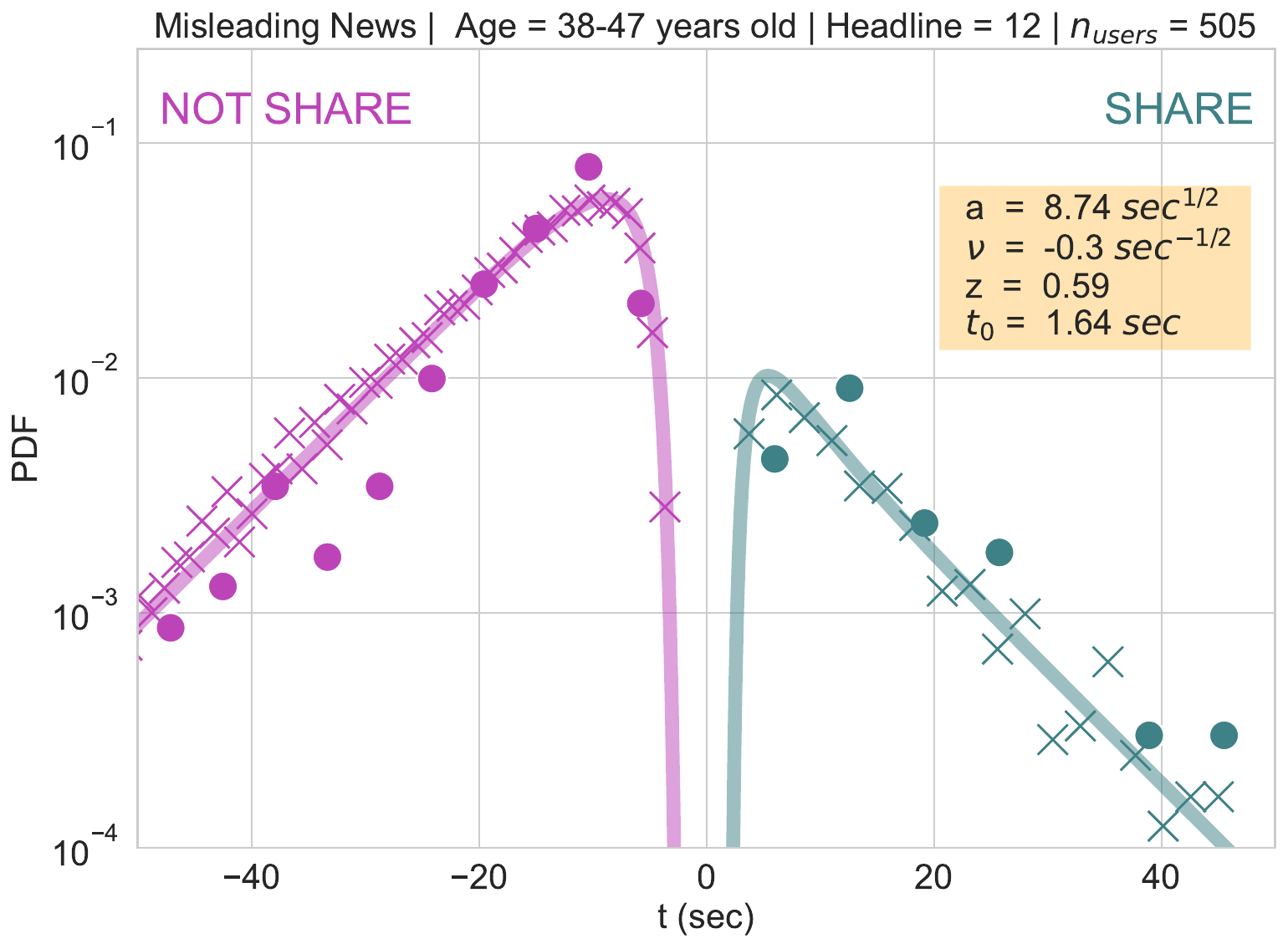}\quad
            \includegraphics[width=.45\textwidth]{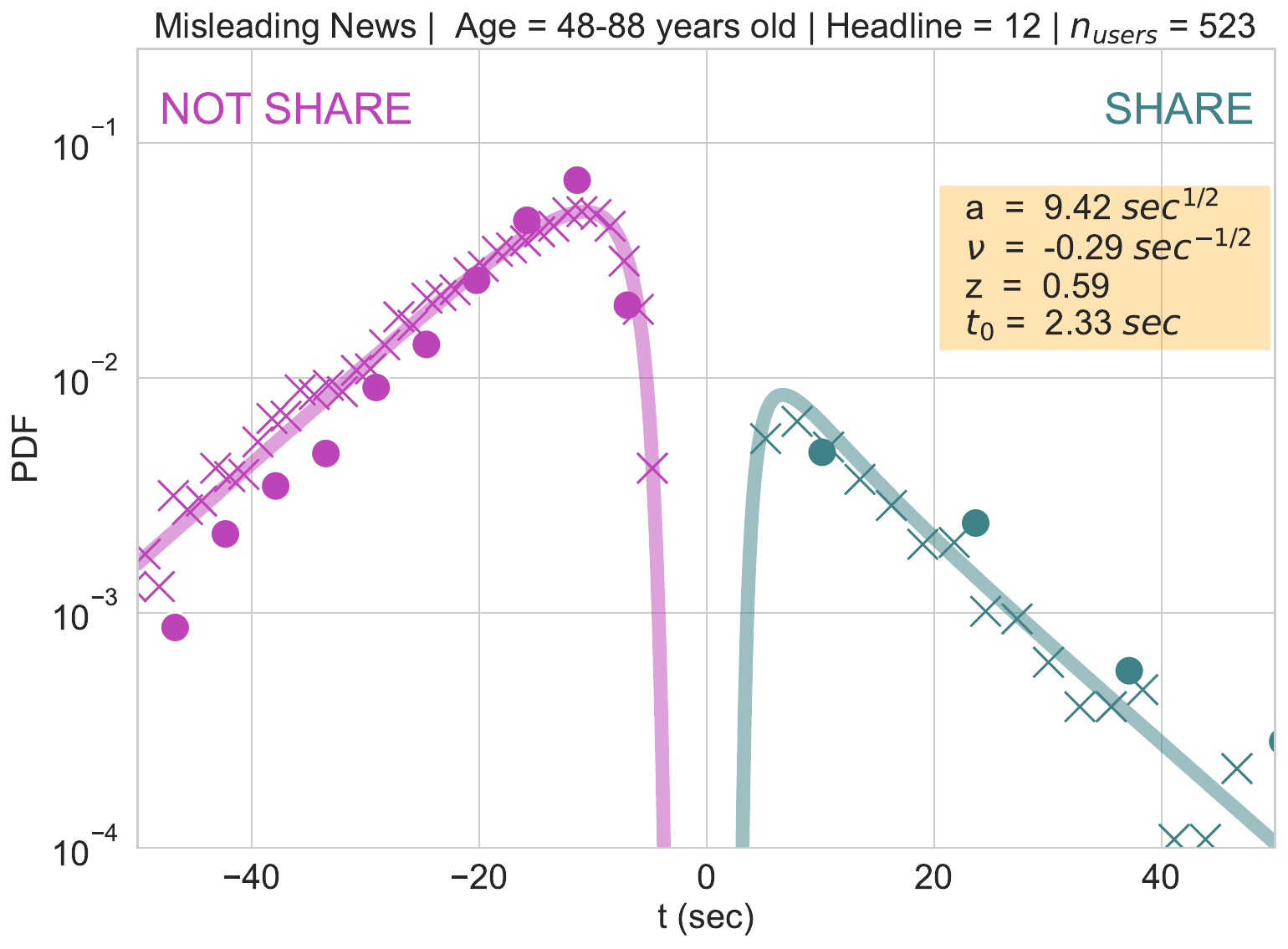}
            \caption{{\bf Headline 12}: Probability distribution of the response time for sharing and not sharing misleading information. Each figure corresponds to different age ranges. The solid line corresponds to theoretical results, dots correspond to empirical data and crosses to stochastic simulations.}
            \label{headline12Fake}
        \end{figure}
        
        \begin{figure}[H]
            \renewcommand{\figurename}{Supplementary Figure}
            \centering
            \includegraphics[width=.45\textwidth]{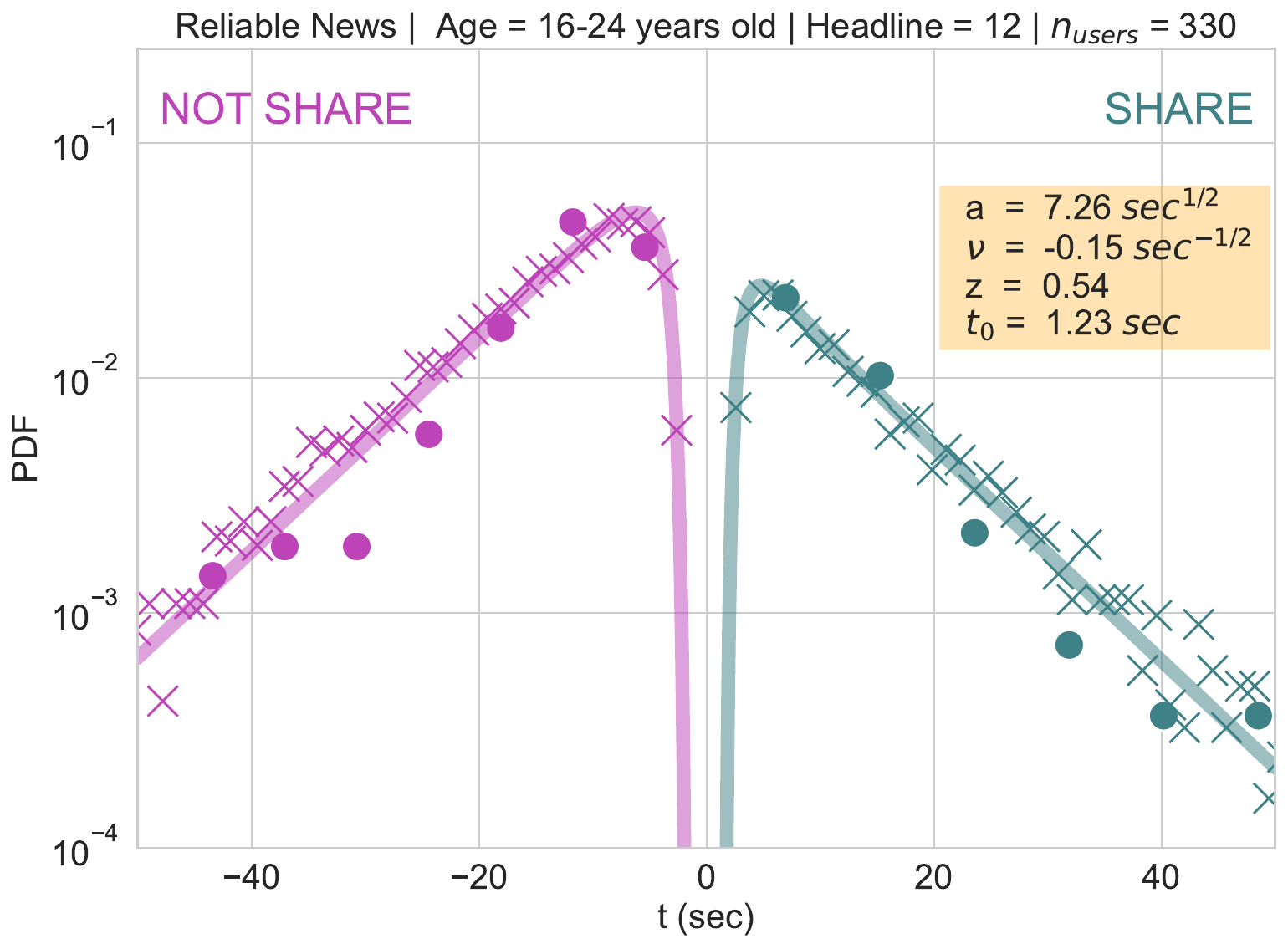}\quad
            \includegraphics[width=.45\textwidth]{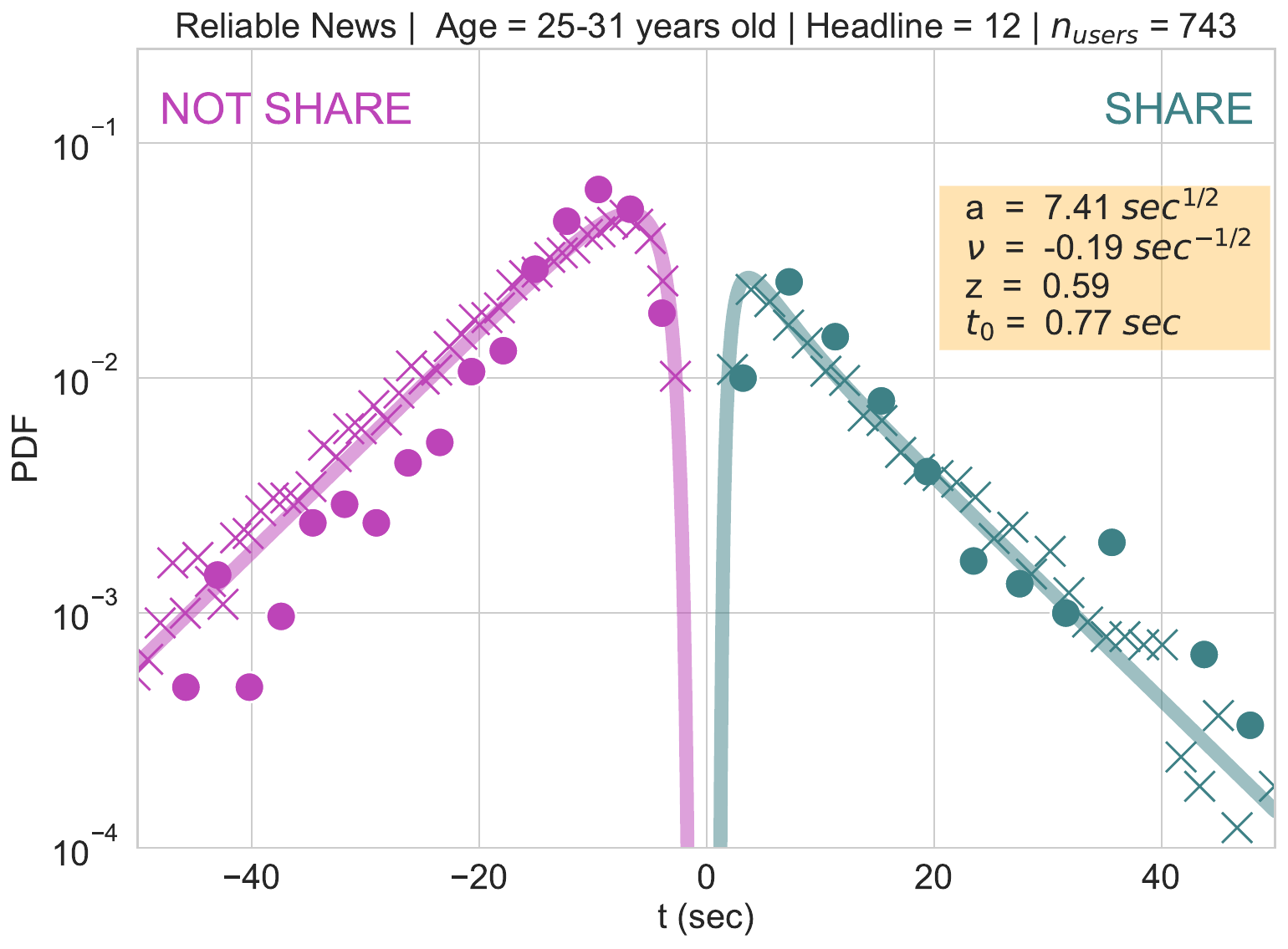}\quad
            \includegraphics[width=.45\textwidth]{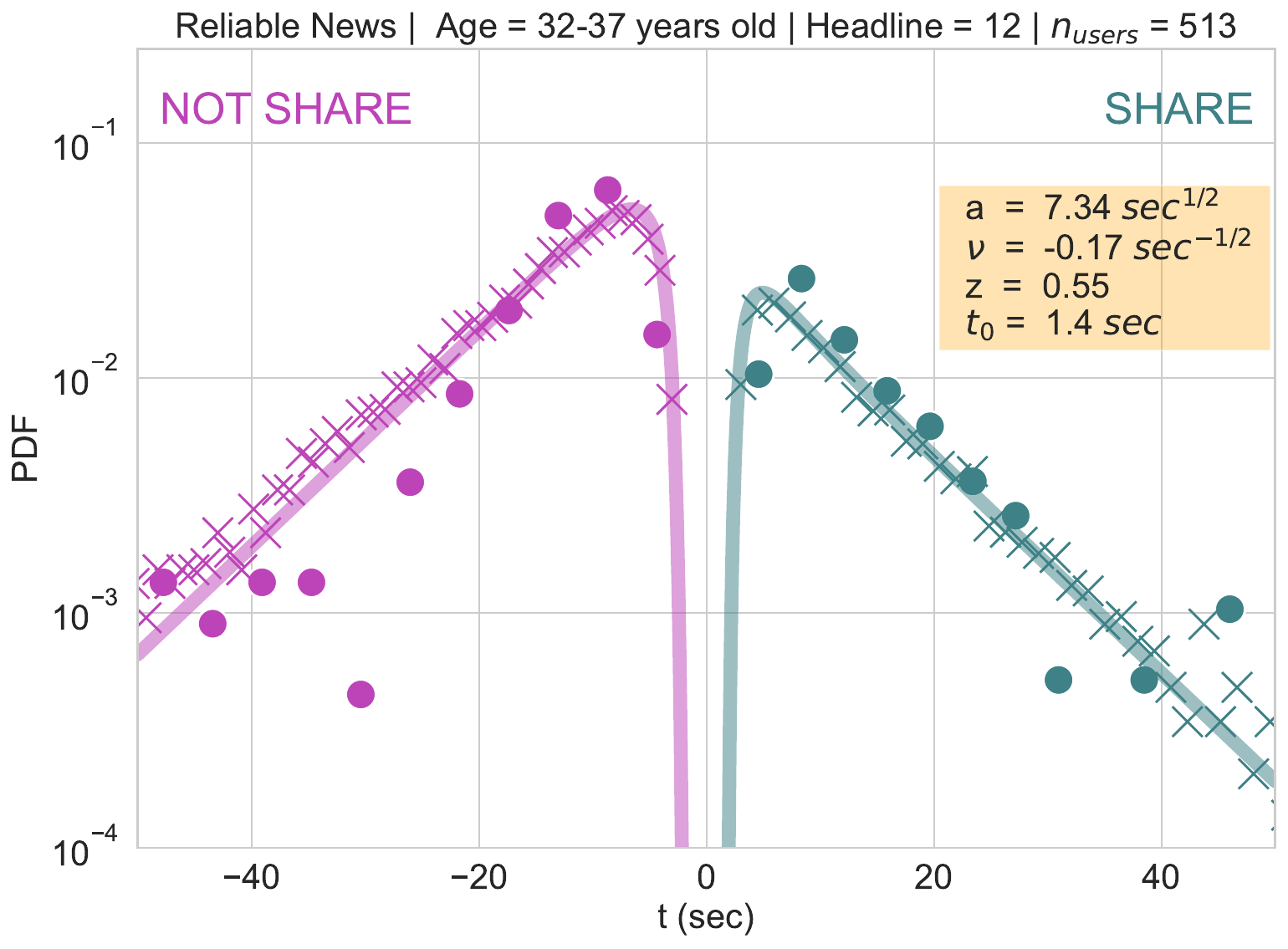}
            \medskip
            \includegraphics[width=.45\textwidth]{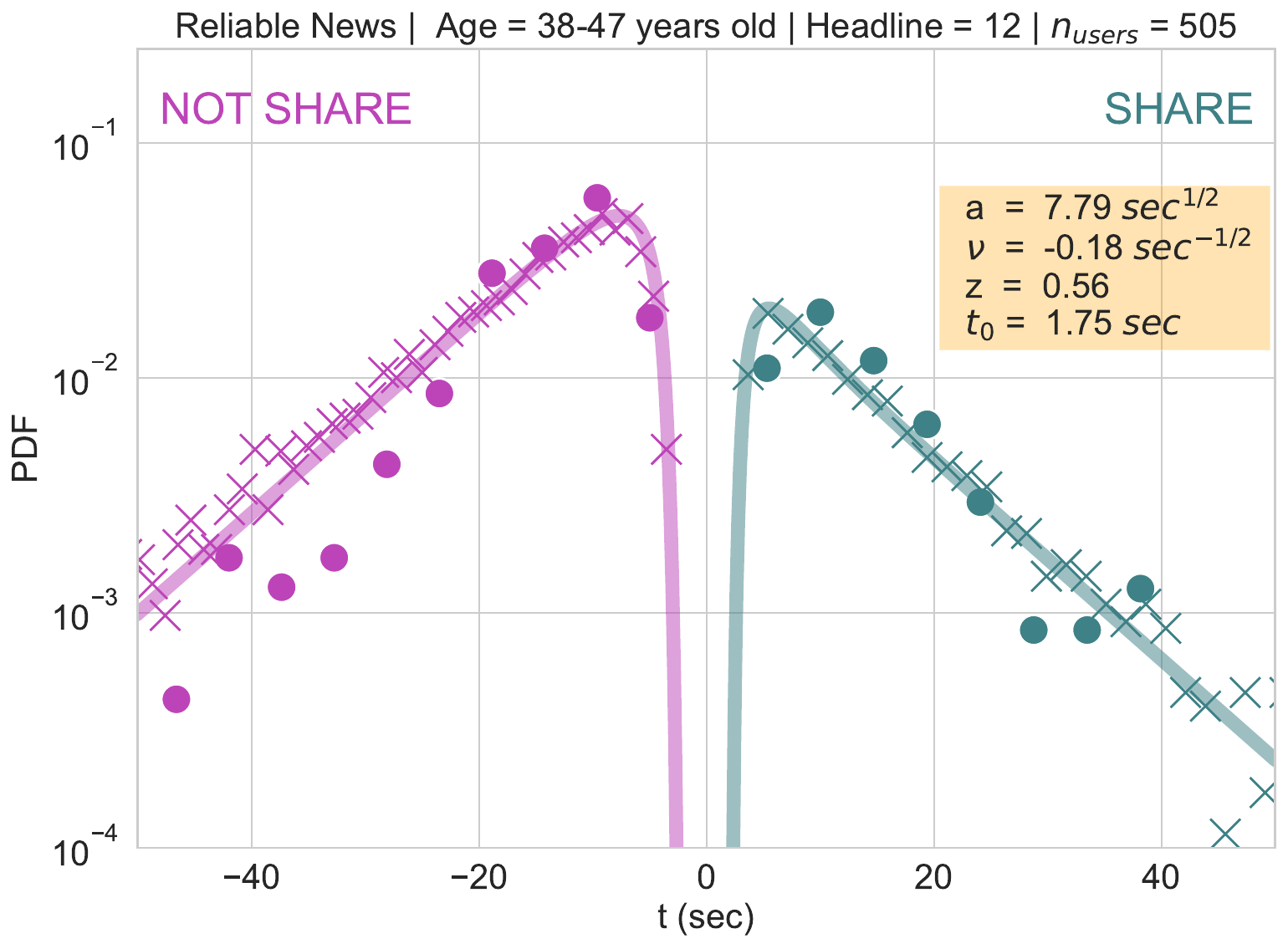}\quad
            \includegraphics[width=.45\textwidth]{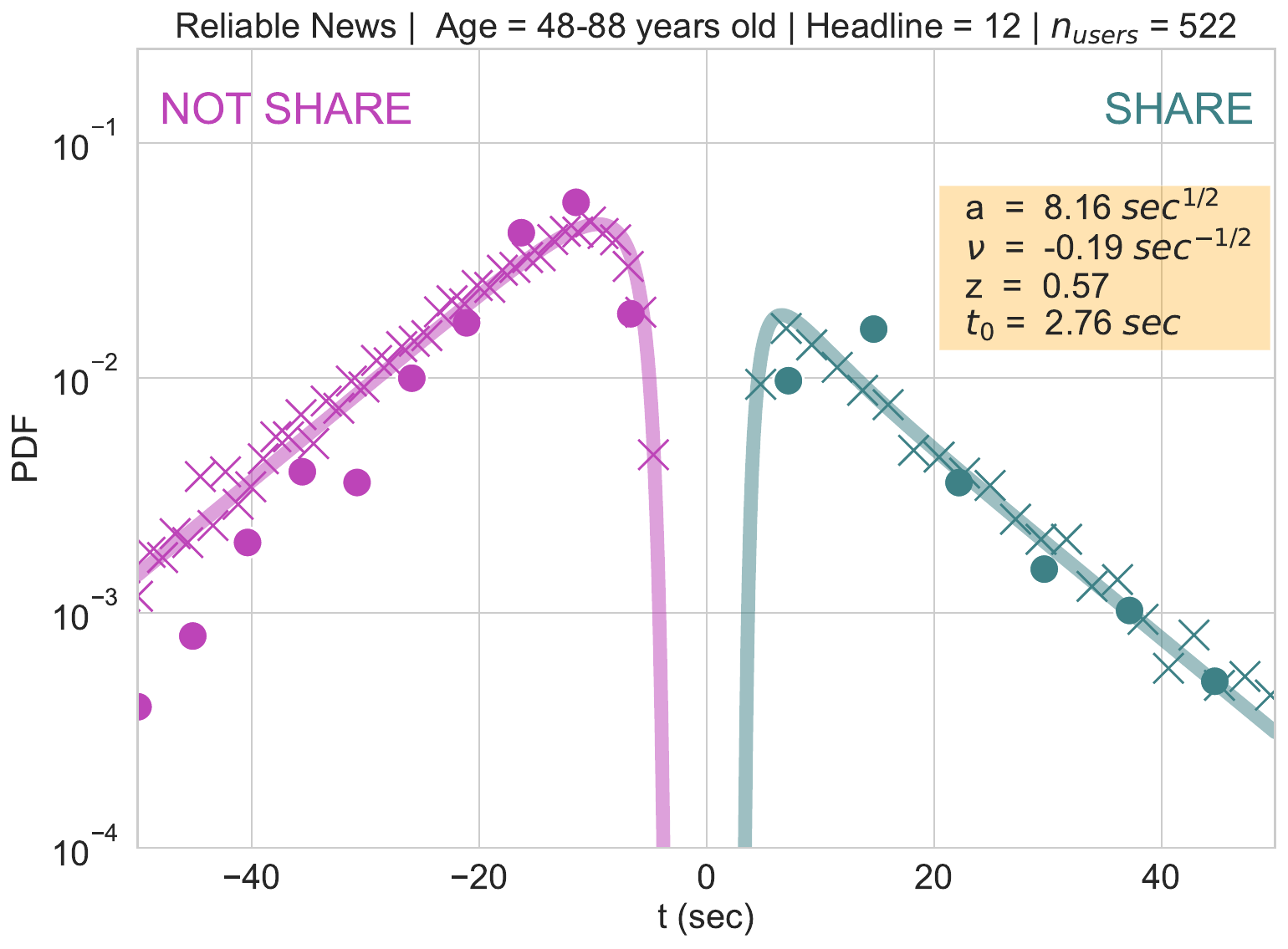}
            \caption{{\bf Headline 12}: Probability distribution of the response time for sharing and not sharing reliable information. Each figure corresponds to different age ranges. The solid line corresponds to theoretical results, dots correspond to empirical data and crosses to stochastic simulations.}
            \label{headline12Real}
        \end{figure}
        
\end{document}